\documentclass{aomart}
\usepackage[OT1]{fontenc}
\usepackage[utf8]{inputenc}
\usepackage{amssymb,mathtools,mathrsfs,microtype}
\usepackage{enumitem,booktabs,longtable,array,tabularx}
\usepackage{bookmark}
\DeclareTextSymbolDefault{\textsection}{OMS}
\hypersetup{hidelinks,pdftitle={The marginal sine-Gordon field theory},pdfauthor={Dyuman Bhattacharya},bookmarksnumbered=true,bookmarksdepth=2}
\numberwithin{equation}{section}
\makeatletter
\newcommand{\equationalias}[1]{\ltx@label{#1}}
\makeatother
\newtheorem{theorem}{Theorem}[section]
\newtheorem{proposition}[theorem]{Proposition}
\newtheorem{lemma}[theorem]{Lemma}
\newtheorem{corollary}[theorem]{Corollary}
\theoremstyle{definition}

\theoremstyle{remark}

\setlist[enumerate]{label=\textup{(\roman*)},leftmargin=2em,itemsep=2pt,topsep=4pt}
\setlist[itemize]{leftmargin=1.5em,itemsep=2pt,topsep=4pt}
\AtBeginDocument{\lefthyphenmin=2\righthyphenmin=3}
\newcommand{\E}{\mathbb E}

\newcommand{\Tr}{\operatorname{Tr}}
\newcommand{\Var}{\operatorname{Var}}
\newcommand{\Cov}{\operatorname{Cov}}

\newcommand{\Spec}{\operatorname{Spec}}

\newcommand{\diag}{\operatorname{diag}}

\DeclareMathOperator{\Gram}{Gram}

\allowdisplaybreaks[2]

\providecommand{\source}[1]{\nolinkurl{#1}}

\ifdefined\pdfmapfile
\pdfmapfile{}
\pdfmapfile{+lm.map}
\pdfmapfile{+cm.map}
\pdfmapfile{+cmextra.map}
\pdfmapfile{+symbols.map}
\pdfmapfile{+euler.map}
\pdfmapfile{+yhmath.map}
\pdfmapfile{+rsfs.map}
\fi
\volumenumber{}\issuenumber{}\publicationyear{}\papernumber{}
\doinumber{}\copyrightnote{}
\fancypagestyle{firstpage}{\fancyhf{}\fancyfoot[C]{\footnotesize\thepage}}

\hypersetup{pdfsubject={Fourth draft}}

\title[Marginal sine--Gordon field theory]{The marginal sine--Gordon field theory}

\author{Dyuman Bhattacharya}
\address{Department of Physics and Astronomy,
University of Waterloo\\
Waterloo, Ontario, Canada}
\email{d7bhatta@uwaterloo.ca}
\givenname{Dyuman}
\surname{Bhattacharya}

\begin{document}

\begin{abstract}
At $\beta_*^2=8\pi$ we construct the two-dimensional sine--Gordon
vacuum from a noncompact real lattice field with nearest-neighbor
kinetic energy, positive cosine activity and zero Dirichlet data.
Its mass $M$ is the lower edge of the energy spectrum above the vacuum. In units of
the length $a_{\rm exit}$ at which the flow leaves weak coupling it obeys
$0<c_-\le M a_{\rm exit}\le c_+<\infty$ on compact positive
endpoint intervals, with uniform constants. A prescribed response
to a smooth background in a fixed square determines the nonlinear
asymptotically free bare parameters at each sufficiently fine cutoff.
In the local relativistic continuum theory the vacuum is unique,
physical dilation realizes every positive mass, and literal
derivatives generate the reflected Hilbert space even for finite
composite and contact families. Variance and exponential moments of a derivative smear rule out a Gaussian law. A connected correlation of some finite even order above two is nonzero on separated supports.
The construction gives universality for the specified finite
nonnegative higher harmonics, the field equation, topological current
and geometric stress tensor with their Ward, trace and contact
identities, and a normalized source-transfer theorem with explicit
finite-scale hypotheses.
\end{abstract}
\maketitle
\tableofcontents
\section{Introduction}
\label{sec:introduction}

In two dimensions with kinetic action $\frac12\int|\nabla\phi|^2$,
the Gaussian dimension of $\cos(\beta\phi)$ reaches two at
$\beta^2=8\pi$. A neutral pair then contributes a logarithmic
ultraviolet divergence to the kinetic energy. This forces the
stiffness to run with the activity. Since the lattice action contains
no quadratic mass term, finding the physical exit length on this
trajectory still leaves a spectral gap on that scale to be proved.

\subsection{The construction and its scope}

Fix $\beta_*^2=8\pi$ and set $h=\beta_*\phi$. The real noncompact
field has on-site potential $2y_\epsilon(1-\cos h)$ and nearest-neighbor
kinetic coefficient $Z_\epsilon>0$. On the nonlinear asymptotically
free branch one measurement in a fixed square chooses both coefficients
at every sufficiently fine cutoff. Integrate fluctuations up to $a_0$
and prescribe the second derivative of the log weight at zero amplitude
of a smooth background. Canonical Gaussian normalization would give
squared charge $8\pi/Z_\epsilon$, tending to $8\pi$ in the ultraviolet.
We keep the field and its period fixed.

Besides the literal derivative $D_\mu=[\partial_\mu\phi]_R$ and
integer vertices $V_n=[e^{in\beta_*\phi}]_R$, we admit each fixed
finite family of their composites, descendants and mixing partners.
Here $R$ specifies the normalization of Section~\ref{sec:model}.
Coincident products have a fixed local subtraction and independent
source data for identity and contact coefficients. All physical
limits first remove the cutoff in a fixed zero-Dirichlet domain and
then exhaust the plane by the selected squares. Spectral limits
refer to the resulting representation.

Use the activity $\rho>0$ of the limiting weak map at $a_0$ to
parameterize the branch. The response in the reference square is
locally invertible in $\rho$, and Section~\ref{sec:ultraviolet}
assigns its trajectory the exit length $a_{\rm exit}(\rho)$.

\begin{theorem}[The marginal vacuum theory]
\label{main:theorem}
For sufficiently small positive $\rho$, the action
\eqref{microscopic-model:action} admits positive, domain-independent
bare functions $Z_\epsilon(\rho)\to1$ and
$y_\epsilon(\rho)\to0$, at every sufficiently fine cutoff, with the
following properties. Uniformity is on compact positive endpoint
intervals and on the fixed source and test families specified in
Section~\ref{sec:model}.
\begin{enumerate}
\item The normalized Schwinger distributions of the local
fields exist, with compatible normalization and contact extensions
under enlargement of a finite field list. Their values and all mixed
reflected Gram matrices converge in the selected state.

\item The complete local fields reconstruct a relativistic vacuum
theory on a common dense invariant domain. Its source, height and
literal-derivative Hilbert spaces are canonically isometric, preserving
the vacuum $\Omega_\rho$ and the physical Hamiltonian $H_\rho$.
The vacuum is unique. The physical mass, defined by
$M(\rho)=\inf\Spec(H_\rho|_{\Omega_\rho^\perp})$, satisfies
\begin{equation}
 0<c_-\le M(\rho)a_{\rm exit}(\rho)\le c_+<\infty.
 \label{main:mass}
\end{equation}
In terms of the microscopic activity $g_\epsilon=c_{\rm lat}y_\epsilon$,
where $c_{\rm lat}=(\pi/2)e^{-2\gamma_E}$ and $\gamma_E$ is Euler's
constant, this comparison gives the mass law
\begin{equation}
 \log\frac{1}{\epsilon M(\rho)}
 =\frac1{g_\epsilon}
       +O\!\left(\log\frac1{g_\epsilon}\right).
 \label{main:clock}
\end{equation}
The same $M(\rho)$ is the infimum of invariant masses in the nonvacuum
joint energy--momentum spectrum. Dilation by $\lambda$ sends the endpoint
to $T_\lambda\rho$ and gives $M(T_\lambda\rho)=\lambda M(\rho)$
whenever both endpoints lie in the admitted charts. A single global
rescaling of the bare family and reference geometry realizes any
prescribed mass $M_*>0$.

\item The derivative field is non-Gaussian. At every admitted endpoint,
some connected derivative distribution of finite even order $n\ge4$
is nonzero on mutually separated compact supports.

\item Fix $N<\infty$ and bounded constants $a_n\ge0$, $2\le n\le N$.
Adding
\[
 2y_\epsilon^2\sum_{n=2}^N a_n
          \sum_{x\in D}[1-\cos(n\beta_*\phi_x)]
\]
to the microscopic action gives the same complete continuum theory
after its own canonical tuning and matching of the physical mass,
literal-field normalization and prescribed composite/contact
convention. The equivalence preserves the vacuum, local fields and
physical translations.

\item The renormalized field equation and topological-current
identities hold in the same local algebra. Finite geometric metric
jets define a symmetric stress tensor, its diffeomorphism and
translation Ward identities, and its local trace relation, with all
mixed and coincidence contacts. The metric variations are smooth,
supported in the interior, and of any fixed finite order at the flat
metric.
\end{enumerate}

The following regularity holds in the parameter $\rho$. At fixed volume
the cutoff approximants converge in $C^3$. At each fixed positive weak
resolution the complete source coefficients have the uniform $C^4$
bounds of Theorem~\ref{source-convergence:cutoff}. Direct differentiation
of the action gives classical fourth derivatives of the fixed-volume
limit, with the volume-uniform bounds of Theorem~\ref{extensive:responses}
on the selected squares and their bounded dilates. The first two
parameter derivatives pass to the selected infinite-volume limit.
\end{theorem}

The bound in \eqref{main:mass} does not determine whether the edge is an isolated one-particle shell. The
leading logarithm in \eqref{main:clock} does not fix the multiplicative
mass constant. The argument leaves open whether the connected
correlation witnessing non-Gaussianity occurs at fourth order. The harmonic perturbations in
the regulator assertion are restricted to the finite nonnegative
sums displayed above.

For Theorem~\ref{main:theorem} the bare family and source convention
are common to Theorems~\ref{arbitrary-cutoffs:allcutoff-theorem},
\ref{source-convergence:cutoff}, \ref{fixed-volume-integration:fixed-box},
\ref{source-admission:local-gram-limits}, \ref{physical-mass:mass-theorem},
\ref{extensive:responses}, \ref{relativistic-reconstruction:full-field-reconstruction},
\ref{physical-scale:mass-coordinate}, \ref{derivative-interaction:derivative-nongaussianity},
\ref{regulator-universality:verified-regulator-universality},
\ref{local-currents:shift-ward} and \ref{stress-and-trace:local-trace}.

\subsection{Relation to earlier work}

Perturbative renormalizability at the endpoint, including the
beta-functional equation, goes back to Nicol\`o and Perfetti
\cite{NicoloPerfetti1989}. Theorem~\ref{main:theorem} takes the
continuum and infinite-volume limits there and locates its positive spectral threshold. For massive sine--Gordon below the endpoint, Benfatto,
Gallavotti and Nicol\`o treated ultraviolet stability in the first
regions of collapse \cite{BenfattoGallavottiNicolo1982}, and
Nicol\`o, Renn and Steinmann obtained finite-volume stability for
$4\pi\le\beta^2<8\pi$ \cite{NicoloRennSteinmann1986}. Dimock and
Hurd studied finite-volume construction below threshold, coupling
analyticity and short-distance asymptotics \cite{DimockHurd1993}.
Their renormalization-group analysis treats ultraviolet asymptotic
freedom below threshold and infrared asymptotic freedom above, keeping
the opposite cutoff \cite{DimockHurd2000}. Their $\beta$ is our squared
frequency. For every real coupling Pelai\v{c} constructs the
finite-volume theory throughout the subcritical range with a massive
Gaussian reference field, Gaussian tails and reflection positivity
\cite{Pelaic2025}. His introduction identifies gaps in Dimock--Hurd,
including the passage from Lemma~6 to (56) in the cluster expansion of
\cite{DimockHurd2000}.

Two probabilistic constructions also concern finite-volume ultraviolet
renormalization. Meyer treats $\beta^2<(6/7)8\pi$ nonperturbatively
and $\beta^2<8\pi$ order by order \cite{Meyer2026}. For a class of
logarithmic covariances with smooth positive decompositions, Bonnefont,
Rajam\"aki and Vargas construct measures on bounded convex domains
throughout the subcritical range \cite{BonnefontRajamakiVargas2026}.
Their singularity $-\log|x-y|$ puts the threshold at $4$, or $8\pi$
here, and their hypotheses exclude the Dirichlet free field on the
interaction domain itself. Screening and identification of the full
local source space at the endpoint require separate arguments.

In the infrared, Fr\"ohlich and Spencer's Kosterlitz--Thouless theorem
applies to two-dimensional Abelian spin systems and the Coulomb lattice
gas \cite{FrohlichSpencer1981}. For each Coulomb-gas inverse temperature
above $8\pi$ in their units, Marchetti and Klein found power-law
external-charge correlations at sufficiently small activity
\cite{MarchettiKlein1991}. D'Alimonte and Lammers obtained analytic
free energy in the disordered regimes of the XY model and lattice
Coulomb gas, together with strong Debye screening for any finite
collection of local Coulomb-gas observables \cite{DAlimonteLammers2026}.
For integer-valued heights and their dual compact spins, the relevant
results include sharpness of localisation and a universal positive gap
in the possible effective temperatures for Lammers's class of height
models \cite{LammersDichotomy2022}, the relation of depinning to slow
dual Villain and XY correlation decay due to Aizenman, Harel, Peled and
Shapiro \cite{AizenmanHarelPeledShapiro2021}, and van Engelenburg and
Lis's alternative proof of the planar XY transition
\cite{vanEngelenburgLis2023}. Their height--spin duality also yields
Green-function bounds and temperature monotonicity of height variances
\cite{vanEngelenburgLis2025}.

For a one-dimensional inverse-square Ising ferromagnet with
translation-invariant couplings $J_n\ge0$, $n^2J_n\to B>0$, the author
proved $2\beta_c Bm_c^2=1$ \cite{BhattacharyaCriticalJump2026}, where
$\beta_c$ is the critical inverse temperature and $m_c$ the critical
spontaneous magnetization. The pressure response to diffuse
interaction deletion recovers the physical squared magnetization.
The proof controls macroscopic interfaces in the Gibbs law and gives
a finite ordering criterion. In \cite{BhattacharyaCriticalRelations2026}
the author relates critical connected correlations to small-field
excess magnetization. The jump is universal. With an exact
inverse-square tail the correlations are inverse-logarithmic and the
ordered cusp square-root. With a relative correction $c/\log(n+e)$,
$c>1$, the logarithmic powers vary continuously and the cusp becomes
linear. This use of physical response, logarithmic defect costs and
finite criteria motivates Section~\ref{sec:pinning}, which proves a
criterion for the noncompact periodic height law. The current
comparison and physical spectral bound are proved in
Section~\ref{sec:screening}.

Two inputs to the endpoint construction are Falco's Gaussian
regulators, local extraction and charged Taylor gains for the Coulomb
gas near the two-dimensional Kosterlitz--Thouless line
\cite{FalcoPressure,FalcoCorrelations}, and Bauerschmidt's spectral
construction \cite{BauerschmidtFRD}, on which our positive continuous
finite-range decomposition is based. The latter puts residual blocking
phases on one physical hierarchy. The nonlinear ultraviolet half-orbit,
its exact compact calibration and the comparison of complete source
histories are constructed here.

Gubinelli and Meyer construct the full-plane model with a massive Gaussian
reference field for $\beta^2<6\pi$, verifying the Osterwalder--Schrader axioms
and non-Gaussianity at sufficiently small nonzero coupling
\cite{GubinelliMeyer2024}. At the free-fermion point Bauerschmidt and
Webb construct planar charge and gradient correlations and identify
them with massive Dirac correlations \cite{BauerschmidtWebb2024}.
At that same frequency, $\beta^2=4\pi$ here, Bauerschmidt, Mason and
Webb construct fractional vertex correlations and identify them with
renormalized determinants of massive twisted Dirac operators
\cite{BauerschmidtMasonWebb2025}. The endpoint uses another ultraviolet
trajectory and requires its own screening estimate. For
$\beta^2<4\pi$ with an interaction cutoff, Fr\"ob and Cadamuro explain
why free normal ordering does not suffice for the stress tensor and
why conservation requires a finite quantum correction
\cite{FrobCadamuro2026}. At the endpoint the comparison is between
the retained kinetic polynomial, geometric metric variation and full
contact extension.

The corrected Osterwalder--Schrader theorem
\cite{OsterwalderSchrader1973,OsterwalderSchrader1975} requires an
order-uniform growth bound on tests flat at the diagonals. We verify
it along with Euclidean covariance, positivity and clustering. The
Hilbert-space identification transfers the mass bound from the
physical source space to the reconstructed local fields.

\subsection{The ideas of the proof}

The main infrared difficulty is to get a bound in the neutral sector
from pinning in the lift sectors. Pinning comes from finite free-box
tests on the positive microscopic law. The passage back to neutral
currents uses long and short rectangles with the same insertion and
caps: a fixed-cavity identity retains the boundary cocycle in this
comparison. Axial summability removes the height's zero-energy atom,
after which association extends the threshold from one square average
to the cyclic space. A derivative vector surviving the limit gives
the opposite mass inequality. Positive comparison identifies the
selected height and derivative spaces. Admitting complete local
sources requires joint reflected-norm approximation there, with the
actual Dirichlet caps and any number of separated repetitions of a
fixed packet.

In the ultraviolet the neutral pair must be extracted from the stable
remainder before the graph transform and weighted half-orbit construction of the asymptotically free branch.
The square-lattice potential kernel fixes its kinetic coefficient in
microscopic units. Positive partial covariance shells and a summable
conjugacy compare different cutoffs at common physical scales. The
same exact integration carries boundary and scalar terms along with
the bulk density. Ordinary source normalization needs only finite
block inverses; the stable and wall terms run forward, and the marked
scalar and contact terms telescope. Taylor subtraction gives positive
ultraviolet powers on each fixed collision forest.

Section~\ref{sec:transfer} gives finite replica and approximation
hypotheses sufficient for source admission. Its response theorem
uses clustering as a separate input. For the extensive derivatives,
including wall bonds, Gaussian observations with matching replica
traces have summable covariance rows. Positive bounds on the actual
replica normalizers let us sum packet errors using screening, in the
prescribed order of limits.

For reconstruction, a radial auxiliary regulator restores rotations,
short-buffer estimates give growth and the chronological derivative
core gives density. Field translations yield the local field identities.
Metric jets retain the Gaussian determinant, source seagulls, scalar
conversion and Weyl contact term.

\section{The model and its observables}
\label{sec:model}
\label{introduction:quantifiers}
\label{microscopic-model:model}

Use units $\hbar=c=1$, Euclidean coordinates $x=(x_0,x_1)$ and
indices $\mu,\nu\in\{0,1\}$. Fix $\beta_* =\sqrt{8\pi}$ and an
operator reference energy $\mu_{\rm ref}>0$. On the real lattice
field $\phi_x$ in a finite rectangle $D\subset\mathbb Z^2$, use the
zero extension outside $D$. Its Dirichlet Laplacian and action are
\begin{equation}
 \begin{aligned}
 A_D\phi(x)&=4\phi_x-\sum_{|e|=1}\phi_{x+e},\\
 S_{\epsilon,D}&=\frac{Z}{2}\phi^\top A_D\phi
 +2y\sum_{x\in D}(1-\cos\beta_*\phi_x)
 +E_\epsilon\epsilon^2|D|.
 \end{aligned}
 \label{microscopic-model:action}
\end{equation}
We require $Z>0$ and $y\ge0$. Every bond meeting $D$, including those
to its zero exterior, occurs once in
$\phi^\top A_D\phi=\sum_{\langle xy\rangle}
(\phi_x-\phi_y)^2$. At physical position $\epsilon x$ the Laplacian is
$L_D=\epsilon^{-2}A_D$. The site activity $y$ is dimensionless (there
is no factor $\epsilon^2$ in the site sum), and $E_\epsilon$ has mass
dimension two. The potential has period $2\pi/\beta_*$ on the
noncompact field. In angular units $h=\beta_*\phi$ the period is
$2\pi$ and the bare stiffness $\kappa_\epsilon=Z_\epsilon/(8\pi)$.

Write $Z_{\epsilon,D}(E)$ for the partition function at vacuum
constant $E$ and $\langle\cdot\rangle_{\epsilon,D}$ for expectation.
The normalized density is
$Z_{\epsilon,D}(E_\epsilon)^{-1}e^{-S_{\epsilon,D}}\prod_{x\in D}d\phi_x$.
For $V$ the prescribed mesh-$\epsilon$ approximation of the vacuum
reference square $(-4/\mu_{\rm ref},4/\mu_{\rm ref})^2$, set
\begin{equation}
 E_\epsilon=\frac{\log Z_{\epsilon,V}(0)}{\epsilon^2|V|}.
 \label{model:vacuum-normalization}
\end{equation}
This fixes the vacuum scalar by $Z_{\epsilon,V}(E_\epsilon)=1$. The compact reference used
to fix the physical family is different (Section~\ref{sec:ultraviolet}).
In a normalized expectation the vacuum scalar cancels, while the
identity coefficients carrying sources belong to the observable
prescription. Existence of the measure and of every polynomial moment
follows from $A_D\succ0$ and $0\le2y(1-\cos u)\le4y$. For bond
reflection positivity use the Fourier representation
\[
 e^{-Z(u-v)^2/2}=(2\pi Z)^{-1/2}\int e^{-p^2/(2Z)}e^{ipu}e^{-ipv}\,dp:
\]
noncrossing factors give reflected half weights, and the crossing
bonds give an integral of squared moduli. Conditioning on the fixed plane gives
site reflection positivity. With $\Theta F=\overline{F\circ\vartheta}$,
we have $\|[F]\|_{\rm OS}^2=\langle(\Theta F)F\rangle\le\langle|F|^2\rangle$.
Reflection positivity here is a property of the complete measure. The
individual Gaussian increments in the calculation need not have this property.

Physical pairings use $\epsilon^2$ in the bulk and $\epsilon$ on the
boundary, so a covariance operator in such a pairing must be distinguished
from the field-coordinate covariance matrix. Our notation is
$A\preceq B$ for positive semidefinite $B-A$, and $G^\dagger$ for
the inverse of $G$ on its range, extended by zero on the kernel.
A superscript $*$ on a matrix or operator denotes the adjoint in the
specified pairing.

\subsection{Fields, finite sources and normalization}
\label{microscopic-model:contract}
The local algebra is generated by
\begin{equation}
 D_\mu=[\partial_\mu\phi]_R,\qquad V_n=[e^{in\beta_*\phi}]_R,
 \quad n\in\mathbb Z,\qquad V_{-n}=V_n^*,\quad V_0=1,
 \label{microscopic-model:algebra}
\end{equation}
and finite composite products. In the prescription denoted by $R$,
the derivative has its literal normalization on a smooth test $f$,
\[
 D_{\mu,\epsilon}(f)
 =\epsilon^2\sum_x f(\epsilon x)\,
       \frac{\phi_{x+e_\mu}-\phi_x}{\epsilon}.
\]
Normal order the free vertices with the bulk covariance
$Z^{-1}\int_0^{(\epsilon\mu_{\rm ref})^{-2}}e^{-tA_\infty}dt$.
Use the Gaussian dimensions $\Delta(D_\mu)=1$ and $\Delta(V_n)=2n^2$ at
the endpoint. When the marks $S$ coalesce with local output
$\mathcal O$, subtract in their relative coordinates through Taylor degree
\begin{equation}
 \omega(S,\mathcal O)=\sum_{i\in S}\Delta_i-\Delta_{\mathcal O}-2(|S|-1)
 \label{microscopic-model:degree}
\end{equation}
if $\omega\ge0$. Here $\Delta_i$ is the Gaussian dimension of the
corresponding input field and $\Delta_{\mathcal O}$ that of the output. In the
fixed relative marked coordinate $r$, localize the Taylor polynomial
with a smooth compactly supported radial cutoff $\chi_{\rm ct}$ equal
to one near zero. This leaves
\begin{equation}
 (W_\omega f)(r)
 =f(r)-\chi_{\rm ct}(\mu_{\rm ref}r)
       \sum_{|\alpha|\le\omega}\frac{r^\alpha}{\alpha!}
                              \partial^\alpha f(0).
 \label{model:contact-prescription}
\end{equation}
For $\omega<0$ leave the test intact. On nested or disjoint partial
diagonals apply the forest subtraction after renormalizing the
source-free subclusters. Set all allowed finite local moments to zero
except for elementary normalizations. Each mixing block contains all
lower partners allowed by the actual symmetries; Gaussian charge
imposes no extra conservation law. Construct the blocks by increasing
dimension (Section~\ref{source-convergence:construction}).

Give repeated occurrences of the same profile distinct labels.
For a finite polarized list $I$ the algebra is
$\mathcal J_I=\mathbb C[\eta_i:i\in I]/(\eta_i^2:i\in I)$ with
monomials $\eta_S=\prod_{i\in S}\eta_i$. Polarization recovers any
fixed finite Taylor order in ordinary sources. Field conjugation
specifies the real structure. A complete source $B_{\epsilon,I}[J]$
has one independent coefficient per required nonempty subset, local
identity and contact outputs included. Its normalized and connected
functionals are
\begin{equation}
 \mathcal Z_{\epsilon,D}[J]
   =\big\langle e^{B_{\epsilon,I}[J]}\big\rangle_{\epsilon,D},
 \qquad W_{\epsilon,D}[J]=\log\mathcal Z_{\epsilon,D}[J].
 \label{model:normalized-functional}
\end{equation}
The unmarked partition function $Z_{\epsilon,D}(E_\epsilon)$ normalizes
the expectation, giving $\mathcal Z_{\epsilon,D}[0]=1$. Write
$\mathcal Z_{D,\rho}$ for the cutoff limit in the box and
$\mathcal Z_\rho$ for the selected plane limit. Nilpotence truncates
exponentials, logarithms and inverses with nonzero constant term.
Unbounded composites therefore require no analytic generating function.

Fix the source list, mixing and contact partners, compact supports,
and any required positive clearance from a wall or reflection plane.
Bound finitely many $C^{R_I}$ seminorms of the tests. Reference
indices and weak domains may depend on the list, but a compatible
enlargement retains every field already defined. Uniformity holds on
compact positive parameter intervals $J_0\Subset J_1$. It is not
asserted as $\rho\downarrow0$, a clearance vanishes, or charges or
composite degrees diverge. Order-uniform reconstruction has its own
quantifier for a fixed alphabet.

The physical limits are successive,
\begin{equation}
 \begin{aligned}
 \epsilon\downarrow0&\quad\hbox{in a fixed physical domain},\\
 D\uparrow\mathbb R^2&\quad\hbox{through selected zero-wall squares},\\
 t\uparrow\infty&\quad\hbox{in the selected representation}.
 \end{aligned}
 \label{model:ordered-limits}
\end{equation}
Domain-uniform estimates use this order. The cutoff threshold may
depend on the fixed domain, and fixed bounded dilates of the squares
give the same selected state. Tall rectangles require the cap
comparison in the screening proof. Remove auxiliary cavities only
after their stated reflected-norm estimates.

\subsection{Exact source calculus}

The same normalized source identity governs local integration,
conditional comparison and parameter response.

\begin{lemma}[Exact normalized calculus]
\label{finite-jet-calculus:jet-naturality}
Let $K$ be a positive integration kernel and
$U=U_0\sum_{S\subseteq I}\eta_Sf_S$, where $U_0>0$, $f_\varnothing=1$,
$0<KU_0<\infty$, and all coefficients are integrable. At each value
of the retained field, write $\nu$ for the kernel tilted by $U_0$.
Then
\begin{equation}
 \nu(dz)=\frac{U_0(z)K(dz)}{KU_0},\qquad
 \frac{KU}{KU_0}=\sum_S\eta_S\,\mathbb E_\nu f_S .
 \label{finite-jet-calculus:posterior-jet}
\end{equation}
Successive integrations, restriction of labels and field-independent
local polynomial source substitutions commute with this formula.
For $U=e^{L_0+B(\eta)}$, let $B_A=[\eta_A]B$ and let $\kappa_\nu$
denote joint cumulants under $\nu$. Then
\begin{equation}
 [\eta_S]\log\frac{KU}{KU_0}
 =\sum_{\pi\in\operatorname{Part}(S)}
       \kappa_\nu(B_A:A\in\pi).
 \label{finite-jet-calculus:complete-cumulant}
\end{equation}
For two labels this reads
\[
 [\eta_1\eta_2]\log\frac{KU}{KU_0}
 =\operatorname{Cov}_\nu(B_1,B_2)+\mathbb E_\nu B_{12},
\]
where $B_i=B_{\{i\}}$ and $B_{12}=B_{\{1,2\}}$. For example, a scalar
contact $c\int f_1(x)f_2(x)\,d^2x$ in $B_{12}$ contributes this same
quantity to the connected coefficient. Division by the source-free
partition function cancels the vacuum constant. The contact survives
because it carries both source labels.
Conditional transport contracts each coefficient in $L^p$, $1\le p\le\infty$.
Parameter derivatives commute whenever the integrals admit differentiation;
for a Gaussian field with covariance $C(\rho)$, write $F_a$ and $F_{ab}$
for its first and second field derivatives. With repeated coordinate
indices summed, the differentiated log integral is
\begin{equation}
 \partial_\rho\log\E_{C(\rho)}e^F
 =\langle\partial_\rho F\rangle+\tfrac12(\partial_\rho C)_{ab}
                  \langle F_{ab}+F_aF_b\rangle.
 \label{finite-jet-calculus:paramforcing}
\end{equation}
\end{lemma}
\begin{proof}
Divide $KU=\sum_S\eta_SK(U_0f_S)$ by $KU_0$. Fubini cancels the
intermediate normalizer in a composition. Algebra homomorphisms commute
with integration and finite inversion; expansion of the logarithm gives
the partition formula. Jensen proves contraction. Gaussian integration
by parts gives the last identity.
\end{proof}

We will repeatedly use two consequences. For a linear source $a$,
\begin{equation}
 \log\E_Ce^{a^\top(\phi+\zeta)+U(\phi+\zeta)}
 =a^\top\phi+\tfrac12a^\top Ca+\log\E_Ce^{U(\phi+Ca+\zeta)}.
 \label{finite-jet-calculus:linearshift}
\end{equation}
For scalar extraction, write the source log weight as $C_j+Q_j$ and use
the normalized tilted map
$\mathcal H_j(Q)=\log\{\E_j e^{F_j^0+Q}/\E_j e^{F_j^0}\}$,
where $F_j^0$ is the source-free log weight. Put $E_j=Q_j(0)$ and let
the exact extraction give
$\mathcal H_j(Q_j)=Q_{j+1}+s_j$. Then
\begin{equation}
 \mathscr A_n=C_0+E_n+\sum_{j<n}s_j,\qquad \mathscr A_j=C_j+E_j.
 \label{finite-jet-calculus:scalar-telescope}
\end{equation}
In particular evaluations of identity and contact columns in $E_n$ are
retained; only the independently generated $s_j$ is estimated as a tail.

\subsection{The finite strip transfer}
\label{finite-transfer:canonical-finite-transfer}
Fix the cutoff and spatial width. The row kernel is
$T(q,q')=e^{-Z\|q-q'\|^2/2-V_s(q)/2-V_s(q')/2}$, with $V_s$ the
original spatial action. The spatial Dirichlet form bounds it by
$T(q,q')\le Ce^{-c(\|q\|^2+\|q'\|^2)}$, hence $T$ is Hilbert--Schmidt.
Gaussian convolution flanked by the two multiplication operators gives
positivity and injectivity. Strict positivity of the kernel yields a
simple top eigenvalue $\lambda_0$ with even eigenvector $\Omega>0$.
By compactness and infinite rank, $0<\lambda_1<\lambda_0$. Thus
\begin{equation}
 H_{\epsilon,W}=-\epsilon^{-1}\log(T/\lambda_0),\qquad
 M_{\epsilon,W}=\epsilon^{-1}\log(\lambda_0/\lambda_1)>0.
 \label{finite-transfer:H}
\end{equation}
The smoothed zero cap $f=T(\cdot,0)$ belongs to $L^2$ and has
positive overlap with $\Omega$, giving
$(T/\lambda_0)^mf\to\langle\Omega,f\rangle\Omega$. One smoothing step
on each side of a fixed polynomial cylinder bounds its kernel by a
polynomial Gaussian integral, so the smoothed cylinder is
Hilbert--Schmidt. Its normalized capped matrix element therefore tends
to the vacuum matrix element. We use this fixed-cutoff limit only for
finite-dimensional comparisons.

\section{Local integration and marginal calibration}
\label{sec:renormalization}

\subsection{Norms and connected estimates}

The irrelevant part of the density must contract under integration, although the
density itself must be represented exactly. For this we weight field growth by a
Gaussian and the number of occupied blocks exponentially. Taking a fixed number of
source derivatives costs at most a polynomial factor.
\label{tree-summation:tree-counting}
Let $b$ be a large odd block factor. A polymer at scale $r=b^j$ is a finite connected
union of $r$-blocks and is called small if it has at most four blocks. Denote the
fixed small-set enlargement by $X^*$, the coarse closure by $\bar X$ and the block count
by $|X|_j$. We take as base analytic norm
\begin{equation}
 \begin{aligned}
 \|K(X)\|_{a,G_j}&=\sup_{\phi\ \text{real}}
 \frac{\sum_{p\ge0}(a^p/p!)\|D^pK(X,\phi)\|_{\rm op}}{G_j(X,\phi)},\\
 \|K\|_{A,a,G}&=\sup_XA^{|X|_j}\|K(X)\|_{a,G_j},
 \end{aligned}
 \label{endpoint-estimates:polymernorm}
\end{equation}
where variations have unit scaled $C^2$ norm, and
\begin{equation}
 \log G_j=\kappa\Bigl(c_1\|\nabla_j\phi\|_{L_j^2(X)}^2
 +c_3\|\nabla_j\phi\|_{L_j^2(\partial X)}^2
 +c_1\sum_{B\subset X}\|\nabla_j^2\phi\|_{\infty,B^*}^2\Bigr).
 \label{endpoint-estimates:regulator}
\end{equation}
In these units $\nabla_j=r\nabla$, the site measure is $r^{-2}$, the boundary measure is $r^{-1}$ and $\kappa=c/\log b$.
For Dirichlet fields we evaluate the norm on the actual odd extension. The
regulator, analytic radius and polymer size must all have their reserves fixed
before we choose the weak coupling radius.

The bulk estimates used here concern the unmarked space with this norm. For each
fixed finite source family, Section~\ref{sec:source-calculus} specifies the corresponding space of
directions, higher derivative quadratic forms and reserves by source degree. The
analytic norm has the same form in that space. The order of $D^p$ counts functional
derivatives independently of the spatial regularity of their directions.

\begin{lemma}[Gaussian and connected-expansion calculus]
\label{endpoint-estimates:regulator-2}\label{endpoint-estimates:majorant}
\label{endpoint-estimates:bounded-germ}\label{tree-summation:tree-sum}
The following estimates will be used with fixed finite source lists.
For $Q\succeq0$, $B=C^{1/2}QC^{1/2}$ and $2\kappa\|B\|<1$,
\begin{equation}
 \E_Ce^{\kappa(\phi+\zeta)^\top Q(\phi+\zeta)}
 \le \exp\!\left(\frac{\kappa(\Tr B+\phi^\top Q\phi)}
                         {1-2\kappa\|B\|}\right).
 \label{endpoint-estimates:regulatorbound}
\end{equation}
For labelled trees on $n\ge2$ vertices with degrees $d_i$,
\begin{align}
 \sum_T\prod_i d_i!&=(n-2)!\binom{3n-3}{n-2},
 \label{tree-summation:tree-factorial-identity}\\
 \sum_{|\alpha|=q}\frac{q!}{\alpha!}\prod_i(d_i+\alpha_i)!
 &=\prod_i d_i!\left(\sum_i(d_i+1)\right)_q.
 \label{tree-summation:mark-allocation-identity}
\end{align}
Here $(a)_q=a(a+1)\cdots(a+q-1)$, with $(a)_0=1$.
Consequently a tree expansion with coefficient $1/n!$, local analytic
factorials and nonnegative edge majorant $K$ satisfying
$\sup_x\int K(x,y)dy\le L$ is bounded by
$N_{\rm root}C_q(Cg)^n$ when its activities have size $g$ and $q$
fixed marks; $N_{\rm root}$ counts the allowed root blocks.
A strict geometric majorant $M\gamma^n$, $\gamma<1$,
also controls every fixed mark, derivative and one-factor comparison,
provided their local operations are bounded and cost only a fixed power
of $n$.

On a bounded complex field neighborhood, let $U=(\prod_BI_B)\Xi$,
where $I_B\ne0$ and $\Xi$ is the compatible gas with
$z(X)=K(X)/\prod_{B\subset X}I_B$. If
\begin{equation}
 \sum_{Y\not\sim X}|z(Y)|e^{(1+d)|Y|}\le |X|\qquad(d>0),
 \label{endpoint-estimates:KP-criterion}
\end{equation}
then its connected logarithm and rooted polynomial moments converge.
If $I_B>0$, $|\log I_B|\le c_0$ and all data are real, then
$U>0$ and $|\log U|\le C|\Lambda|$.
\end{lemma}
\begin{proof}
Completing the Gaussian square produces $\det(1-2\kappa B)^{-1/2}$ from the determinant and $\kappa Q+2\kappa^2QC^{1/2}(1-2\kappa B)^{-1}C^{1/2}Q$ as the
background precision. The claimed bound follows from $-\log(1-t)\le t/(1-t)$ and $QCQ\preceq\|B\|Q$. In the tree sum
set $k_i=d_i-1$. Pr\"ufer's formula gives $(n-2)![z^{n-2}](1-z)^{-2n}$, and extracting the coefficient of $z^q$ in
$\prod_i(1-z)^{-d_i-1}$ gives the mark identity. Successive leaf integrations bound the $n-1$ edge
integrals by $L^{n-1}$. Both the resulting binomial coefficient and the fixed polynomial
are at most exponential in $n$. For additional derivatives of any fixed order we
use $\sup_n n^p(\gamma/\gamma')^n<\infty$, with $\gamma<\gamma'<1$.
Cauchy's estimate on a smaller ball handles unmarked derivatives. For a
comparison, telescope the product one factor at a time, counting the full
covariance derivative of a Gaussian expectation as one changed factor. Apply the
Koteck\'y--Preiss criterion~\cite[Section 2]{KoteckyPreiss1986} to the compatible gas. Given $|z(Y)|\le\eta A^{-|Y|}e^{c|Y|}$, at most $C_0^n$
polymers of a given block count pass through one block. Choosing $A$ and then $\eta$
verifies the criterion. Along $tz$, $0\le t\le1$, the partition function stays real and
never crosses zero, hence stays positive.
\end{proof}

\begin{lemma}[Uniform parameter calculus]
\label{endpoint-estimates:uniform-calculus}
A uniformly contracting $C^r$ map of a common complete invariant ball
has a unique $C^r$ parameterized fixed
point; changing the map by $O(\delta)$ in $C^r$ changes that point by
$O(\delta)$ when one additional derivative is bounded. A finite triangular
system with diagonal norms below $\theta<1$ is contracting in a fixed
weighted norm. Its forced history retains the initial term:
\begin{equation}
 e_n\le\theta^{n-j}e_j+\sum_{k=j}^{n-1}\theta^{n-1-k}d_k.
 \label{endpoint-estimates:retarded-bound}
\end{equation}
If $L\ge1$, $\|A_j\|\le L$ and
$\|\partial^mA_j\|\le C_mw_j$, $\sum_jw_j\le W$, every fixed derivative
of $A_{n-1}\cdots A_j$ is bounded by $C_{r,W}L^{n-j}$. Finally,
on $J\Subset J'$, $\|f\|_{C^0(J')}\le d$, $\|f\|_{C^r(J')}\le M$ imply
\begin{equation}
 \|f\|_{C^k(J)}\le C(d+d^{1-k/r}M^{k/r}),\qquad k<r.
 \label{endpoint-estimates:interpolation}
\end{equation}
\end{lemma}
\begin{proof}
Successive differentiation of the fixed-point equation requires the inverse of $1-D_qR$
at each order. Its Neumann series gives the assertion by induction~\cite[Sections I.5.2--I.5.3]{Hamilton1982}. For the
triangular system, successive coordinate weights make the strictly triangular part
arbitrarily small. In the ordered Leibniz sum each degree is bounded by a
polynomial in $\sum w_j$ because $\sum_{i_1<\cdots<i_m}\prod w_{i_l}\le W^m/m!$. Finally, Taylor differences on a mesh of size $s$
bound the interior derivative by $C(ds^{-k}+Ms^{r-k})$. Take $s=(d/M)^{1/r}$ or the fixed interior margin. The
same proofs hold for Banach-valued maps.
\end{proof}

\subsection{The prepared local operation}
\label{polynomial-operation:section}\label{polynomial-operation:section-2}

The positive finite-range shells constructed in \cite[Appendix A.2 and Lemma 3.1]{FalcoPressure} sum to the massless resolvent
when the terminal covariance is included. They obey
\begin{equation}
 \begin{aligned}
 \Gamma_j(x)&=0\quad (|x|_\infty\ge b^{j+1}/2),\\
 |\nabla^\alpha\Gamma_j(x)|&\le C_\alpha b^{-j|\alpha|},\qquad
 \Gamma_j(0)=\frac{\log b}{2\pi}+O(b^{-j/4}).
 \end{aligned}
 \label{endpoint-estimates:covariance-interface}
\end{equation}
Each shell splits into positive finer shells whose ranges increase geometrically.
We use the regulator merger and Gaussian bound of \cite[Lemmas 5.1--5.2]{FalcoPressure} and the bounded neutral and
charge-one extraction with strict contraction of the free remainder from \cite[Lemmas 3.3--3.4, Section 6.2]{FalcoPressure}.
These estimates apply to one local operation on the even, periodic,
square-covariant bulk activities in \eqref{endpoint-estimates:polymernorm}. They give $\E_jG_j(X)\le2^{|X|_j}G_{j+1}(\bar X)$ and $\|\mathcal L_j\|\le\theta<1$ and leave a strict
exponential size reserve on reblocking. We shall construct the trajectory from
this single-step operation.

\subsection{The exact operation and its bounds}

In lattice units define the local density
\[
 V_j(s,z,B)=\frac{s}{2}\sum_{x\in B}\sum_\mu(\partial_\mu\phi_x)^2
       +zb^{-2j}\sum_{x\in B,\sigma=\pm1}e^{i\sigma\beta_*\phi_x}.
\]
We sum over $\pm e_0,\pm e_1$ with weight $1/2$ in each direction. The log density has a plus sign
on $s$, so the nominal stiffness is $1-s$. Set $\ell_j=b^2e^{-4\pi\Gamma_j(0)}$, $P_j=\prod_{k<j}\ell_k$ and $L_j(s,z)=(s,\ell_jz)$. The products $P_j$
stay bounded above and away from zero, with exponential convergence to $P_\infty>0$.

For a centered bulk shell field $\zeta$ with covariance $\Gamma_j$,
define the linear kinetic vacuum density
\[
 e_j^{\rm kin}(s)
 :=\frac{s}{2}\sum_\mu \E_j[(\partial_\mu\zeta_0)^2],
\]
using the same direction weights as in $V_j$. For $g=(s,z)$ and
a coarse block $D$, with $V_j(g,D)$ summed over its fine blocks,
\[
 \E_j V_j(g,D;\phi'+\zeta)
 =V_{j+1}(L_jg,D;\phi')+e_j^{\rm kin}(s)|D|.
\]
The volume $|D|$ counts original lattice sites. Evenness and the
second-difference shell bound give
$|e_j^{\rm kin}(s)|\le C|s|b^{-2j}$.

For the density projection $\mathsf T_j$, take the period average at a constant field, the
square average of the second affine variation and both charge-one Fourier
coefficients. Normalize it by $\mathsf T_j(e|B|+V_j(s,z,B))=e|B|+V_j(s,z,B)$. Summing over ordered pairs of blocks with $\overline{B_0\cup B_1}=Y$,
define
\[
 Q_j(g,Y)=\frac12\sum_{B_0,B_1}
             \Cov_j(V_j(g,B_0),V_j(g,B_1)),\qquad g=(s,z).
\]
Start the quadratic preparation at $W_1=0$ and iterate
\begin{equation}
 W_{j+1}(L_jg,D)=
 (1-\mathsf T_{j+1})\left(\E_jW_j(g,D)
                   +\sum_{Y\supset D}\frac{Q_j(g,Y)}{|Y|_{j+1}}\right).
 \label{polynomial-operation:preparation-recursion}
\end{equation}
The projected part is $e_{2,j}(g)|D|+V_{j+1}(q_j(g),D)$. The recursion propagates the microscopic initial
condition and keeps the linear transport at every shell. In particular a
contribution born at $n$ with $k$ activity factors is multiplied by $(P_n/P_j)^k$, $k=0,1,2$.

The block factor $I_j(g,B)=e^{V_j(g,B)}(1+W_j(g,B))$ is polynomial in $W_j$. With this choice the density is exactly
\begin{equation}
 e^{E_j|\Lambda|}\sum_{X\subset\Lambda}
 I_j(g,\Lambda\setminus X)\prod_{Y\in\mathcal C_j(X)}K_j(Y).
 \label{polynomial-operation:eq-R-15}
\end{equation}
The factor $I_j(g,X)$ is the product over the blocks of its argument, and $\mathcal C_j(X)$ is the set
of connected components. Project $K$ linearly onto its neutral Taylor jet through
degree two and its constant charge-one jets, averaging over the anchor. Divide
each small-polymer extraction equally among its blocks, with density coefficients
$B_jK$ and $e_{1,j}(K)$. This gives the extraction in \cite[eqs. (4.15)--(4.19)]{FalcoPressure}, whose kinetic coefficient is half
the trace per site of the affine Hessian. Set
\[
 \begin{aligned}
 g^+&=L_jg+B_jK+q_j(g),\\
 \mathcal E_j&=e_j^{\rm kin}(s)+e_{1,j}(K)+e_{2,j}(g),\\
 E_{j+1}&=E_j+\mathcal E_j.
 \end{aligned}
\]
The coupling $s$ is evaluated at the input, and $E_j$ denotes the unmarked bulk
scalar of \eqref{polynomial-operation:eq-R-15}. Extract the linear kinetic scalar directly. Since its allocation
occupies a single block, centering makes its contribution to $J$ vanish. The
connected quadratic preparation, which defines $Q_j$, $W_j$, $q_j$ and $e_{2,j}$, is separate.

To write the exact operation as a finite sum, let $\widetilde Q(D,Y)$ be the linear and quadratic
extractions after equal allocation. Reblock the activity and center the extraction
by setting
\begin{align*}
 \widetilde K(Y)&=\sum_{\bar X=Y} I_j(Y\setminus X)
                      \prod_{Z\in\mathcal C_j(X)}K(Z),\\
 J(D,Y)&=\widetilde Q(D,Y)
             -\mathbf1_{Y=D}\sum_{Y'\supset D}\widetilde Q(D,Y').
\end{align*}
Centering gives $\sum_YJ(D,Y)=0$. With $\phi=\phi'+\zeta$, put
$\mathcal P(D)=I_j(g,\phi,D)-e^{\mathcal E_j|D|}I_{j+1}(g^+,\phi',D)$
and $\mathcal R(Y)=\widetilde K(\phi,Y)-\sum_{D\subset Y}J(\phi',D,Y)$.
Then
\begin{equation}
 K^+(Y')=\sum
 e^{-\mathcal E_j|W|}I_{j+1}(g^+,Y'\setminus W)
 \E_j[\mathcal P^Z\mathcal R^{X_1}]
 \prod_{Y\in\mathcal C_{j+1}(X_0)}J(D_Y,Y).
 \label{polynomial-operation:literal-operation}
\end{equation}
Set $W=X_0\cup X_1\cup Z$. Sum over separated coarse polymers $X_0,X_1$, disjoint $Z\subset Y'\setminus(X_0\cup X_1)$, small components
$Y$ of $X_0$ and a block $D_Y\subset Y$ for each such component. These choices must satisfy
$Y'=X_1\cup Z\cup\bigcup_YD_Y^*$. A power with exponent $Z$ is a product over blocks, one with exponent $X_1$ a
product over components. On expanding the sum, terms missing their assigned
enlargement cancel by the zero-sum identity. The remaining components factor by
finite range. This gives the exact operation without dividing by $I_j$.

\begin{proposition}[Complete weak map]
\label{polynomial-operation:smooth}\label{polynomial-operation:covariance}
\label{polynomial-operation:contraction}
Choose sufficiently large $b$, admissible regulator reserves and then
sufficiently large $A$. For a sufficiently small weak radius, the
operation is analytic in a fixed complex ball and obeys
\begin{align}
 s^+&=s-a_jz^2+F_jK,\qquad
 z^+=\ell_j(z-b_jsz+M_jK),\nonumber\\
 K^+&=\mathcal L_jK+\mathcal R_j(g,K),\nonumber\\
 \|\mathcal R_j\|&\le C(|g|^3+|g|\|K\|+\|K\|^2),
 \label{polynomial-operation:prepared}\\
 |F_jK|+|M_jK|&\le CA^{-1}\|K\|,\nonumber\\
 \|D_g\mathcal R_j\|&\le C(|g|^2+\|K\|),\qquad
 \|D_K\mathcal R_j\|\le C(|g|+\|K\|).
 \label{polynomial-operation:import}
\end{align}
All fixed derivatives are uniform. The prepared recursion applies
for $j\ge1$. The first integration has the actual data
\begin{equation}
 \begin{aligned}
 E_1&=E_0+e_0^{\rm kin}(s_0),\\
 s_1&=s_0,\qquad z_1=\ell_0y,\qquad
 K_1=O((|s_0|+y)^2).
 \end{aligned}
 \label{polynomial-operation:preparation}
\end{equation}
Here $E_0$ is the microscopic unmarked scalar. Since $W_1=0$,
all remaining quadratic terms, including their scalar parts,
are retained in $K_1$.
Here $a_j\to a_\infty>0$, $b_j\to2\log b$ exponentially. In a true
Dirichlet box, after the separate extraction of its local scalar
terms, the boundary activity obeys
\begin{equation}
 \begin{aligned}
 B^+&=\mathcal L_\partial B+Q_\partial(g,K)+N_\partial(g,K,B),\\
 \|\mathcal L_\partial\|&<1,\qquad
 \|Q_\partial\|\le C(|z|+|g|^2+\|K\|),
 \end{aligned}
 \label{polynomial-operation:nonlinear}
\end{equation}
where $\|N_\partial\|\le C[(|g|+\|K\|)\|B\|+\|B\|^2]$.
\end{proposition}
\begin{proof}
\emph{Preparation and local factors.} Gaussian characters compute all
coefficients in \eqref{polynomial-operation:preparation-recursion}:
\[
 \E[P(\phi+\zeta)e^{i\sigma\beta_*\zeta_y}]
 =e^{-4\pi\Gamma_j(0)}\E P(\phi+\zeta+i\sigma\beta_*\Gamma_j(\cdot-y)).
\]
For an age-$j-n$ neutral cosine contribution its kernel has support
$|x|\lesssim b^n$, size $Cb^{-4n}$ and third moment $Cb^n$.
Subtracting its constant and affine quadratic parts leaves
$C|x|^3b^{-3j}(1+H_{j,B}^3)$, where
$H_{j,B}=\max_{k=1,2}\|\nabla_j^k\phi\|_{\infty,B^*}$.
The block volume supplies $b^{2j}$, giving $Cb^{-(j-n)}$.
The charge-two contribution has propagation $b^{-8(j-n)}$ and total
gain $b^{-6(j-n)}$; the mixed component has
$b^j\sum|w_d|+b^j\sum|w_e||x|\le Cb^{-(j-n)}$ from
$|\nabla\Gamma_n|\le Cb^{-n}$ and charge-one propagation
$b^{-2(j-n)}$. The kinetic component has
$b^{-j}\sum|\nabla^2\Gamma_n(x)||x|\le Cb^{-(j-n)}$.
The same Taylor remainders control the full analytic jet. Summation gives
\begin{equation}
 |V_j(g,B)|_a\le C|g|(1+H_{j,B}^2),\qquad
 |W_j(g,B)|_a\le C|g|^2(1+H_{j,B}^3).
 \label{polynomial-operation:eq-R-11}
\end{equation}
Use the Gaussian polynomial reserve for $1+H^3$. At small $|g|$ the product of all $b^2$
fine factors in a coarse block is bounded by $2(G_j^{\rm str})^{1/4}$. Another quarter of the reserve
controls both the subtraction of one from this product and its coupling
derivatives.

\emph{Summation and cancellation.} Reblocking satisfies, for some
$\eta>0$,
$(1+2\eta)|\bar X|\le|X|+8(1+2\eta)|\mathcal C_j(X)|$.
It bounds $\mathcal P,\mathcal R,J$ by their regulators times,
respectively, $C_A\varepsilon A^{-(1+\eta)}$,
$C_A\varepsilon^2A^{-(1+\eta)|Y|}$ and
$C_A\varepsilon^2A^{-(1+\eta)|D^*|}$ on
$|g|\le\varepsilon$, $\|K\|\le\varepsilon^2$.
Indeed there are at most $2^{b^2n}$ fine subsets of a size-$n$ coarse
polymer; choose $(2C_0)^{b^2}A^{-\eta}<1/4$ and
$\varepsilon^2A^{8(1+2\eta)}<1/4$. In
\eqref{polynomial-operation:literal-operation} merge the disjoint fine
regulators, integrate once, then merge exterior coarse factors.
The result contains one output regulator. At most $16^n$ assignments
and the Gaussian cost $2^{(b^2+2)n}$ are absorbed by the remaining
$A^{-\eta n}$. Every nonempty term has a small factor. This gives a
strict analytic majorant. At degrees one and two,
$I_j=1+V_j+V_j^2/2+W_j+O(g^3)$.
For the aggregate $V=V_j(g,D)$, put
$V_+=V_{j+1}(L_jg,D)$ and $c=e_j^{\rm kin}(s)|D|$. Then
\[
 \E_jV=V_++c,\qquad
 \frac12\E_jV^2
 =\frac12(V_++c)^2+\frac12\Cov_j(V,V).
\]
The factor $e^c e^{V_+}$ supplies the linear contraction and the
quadratic products $cV_++c^2/2$. The remaining quadratic connected
term is the covariance defining $Q_j$. The zero-sum allocation
cancels the two-block covariance, and
\eqref{polynomial-operation:preparation-recursion} cancels the
one-block quadratic remainder. The term linear in $K$ is the
imported $\mathcal L_j$. Cauchy estimates now give the three
remainder bounds in the statement. Comparing each transported
shell with its untransported counterpart costs
$\sum_{n<j}b^{-(j-n)}b^{-n/4}=O(b^{-j/4})$, including the omitted
initial shell. Thus the rebuilt coefficients have the stated limits.

\emph{Dirichlet compression and boundary.} On the doubled torus let
$J$ be the extension odd in both coordinates, $E=J/2$, and $P=EE^*$.
Then $E^*E=1$, $A_TE=EA_D$. Periodic shells commute with $P$, and
\begin{equation}
 C_j^D=E^*C_j^TE,\qquad EC_j^DE^*=PC_j^TP\preceq C_j^T,\qquad
 JC_j^DJ^*\preceq4C_j^T.
 \label{polynomial-operation:domination}
\end{equation}
Compress the massive resolvent sum and let its auxiliary mass decrease
to zero; the odd sector has no constant mode. The four-image formula
preserves range and derivative bounds and equals the bulk kernel off
the wall collar. Convexity of the regulator and conditional Jensen
transfer its Gaussian bound. The regulator coefficient is chosen small
enough for the factor four in \eqref{polynomial-operation:domination}.

For a true box $D$, write the kinetic contribution of a rooted block
$B$ as $s\phi^\top A_{D,B}\phi/2$, using the actual bond allocation,
including bonds to the zero exterior, so that
$\sum_B A_{D,B}=A_D$. Its linear Gaussian scalar and wall correction are
\[
 \begin{aligned}
 c^D_{j,B}(s)&=\frac{s}{2}\Tr(A_{D,B}C_j^D),\\
 c^{\partial,D}_{j,B}(s)
 &=c^D_{j,B}(s)-e_j^{\rm kin}(s)|B|.
 \end{aligned}
\]
Thus $\sum_B c^D_{j,B}(s)=s\Tr(A_DC_j^D)/2$.
The wall correction vanishes off the finite-range wall collar
and is retained as a boundary-colored unmarked scalar.
Every scalar exponential in the true-box local operation uses
the sum of the actual bulk and wall allocations on its support.
The boundary activity and its norm bounds refer to the remainder
after these constants have been extracted and stored.

For an even small boundary activity, subtract its integrated constant.
Pinned directions and path integration to the wall give a local
$b^{-2}$ Taylor gain. Bounding the residual polynomial by the Gaussian
weight contributes $C(\log b)^{3/2}$. There are $O(b)$ anchors per coarse face block,
so the homogeneous boundary norm is
$C(\log b)^{3/2}b^{-1}+CA^{-\eta}<1$; corners are better.
For a boundary-collar point at distance $1\le d\le Cb^{j+1}$,
the shell attenuation is
$e^{-4\pi C_j^D(x,x)}\le C\min(1,(b^j/d)^2)$.
Multiplication by the pinned gain $(d/b^{j+1})^2$ and summation of
sites gives a first-order forcing $O(|z|)$. Color a term whenever it
contains a wall image, clipped support or boundary factor. Finite
range preserves that color, and every remaining nonlinear colored
term contains $gB$, $KB$ or $B^2$, proving the boundary equation.
\end{proof}

Both stable equations have their exact ordered histories; in particular
\begin{equation}
 \begin{aligned}
 \|B_j\|&\le\theta_\partial^j\|B_0\|+
 C\sum_{k<j}\theta_\partial^{j-1-k}(|z_k|+|g_k|^2+\|K_k\|),\\
 \|K_j\|&\le\theta_K^{j-1}\|K_1\|+
 C\sum_{k<j}\theta_K^{j-1-k}|g_k|^3.
 \end{aligned}
 \label{polynomial-operation:history}
\end{equation}
Higher spatial derivatives use
$Q_{j,k}=b^{(2k-2)j}\sum_{B,x\in B^*}|\nabla^k\phi_x|^2$, $k\ge2$.
Half of the coarse background
coefficient survives reblocking as a reserve. On fine subscales of range $t$ the
covariance norm is bounded by $C_kb^{(2k-2)j}\sum_{b^j\le t\le b^{j+1}}t^{2-2k}\le C_k$ and its trace by $C_k|X|_j$. The ranges in the sum form
a geometric sequence with fixed ratio greater than one. The Gaussian estimate
costs $e^{C_k\kappa_k|X|_j}$ when the coefficients lie below fixed positive bounds independent of the
grouped block factor. Each source space involves only finitely many $k$.
H\"older's inequality therefore combines all these forms with the base regulator,
even for positive coefficients equal to fixed multiples of $(1+\log B)^{-1}$. This applies
through order $R+2$ in the finite-source norm, including source degree zero, and
leaves a fixed fraction of every coarse background coefficient available after
reblocking.

\subsection{Calibration and reduction}

\begin{theorem}[Canonical marginal coordinates]
\label{polynomial-operation:graph}\label{polynomial-operation:calibration}
On the forward graphs of the actual preparation there are scale-prescribed
coordinates $(x,v)$ in which, with $\ell_b=\log b$,
\begin{equation}
 \begin{aligned}
 x^+&=x-\ell_bv^2+O(|x|v^2+v^4),\\
 v^+&=v-\ell_bxv+O((x^2+v^2)|v|).
 \end{aligned}
 \label{polynomial-operation:normalform}
\end{equation}
The graphs and maps have every prescribed finite differentiability
order after a fixed choice of weak radius. The Gaussian axis is fixed
pointwise. The microscopic normalization is
\begin{equation}
 \begin{aligned}
 c_{\rm lat}&=\frac\pi2e^{-2\gamma_E},\qquad
 \frac{2a_\infty P_\infty^2}{\ell_b}=c_{\rm lat}^2,\\
 v_1&=c_{\rm lat}y+O((|s_0|+y)y),\\
 x_1&=2s_0+O((|s_0|+y)^2).
 \end{aligned}
 \label{polynomial-operation:match}
\end{equation}
Off the graph the stable displacement contracts uniformly.
\end{theorem}
\begin{proof}
\emph{The quadratic coefficient.} For any covariance $C$ the degree-two
Taylor coefficient of $\log\E_Ce^{2y\sum\cos\beta_*\phi}$ is
\[
 y^2\sum_{x,z}e^{-4\pi(C_{xx}+C_{zz})}
 \{(e^{8\pi C_{xz}}-1)\cos\beta_*(\phi_x-\phi_z)
 +(e^{-8\pi C_{xz}}-1)\cos\beta_*(\phi_x+\phi_z)\}.
\]
The sum is ordered. Its neutral affine projection in the action is
$\delta Z=(\beta_*^2y^2/2)\sum_r|r|^2k(r)$.
For $A=4I-\text{adjacency}$ the potential kernel is
$a(r)=(\log|r|+\gamma_E+\frac32\log2)/(2\pi)+O(|r|^{-2})$
\cite[Theorem 4.4.4]{lawler}; the cited simple-walk kernel is $4a$.
Consequently $e^{-8\pi a(r)}=\kappa_{\rm lat}|r|^{-4}(1+O(|r|^{-2}))$,
$\kappa_{\rm lat}=e^{-4\gamma_E}/64$.
For $C_{<N}=\sum_{j<N}\Gamma_j$ and $R=b^N$,
$0\le a(r)-C_{<N}(0)+C_{<N}(r)\le C|r|^2/R^2$ on $|r|\le R$.
This follows by evenness, the second-difference shell bounds and
positive definiteness. Also $e^{-8\pi C_{<N}(0)}=O(R^{-4})$.
Both replacement errors contribute $O(1)$ to the weighted sum, hence
\[
 \sum_r|r|^2\bigl(e^{-8\pi(C_{<N}(0)-C_{<N}(r))}
                       -e^{-8\pi C_{<N}(0)}\bigr)
 =2\pi\kappa_{\rm lat}\log R+O(1).
\]
The exact map gives $z_j^{[1]}=P_jy$ and
$K_j^{[2]}=\mathcal L_{j-1}\cdots\mathcal L_1K_1^{[2]}$, whose
summed feedback is $O(y^2)$. The preparation has zero affine kinetic
projection. The remaining projection of the connected logarithm grows
at most polynomially in the polymer size, so exponential decay in that
size makes its sum finite. Comparing
$s_N^{[2]}=-a_\infty P_\infty^2Ny^2+O(y^2)$ with the preceding moment
proves \eqref{polynomial-operation:match}.

\emph{Graph and physical axis.} Transport the microscopic initial graph to the
bounded, scale-dependent coordinates $x_G=2s/(1-s)$ and $w_j=c_{\rm lat}z/P_j$. The ordinary free derivative
becomes exactly $I$. Denote the full map in these coordinates by $(f_j,p_j)$ and extend it
with a smooth cutoff acting only on the two finite-dimensional couplings and
preserving activity parity. For a small graph $K=H_j(g)$ the base map $q_j(g)=f_j(g,H_j(g))$ satisfies $\|Dq_j-1\|<1/2$.
Its graph transform reads
\begin{equation}
 H_{j+1}=p_j(\cdot,H_j(\cdot))\circ q_j^{-1}.
 \label{polynomial-operation:graphtransform}
\end{equation}
This starts from the true $H_1$. At derivative order $r$, the highest
input is $(D_Kp_j-DH_{j+1}D_Kf_j)D^rH_j$ and the highest output is
$D^rH_{j+1}(Dq_j)^{\otimes r}$. Therefore
\begin{equation}
 \begin{gathered}
 M_r^+\le(\theta+C\eta)(1-C\eta)^{-r}M_r+
 C_r\mathcal P_r(M_0,\ldots,M_{r-1}),\\
 (\theta+C\eta)(1-C\eta)^{-r}<1.
 \end{gathered}
 \label{polynomial-operation:graphr}
\end{equation}
Triangular induction bounds the chosen finite jet and gives
$H_j(0)=DH_j(0)=0$, $H_j^{[2]}=\mathcal L_{j-1}\cdots\mathcal L_1H_1^{[2]}$.

On a bounded field neighborhood the connected logarithm exists by
\eqref{endpoint-estimates:KP-criterion}. Its period-averaged affine
kinetic projection is $\sigma_j=s+O(\|K\|+|g|^4)$: $W_j$ has zero
projection and $\log(1+W_j)-W_j=O(g^4)$. Rooted size moments bound
the affine derivatives. At $y=0$ the exact Gaussian log is
$\tfrac12\phi^\top s_0A(1-s_0C_{<j}A)^{-1}\phi$ plus a constant,
whose $|p|^2$ coefficient is exactly $s_0$. Since
$\|AC_{<j}\|_{\ell^1}\le C$, its Neumann series stays in the weak
space. Equivalently the stable history and the projection estimate
prevent a first exit for sufficiently small $s_0$. Thus
$\xi=2\sigma_j/(1-\sigma_j)$ is an exactly fixed Gaussian coordinate.

Keep $w=c_{\rm lat}z/P_j$. The exact identity
$x_G^+-x_G=2(s^+-s)/[(1-s)(1-s^+)]$, $x_G=2s/(1-s)$,
shows that the graph map in $(\xi,w)$ has quadratic coefficients
$\widetilde a_j,\widetilde b_j=\ell_b+O(b^{-j/4}+\theta^j)$.
The changes
$x=\xi-\sum_{k\ge j}(\widetilde a_k-\ell_b)w^2$,
$v=w-\sum_{k\ge j}(\widetilde b_k-\ell_b)\xi w$
remove those defects. The kinetic increment is even in $v$ and
vanishes on $v=0$; the activity increment is odd. Taylor's theorem
now gives \eqref{polynomial-operation:normalform} with its stated powers.
To estimate the displacement $R=K-H_j(g)$ from the graph, write
$\Psi_j=p_j-H_{j+1}\circ f_j$. Graph invariance says that
$\Psi_j(g,H_j(g))=0$, while $R^+=\Psi_j(g,K)$. Thus
\begin{equation}
 \begin{aligned}
 R^+&=\int_0^1 D_K\Psi_j\bigl(g,H_j(g)+tR\bigr)R\,dt,\\
 \|R^+\|&\le(\theta+C\eta)\|R\|.
 \end{aligned}
 \label{polynomial-operation:offgraphstable}
\end{equation}
\end{proof}

Freeze this construction radius and choose a smaller exit $\eta_0$.
On the cone $1/2\le -x/v\le2$, $v>0$, writing $a=-x/v$ gives
$a^+=a+\ell_b(1-a^2)v+O(v^2)$ and
$\partial_aa^+=1-2\ell_bav+O(v^2)$. Its endpoints point inward, and
$\ell_bv^2/4\le v^+-v\le3\ell_bv^2$. The preparation with
$s_0=-c_{\rm lat}y/2$ enters this cone. From $v\ge\eta_*>0$ it takes
at most $1+(1+\ell_b\eta_0/4)/(\ell_b\eta_*/4)$ further steps to exit.
Along such an orbit monotonicity and the exact history give
\begin{equation}
 \|K_j\|\le C\theta_K^{j-1}v_1^2+Cv_j^3.
 \label{polynomial-operation:conehistory}
\end{equation}
This bounded continuation will allow each finite source family to
start in its own smaller weak domain.

\section{Local sources and contact extensions}
\label{sec:source-calculus}
\label{local-source-operation:section}

A product of local fields can generate lower-dimensional fields when
its insertion points collide. Its normalization must therefore be
specified on a finite module of local outputs, together with the
complementary remainder. The output module is finite by dimension
counting, but its coefficients need not contract. We construct an
ordinary inverse on this module and a forward contraction on the
complement. Independent scalar and wall outputs are retained throughout.

Fix a finite list $I$ of size $N$. For each nonempty subset $S\subseteq I$,
write $\sigma_S=2+\sum_{i\in S}(\Delta_i-2)$. A collision with $m$
integrated dimension-two interactions has degree
$\sum_i\Delta_i+2m-\Delta_\mathcal O-2(|S|+m-1)=\sigma_S-\Delta_\mathcal O$.
Hence its physical output list is the finite set
\begin{equation}
 \mathcal L_S=\{(q,d,\hbox{tensor/profile indices}):2q^2+d\le\sigma_S\}.
 \label{local-source-operation:physical}
\end{equation}
The scaled profile is $J_{j,i}(u)=a_j^{2-\Delta_i}J_i(a_ju)$,
$a_j=\epsilon b^j$; a grouped step $B=b^m$ therefore contributes
$B^{\sigma_S}$, including its anchor volume.

Take $\alpha_{\rm H}=1/2$ and an integer $R\ge4$ exceeding every retained source Taylor order by
at least two. This choice includes the unmarked neutral jet through degree
two.
On a block enlargement of scale $r$ put
\[
 \|h\|_{r;R,\alpha_{\rm H}}
 =\sum_{|\alpha|\le R}r^{|\alpha|}
       \|\partial^\alpha h\|_\infty
  +\sum_{|\alpha|=R}r^{R+\alpha_{\rm H}}
       [\partial^\alpha h]_{C^{0,\alpha_{\rm H}}}.
\]
Complete the smooth restrictions on the block enlargement in this H\"older
norm to obtain the separable little H\"older space $\mathfrak h_r^{R,\alpha_{\rm H}}$ for field directions
and ordinary profiles. Use odd Dirichlet field extensions and the prescribed
smooth profile extensions. For lattice data take divided differences on
fixed stencil enlargements, forming H\"older quotients only at distinct
sites. The kernel and contact estimates require separate integer seminorms
$C^{R_I}$ for coefficient profiles, with $R_I\ge R+3$ for metric labels. Field directions
have order $R$ throughout. Taking $\mathfrak h_r^{R,\alpha_{\rm H}}$ in \eqref{endpoint-estimates:polymernorm} sums over $p\ge0$ independently of
that spatial order $R$.

Add $Q_{j,k}$ to the base regulator for each $2\le k\le R+2$. At source degree $m$ set
\begin{equation}
 \begin{aligned}
 \kappa_m&=\kappa_I(1+m/N),&
 \kappa_I&\asymp(1+\log B)^{-1},\\
 \kappa_{k,m}&=\kappa_{k,0}+m\delta\kappa_k
              =\kappa_I(c_k+m d_k),&
 2&\le k\le R+2,\qquad c_k,d_k>0 .
 \end{aligned}
 \label{local-source-operation:gradedregulator}
\end{equation}
Choose $c_k,d_k$ positive within the Gaussian, odd-extension and finite H\"older
reserves, then fix the weak radius. Thus the unmarked coefficients ($m=0$) are
positive through $R+2$. The analytic radius
includes twice the $\mathfrak h_{a_j}^{R,\alpha_{\rm H}}$ norm of the pinned
imaginary shift $\beta_*(\Gamma_j(\cdot-x_0)-\Gamma_j(0))$, as well as the
fixed radius reserves used for the prescribed parameter derivatives.
Let
$G_{j,S}$ denote this regulator, including the higher quadratic forms.
The marked norm is
\begin{equation}
 \|K_S(X)\|_{j,I}=\sup_{\phi,J_S}
       |K_S(X;J_S,\phi)|_a/G_{j,S}(X,\phi),
 \label{local-source-operation:localnorm}
\end{equation}
with the preceding unit profile seminorms and polymer weight $A^{|X|_j}$.
All later integer test indices $R_I$ are chosen to dominate these fixed
profile seminorms. The higher regulator coefficients in source degree
zero are included when the unmarked body is reestimated in this space.

The two extra regulator derivatives supply the background estimate needed
by the analytic Taylor bounds. In physical coordinates write
\[
 \begin{aligned}
 E_r&=\int_X|\nabla\phi|^2, &
 H_r&=\sum_{B\subset X}r^4\sup_{B^*}|\nabla^2\phi|^2,\\
 Q_{r,k}&=r^{2k-2}\sum_{B\subset X}\int_{B^*}|\nabla^k\phi|^2.
 \end{aligned}
\]
At finite cutoff the integrals mean the physical lattice quadrature;
these are the rescaled forms introduced above. On each fixed small enlargement let
$\mathfrak H_{r,R,\alpha_{\rm H}}(\phi)$ denote the preceding seminorm with its
constant-field part omitted. Scaled Poincar\'e and Sobolev inequalities give
\begin{equation}
 \mathfrak H_{r,R,\alpha_{\rm H}}(\phi)^2
 \le C_I\left(E_r+H_r+\sum_{k=2}^{R+2}Q_{r,k}\right).
 \label{local-source-operation:sobolev-reserve}
\end{equation}
Indeed, after rescaling to a unit enlargement, the right-hand side controls
$\phi$ modulo constants in $H^{R+2}$. In two dimensions
$H^{R+2}$ embeds into $C^{R,\alpha_+}$ for every $\alpha_+<1$; choose
$\alpha_{\rm H}<\alpha_+<1$ to land in little $C^{R,\alpha_{\rm H}}$. Here $E_r$ controls the gradient average and the
Hessian term extends this control to the fixed enlargement. Odd extension
costs a fixed constant. Polynomial reproduction in the lifting below
transfers the bound to resolved lattice stencils uniformly.

Consequently every unused regulator fraction absorbs any fixed polynomial:
\[
 \sup_{H\ge0}(1+H)^p e^{-c_I\kappa_I H^2}
       \le C_{I,p}(1+\log B)^{p/2}.
\]
If a local coefficient overlaps the input, pay for its polynomial
growth with the source-degree increments $\kappa_I|S|/N$ and $|S|\delta\kappa_k$. Each nonempty degree
increase makes both increments positive. The fixed list allows at most $N$
increases and the degree-zero reserve is untouched.
\label{local-source-operation:overlapreserve}
For full activities the exact gas uses disjoint polymers. On merging
supports, integrate the joint Gaussian with each regulator counted once. The
preceding overlap estimate gives no unrestricted pointwise bound for the
product of full Gaussian envelopes.

\begin{proposition}[Local projection and graded transport]
\label{local-source-operation:projector}\label{local-source-operation:graph}
Every finite list has a compatible lattice-local projection, a uniformly
contracting ordinary stable complement and an exact invertible
polynomial transport of its physical coefficients. Its inverse has at
most a fixed arbitrarily small power loss in physical scale. Independent
marked-wall coefficients are transported forward.
\end{proposition}
\begin{proof}
\emph{A true finite projector.} Construct a smooth cardinal lifting reproducing polynomials through
order $R+2$. Blend degree-$(R+2)$ Lagrange interpolants with a fixed local
partition of unity, symmetrize by reflections, then take tensor products. For a fixed source family the stencil is finite, and cardinal interpolation gives $S_\epsilon I_\epsilon=1$. Because the
derivative coefficients annihilate lower polynomials, Newton differences
bound the lifted derivatives up to $R+2$ by the corresponding finite
differences in local $L^2$ and supremum norms. For $\mathfrak h_r^{R,\alpha_{\rm H}}$ use two ranges. Below
one mesh length the next derivative suffices by $\epsilon^{\alpha_{\rm H}-1}|x-y|\le|x-y|^{\alpha_{\rm H}}$. Above that length apply
the discrete H\"older bound to the $R$th differences. This works on all
resolved meshes with the same constants and respects odd extension.

For $M<R$ use Taylor's integral remainder up to derivative order
$M$ and direct rescaling for higher derivatives and the top
H\"older seminorm. This gives
\begin{equation}
 \|f-T_Mf\|_{\mathfrak h_{a_j}^{R,\alpha_{\rm H}}}
 \le C_I B^{-M-1}
       \|f\|_{\mathfrak h_{a_{j+1}}^{R,\alpha_{\rm H}}}.
 \label{local-source-operation:spatialrem}
\end{equation}
This profile estimate uses derivatives up to $R$. The two further
orders are for the background regulator and its lifting bound.

For charge $q$ embed derivative-weight-$M$ polynomials by
\[
 \iota_Xp=|X|^{-1}\sum_{x\in X}e^{iq\beta_*\phi(x)}p(z(x),u(x)),
\]
where $z_\alpha=a_j^{|\alpha|}\partial^\alpha\phi$ and
$u_{i,\alpha}=a_j^{|\alpha|}\partial^\alpha J_i$ at block scale $a_j$.
Use the scaled coordinate $y=(x-x_0)/a_j$ for the polynomial tests
\begin{equation}
 \phi_t(x)=t_0+\sum_{1\le|\alpha|\le M}
 t_\alpha\frac{y^\alpha}{\alpha!},\qquad
 J_{i,u}(x)=\sum_{|\alpha|\le M}
 u_{i,\alpha}\frac{y^\alpha}{\alpha!}.
 \label{local-source-operation:polytests}
\end{equation}
Divide out $e^{iq\beta_*t_0}$ and denote the moments through weight $M$ by $\mathcal J_X$. Ordered by
derivative weight, $\mathcal J_X\iota_X=1+N_X$ with $N_X$ uniformly bounded and strictly weight-raising. The
finite triangular inverse $P_X=(1+N_X)^{-1}\mathcal J_X$ defines the projector $\operatorname{Loc}_X=\iota_XP_X$. All prescribed moments of
$(1-\operatorname{Loc}_X)F$ vanish for every $F$. Polynomial reproduction gives this identity directly on
the lattice without inverting a spatial average. Indeed $N_X^{M+1}=0$, since replacing $t_\alpha$
by $t_{\alpha+\beta}y^\beta/\beta!$ for $\beta\ne0$ raises weight, as does each nonconstant anchor-character term.
The leading terms reproduce the polynomial, so $(1+N_X)^{-1}=\sum_{\ell=0}^{M}(-N_X)^\ell$ on the finite module. Higher
interpolation order in the proof leaves earlier microscopic source columns equal
to the same sampled lattice functionals with their fixed stencils.

\emph{The charge gain.} Set $d(q)=q^2$ for $|q|\le1$ and
$d(q)=2|q|-1$ otherwise. Extract through
$M_{S,q}=\lfloor\sigma_S-2d(q)\rfloor$ if this is nonnegative.
The capped Gaussian shift~\cite[Appendix B.3, eq. (B.11)]{FalcoCorrelations}
with $f=\beta_*q\delta_{x_0}$ for $|q|\le1$ and
$f=\beta_*\operatorname{sgn}(q)\delta_{x_0}$ otherwise gives
\begin{equation}
 \begin{aligned}
 \E_jF_q(\phi+\zeta)&=e^{-\beta_*^2d(q)\Gamma_j(0)/2}
 \E_j[e^{-i(\zeta,f)}F_q(\phi+\zeta+i\delta\psi)],\\
 \delta\psi_x&=(\Gamma_jf)_x-(\Gamma_jf)_{x_0}.
 \end{aligned}
 \label{local-source-operation:contour}
\end{equation}
The pinned shift is bounded in $\mathfrak h_{a_j}^{R,\alpha_{\rm H}}$ uniformly
in the grouped factor: on a small support its value follows from the
gradient kernel bound, and its top H\"older seminorm follows from the
next fixed kernel derivative. The capped choice makes this bound independent
of $q$. The shift uses half the analytic radius. Moreover, for a coarse
direction $h$,
\[
 \|h-h(x_0)\|_{\mathfrak h_{a_j}^{R,\alpha_{\rm H}}}
       \le C_I B^{-1}
          \|h\|_{\mathfrak h_{a_{j+1}}^{R,\alpha_{\rm H}}}.
\]
Take $B$ large enough that $C_I/B\le1/4$. After the half-radius
imaginary shift, a quarter-radius therefore remains for the fixed
Taylor--Cauchy reserves, including the prescribed parameter jets.
The corresponding background bound has the same factor and the coarse
seminorm $\mathfrak H_{a_{j+1},R,\alpha_{\rm H}}(\phi)$, which is controlled by
\eqref{local-source-operation:sobolev-reserve}.

First remove the constant-field character. Expand in the nonconstant
profiles and pinned field increment. Distinct profile labels enter
multilinearly, so the profile expansion terminates. Taylor-expand the field functional at $\zeta+i\delta\psi$ to the remaining total degree and put the spatial Taylor polynomial
in each retained field and profile argument. In weight $w$ each jet
contributes $B^{-|\alpha|}$. Every discarded term has $w\ge M+1$ and therefore gains $B^{-M-1}$, and \eqref{local-source-operation:spatialrem} gives the
same bound for a spatial replacement error.

To sum functional derivatives, give a derivative on the pinned increment its
$B^{-1}$ and keep the complementary powers on the other Taylor factors.
Derivatives on the character sum to $e^{C_Ia|q|}$. Absorb the fixed coarse-background
polynomial with the unused Gaussian fraction in \eqref{local-source-operation:sobolev-reserve}. For the actual
projector use agreement of the anchored representative with $\iota_XP_X$ in moments
through $M$. Expanding the finite templates and anchor characters to this
order gives the same weighted error for their difference. The triangular
inverse contributes only constants and a fixed polynomial in $|q|$ absorbed
by $C_I^{1+|q|}$. We can therefore use the actual projector. Substituting $\Gamma_j(0)=(\log B)/(2\pi)+O(1)$ in the
capped-shift prefactor gives
\begin{equation}
 \|(1-\operatorname{Loc}_{q,M})\E_jF_q\|_{j+1,I}
 \le C_I^{1+|q|}(1+\log B)^{c_I}
       B^{-2d(q)-M-1}\|F_q\|_{j,I}.
 \label{local-source-operation:localizationbound}
\end{equation}
The finite polynomial absorptions determine $c_I<\infty$. Both sides use directions in
$\mathfrak h^{R,\alpha_{\rm H}}$ and background order $R+2$. Spare exponential weight absorbs larger-support
size moments. After multiplication by $B^{\sigma_S}$ the gain is at least $B^{-1}$, with
summable unextracted charges. Choose $B$, then $A$, then the sourcewise weak
radius. The canonical initial norm in this stronger space is finite.
Iteration gives
\begin{equation}
 \|K_j\|_{A_I}\le\theta^j\|K_0\|_{A_I}
       +C_I\sum_{k<j}\theta^{j-1-k}|g_k|^3.
 \label{local-source-operation:strongbody}
\end{equation}
Intermediate master steps are finitely many bounded operations.

\emph{The physical/stable split.} For positive-order derivative jets
let $D_j,c_j$ be the derivative--derivative and derivative--field
covariances of the actual tail $\sum_{k\ge j}\Gamma_k$, with current-scale
weights. Both converge geometrically. The finite Wick change
$\exp[-D_j:\partial_z^2/2-iq\beta_*c_j\cdot\partial_z]$
cancels the shell jet covariance. A further bounded scalar tail factor
cancels the summable defect of $\Gamma_j(0)$ from
$(\log B)/(2\pi)$. In this basis a monomial of derivative degree $d$
has exact free multiplier $B^{\sigma_S-2q^2-d}$. The physical
eigenvalues are at least one; the surplus extracted coordinates have
true gap at least one. Their matrix with the infinite remainder is
upper triangular, so a fixed relative weight makes the combined stable
norm contracting. Interaction perturbs it by $O_I(\eta)$.

Write physical and stable coefficients as $(\lambda_S,r_S)$. The exact
top-degree map has blocks $A_S,B_S,C_S,D_S$ plus proper-subset polynomials,
with
\begin{equation}
 \|A_S^{-1}\|\le1+C_I\eta,\quad \|C_S\|\le C_I\eta,\quad
 \|D_S\|\le\theta+C_I\eta,\quad \|B_S\|\le C_I.
 \label{local-source-operation:blocks}
\end{equation}
Invert in increasing subset order, with all proper-subset contacts already known
at each step. For a subset $S$ assign weight $w_S=\delta^{|S|^2}$. Any proper partition then
satisfies
\begin{equation}
 w_S/\prod_{B\in\pi}w_B
 =\delta^{2\sum_{B<C}|B||C|}\le\delta^2.
 \label{local-source-operation:grading}
\end{equation}
The forward graph $r=H_j(\lambda)$ is defined by $H_{j+1}=Q_jT_j^{-1}$,
where $T_j,Q_j$ are its physical and stable outputs. Its linear part is
$O_I(\eta)$, so the inverse physical block is $1+O_I(\eta)$.
For each proper-partition coefficient the highest operator is the
stable block minus the graph correction, times at most $N$ inverse
physical blocks. Choose
\begin{equation}
 (\theta+C_I\eta)(1+C_I\eta)^N<\theta_1<1,
 \label{local-source-operation:jetgap}
\end{equation}
then $\delta$ small. Triangular induction and the retarded estimate
bound all coefficients, including the true initial graph. On the finite
space $\mathcal M_I$ of label-compatible monomials the inverse is substitution
$U_jF=F\circ T_j^{-1}$, with
\begin{equation}
 \|U_j\cdots U_{n-1}\|\le L_I^{n-j},\qquad
 L_I=1+C_I(\eta+\delta^2),\qquad L_I^{n-j}\le C(a_n/a_j)^{a_I},
 \label{local-source-operation:backward}
\end{equation}
where the total normalization loss, including output blocks, is chosen
\begin{equation}
                         0<a_I<1/(2N).
 \label{local-source-operation:anomalousloss}
\end{equation}
Gaussian proper-subset contractions are included in this polynomial
cocycle. Independent odd marked-wall slots are not inverted.
\end{proof}

\begin{proposition}[Contact and spatial bounds]
\label{local-source-operation:anchored}
The preceding operation gives the prescribed contact extension and
one common exponential spatial rate for all fixed finite lists.
\end{proposition}
\begin{proof}
Partition each profile into blocks. A size-$n$ connected support has
at most $C_In^N$ marked assignments. The geometric decay of the unmarked
activities makes their sum finite, since
\begin{equation}
                         \sum_{n\ge1}n^N\gamma^n<\infty,\quad\gamma<1.
 \label{local-source-operation:pinning}
\end{equation}
The finite-algebra product is $(FG)_S=\sum_{A\sqcup B=S}F_AG_B$;
the chain rule sums partitions of $S$. This retains every nested or
disjoint contact, and a fusion tree has at most $N-1$ branching vertices.
Fix the label set $S$ and output $\mathcal O$ for the next estimates.
Write $\omega=\sigma_S-\Delta_\mathcal O$ and
$\mathcal K_j=\mathcal K_{j,S,\mathcal O}$ for their shell kernel.
Restoring physical dimensions gives the distributional shell bound
\begin{equation}
 |\mathcal K_j(f)|
 \le C_Ia_j^{-\omega-a_I}\|f(a_j\cdot)\|_{C^{R_I}}.
 \label{local-source-operation:shelltest}
\end{equation}
Separated profiles extend to general relative tests by a product Fourier
series with enough derivatives to be absolutely summable. If $\omega\ge0$,
\begin{equation}
 |\mathcal K_j(f-T_\omega f)|
                       \le C_fa_j^{1-a_I};
 \label{local-source-operation:contactgain}
\end{equation}
if $\omega<0$, its integer value is at most $-1$ and no subtraction
is needed. The budget \eqref{local-source-operation:anomalousloss}
keeps a positive power through the finite contact forest. Auxiliary
outputs vanish with their true positive gap and stay paired with their
stable remainders. For this $S$ and $\mathcal O$, let $c_{j,\alpha}$
denote the contact coefficient for the multi-index $\alpha$. Converting
to the chosen physical cutoff gives
\begin{equation}
 c_{j,\alpha}
 =\mathcal K_j\!\left[(1-\chi_{\rm ct}(\mu_{\rm ref}r))\frac{r^\alpha}{\alpha!}\right],
 \quad |\alpha|\le\omega,
 \label{local-source-operation:schemedifference}
\end{equation}
and its connected support must span distance $c/\mu_{\rm ref}$, whence
\begin{equation}
 |c_{j,\alpha}|
 \le C_Ia_j^{|\alpha|-\omega-a_I}e^{-c_I/(\mu_{\rm ref}a_j)}.
 \label{local-source-operation:schemetail}
\end{equation}
Proper-subset induction therefore fixes the finite parts without
assuming convergence of a new full-subset contact.

Retain the labelled support history under every integration and
projection, including scalar allocation. For $n$ blocks its connecting
tree length obeys
\begin{equation}
 \tau(X,z_S)\le a_j(c_{\rm geom}n+R_I).
 \label{local-source-operation:eq-AS-2a}
\end{equation}
The unmarked sum decays as $e^{-4\sigma_0 n}$. Placing the finitely
many labels adds a polynomial in $n$, so the complete coefficient is bounded by
\begin{equation}
 C_Ia_j^{-P_I}(1+n)^{p_I}e^{-4\sigma_0n}\prod_i\|J_i\|_{C^{R_I}}.
 \label{local-source-operation:eq-AS-3}
\end{equation}
For $\kappa_0\le\sigma_0/c_{\rm geom}$, summing all but one anchor gives
\begin{equation}
 \sup_{z_{i_0}}\sum_{z_{S\setminus i_0},X}
 e^{\kappa_0\tau(X,z_S)/a_j}\|\mathcal F_{j,X,S}\|_{j,I}^{\rm wk}
 \le C_Ia_j^{-P_I}\prod_i\|J_i\|_{C^{R_I}}.
 \label{local-source-operation:eq-AS-3a}
\end{equation}
Scalar and boundary colors retain the same connecting tree. In particular
for $L\ge a_n$,
\begin{equation}
 \sum_{j\le n}a_j^{-P}e^{-\kappa_0L/a_j}
 \le C a_n^{-P}e^{-\kappa_0L/(2a_n)},
 \label{local-source-operation:eq-AS-5}
\end{equation}
by comparison with $\sum_{k\ge0}b^{Pk}e^{-\kappa_0b^k/2}$.
After the sourcewise small section the cone estimate permits only a
bounded number of master steps to the common exit. Decreasing $A_I$
and increasing the regulator coefficient weakens the norm, so these
steps preserve all bounds. Before a wall at distance $d$ is reached,
its history has the summable factor $a_j^{-P}e^{-cd/a_j}$; afterwards
only finitely many boundary-sensitive steps remain.
\end{proof}

\section{The canonical ultraviolet family}
\label{sec:ultraviolet}

The calibrated map determines a candidate ultraviolet branch only after
its true microscopic preparation has been included. Two issues enter at
this point. The branch is nonlinear, so its tangent ray does not fix the
bare family. Moreover, a general ratio $a_0/\epsilon$ has a residual
blocking phase. We resolve both issues on one physical hierarchy: a
positive continuous covariance decomposition gives an exact operation
at every cutoff, and the summable difference between these operations
is removed by a conjugacy.

The base comparison includes the marginal pair, quadratic preparation,
stable remainder, boundary and covariance records. The accumulated
source-zero scalar is factored from its norm, since that scalar need not
converge. Its increments remain in the complete operation, and its full
history remains in every exact reconstruction identity.

This scalar history includes the linear kinetic contractions and
their actual wall corrections. Parameter and geometric coefficients
are formed from the full finite-cutoff labelled record, including
these contractions. Only scalar factors common to the numerator
and its specified normalizer are cancelled; identity and contact
coefficients not common to those two factors remain in the record.

\subsection{Positive scales and a common operation}

Fix $a_0>0$, an odd integer subscale factor $\gamma\ge3$ and a sufficiently
large odd master factor $b=\gamma^m$. Put
\begin{equation}
                         a_k=a_0b^k,\quad k\in\mathbb Z.
 \label{arbitrary-cutoffs:physical-grid}
\end{equation}
Choose an even nonnegative Schwartz function $\varphi$ with smooth
Fourier transform supported in $[-1,1]$ and
$\int_0^\infty u\varphi(u)du=1$. With
$W_t^*(\lambda)=\sum_{n\in\mathbb Z}
\varphi(t\arccos(1-\lambda/2)-2\pi nt)$, define, in the physical
pairing $\epsilon^2\sum_x$,
\begin{equation}
 C_\epsilon[a,d]=\frac{\epsilon^2}{3}
 \int_{a/(c_{\rm fr}\epsilon)}^{d/(c_{\rm fr}\epsilon)}
 tW_t^*(A_D/3)dt,\qquad Q_\epsilon(a)=C_\epsilon[a,\infty],
 \label{arbitrary-cutoffs:cov-int}
\end{equation}
where the fixed divisor $c_{\rm fr}>32\gamma$ includes the template
collars. Its matrix divided by $\epsilon^2$ is the field covariance.

\begin{proposition}[Common physical-scale operation]
\label{arbitrary-cutoffs:covariance-lemma}\label{arbitrary-cutoffs:input-transfer}
\label{ultraviolet-trajectory:common-theorem}
The operators in \eqref{arbitrary-cutoffs:cov-int} are positive,
additive in their scale intervals, have range at most $d/c_{\rm fr}$,
and satisfy $C_\epsilon[0,\infty]=\epsilon^2A_D^{-1}$.
Their continuum limits are
\begin{equation}
 C_0[a,d]=\frac13\int_{a/c_{\rm fr}}^{d/c_{\rm fr}}
 r\varphi(r\sqrt{-\Delta_D/3})dr.
 \label{arbitrary-cutoffs:cov-cont}
\end{equation}
At fixed positive scale, every prescribed finite kernel derivative and
projective covariance norm has error $O((\epsilon/a)^\nu)$ for some
$\nu>0$, including the full retained tail in a fixed Dirichlet box.
On nested physical blocks the complete weak operation consequently has
one limiting map, independent of residual phase, with
$\|\mathcal F_{\epsilon,j}-\mathcal F_0\|_{C^k(s\to w)}
\le C_{I,k}(\epsilon/a_j)^\nu$ for every fixed source list and order.
Its actual forward graphs have uniform $C^{10}$ bounds and converge in
$C^8$ at a positive exponential rate in the number of resolved scales.
\end{proposition}
\begin{proof}
\emph{The spectral construction.} We use the Fourier convention
\[
 \widehat\varphi(\xi)=(2\pi)^{-1}\int e^{-i\xi u}\varphi(u)du.
\]
Poisson summation gives
$W_t^*(\lambda)=t^{-1}\sum_{|k|\le t}\widehat\varphi(k/t)
T_k(1-\lambda/2)$. This is a polynomial of range at most $t$.
Positivity follows from the original sum. For
$\lambda=4\sin^2(\omega/2)$, the identity
$\lambda^{-1}=\sum_n(\omega-2\pi n)^{-2}$ gives
$\int_0^\infty tW_t^*(\lambda)dt=\lambda^{-1}$.
Since $\operatorname{Spec}(A_D/3)\subset(0,8/3]$, the factor $1/3$
gives the exact inverse. The kernel estimates of
\cite[Theorem 1.2, Lemma 2.7, Proposition 3.1 and (1.35)--(1.38)]{BauerschmidtFRD}
bound fixed rescaled derivatives and their lattice errors by a positive
power of $\epsilon/a$. Cardinal lifting transfers these bounds; Fourier
expansion on two block enlargements with sufficiently many derivatives
gives the projective tensor estimate. The sine spectral basis and
Schwartz decay sum the retained tail in a fixed box. Odd compression
and an $O(\epsilon)$ affine wall adjustment preserve these estimates.
The continuum formula has exact two-dimensional scale covariance.

\emph{Norms.} Use stronger and weaker versions of
\eqref{endpoint-estimates:polymernorm}: the stronger has larger analytic
radius and size weight and smaller admissible regulators. In physical
coordinates the common regulator is
\begin{equation}
 G_{r,S}=\exp\{\kappa_S(c_1E_r+c_3T_r+c_1H_r)
                         +\sum_{k=2}^{R+2}\kappa_{k,S}Q_{r,k}\},
 \label{ultraviolet-trajectory:common-regulator}
\end{equation}
where $E_r=\int_X|\nabla\phi|^2$, $T_r=r\int_{\partial X}|\nabla\phi|^2$,
$H_r=\sum_{B\subset X}r^4\sup_{B^*}|\nabla^2\phi|^2$ and
$Q_{r,k}=r^{2k-2}\sum_B\int_{B^*}|\nabla^k\phi|^2$.
Here the spatial direction order remains $R$: the sum through $R+2$
is the separate Sobolev regulator reserve. All its coefficients have the
positive degree-zero baseline and the grading in
\eqref{local-source-operation:gradedregulator}. The completion of smooth
local cylinder functionals in the analytic norm over
$\mathfrak h_r^{R,\alpha_{\rm H}}$ defines the activity Banach spaces. Retained
Gaussian fields at positive resolution are smooth and belong to this
separable little H\"older space. The stronger and weaker norms use the
same fixed orders and exponent, with their previously specified radius,
size-weight and regulator reserves.
In the continuum norm the real-background supremum is first taken on
smooth fields, then on their closure in the norm defined by
$E_r+H_r+\sum_{k=2}^{R+2}Q_{r,k}$ modulo constants, with any constant
coordinate kept separately. Analytic jets extend to little H\"older
directions by their common radius. Taylor series on balls about smooth
centers agree on their smooth dense overlaps, so these local extensions
are consistent. No finite regulator is required of every little H\"older
background. Positive-resolution Gaussian fields have all the finite
regularities used here. Sampling and cardinal lifting satisfy $S_\epsilon I_\epsilon=1$,
so lifted Gaussian integration is exact. Sampling bounds follow from
Newton differences and the alias estimate
$|d_\epsilon(p)|^{2k}\sum_\ell(1+|p+2\pi\ell/\epsilon|^2)^{-k}\le C_k$
for $k>1$. For the gradient, use the Hessian term in the regulator.

Polynomial reproduction through $R+2$ covers the last two regulator forms. The
uniform little H\"older lifting estimate (Section~\ref{sec:source-calculus}) puts smooth-kernel
comparisons in the same spaces. Interpolating a positive-power $C^R$ error
against a uniform $C^{R+1}$ bound retains a positive power in $C^{R,\alpha_{\rm H}}$. Additional
fixed derivatives give the projective tensor estimate. All source-family
indices are independent of the ultraviolet cutoff.
More explicitly, on two rescaled block enlargements expand the smooth
kernel difference in a tensor-product Fourier series. A Fourier mode has
$\mathfrak h^{R,\alpha_{\rm H}}$ norm bounded by
$C(1+|n|)^{R+\alpha_{\rm H}}$. Taking more than
$R+\alpha_{\rm H}+2$ derivatives in each endpoint makes the weighted
coefficient sum absolutely convergent. The stipulated fixed mixed-kernel
error therefore gives the same positive-power projective error in the
new direction spaces.

At each positive $\gamma$-substep, integration by parts bounds the
gradient cross term by boundary and Hessian energies and a positive
noise form, leaving the background gradient coefficient one. A small
boundary coefficient and reblocking restore those reserves; translating
a small polymer into a disjoint part of its enlargement leaves a fixed
fraction of the background weight unused. Noise covariance norm is
bounded and its trace is $C|X|_r$. The Gaussian determinant estimate
therefore proves the same regulator bound. For $k\ge2$ the noise norm
uses $r^{2k-2}\int_r^{br}t^{2-2k}dt/t\le C_k$ and reblocking gains
$b^{2k-2}$. These estimates retain the source grading and odd factor four.

\emph{Exact physical-block projection.} Assign each lattice site and
bond once to the half-open physical square containing its root; coarse
squares are exact unions of children. The readout is taken from the
intrinsic rooted functional-derivative kernels of the complete bounded
connected germ, before any allocation among its individual $W,K$
entries. In the neutral quadratic sector use the second relative
coordinate moments of the period-averaged Hessian; in the charge-one
sector use the corresponding Fourier coefficient of the rooted first
derivative. These are bounded moment tests: expanding their relative
coordinate powers gives smooth polynomial field directions, and the
resulting support powers are absorbed by the size reserve.

Let $V$ be the unmarked physical marginal space (the physical pair), and let
$\mathcal I=\sum_D i_D^\chi:V\to\{\text{density functionals}\}$
be the homogeneous injection given by a smooth partition $\chi_D$.
Write $\widetilde J_B$ for the raw rooted tests and set, at the actual
cutoff and block phase,
\[
 M_B=\widetilde J_B\mathcal I,\qquad
 J_B=M_B^{-1}\widetilde J_B.
\]
Their continuum moment matrices are nonsingular. Uniform quadrature
convergence makes $M_B^{-1}$ bounded once $r/\epsilon\ge K$, and
\[
                         J_B\mathcal I=\operatorname{id}_V
\]
holds exactly, including for noncommensurate physical blocks. Choose
the partition equal to one on smaller disjoint cores. Invert the actual
finite moment matrices of core-supported templates to obtain
$J_Ce_B=\delta_{CB}\operatorname{id}_V$; the cores include the fixed
lattice stencil collars. Setting
\[
 i_D^\sharp=i_D^\chi+
       \sum_Be_B(\delta_{BD}\operatorname{id}_V-J_Bi_D^\chi)
\]
therefore gives
\[
 J_Ci_D^\sharp=\delta_{CD}\operatorname{id}_V,\qquad
 \sum_Di_D^\sharp
 =\mathcal I+\sum_Be_B(\operatorname{id}_V-J_B\mathcal I)
 =\mathcal I.
\]
Only bounded neighborhoods occur in these sums. Subtract the moments on fine
connected supports before reblocking to retain the Taylor and charge gains. Exact
zero-sum allocation gives the subtraction. Its nonlinear correction is contractive
since its derivative is $O(\eta)$.

Temporary local marginal arrays are determined, not additional bare
parameters. Their complete projection has the form
$\jmath g+q+R(g,q,K)$, with
$\|D_qR\|\le C\eta$,
$\|D_KR\|\le\delta_A+C\eta$, and
$\|R(g,0,K)\|\le C|g|^4+(\delta_A+C|g|)\|K\|+C\|K\|^2$.
Here $\jmath g$ is the constant array with value $g$, and
$\delta_A\to0$ bounds the unextracted large-polymer roots.
The equation $q=-R(g,q,K)$ defines the uniform local inverse
$q=Q_r(g,K)$ in the block $\ell^\infty$ norm.
On a canonical uncolored bulk history the intrinsic rooted kernels of
the complete germ are lattice-translation invariant. Their retained
root moment vector is therefore independent of the root, with square
symmetry reducing its unmarked marginal part to the physical pair.
The raw readout in block $B$ is $M_B$ times this common vector, so the
exact normalization above makes the physical output homogeneous even
when different physical blocks contain different numbers of lattice
roots. Uniqueness identifies the actual arrays with the local inverse. For graph estimates, extend off these
histories by applying a fixed norm-one projection of the output physical
array onto the homogeneous pair and then reconstructing $Q_r$.
This extension equals the exact operation on canonical histories and
leaves all marked arrays unchanged; no physical reconstruction is
asserted for artificial nonuniform inputs. The stable gap is
$\theta+C\delta_A+C\eta<1$. Marked arrays and wall colors remain
independent complete data.

\emph{Comparison.} Use one common continuum background $\phi$ and the
lifted lattice covariance $\widehat C_\epsilon=I_\epsilon C_\epsilon
I_\epsilon^*$. On a canonical sampled cylinder
$F_\epsilon=\widetilde F_\epsilon\circ S_\epsilon$, the expectation
$\mathbb E_{\widehat C_\epsilon}F_\epsilon(\phi+\zeta)$ agrees exactly
with the lattice expectation at background $S_\epsilon\phi$, since
$S_\epsilon I_\epsilon=1$.
At this fixed background convex Gaussian interpolation pairs the covariance
difference with $D^2F/2$, bounded by the projective norm and an analytic-radius
reserve. The identity extends from smooth cylinders by the uniform regulator
bound, without needing convergence of $I_\epsilon S_\epsilon$ to the identity in operator norm on the
same little H\"older space. Age-$m$ preparation terms cost $Cb^{-m}(\epsilon/a_{j-m})^\nu$, summable for $0<\nu<1$.
The same comparison holds for template quadratures and their finite inverses.
Changing one factor at a time in the strict connected majorant gives the complete
estimate including finite source jets, and the bounded array inverse preserves it.
For the graph choose a radius with
\begin{equation}
                  (\theta+C\eta)(1-C\eta)^{-r}<1,\quad r\le10.
 \label{ultraviolet-trajectory:ten-jet-gap}
\end{equation}
The differentiated graph recurrence, with two spare orders for moved
inverse arguments, yields
\begin{equation}
 e_{j+1}\le qe_j+C(\epsilon/a_j)^\nu,\quad q<1,\qquad
 \|H_j-H_0\|_{C^8(w)}\le C\sigma^j,\quad\sigma<1.
 \label{ultraviolet-trajectory:graph-difference-recursion}
\end{equation}
The retarded sum keeps the microscopic preparation and boundary forcing. Physical
source coefficients and scalar histories still have their polynomial transport.
\end{proof}

\subsection{The nonlinear orbit and its microscopic realization}

\begin{theorem}[Canonical ultraviolet family]
\label{ultraviolet-trajectory:trajectory-theorem}
\label{arbitrary-cutoffs:allcutoff-theorem}
The limiting map has an invariant $C^6$ one-sided curve
$p_\rho=(\mathfrak h(\rho),\rho)$, $\mathfrak h(\rho)=-\rho+O(\rho^2)$,
with a unique ultraviolet half-orbit in the weighted class below.
It satisfies $v_n\sim(\ell_b|n|)^{-1}$ as $n\to-\infty$.
Every sufficiently small cutoff admits positive canonical bare data
$Z_\epsilon\to1$, $y_\epsilon\to0$ realizing this curve. Their complete
base records have one common limit at every fixed positive resolution,
uniformly in residual phase, through four endpoint derivatives on compact
positive intervals. With $g_\epsilon=c_{\rm lat}y_\epsilon$,
\begin{equation}
 \log(a_0/\epsilon)=g_\epsilon^{-1}+O(\log(1/g_\epsilon)),\qquad
 1-Z_\epsilon=-g_\epsilon/2+O(g_\epsilon^2).
 \label{arbitrary-cutoffs:all-clock}
\end{equation}
\end{theorem}
\begin{proof}
\emph{The half-orbit.} Put $T=(-x+2v)/3$, $U=x+v$.
The limiting normal form is
$T^+=T+\ell_bT^2-\ell_bU^2/9+O((|T|+|U|)^3)$,
$U^+=(1-\ell_bT)U-\ell_bU^2/3+O((|T|+|U|)^3)$.
For $n\le0$ set $q_n=(\rho^{-1}-\ell_bn)^{-1}$,
$t_n=(\ell_b\rho)^{-1}-n$, $L_n=1+\log(\rho/q_n)$ and
$T_n=q_n+e_n$, $U_n=u_n$. Write $w_n=(t_n+1)(t_n+2)$.
For forcing terms $F_k^U,F_k^T$, the linear Green problem is
\begin{align*}
 u_n&=\sum_{k<n}\frac{t_n}{t_{k+1}}F_k^U,\\
 e_n&=-\frac{w_0}{3w_n}u_0
       -\sum_{k=n}^{-1}\frac{w_{k+1}}{w_n}F_k^T .
\end{align*}
It imposes $u_n=O(q_n^2)$ and $e_0=-u_0/3$, hence $v_0=\rho$.
The excluded homogeneous normal solution grows like $t_n$.
In the norm $\sup_n\max(|e_n|/(q_n^2L_n),|u_n|/q_n^2)$,
forcing $Mq_n^3$ has bounded response and forcing $Mq_n^4L_n^2$
has norm at most $C\rho M$. These follow by summing powers of
$t_n^{-1}$; in particular
$\sum_{k<n}q_k^4\le Cq_n^3$, $\sum_{k=n}^{-1}q_k\le CL_n$ and
$\sup_{q\le\rho}q(1+\log(\rho/q))^2\le C\rho$.
The reference defect is cubic, with tangent contribution
$-\ell_b^2q_n^3/(1-\ell_bq_n)$. On a fixed norm ball the nonlinear
difference is $Cq_n^4L_n^2$ times the difference norm. Small $\rho$
therefore gives a strict contraction. It yields $u_n=O(q_n^2)$,
$e_n=O(q_n^2L_n)$; shifting and uniqueness give invariance.

For endpoint regularity, the polynomial
$P_m(v)=-v+\sum_{k=2}^ma_kv^k$ has invariance defect $O(v^{m+2})$
when coefficients are solved successively. The homological denominator
is $(m+1)\ell_b\ne0$. The normal remainder obeys
$e_{n+1}=(1-\ell_bv_n+O(v_n^2))e_n+O(v_n^{m+2})$.
Its retarded product is $Cv_k/v_n$ and its differentiated tangent
product is $C(v_k/v_n)^2$. The same sums give
$|\partial_v^r(\mathfrak h-P_m)|\le C_{m,r}v^{m+1-r}$ for $r\le m$.
Taking $m=6$ in these estimates gives six continuous one-sided
derivatives at zero.

\emph{Summable nonautonomy.}
\label{ultraviolet-trajectory:nonautonomous-conjugacy-2}
\label{arbitrary-cutoffs:uniform-conjugacy}
Write $f_0$ for the limiting marginal map and $f_{\epsilon,j}$ for the
scale-dependent maps in this comparison. Their $C^8$ defects are
$O(\sigma^j)$, and they are uniformly bounded in $C^{10}$. Extend
the near-identity maps by a common smooth coupling cutoff and set
$c_j^{(n)}=f_0^{-n}\circ f_{\epsilon,j+n-1}\circ\cdots\circ f_{\epsilon,j}$.
Here the cutoff is fixed in $c_j^{(n)}$. Consecutive terms differ
by one defect, with $C^r$ norm at most
$C_r(1+n)^{C_r}e^{C_r\eta n}\sigma^{j+n}$, $r\le6$.
Taking $\eta$ smaller than the exponential margin gives a convergent
series and
$c_{j+1}\circ f_{\epsilon,j}=f_0\circ c_j$. Equal linear jets imply
$c_j(g)=g+O(\sigma_1^j|g|^2)$, with the corresponding differentiated
bounds. Any common invariant Gaussian axis and any common symmetry,
including activity parity when present, are preserved. Uniform
inversion holds on a smaller operative ball.

Choose the unique $N$ with
\begin{equation}
 r_0=a_0b^{-N},\qquad K\epsilon\le r_0<Kb\epsilon,\qquad r_j=r_0b^j.
 \label{arbitrary-cutoffs:prefix}
\end{equation}
Integration over $C_\epsilon[0,r_0]$ gives the microscopic preparation $\mathcal P_\epsilon$. Throughout the compact
phase interval both its range and its site count are uniformly bounded in lattice
units. For the bulk calculation denote the field-coordinate covariance by $\Gamma_{<r}=\epsilon^{-2}C^{\rm bulk}_\epsilon[0,r]$ at
cutoff $\epsilon$.

Already in this first preparation, the linear kinetic scalar in a true box $D$ is
\[
 \frac{s}{2}\Tr(A_D\Gamma_{<r_0}^D),\qquad
 \Gamma_{<r_0}^D=\epsilon^{-2}C_\epsilon^D[0,r_0].
\]
Record both its bulk and its wall allocation. Do this on each positive partial
shell with that shell's field-coordinate covariance and input coupling. The
variance factor and coordinate change are
\begin{align}
 P_\epsilon(r)&=(r/\epsilon)^2
                 e^{-4\pi\Gamma_{<r}(0,0)},
 \label{arbitrary-cutoffs:P}\\
 v_0&=c_{\rm lat}y+O((|s|+y)y),\qquad x_0=2s+O((|s|+y)^2).
 \label{arbitrary-cutoffs:prefix-jacobian}
\end{align}
To first order in the activity the dimensionless block coefficient is $z_\epsilon(r)=P_\epsilon(r)y$. The
exact coordinate $v=c_{\rm lat}z/P_\epsilon(r)$ cancels successive variance factors, so the free multiplier
is one and \eqref{arbitrary-cutoffs:prefix-jacobian} carries no residual phase. In logarithmic scale measure the
scale-density diagonal equals $1/(2\pi)+O(\epsilon/r)$. Hence the limit
\[
 P_{\rm fr}:=\lim_{r/\epsilon\to\infty}P_\epsilon(r)>0
\]
exists independently of phase, for this spectral shape. Other decompositions may
have different variance constants. For the square-lattice potential kernel we have
\[
 0\le a_{\rm lat}(x)-
 [\Gamma_{<r}(0,0)-\Gamma_{<r}(0,x)]
 \le C|\epsilon x|^2/r^2 .
\]
The last inequality follows from the retained-covariance quadratic-form bound. It
gives the cumulative neutral-pair moment the logarithmic coefficient in \eqref{polynomial-operation:match}.
Rebuilding the actual quadratic memory thus recovers the same $c_{\rm lat}$ and quadratic
marginal coefficients. At higher orders we compute the entire limiting map $f_0$ in
the present spectral scheme and make no identification across decompositions.
Since $\mathcal P_\epsilon$ has Jacobian $\operatorname{diag}(2,c_{\rm lat})$, its inverse is uniform. Take as bare pair
\[
 (s_\epsilon,y_\epsilon)=\mathcal P_\epsilon^{-1}
                  c_{\epsilon,0}^{-1}f_0^{-N}p_\rho.
\]
Its exact prepared marginal point at step $j$ is
$c_{\epsilon,j}^{-1}f_0^{j-N}p_\rho$; all stable and boundary data are
generated forward. Positivity follows near the outgoing ray. Along
the invariant curve,
$v^+-v=\ell_bv^2+O(v^3)$ and $1/v-1/v^+=\ell_b+O(v)$.
Summation proves \eqref{arbitrary-cutoffs:all-clock}; the bounded phase
changes only its finite part. At physical level $a_k$ the autonomous
point is exactly $f_0^kp_\rho$, and $\sigma_1^{N+k}\le C(\epsilon/a_k)^\nu$.
Graph convolution and kernel comparison thus give the common complete
base limit and, on $\delta\le r\le a_+$, error
$C\delta^{-p}[(\epsilon/\delta)^\nu+(\epsilon'/\delta)^\nu]$.
Positive partial shells give all intermediate resolutions.

\emph{Four derivatives.}
\label{ultraviolet-trajectory:af-four-derivatives}
Let $F$ be the activity map on the invariant curve and $q=F^{-1}$.
The inverse iterates $v_k=q^k(\rho)$ satisfy $v_k\asymp_J(1+k)^{-1}$.
Write $\dot v_k=\partial_\rho v_k$. Then
$\dot v_{k+1}=q'(v_k)\dot v_k$ and
$q'(v_k)/(v_{k+1}/v_k)^2=1+O(v_k^2)$, so $\dot v_k\asymp_Jv_k^2$.
Divide each higher differentiated recurrence by $\dot v_{k+1}$.
Its leading term is $\partial_\rho^m v_k/\dot v_k$, and every other term contains at
least two derivatives of $v_k$. Induction bounds its increment by
$C_m\dot v_k$, a summable sequence. For a fixed reference length
$a_*>0$ and $1\le m\le4$, this gives
\begin{equation}
 |\partial_\rho^m(1-Z_\epsilon,y_\epsilon)|\le C_{J,m}t_\epsilon^2,\qquad
 t_\epsilon=[1+\log(a_*/\epsilon)]^{-1},\qquad
 \sum_kv_k^2<\infty.
 \label{ultraviolet-trajectory:bare-four-derivatives}
\end{equation}
The $C^6$ conjugacies and preparation inverse preserve the estimate.
The graph has two spare orders, so the complete fixed-window base
comparison holds through four derivatives. The cutoff fixes the integer
step count; it is never differentiated.
\end{proof}

\subsection{The compact physical normalization}
\label{arbitrary-cutoffs:exact-scalar-allcutoffs}
In the square $D_0=(-4a_0,4a_0)^2$ fix a nonzero real smooth $f$ supported in $|x|<a_0$. Write $A_f=\int f^2$ and
$E_f=\int|\nabla f|^2$. After integrating the modes up to $a_0$ in this square, let $F_{\epsilon,a_0}$ denote the
complete logarithm. We use its curvature on the sampled background $u f_\epsilon$ as
reference datum,
\[
 \mathcal A_\epsilon(\rho)
 =-\frac{a_0^2}{2\beta_*^2 A_f}
   \left.\frac{d^2}{du^2}F_{\epsilon,a_0}(uf_\epsilon;\rho)\right|_{u=0}.
\]
The connected-germ lower bound and the preceding comparison give a
$C^4$ limit and $C^3$ convergence, with uniform $C^4$ bounds. At zero,
\[
 \mathcal A(0)=0,\qquad
 \mathcal A'(0)=P_{\rm fr}/c_{\rm lat}+a_0^2E_f/(4\beta_*^2A_f)>0.
\]
Indeed the first-order kinetic and vertex Hessians are respectively
\[
 -\rho E_f/2\qquad\hbox{and}\qquad
 -2\beta_*^2P_{\rm fr}\rho A_f/(c_{\rm lat}a_0^2).
\]
The source clearance excludes a first-order wall contribution.
On compact positive intervals with $\mathcal A'>0$, the unique inverse
$\rho_\epsilon(a)$ satisfies $\mathcal A_\epsilon(\rho_\epsilon(a))=a$
exactly and converges in $C^3$ with uniform $C^4$ bounds. This same bare
pair is used in every target volume.

\subsection{Weak exits at fixed physical scales}

Choose an activity threshold $\eta_{\rm e}$ above the compact interval of endpoints, low
enough that even the $O(\eta_{\rm e}^2)$ overshoot lies strictly inside the fixed weak domain.
Write $v_k(\rho)$ for the activity of $f_0^kp_\rho$ at $a_k=b^ka_0$. The scale where the trajectory first
leaves the smaller weak region is given by
\begin{equation}
 K(\rho)=\min\{k\ge0:v_k(\rho)\ge\eta_{\rm e}\},
 \qquad a_{\rm exit}(\rho)=b^{K(\rho)}a_0.
 \label{ultraviolet:exit-definition}
\end{equation}
The first crossing can jump. We therefore differentiate the physical family at a
fixed output scale on each patch of a finite atlas.

\begin{proposition}[Fixed-scale exit atlas]
\label{ultraviolet:exit-atlas}
Each compact positive endpoint interval has a finite cover by patches
with fixed output lengths $a_k$, $0\le k\le K_*$. On each patch the
limiting output activity lies in one fixed enlargement of the exit
slab, still inside the weak domain. The same holds at every
sufficiently fine cutoff. The complete records at these outputs have
the preceding finite-window regularity and convergence. Their stable,
quadratic and even-wall components obey
\begin{equation}
 \|K_j\|\le C\eta_{\rm e}^3+C\theta_K^{j-1}t_\epsilon^2,
 \qquad \|W_j^\sharp\|\le C\eta_{\rm e}^2,
 \qquad \|B_j\|\le C\eta_{\rm e}.
 \label{ultraviolet:exit-records}
\end{equation}
Here $j$ counts steps from the actual microscopic preparation and
$t_\epsilon$ is defined in
\eqref{ultraviolet-trajectory:bare-four-derivatives}. The output
lengths and $a_{\rm exit}$ are uniformly comparable on the cover.
All boundary, covariance and scalar histories are retained.
For a complete record in a true box, its output precedes the
true-box terminal scale and its sources retain the prescribed
positive support clearances. The bulk exit length is independent
of that containing box.
\end{proposition}
\begin{proof}
In the strict outgoing cone the calibrated increments satisfy $c_*v_k^2\le v_{k+1}-v_k\le C_*v_k^2$. Telescoping
reciprocals from the compact positive entering interval up to $\eta_{\rm e}$ bounds the step
count by $K_*<\infty$, with overshoot $O(\eta_{\rm e}^2)$. Fix the iterate at an endpoint. By continuity it
stays in a slightly enlarged exit slab on a neighborhood of that endpoint.
Shrinking the neighborhoods to closed patches gives a finite subcover, and the
finite-window comparison puts sufficiently fine cutoffs in the same slab on each
patch. Forward iteration of the stable and wall equations gives \eqref{ultraviolet:exit-records} with the
homogeneous preparation term included. Fixed finite compositions preserve the
parameter bounds, with the exact intervening operation as transition on overlaps.
Transfer the cover to the prescribed datum by the compact physical inverse. The
bounds $a_0$ and $b^{K_*}a_0$ on all lengths give comparability without a derivative of the
first-crossing integer.
\end{proof}

\subsection{Smoothing of retained translations}
The field equation requires the part of a smooth translation retained after
ultraviolet integration. To determine its locality and physical normalization,
take a lattice rectangle with both physical sides at least $c_0r$ and split $G=A_D^{-1}=C+Q$ at
scale $r$ into prefix and tail. Then $R_C=QA_D=1-CA_D$ has range $O(r)$, and its field-coordinate
matrix satisfies the physical finite-difference bounds
\begin{equation}
 |\nabla_{\epsilon,x}^kR_C(x,y)|\le C_k\epsilon^2r^{-2-k},\qquad
 \sup_x\sum_y|\nabla_{\epsilon,x}^kR_C(x,y)|\le C_kr^{-k}.
 \label{retained-smoothing:eq-CG-25}
\end{equation}
Apply $A_D$ to the smooth retained kernel and use its range for the
row bound. Since $0\preceq Q\preceq G$,
$0\le\xi^\top A_DR_C\xi\le\xi^\top A_D\xi$.
The common kernel estimate also proves convergence on sampled smooth
tests, with support in a fixed enlargement. Constants may depend on
$c_0$.

\section{Renormalized observables and the continuum limit}
\label{sec:continuum-observables}
\subsection{The compatible source realization}

We now fix the finite normalization data and construct their actual
microscopic sources. The inverse is taken before restriction to a box.
This gives a common bulk prescription, while the exact integration
retains the response to each true Dirichlet wall. The main convergence
point is the scalar history: its local identity evaluation is kept
explicitly, and only its Taylor-subtracted remainder is estimated as an
ultraviolet tail.
\label{source-convergence:construction}

Choose operator references by dimension. If the physical exit lengths
lie in $[a_-,a_+]$ and the cone permits at most $T_D$ steps after the
dimension-$D$ block reaches its sourcewise weak section, use
$a_D=a_-b^{-(T_D+2)}$. The orbit is still in that section at $a_D$.
Set the new principal block there to the identity and its lower
coefficients to zero after the free Wick conversion; transport earlier
columns without redefining them. Positive partial shells handle references
between master scales. The finite triangular contact correction imposes
the fixed $\chi_{\rm ct}$ convention. Every finite list has a sufficiently
fine cutoff threshold resolving its finitely many references.

Let $\iota_{\epsilon,I}$ be the microscopic density embedding with zero
ordinary stable slot before preparation. Its coefficients are
\begin{equation}
 [\iota_{\epsilon,I}(c)[J]]_S=
 \epsilon^2\sum_{x,\nu}c^\epsilon_{S,\nu}(x)e^{iq_\nu\beta_*\phi_x}
 P_\nu((\nabla_\epsilon^\alpha\phi)_x)
                \prod_{i\in S}(\nabla_\epsilon^{\gamma_{i,\nu}}J_i)_x.
 \label{source-convergence:bare-local}
\end{equation}
The templates, anchors and stencils fix the displayed spatial dependence;
these are finite ordinary coefficient blocks, not arbitrary site arrays.
If $P_{R,0}^\epsilon$ is their exact ordered transport to the references,
including preparation and partial shells, and $\mathcal C_{R,I}$ their
fixed targets, define
\begin{equation}
 B_{\epsilon,I}=\iota_{\epsilon,I}(P_{R,0}^\epsilon)^{-1}\mathcal C_{R,I}.
 \label{source-convergence:injection}
\end{equation}

\begin{theorem}[Complete local realization and convergence]
\label{source-convergence:local-injection}
\label{source-convergence:ordinary-left-inverse}
\label{source-convergence:cutoff}\label{source-convergence:controlled-source-transfer}
The source \eqref{source-convergence:injection} is local, compatible
under label restriction and earlier-block inclusion, and preserves the
declared conjugations and reflections. Its ordinary readout is exactly
$\mathcal C_{R,I}$. Restrict this same source to every true box containing
its stencil enlargement. At each fixed positive weak resolution its
complete local, stable, wall, scalar and contact record has uniform
$C^4$ bounds and converges in $C^3$ on compact positive interior parameter
intervals, along arbitrary cutoffs. The same conclusion applies to any
compatible local injection with the backward, reference-data and contact
bounds used below; an initial independent scalar must have its own
convergent modulus.
\end{theorem}
\begin{proof}
\emph{Local inversion and its derivatives.} Solve the ordinary inverse
subset by subset. It changes only the coefficients of the original
embedding, so preserves locality and exact symmetries; uniqueness gives
restriction compatibility and the readout identity. On the canonical
base, for $1\le m\le4$,
\begin{equation}
                 \|\partial_\rho^mX_\epsilon(a_j)\|\le C_{I,m}v_j^2.
 \label{source-convergence:source-base-jet}
\end{equation}
Differentiate the finite graph equations using their stable gap and two
spare orders. The inverse-substitution matrices therefore satisfy
\begin{equation}
 \|U_j\|\le L_I,\qquad \|\partial_\rho^mU_j\|\le C_{I,m}v_j^2,\qquad
 \|\partial_\rho^m(U_j\cdots U_{n-1})\|\le C_{I,m}L_I^{n-j}.
 \label{source-convergence:local-inverse-jets}
\end{equation}
\label{source-convergence:source-product-jets}
The ordered-product bound follows from $\sum_jv_j^2<\infty$.
Taking up to four parameter derivatives leaves the scale exponent
of the inverse transport unchanged.

\emph{Exact cancellation of the free normalization.}
\label{source-convergence:gaussian-source-conversion}
Write $C=A_D^{-1}$, $s=1-Z$, $\Gamma=C-Q_r$, and
$C_H=\int_0^{(\epsilon\mu_{\rm ref})^{-2}}e^{-tA_D}dt=C-Q_H$ in field
coordinates. For a finite vector of field/derivative jets let
$\mathcal W_B=\exp(-B_{ab}\partial_a\partial_b/2)$; on a polynomial
times a vertex its polynomial part is finite. Complete the exact kinetic
tilt $w_s=e^{s(\phi+\zeta)^\top A_D(\phi+\zeta)/2}$. The mean of
$\phi+\zeta$ under the tilt is $M\phi$, and its covariance is $K$, where
\[
 M=(1-s\Gamma A_D)^{-1},\qquad
 K=\Gamma^{1/2}(1-s\Gamma^{1/2}A_D\Gamma^{1/2})^{-1}\Gamma^{1/2}.
\]
With these two quantities the normalization becomes
\begin{equation}
 \begin{aligned}
 \frac{\E_\Gamma[w_s(\mathcal W_{C_H/Z}F)(\phi+\zeta)]}{\E_\Gamma w_s}
 &= (\mathcal W_R F)(M\phi),\\
 R&=Z^{-1}\{Q_r(Z+sA_DQ_r)^{-1}-Q_H\}.
 \end{aligned}
 \label{source-convergence:gaussian-normalized-conversion}
\end{equation}
The identity follows from $(C/Z-K)(1-sA_D\Gamma)=Q_r/Z$ and does
not assume commuting covariance matrices. The bulk heat convention adds
$(C_H^\infty-C_H)/Z$ to $R$. This wall difference is smooth at positive
clearance. Indeed a shifted Fourier contour gives the lattice heat bounds
$C_mt^{-1-m/2}e^{-cd/\sqrt t}$ for $t\ge\epsilon^2$ and
$C_m\epsilon^{-2-m}e^{-cd/\epsilon}$ below it; the nontrivial rectangular
images have distance at least $cd$. Integration gives
$C_me^{-c_md/\sqrt T}$ for a finite heat horizon $T$ and $d\ge\sqrt T$.
At fixed clearance the same calculation is finite. In kinetic resolvent
differences, a path reaching the wall needs $cd/r$ finite-range factors;
the small-$s$ Neumann series gives the corresponding exponential tail.
Thus $R$, the smooth part of $M$ and their finite inverses have four
bounded parameter derivatives at fixed physical reference scales.
For several primitive labels use block-diagonal input Wick covariances;
the joint $K$ cancels self-contractions and retains the off-diagonal
contact contractions. Determinants cancel in the quotient before
differentiation. At nonzero interaction the Wick change is still an exact
source-chart identity, so complete integration cancels it with the
ordinary normalization inverse.

\label{source-convergence:protected-derivative}
For a derivative singleton, \eqref{finite-jet-calculus:linearshift}
returns the literal linear source plus $DW_C[Ca]$, $a=-\operatorname{div}f$.
The high covariance is analytic at zero momentum: a neutral correction
has at least two source derivatives and parity supplies a field derivative;
charged corrections already exceed dimension one. The principal coefficient
is therefore one. The unique ordinary inverse returns the literal
derivative, transporting the other responses forward.

\emph{Wall terms and scalar history.}
\label{source-convergence:wall}
Compare the records $X^D_n$ (true box) and $Y_n$ (bulk) for one bare source. In
top degree write $M_j^D,M_j^{\rm bulk}$ for the respective one-step maps and $M^D_{n,j+1}$ for ordered
propagation from $j+1$ to $n$. Telescoping gives
\[
 X^D_n-Y_n=\sum_{j<n}M^D_{n,j+1}(M_j^D-M_j^{\rm bulk})Y_j,
\]
with proper-subset forcing treated inductively. Every term has a source
mark and a wall distinction; its support history joins them. Hence
\begin{equation}
 C_{I,r}a_j^{-P_I}e^{-\kappa d(D)/a_j}\prod_i\|J_i\|_{C^{R_I}}
 \label{source-convergence:wall-shell}
\end{equation}
bounds its shell contribution. Summing gives $C_{I,r}e^{-\kappa d(D)/(2r)}$
for $d(D)\ge r$; shorter positive clearances leave finitely many later
steps. Four derivatives retain the same power by
\eqref{source-convergence:local-inverse-jets}.

The ordinary prefix cancels in its original order:
\begin{equation}
 P_{n,0}P_{R,0}^{-1}=
 \begin{cases}P_{n,R},&n\ge R,\\P_{R,n}^{-1},&n<R.
 \end{cases}
 \label{source-convergence:ordered-normalization-cancellation}
\end{equation}
For the complete logarithm retain
\begin{equation}
 \begin{gathered}
 F_j(J)-F_j(0)=C_j[J]+Q_j(\phi;J),\\
 E_j[J]=Q_j(0;J),\qquad\mathcal T_j=Q_j-E_j,\qquad\mathscr A_j=C_j+E_j.
 \end{gathered}
 \label{source-convergence:source-zero-evaluation}
\end{equation}
Thus $F_j(0)$ sets the source to zero, whereas $E_j$ sets the field
to zero and keeps the source. The tilted operation and its scalar telescope are
\begin{align}
 \mathcal H_j(Q_j)&=Q_{j+1}+s_j,\quad C_{j+1}=C_j+s_j,
 \label{source-convergence:source-tilted-scalar-map}\\
 \mathscr A_n&=E_n+\sum_{j<n}s_j.
 \label{source-convergence:source-scalar-coboundary}
\end{align}
Before preparation $C_0=0$. Assign the local identity and contact evaluations to $E_j$
and the Taylor-subtracted scalar to $s_j$. Each vertex of the contact forest
satisfies
\begin{equation}
 |\partial_\rho^k\mathcal K_{j,S,\mathcal O}(f-T_\omega f)|
 \le C_{I,k}a_j^{1-a_I}\|f\|_{C^{R_I}},\qquad0\le k\le4.
 \label{source-convergence:differentiated-contact-gain}
\end{equation}
The dimension-two parameter insertions have at most polynomially many placements
in the number of blocks. The exponential block weight sums them. With the contact
cutoff fixed, the scalar sum below $\delta$ is $O(\delta^{\eta_I})$ in $C^4$ plus the wall tail. The local
evaluation $E_j$ is excluded from this tail bound.

\label{source-convergence:positive-local-core}
On a small fixed complex field neighborhood $|W_B|<1/2$ and
$|\log I_B|\le C\eta$. The connected criterion gives
$U_j\ge e^{-C N_r(D)}>0$ on the real neighborhood, uniformly in cutoff,
where $N_r(D)$ counts the $r$-blocks meeting $D$.
Finite logarithmic jets and $E_n$ are consequently controlled there.
No condition $\eta M_{D,r}\ll1$ is required.

\emph{Cutoff passage.} Let $t_\epsilon=[1+\log(a_*/\epsilon)]^{-1}$,
$u_\epsilon=t_\epsilon(1+\log(1/t_\epsilon))^2$.
Above $\delta$ the normalized prefix is a finite physical window.
Write $e_\epsilon(\delta)=u_\epsilon+(\epsilon/\delta)^{\nu_I}$ for
the two cutoff-dependent errors. Telescoping the common operation
and applying the exact reference conversion gives
\[
 \|X^D_\epsilon-X^D_{\epsilon'}\|_{C^0(J_1)}
 \le C\{\delta^{\eta_I}+\delta^{-P_I}
 [e_\epsilon(\delta)+e_{\epsilon'}(\delta)+e^{-c_Id(D)/\delta}]\}.
\]
The low scalar tail was proved independently; its increments on the
window depend on $Q_j$, not old $C_j$, so this argument is noncircular.
Choose $\delta$ to balance the algebraic terms. The result is a modulus
$d_{\epsilon,\epsilon'}\to0$ with uniform $C^4(J_1)$ bounds. On
$J\Subset J_1$, interpolation gives $C^k$ error
$Cd_{\epsilon,\epsilon'}^{1-k/4}$ for $k\le3$, in particular the
square-root $C^2$ rate. Exact scalar calibration composes with the
uniform $C^4$ inverse $\rho_\epsilon(a)$ and preserves this conclusion.
An alternative injection with the stated hypotheses follows the same
proof, adding its actual reference and initial-scalar moduli.
\end{proof}

\subsection{Finite metric sources}
\label{metric-source-prescription:renormalized-metric-completion}
At one weak physical reference resolve the identity, the kinetic tensor
and the dimension-two fundamental cosine. We write the latter two as
\[
 \mathcal K_{\mu\nu}=[\partial_\mu\phi\,\partial_\nu\phi]_R,\qquad
 V_{\cos}=\tfrac12\mu_{\rm ref}^2(V_1+V_{-1}).
\]
Project the complete limiting flat action at the fixed reference to obtain $Z_R(\rho)$ and
$\lambda_c(\rho)$, with $Z_R(0)=1$ and $\lambda_c(0)=0$. Set the scalar coefficient $e=0$. For $C^4$ regularity use the
bound $C\theta^m+C\sum_{l<m}\theta^{m-1-l}d_{\epsilon,\epsilon'}(aB^{l-m})$ on the ordinary four-jet graph difference, taking the cutoff limit at
fixed $m$ before sending $m\to\infty$. The finite reference windows and their exact Wick
conversions preserve this regularity independently of the $C^3$ theorem for the
complete source.

For $g=\delta+k$ with compact interior support let $q(k)$ be the finite
Taylor jet of
\[
 -\int\!\left\{\frac{Z_R}{2}
 [\sqrt{\det g}\,g^{\mu\nu}-\delta^{\mu\nu}]\mathcal K_{\mu\nu}
 +\lambda_c[\sqrt{\det g}-1]V_{\cos}\right\}d^2x.
\]
Then the retained metric prescription is
\begin{equation}
                  B^{\rm met}_\epsilon[k,J]=B_\epsilon[J+q(k)].
 \label{metric-source-prescription:metric}
\end{equation}
Substitution in the finite source algebra gives locality, all mixed contacts and
$C^3$ convergence as the cutoff is removed. The geometric action itself contains $Z\epsilon^2\sum_x\sqrt g\,g^{\mu\nu}\nabla_\mu\phi\nabla_\nu\phi/2$
and $\sum_x\sqrt g[2y(1-\cos\beta_*\phi)+E_\epsilon\epsilon^2]$. To realize \eqref{metric-source-prescription:metric}, add to the log weight the local counterterm
\[
 \mathcal C^{\rm met}_\epsilon[k,J]
 =B_\epsilon[J+q(k)]-B_\epsilon[J]
             +S_\epsilon[\delta+k]-S_\epsilon[\delta].
\]
This vanishes for $k=0$. Stress and Ward identities also require the geometric conversion, including determinants and contacts. For parameter and geometric variations
differentiate the complete action, source realization and scalar normalization,
inserting $\tfrac12\dot C:D^2$ when the covariance varies.

\subsection{Fixed-volume integration of the complete source}

\begin{lemma}[Positive reweighting]
\label{positive-cosine-laws:positive-stability}
Let $\mu_n\Rightarrow\mu$ on a separable Banach space, and let continuous
$U_{n,a}>0$ converge, uniformly on norm balls and compact parameter sets,
to $U_a\ge0$, continuously in $a$. Suppose the normalized laws
$\nu_{n,a}=U_{n,a}\mu_n/Z_{n,a}$ are uniformly tight, the densities are
uniformly bounded on norm balls, and $U_{n,a}\ge c>0$ on one open ball
of positive $\mu$-mass. Then $Z_{n,a}\to\int U_a\,d\mu>0$ uniformly,
and $\nu_{n,a}$ converge uniformly weakly. Uniform integrability of
their $p$th norm powers gives $W_p$ convergence. The limits depend
continuously on $a$ in total variation, and in $W_p$ under that moment
condition.
\end{lemma}
\begin{proof}
Portmanteau gives the lower bound on $Z_{n,a}$. Put at least half the
normalized mass in a fixed norm ball; its density bound gives the upper
bound. Tightness now gives uniform tails for the unnormalized integrals,
since $\int_{B_R^c}U_{n,a}d\mu_n=Z_{n,a}\nu_{n,a}(B_R^c)$.
On a Skorokhod coupling \cite[Theorem~6.7]{Billingsley1999} the fields converge almost surely and are
eventually bounded; uniform convergence on balls gives convergence of
the density values, and the tail bound gives $L^1$ convergence. Compactness
of the density family's image gives parameter uniformity. The moment
criterion \cite[Theorem~7.12]{Villani2003} gives $W_p$ convergence. With the reference fixed to $\mu$,
the same $L^1$ argument gives total-variation continuity in $a$.
\end{proof}

\begin{theorem}[Physical fixed-volume source limit]
\label{fixed-volume-integration:fixed-box}
In each true Dirichlet rectangle, the complete normalized sources and
connected coefficients have an arbitrary-cutoff $C^3$ limit, with fixed
contacts and the preceding $C^2$ rate. All finite reflected Gram matrices
converge and remain positive. The actual retained source-zero laws
converge in every finite $W_p$ on each fixed $C^R$ space, and their
normalizers stay bounded above and away from zero. Constants may depend
on the box and positive physical resolution.
\end{theorem}
\begin{proof}
Take a weak resolution $a$ below all the reference and clearance scales of the
finite list. Include the finite H\"older moment order $p>1$ in the proof norm,
keeping the bare family and its normalization fixed. The stronger canonical
preparation estimate permits smaller regulator coefficients. Choose these first,
then enlarge the grouped block factor and reduce the weak radius. By the scaled
trace and Sobolev bounds, $\log G$ is at most $\psi^\top T\psi/2$. The retained multiplier $q_a(\lambda)\le C_M\lambda^{-1}(1+a^2\lambda)^{-M}$ gives
\[
 \|T^{1/2}Q_aT^{1/2}\|\le C\kappa,\qquad
 \Tr(T^{1/2}Q_aT^{1/2})\le C\kappa|X|_a.
\]
These follow from the bounded multipliers
$u^{k-1}(1+u)^{-M}$, bounded overlap and the odd extension. With
$pC\kappa<1$, \eqref{positive-cosine-laws:regulator-determinant} gives
$\E_\nu G(X)^p\le e^{c_p|X|_a}$ uniformly also under added centered
precision. Taking $c_p/p$ smaller than the unused size decay makes Minkowski's
inequality applicable to every anchored representative, difference and
controlled derivative, with the original test seminorms.

Write the exact finite circle-product kernel as $e^{S_\epsilon}\mathbf U_\epsilon$, so only the unmarked
scalar is factored out. All marked scalars belong to $\mathbf U$. At fixed box and
resolution, partition the degrees among the finitely many blocks and merge
their regulators. For $0\le k\le3$ this gives
$|\partial_\rho^k U_{\epsilon,S}(\psi)|\le
C_{D,I,a}\mathcal G_{D,a}(\psi)\prod_i\|f_i\|_{C^{R_I}}$,
where $\mathcal G$ has a common Gaussian $p$th moment by the same
trace estimate. At this fixed box and scale write
$N_{\epsilon,S}=\E_{Q_\epsilon(a)}U_{\epsilon,S}$ for each label set. To bound $N_{\epsilon,\varnothing}$ from below, integrate the positive
bounded-field core over its small support neighborhood, which has positive
centered Gaussian probability. The common envelope gives the upper bound and
uniform integrability. Finite reconstruction passes complete-record
convergence to the kernel. To compare the cutoff covariances use
\[
 (\E_Q-\E_{Q'})F=\tfrac12\int_0^1
 \E_{(1-t)Q'+tQ}[D^2F:(Q-Q')]dt.
\]
Prove the identity first on finite sine projections with the common
quadratic Sobolev envelope. The projective covariance norm and dominated
convergence allow passage to the full field. The reference covariance is independent
of $\rho$, so this also proves convergence through three parameter
derivatives. Differentiating a varying covariance $r$ times requires
bounds on $2r$ additional field derivatives. The normalized functional is
\[
 \mathcal Z_\epsilon[\eta]
       =\frac{\sum_S\eta_S N_{\epsilon,S}}{N_{\epsilon,\varnothing}}.
\]
Finite inversion and logarithm preserve the full contact algebra and
the $C^3$ estimates. Applying them to all entries of a reflected list
passes reflection positivity to the limit. Positive clearance prevents
subtraction stencils crossing the reflection plane.

For the actual retained law combine the conditional proxy \eqref{positive-cosine-laws:retained-domination} with smoothness of
the retained kernel to obtain $\E e^{c_s\|\psi\|_{H^s}^2}\le C_{D,a,s}$ at each fixed $s$ and sufficiently small $c_s>0$.
Compact Sobolev embedding gives tightness and all finite norm moments. Density
convergence on bounded fields and the positive core thus give the remaining
fixed-volume assertions by the reweighting lemma. Volume-uniform reflected norms
follow later from source admission.
\end{proof}

\subsection{Continuous physical dilations}

\begin{proposition}[Recognition and dilation action]
\label{canonical-dilations:recognition}\label{canonical-dilations:action}
Physical scaling $h'(x)=h(\lambda x)$ carries the source-free continuum
family at $\rho$ to the canonical family at $T_\lambda\rho$. Its complete
normalized retained functionals and actual generated histories agree.
The maps form a continuous local action
$T_1=1$, $T_\lambda T_\mu=T_{\lambda\mu}$, with $T_b=F$, where $F$
is the activity map on the ultraviolet curve. There are no positive weak
fixed points. Every compact parameter interval is one orbit segment,
with a continuous compact positive range of dilation factors. At fixed
$\lambda$, the endpoint map is at least $C^2$ in $\rho$.
\end{proposition}
\begin{proof}
Use an Abel coordinate for the nonlinear activity map to measure the finite clock
shift from the quadratic rescaling error in the microscopic preparation.
For $F(v)=v+\ell_bv^2+av^3+O(v^4)$ set $\Phi_0(v)=-1/(\ell_bv)+(1-a/\ell_b^2)\log v$ and $r(v)=\Phi_0(F(v))-\Phi_0(v)-1=O(v^2)$. In the series $\Phi(v)=\Phi_0(v)-\sum_{l\ge1}r(F^{-l}v)$ the inverse-iterate bounds give
convergence with two derivatives on positive compact intervals. It obeys $\Phi(F(v))=\Phi(v)+1$, $\Phi'=1/(\ell_bv^2)+O(v^{-1})>0$.
In particular, the differentiated tail has bound $Cv^{-2}\sum_l(F^{-l}v)^3=O(1)$.

In coordinates $u=x-\mathfrak h(v)$ the full limiting map gives
$v^+=F(v)+O(v|u|)$,
$u^+=(1-\ell_bv+O(v^2+|u|))u$.
The cone $|u|\le Kv^2$ is invariant at small radius, and
$|u_j|v_j\le C|u_0|v_0$. Therefore
$|\Phi(v_j)-\Phi(v_0)-j|\le C|u_0|/v_0^2$:
the error per step is $C|u_j|/v_j$ and
$\sum_jv_j^{-2}\le Cv_0^{-3}$. A quadratic initial mismatch thus
produces a bounded clock displacement and a vanishing terminal normal
coordinate. Read the endpoint sufficiently far into the ultraviolet
that this entire bounded clock interval stays weak.

Use the same bare pair at $\epsilon' =\epsilon/\lambda$. The new initial point of the autonomous map differs
by $O(v_0^2)$ because both preparation Jacobians equal $\operatorname{diag}(2,c_{\rm lat})$ and the conjugacies have
identity linear jets. On compact dilation intervals the step counts differ by a
bounded integer. Hence all subsequential endpoints lie on the AF curve. Graph
comparison and the retarded boundary history determine their complete canonical
records. The lattice covariance scales exactly,
\[
 \frac{C_{\epsilon/\lambda,D/\lambda}[a/\lambda,d/\lambda]}
 { (\epsilon/\lambda)^2}
       =\frac{C_{\epsilon,D}[a,d]}{\epsilon^2}.
\]
The rescaled high prefix therefore has a unique full limit. The injective
compact reference determines its endpoint $T_\lambda\rho$, with $C^2$ parameter regularity
by smooth inversion. Lattice scaling composes exactly and $\lambda=b$ adds one step
without changing microscopic phase, giving $T_b=F$.

To get continuity in $\lambda$ directly in the Abel coordinate, extend $D_t=\Phi T_{e^t}\Phi^{-1}$ from a
negative half-line by $D_t(q+1)=D_t(q)+1$. These homeomorphisms are increasing because $D_t=D_{t/2}^2$, and
the clock bound gives local boundedness in $t$. The integer-part almost-additivity
bounds give the translation number $\tau(D)=\lim_n(D^n(q)-q)/n$ with error at most two. Its additivity for
commuting maps yields $\tau(D_t)=t/\log b$. For $t>0$ there can be no fixed point and $D_t(q)>q$. Each orbit
increases strictly with $t$. A jump would then propagate to every time by the
group law, but a monotone function has only countably many discontinuities.
Two-sided monotone bounds yield joint continuity. Finally $D_{\log b}(q)=q+1$ implies that every
orbit covers the Abel axis.
\end{proof}

\section{Normalized transfer and response theorems}
\label{sec:transfer}
\label{normalized-source-theorems:reusable-section}

To bound a normalized source one must compare its numerator with the
partition function. For a complete jet, one comparison covers ordinary
insertions, nonlinear contacts and differences of source constructions.
The theorem uses only finite integral identities and estimates, verified
below for the original positive sine--Gordon laws.

We label each occurrence separately, including repetitions. At fixed
complete list and local test geometry the constants are uniform in the
containing domain and all sufficiently fine cutoffs. The cutoff
threshold may depend on the domain.

\begin{theorem}[Normalized transfer and source admission]
\label{normalized-source-theorems:reusable-normalized-transfer}
Suppose that for each $\alpha=(\epsilon,D,\rho)$ the finite positive
law has a conditional representation with integration measure $\mu$.
Write $\mu(f)=\int f\,d\mu$ and use the $L^2(\mu)$ norm. At fixed
$\alpha$ the conditional amplitudes $u_F$ and $v$ satisfy
\begin{equation}
 \langle F\rangle_\alpha
  =\frac{\mu(u_Fv)}{\mu(uv)},
 \qquad u_1=u>0,\quad v>0,
 \label{normalized-source-theorems:reuse-factorization}
\end{equation}
where $F\mapsto u_F$ is linear. Any quotient used to obtain
this formula must preserve the complete insertion. Let
$R_S=[\eta_S\bar\eta_S]\mathcal R_\alpha$ be the coefficient
with the labels $S$ in both replicas. Assume
\begin{align}
 p_\alpha&:=\frac{\mu(uv)^2}{\|u\|_2^2\,\|v\|_2^2}
 \ge p_*>0,\nonumber\\
 Z_\alpha^{\rm rep}&:=\|u\|_2^2
 \ge z_*e^{E_\alpha},                                      \label{normalized-source-theorems:reuse-affinity}\\
 \|u_{F_S}\|_2^2
 &=e^{E_\alpha}\int R_S\,d\gamma_{Q_\alpha},                 \label{normalized-source-theorems:reuse-replica-record}\\
 \int |R_S|\,d\gamma_{Q_\alpha}
 &\le C_S\prod_{i\in A}d_{i,\alpha}^2
                  \prod_{i\in S\setminus A}M_{i,\alpha}^2.  \label{normalized-source-theorems:reuse-record-bound}
\end{align}
Here $A\subseteq S$ is the set of error labels. The replica lower
bound and the marked identity use the same unmarked scalar $E_\alpha$.
Every source-bearing scalar remains in $\mathcal R_\alpha$.
The hypotheses apply to all needed sublists,
mixed switches and reflected copies. Then
\begin{equation}
 |\langle F_S\rangle_\alpha|
 \le\sqrt{C_S/(p_*z_*)}\prod_{i\in A}d_{i,\alpha}
                         \prod_{i\in S\setminus A}M_{i,\alpha}.
 \label{normalized-source-theorems:reuse-normalized-bound}
\end{equation}
The estimate survives bounded local substitutions of complete jets
that preserve the stated error grading. For a reflection-positive law,
two switches $E=F-P$ and $\Theta E$ in separated time halves give
\begin{equation}
                 0\le\langle\Theta E\,E\rangle_\alpha
                         \le C d_\alpha^2.
 \label{normalized-source-theorems:reuse-own-gram}
\end{equation}

\label{normalized-source-theorems:reusable-source-admission}
Suppose a microscopic core has numerical moment and reflected-Gram
Cauchy bounds uniformly on each stipulated bounded test family, first as $\epsilon\downarrow0$ in each fixed $D$, then
along the selected exhaustion. If each complete coefficient $F$ has
core comparators $P_r(F)$ chosen before $D,\epsilon$, with
\begin{equation}
 d_{r,\epsilon,D}\le C_Ir^{\nu_I}+\omega_{r,D}(\epsilon),
 \qquad \nu_I>0,\quad\omega_{r,D}(\epsilon)\longrightarrow0,
 \label{normalized-source-theorems:reuse-approximation-modulus}
\end{equation}
then its complete values and mixed Grams have unique cutoff-first,
selected-volume limits. The completed source OS quotient embeds canonically in
the completed core quotient, with
$\|[F]-[P_r(F)]\|\le C_Ir^{\nu_I}$. The embedding is onto exactly
when these comparator limits contain a dense core. Writing $M_{D,r}$
for the cutoff limit of $\langle P_r(F)\rangle_D$, one has
\begin{equation}
 |\langle F\rangle_D-\langle F\rangle_\infty|
 \le2C_Ir^{\nu_I}+|M_{D,r}-M_{\infty,r}|.
 \label{normalized-source-theorems:reuse-source-volume-modulus}
\end{equation}
All assertions hold in the stipulated bounded-test seminorms.
\end{theorem}
\begin{proof}
Conditional Cauchy--Schwarz gives $|\langle F\rangle_\alpha|^2\le
p_\alpha^{-1}\|u_F\|_2^2/Z_\alpha^{\rm rep}$. Cancellation of the common scalar
yields the estimate. For the complete two-switch coefficient this is the
Gram bound, whose sign follows by reflection positivity. The coefficient
chain rule for a finite substitution is a finite sum, so the estimate
extends to each resulting contact column.

At two cutoffs with $r,D$ fixed, comparison through $P_r(F)$ bounds the
difference of the source values by two approximation errors and the core
Cauchy error. Let both cutoffs tend to zero and then let $r$ tend to
zero. Comparing two domains after cutoff removal gives the second Cauchy
bound and the displayed modulus. In the completed core quotient, $P_r-P_s=(P_r-F)+(F-P_s)$ and
the Gram estimate imply that $[P_r(F)]$ is Cauchy. Polarizing the mixed-switch bounds identifies its limiting inner products with the complete source Grams. The isometry therefore has
the stated density criterion and is independent of the comparators.
\end{proof}

\subsection{Finite certificates for the hypotheses}
\label{normalized-source-theorems:reuse-affinity-certificate}
For the three replica partitions $Z=(Z_{II},Z_{IE},Z_{EE})$ put $p(Z)=Z_{IE}^2/(Z_{II}Z_{EE})$. Suppose restoration of the
collar gives $Z_s^{\rm full}=v_sr_sZ_s^{\rm del}$ for each replica $s$. The bounds $p(Z^{\rm del})\ge c_0$, $e^{-L}\le r_s\le1$ and $|2\log v_{IE}-\log v_{II}-\log v_{EE}|\le L'$ imply $p(Z^{\rm full})\ge c_0e^{-2L-L'}$,
to be multiplied by any further exact positive comparison from folding.
Convergence of the individual $v_s$ is unnecessary. For a Gaussian
regulator,
\begin{equation}
 \int e^{c\|L\psi\|^2}d\gamma_Q
 =\det(1-2cLQL^*)^{-1/2}
 \le\exp\{cKN/(1-2cK)\}
 \label{normalized-source-theorems:reuse-gaussian-moment}
\end{equation}
whenever $\|LQL^*\|\le K$, $\operatorname{Tr}LQL^*\le KN$ and $2cK<1$. Diagonalization works also for
singular $Q$.

Lemma~\ref{connected-expansions:unified-tree-count} gives \eqref{normalized-source-theorems:reuse-record-bound} using the local block-Cauchy factorials, summable edge
rows and exact temporary-label factor, together with a joint bound on
the actual unmarked reconstruction and the merged Gaussian regulators.
Each replica contributes one copy of every prescribed source or error
weight.

\subsection{From local comparison to extensive response}

Differentiating a parameter inserts the action throughout the volume. To
sum the resulting connected components, we anchor each at an observable
and use quantitative factorization of separated packets. For
sine--Gordon the required clustering comes from screening, after the
proof of the physical mass.

\begin{theorem}[Anchored responses]
\label{normalized-source-theorems:reusable-response}
Fix $q_0\ge2$ and $J_0\Subset J_1$. Suppose a physical partition of
unity localizes the actual $C^{q_0}$ parameter family into labels
$t_b$, whose uniform diagonal is exactly $\rho\mapsto\rho+t$,
including boundary bonds. Assume complete source/action coefficients
through $q_0$ action labels satisfy the preceding packet bounds and
are directly Cauchy at fixed box through that order, locally uniformly
in $\rho$, with continuous limits. For two separated groups of
clipped core comparators at approximation scale $r$ and clipping
height $K$, let $d$ be the distance between the groups. Assume, after cutoff removal,
\begin{equation}
 |\operatorname{Cov}_D(P_{A,r,K},P_{B,r,K})|
 \le Cr^{-P}(1+K)^Pe^{-md},
 \label{normalized-source-theorems:reuse-core-cluster}
\end{equation}
uniformly in selected domains, with packet error $Cr^\nu$ and
clipping error $Cr^{-P}e^{-cK^2}$, where $m,c,\nu>0$.
Then every fixed complete observable bundle $F_D$ has a
domain-uniform $C^{q_0}(J_1)$ bound. If
$\|F_D-F_\infty\|_{J_1}\le d_D\to0$, then
\begin{equation}
 \|\partial_\rho^qF_D-\partial_\rho^qF_\infty\|_{J_0}
 \le C_qd_D^{1-q/q_0},\qquad 1\le q<q_0.
 \label{normalized-source-theorems:reuse-response-rate}
\end{equation}
No convergence of the selected $q_0$th derivative is asserted.
\end{theorem}
\begin{proof}
The factorization error is
$C[r^\nu+r^{-P}(1+K)^Pe^{-md}+r^{-P}e^{-cK^2}]$.
Choose $r=r_0e^{-\zeta d}$ with $\zeta P<m/2$ and $K^2=Md$
with $cM>\zeta P+1$. This is $Ce^{-m'd}$. For a fixed number
$n$ of packets, removing a longest edge of a minimum spanning tree
separates two groups by at least their diameter divided by $n-1$,
up to fixed region widths. The mixed cumulant of the product of
their group laws is zero. Telescoping the finite moment--cumulant
formula therefore gives $|\kappa_D(O_1,\ldots,O_n)|\le
C_ne^{-c_n\operatorname{diam}(O_1,\ldots,O_n)}$.
Its sum over all but one packet position converges uniformly;
coincident contacts are merged into one region.

For $B\subseteq[q]=\{1,\ldots,q\}$ write
$\partial_B=\prod_{i\in B}\partial_{t_i}$, with
$\partial_\varnothing$ the identity. For nonempty $B$ set
$A_B=-\partial_B S$. Differentiation of normalized integrals gives
\begin{equation}
 \partial_{[q]}\langle O\rangle
 =\sum_{S_0\subseteq[q]}
   \sum_{\pi\in\operatorname{Part}([q]\setminus S_0)}
   \kappa\bigl(\partial_{S_0}O,(A_B)_{B\in\pi}\bigr).
 \label{normalized-source-theorems:reuse-polarized-response}
\end{equation}
Each new derivative either differentiates an argument or appends a
score, which proves the identity by induction. The normalized generating
logarithm cancels the parameter-only logarithm. Each surviving component
thus has an observable label at which to anchor the sum of the
exponential cumulant bound. The sum is uniform in the volume and
includes every contact and source-bearing scalar. At fixed box, direct
Cauchy convergence and repeated application of the fundamental theorem
of calculus identify the classical derivatives through order $q_0$.

For $f=F_D-F_{D'}$, an interior finite-difference stencil and Taylor's
integral remainder give
$\|f^{(q)}\|\le C_q(ds^{-q}+Ks^{q_0-q})$ when
$\|f\|\le d$ and $\|f^{(q_0)}\|\le K$.
Take $s=d^{1/q_0}$ within the fixed interval clearance. The resulting
Cauchy limits are derivatives by the same integral identities, proving
the stated rate.
\end{proof}

For $q_0=4$, set $H_D=F_D-F_\infty$. In the application below,
$\mathfrak b$ generates dilations of the endpoint and $\mathscr D_\rho$
generates their action on the sources. Suppose the domain cocycle
independently gives
\[
\begin{aligned}
 B_{D,1}F_D&=\mathfrak b H_D'-\mathscr D_\rho H_D,\\
 B_{D,2}F_D&=\mathfrak b^2H_D''-L_\rho H_D,\qquad
 L_\rho=\mathscr D_\rho^2+\mathfrak b\,\partial_\rho\mathscr D_\rho
                              -\mathfrak b'\mathscr D_\rho .
\end{aligned}
\]
The two domain responses are then bounded by $C(d_D^{3/4}+d_D)$
and $C(d_D^{1/2}+d_D)$, respectively. The local metric identities
require the separate geometric construction below.

\section{Pinning and the selected state}
\label{sec:pinning}
\label{physical-theory:section}
\label{physical-theory:pinning-selection-section}

The bounded-domain ultraviolet law has real height $h=\beta_*\phi$: period $2\pi$ belongs
to the interaction, while $h$ remains noncompact as the Dirichlet wall
recedes. Finite pinning bounds moments of block averages uniformly. These
determine the selected law and its cyclic Osterwalder--Schrader space. A
separate contour estimate gives the axial summability that excludes a
zero-energy atom in screening.

Throughout, the endpoints lie in a fixed compact positive interval. We
remove the cutoff in each fixed zero-Dirichlet rectangle before taking a
cofinal exhaustion by these rectangles. Unless specified, constants may
depend on the fixed source list, test seminorms and positive support
clearances and are uniform in the containing rectangle and sufficiently
fine cutoff. We use the strict contraction, Gaussian regulator and
support reserves of Proposition~\ref{local-source-operation:graph}, Theorem~\ref{source-convergence:cutoff} and Corollary~\ref{massive-source-integration:massive-stop}.

\subsection{A finite pinning criterion and its physical realization}
\label{physical-theory:pinning}

Write $A_B$ for the graph Laplacian of a finite connected graph $B$,
with only its internal nearest-neighbor edges, and put
$k=\beta_*^2/Z$. In this subsection lattice pairings use counting
measure. Removing the common noncompact translation gives the free
law
\[
 Z_B^{-1}e^{-(u,A_Bu)/(2k)+2y\sum_x\cos(t+u_x)}
       \,dt\prod_{x\ne o}du_x,
 \qquad u_o=0,\quad -\pi\le t<\pi.
\]
The root $o$ fixes the decomposition $h_x=t+u_x$. Real neutral observations are $(h,f)$
with $\sum_xf_x=0$, and a character $e^{i(h,s)}$ is well defined whenever $\sum_xs_x\in\mathbb Z$. The free law
integrates the common phase without choosing a height. Indeed, on adding
precision $\eta I$ and sending $\eta\downarrow0$, the periodized common Gaussian tends to
Haar measure and the root-pinned Gaussian dominates every fixed neutral
exponential. A cut gives the same limit separately on each component.

The positive-law inequalities of Proposition~\ref{positive-cosine-laws:phase}
give, for neutral $f$ and the inverse $G_B$ of $A_B$ on $\mathbf1^\perp$,
\begin{equation}
 \frac12\operatorname{Var}\langle h,f\rangle\le\log\E e^{(h,f)}
 \le\frac k2(f,G_Bf),\qquad
 (f,G_Bf)=\min_{\operatorname{div}w=f}\sum_e w_e^2.
 \label{physical-theory:comparison}
\end{equation}
Both quantities decrease under an increase in centered positive
precision and increase under a cut if the test is neutral on every
released component. For a proper law any positive semidefinite precision
may be added. Projection onto gradients gives the flow identity.
Conditional on a fixed nonempty trace, the original nearest-neighbor law
has nonnegative mixed log-density derivatives, given by the edge
couplings. The lattice condition gives association on truncated
coordinate boxes, and Gaussian tails let us pass to the full conditional
law.

For comparison with the integer charge weights retained, let $I_n(2y)$ denote
the $n$th Fourier coefficient of $e^{2y\cos t}$. A profile has total charge $Q(f)=\sum_x f_x$, and
an integer charge configuration has weight $b_y(q)=\prod_x I_{q_x}(2y)$. We write $\mathcal E_B(f)=(f,G_Bf)$ for the
energy of a neutral profile $f$.
The characteristic function $w_B(s)=\E e^{i(h,s)}$ is
\begin{equation}
 w_B(s)=
 \frac{\displaystyle\sum_{Q(q)=Q(s)}b_y(q)\,e^{-k\mathcal E_B(q-s)/2}}
      {\displaystyle\sum_{Q(q)=0}b_y(q)\,e^{-k\mathcal E_B(q)/2}}.
 \label{physical-theory:charge-ratio}
\end{equation}
For $y>0$ this lies in $(0,1]$. A larger neutral quadratic form
decreases the ratio and does not increase the centered neutral
charge covariance. At fixed precision, the shifted-sector charge
covariance dominates the centered one. Along a neutral direction $f$,
\begin{equation}
 \begin{gathered}
 D_f^2\log w_B(s)\ge-\operatorname{Var}\langle h,f\rangle,\\
 \big|\sqrt{-\log w_B(s+f)}-\sqrt{-\log w_B(s)}\big|
 \le\sqrt{\operatorname{Var}\langle h,f\rangle/2}.
 \end{gathered}
 \label{physical-theory:phase-shift}
\end{equation}
Here is a direct reduction to the lattice Gaussian comparison, avoiding
any relabelling of Bessel weights. For $N>y$, attach $N$ integer
leaves to each site, put $t_N=2\log(N/y)$, and let $E$ sum the
leaves at each site. In $\mathbb R^{N|B|}\oplus\mathbf1^\perp$ use
the lattice and displacement
\[
 \mathcal L_N=\{(v,Ev):v\in\mathbb Z^{N|B|},\ \sum v=0\},
 \qquad x_s=(v_0,Ev_0-s),\quad \sum v_0=Q(s).
\]
The second coordinate of $x_s$ is neutral. With precision
$\operatorname{diag}(t_NI,kG_B)$, the coset sum is
\[
 \rho_N(x_s)=\sum_{\sum v=Q(s)}
 e^{-t_N\|v\|^2/2-k\mathcal E_B(Ev-s)/2}.
\]
Increasing the neutral form therefore changes the precision at a fixed
ambient displacement. The coset comparisons in
Lemma~\ref{positive-cosine-laws:lattice-comparison} apply to
$\rho_N(x_s)/\rho_N(0)$, to its neutral profile derivatives, and
to the covariance of $Ev$. The aggregate leaf coefficients converge
in $\ell^1$ with every fixed polynomial moment to $b_y(q)$.
The quadratic factors are at most one, and the centered denominator
has a positive zero-charge term. Thus these ratios, derivatives and
charge covariances converge to their counterparts in
\eqref{physical-theory:charge-ratio}. All leaf limits are taken at
fixed graph and $k,y$; no uniform estimate in leaf degree is needed.
The same construction gives the midpoint inequality
$w(s)^2w(t)^2\le w(s+t)w(s-t)$ whenever both profiles have integer
total. For the last bound in
\eqref{physical-theory:phase-shift}, put $F(t)=\log w(s+tf)\le0$.
The lower Hessian bound implies $|F'|^2\le-2VF$, $V=\operatorname{Var}\langle h,f\rangle$;
integrate, then interchange the endpoints. The same bound holds with
$V$ replaced by the flow upper bound in \eqref{physical-theory:comparison}.

\begin{proposition}[Finite tests for pinning]
\label{physical-theory:finite-pinning}
Let $B_m=\{0,\ldots,m-1\}^2$ be the free square, $u_m$ its uniform probability,
$g_m(x)=x_1/m$, and $\bar\rho_m$ the average of its four face
probabilities. A face probability assigns equal mass to the sites of
one face. Write $\bar h_i=m^{-2}\sum_{x\in B_i}h_x$ for a block
average, and let $P_m$ be the free rectangle obtained by joining two
adjacent $m$-squares, with their interface bonds retained. Fix
$k_{\max}>0$, $0<\delta<8\pi$, $z_*>0$,
$0<d_0<\pi/10$ and $0<p<1$. There are finite $M,L$ such that
\[
 \begin{gathered}
 0<k\le k_{\max},\qquad y>0,\qquad m_0\ge M,\\
 \operatorname{Var}_{B_{m_0}}\langle h,A g_{m_0}\rangle\le8\pi-\delta,
 \qquad w_{B_{m_0}}(\bar\rho_{m_0})\ge z_*
 \end{gathered}
\]
imply the following. At some $m=rm_0$, with integers $1\le r,\ell\le L$, the
free-square phase and adjacent-free-pair Laplace tests satisfy
\begin{equation}
 1-w_{B_m}(\ell u_m)<\delta_{\rm ph},\qquad
 \log\E_{P_m}e^{\tau\ell(\bar h_L-\bar h_R)}<1,
 \label{physical-theory:free-tests}
\end{equation}
where $\delta_{\rm ph}>0,\tau<\infty$ depend only on $d_0,p$.
Declaring a cell bad when
$\operatorname{dist}(\ell \bar h_i,2\pi\mathbb Z)\ge d_0$ or
$|\bar h_i-\bar h_j|>d_0/\ell$ for a neighbor, one has
$\Pr(S\subset\mathcal B)\le p^{|S|}$ for every prescribed set.
This holds on tiled tori, free squares (omit missing-neighbor tests),
and proper zero-wall laws, including arbitrary centered PSD pins and
zero-extended boundary blocks with their full denominators.
\end{proposition}
\begin{proof}
The threshold $8\pi$ comes from counting sites for a unit charge. In a
square of $G^2$ cells any central cell is available. If $a$ is the limiting
affine variance, the energy penalty divides the weight of a placement by
at most a factor of order $G^{a/(4\pi)}$. For $0<a<8\pi$ the multiplicity gain thus grows
with $G$. The identification of $a$ precedes the comparison \eqref{physical-theory:charge-amplification}.

The inverse-square Ising jump proof \cite{BhattacharyaCriticalJump2026} likewise identifies a physical
response before comparing the logarithmic defect cost with the number of
possible locations. Its response is the square of the spontaneous
magnetization. For the continuous height it is the affine variance, and
the comparison uses Gaussian-coset inequalities, replica identities and
radial-flow estimates.

\emph{Neutral macroscopic response.}
For a fixed free law let $a_m=\operatorname{Var}\langle h,Ag_m\rangle$.
The identity $A_{B_m}g_m=\rho_R-\rho_L$ is exact, including the
corners, and $(g_m,A_{B_m}g_m)=1-m^{-1}$. Cut an $M$-square into
$m$-squares and residual strips. In the affine test of slope $M^{-1}$,
the omitted seams and strips contain $O(M^2/m+mM)$ edges, each
with flow coefficient $O(M^{-1})$. Their squared cost is at most
$C(m^{-1}+m/M)$. Cutting, independence of the released cells and
\eqref{physical-theory:comparison} therefore give
\begin{equation}
 \sqrt{a_M}\le\sqrt{a_m}+C\sqrt{k(m^{-1}+m/M)}.
 \label{physical-theory:affine-cutting}
\end{equation}
First let $M\to\infty$, then let $m$ run through a liminf sequence.
The resulting limsup--liminf inequality proves $a_m\to a\in[0,k]$.
Reflection and quarter-turn symmetry make the two-coordinate affine
covariance scalar.

For a fixed axis rectangle $D$, let $D_m$ be a lattice approximation
after rescaling by $m^{-1}$, and write
$X_{D,m}(g)=(h,A_{D_m}g(\cdot/m))$. Every law in this paragraph
is the free law of its displayed rectangle. Cutting into small
macroscopic squares gives the affine upper bound
$a\int_D|\nabla g|^2$. For the reverse bound, tile a much larger
square by copies of $D_m$. Its saturated affine variance is at most
the sum of the rescaled free-rectangle variances, up to seam and
residual-strip flow errors. Let the number of copies tend to infinity
and then $m\to\infty$. This forces the matching lower bound for
every affine $g$ on $D$. Approximation of a smooth gradient by constants
on finitely many small rectangles now gives
\begin{equation}
 \limsup_{m\to\infty}\operatorname{Var}X_{D,m}(g)
       \le a\int_D|\nabla g|^2.
 \label{physical-theory:macroscopic-upper-energy}
\end{equation}
Indeed the squared flow cost of the gradient approximation is bounded
by its $L^2$ error, while a fixed partition has seam cost $O(m^{-1})$.
The same bound holds for continuous piecewise-smooth shifts and for
observations formed from their gradients on polygonal edge sets.

Call $g$ saturated on $D$ when equality holds in
\eqref{physical-theory:macroscopic-upper-energy}. Saturated shifts
form a linear class, closed in the gradient $L^2$ norm. For linearity,
take a subsequential covariance matrix for any finite list of shifts.
Its difference from $a$ times the Dirichlet-energy matrix is positive
semidefinite; a zero diagonal forces the corresponding row to vanish.
For closure, use the bare flow bound on the difference of two shifts.

For a finite rectangular partition one obtains a lower bound on every
piece. Indeed, if $g$ is saturated on $D$ and $D=\bigcup_jD_j$, its observation differs
from the sum of the internal-edge observations on the $D_j$ by a flow of
vanishing cost. Cutting the seams and applying \eqref{physical-theory:macroscopic-upper-energy} gives
\[
 \begin{aligned}
 a\int_D|\nabla g|^2
 &\le\liminf_m\sum_j\operatorname{Var}_{D_{j,m}}
                 X_{D_j,m}(g|_{D_j})\\
 &\le\limsup_m\sum_j\operatorname{Var}_{D_{j,m}}
                 X_{D_j,m}(g|_{D_j})\\
 &\le a\sum_j\int_{D_j}|\nabla g|^2.
 \end{aligned}
\]
The energies
at the two ends agree. Each of the finitely many variance deficits has
nonnegative liminf, so every deficit tends to zero. Restriction from an
enclosing rectangle thus gives saturation in the free law of the smaller
rectangle itself.

For a nonaffine shift, start with a rectangle invariant under axis or
diagonal reflection, a reflection-odd saturated shift $g$ and a free root
on the fixed vertex trace. Conditioning on the whole trace, common phase
included, makes the open halves independent. The shift vanishes on the
trace. Thus their edge observations $X_+,X_-$ obey $X_++X_-=X_{D,m}(g)$ and have opposite
conditional means by reflection. In particular,
\[
 \operatorname{Cov}(X_+,X_-)
 =-\operatorname{Var}\bigl(\E[X_+\mid\text{trace}]\bigr)\le0.
\]
Since the separate
upper-energy bounds sum to the saturated total energy, both must
saturate and the covariance must tend to zero. This gives saturation of
the half restriction $g\mathbf1_+$ and its reflected fold on the original
rectangle. Restriction by the partition argument gives the same result
on every enclosed rectangle.

Axis folds give $(x_1-t)_+$ on arbitrary rectangles: perform the fold
in a larger rectangle symmetric about $x_1=t$, then restrict. Linear
combinations and gradient approximation give all smooth functions of
one coordinate. Next perform a diagonal fold of $F(x)-F(y+t)$ in
a square symmetric about $x=y+t$ and restrict again. This yields
$(F(x)-F(y+t))\mathbf1_{x>y+t}$ on the requested rectangle; the
requested rectangle need not itself have that reflection symmetry.
For a smooth $g$, extend it smoothly to an enclosing rectangle and
choose $F_t'(x)=-g_{xy}(x,x-t)$. Integrate over a fixed finite
interval containing all differences $x-y$ on the rectangle. Distributionally,
\[
 \partial_x\partial_y\int
    (F_t(x)-F_t(y+t))\mathbf1_{x>y+t}\,dt=g_{xy}(x,y).
\]
The remaining difference is a sum of one-coordinate functions.
Riemann sums for the integral converge in gradient $L^2$, proving
saturation of $g$. Use aligned lattice reflection traces first;
rounding them by $O(m^{-1})$ changes the gradient only in a vanishing
strip and by $O(m^{-1})$ elsewhere. The flow bound removes this error.
For rounded target walls, take an aligned enclosing rectangle and use
the same partition squeeze before removing its $O(m^{-1})$ strip.

Smooth shifts have covariance $a\int\nabla g\cdot\nabla g'$. Cut into cells of macroscopic side $r$.
The single-cell variance is $O(r^2)$, and Gaussian domination gives $\log\E e^{tX}=t^2\operatorname{Var}X/2+O_t(r^3)$.
For $O(r^{-2})$ independent
cells the logarithms add with total error $O_t(r)$. In the microscopic limit the
interface flow has vanishing energy and can be dropped by H\"older. The
cut upper bound then has the quadratic term of \eqref{physical-theory:comparison} applied to $tX$. As $r\downarrow0$ the
two bounds squeeze the transform to its Gaussian value for every linear
combination.

For a bounded zero-mean density use finite Neumann cosine sums
approximating it in $L^2$. The inverse-Laplacian shifts of these modes are
smooth Neumann functions. Their sampled discrete Laplacians approach the
modes in rescaled density $L^2$, with boundary discrepancy on a strip of
vanishing area. Discrete Neumann Poincar\'e converts the residual density
error to a flow estimate uniform in lattice size. This gives joint
Gaussian convergence for the density tests as for the smooth shifts. For
face tests first move the probability into a strip of width $r$ at cost
$O(r)$. All fixed moments converge as well.

\emph{Why $0<a<8\pi$ is impossible.}
For two independent copies of the height field, put $S=h^{(1)}+h^{(2)}$,
$D=h^{(1)}-h^{(2)}$, and $S_X=X(h^{(1)})+X(h^{(2)})$ for a
neutral linear observation $X$. Write $\kappa$ for the joint cumulant.
For a second neutral linear observation $Y$ and a linear phase $L$
with integer total charge, the continuous-law replica inequalities are
\[
 \begin{gathered}
 |\E[(S_XS_Y-2\operatorname{Cov}(X,Y))e^{iL(D)}]|
 \le\sqrt{2\kappa(X,X,Y,Y)},\\
 \kappa(X,X,Y,Y)^2\le\kappa(X,X,X,X)\kappa(Y,Y,Y,Y).
 \end{gathered}
\]
To prove them, add a proper Gaussian mass, use the finite comb of
Appendix~\ref{app:positive-comparison}, and restrict the real cores
to a mesh. The joint core--leaf field is then a centered Gaussian
on a lattice $\mathcal L$. The map from the two independent copies
to $S,D$ gives a bijection onto pairs with the same class in
$\mathcal L/(2\mathcal L)$. Its quadratic energy is the sum of
the two half-precision energies. Conditional on that class, $S,D$
are independent and have the same law; the class is symmetric under
sign reversal. Set $V_{XY}=\E[S_XS_Y\mid\text{class}]$. Expanding
the independent original copies gives
\[
 \begin{gathered}
 \E V_{XY}=2\operatorname{Cov}(X,Y),\qquad
 \E V_{XY}^2=2\operatorname{Cov}(X^2,Y^2),\\
 \operatorname{Var}V_{XY}
 =\operatorname{Cov}(V_{XX},V_{YY})
 =2\kappa(X,X,Y,Y).
 \end{gathered}
\]
Conditional independence and $|e^{iL(D)}|=1$ give the first
inequality by Cauchy--Schwarz; applying it to $V_{XX},V_{YY}$
gives the second. Remove the core mesh and then the comb at fixed
proper mass and graph. Finally remove the mass. Neutral Gaussian
moment bounds give uniform integrability of the displayed products,
and integer total charge makes the phase invariant under the common
$2\pi$ translations of each released component. Only the numerical
moment inequalities pass to the free law; no parity variable is
assigned to the limiting real field.

Let $\rho_R$ be the uniform probability on the right face of the
square. Take $X=(h,Ag_m)$ and $Y=(h,\rho_R-u_m)$, with centered $g_m$.
Their limiting covariance is $a/2$, since
$(g_m,\rho_R-u_m)=(m-1)/(2m)$ before centering $g_m$ and is
unchanged by centering. If a nonzero mean harmonic had
a positive subsequential limit, take its locally uniform entire
transforms $F_m(t,u)=\E e^{i\ell(h,u_m)+itX+iuY}$ converging
to $F(t,u)$. Gaussian
convergence makes the fourth cumulant of $X$ vanish; the bounded
fourth moment of $Y$ and the replica inequality remove the mixed
cumulant. The identity
$\E[S_XS_Ye^{iL(D)}]=-2(F_mF_{m,tu}-F_{m,t}F_{m,u})$
and its $X,X$ version imply $(\log F)_{tt}=-a$ and
$(\log F)_{tu}=-a/2$ near the origin. Analytic continuation gives
\begin{equation}
 F(t,u)=F(0,u)e^{-at^2/2-atu/2}.
 \label{physical-theory:replica-PDE}
\end{equation}
Reflection gives $F_t(0,0)=0$; positivity and
\eqref{physical-theory:phase-shift} give $F(0,\ell)>0$.
Join two squares and group the neutral charge sum by their totals
$Q,-Q$. In this two-cell calculation $\rho_L,\rho_R$ denote the
probabilities on the facing sides of the left and right cells.
Internal minimum-energy flows to the common faces, parallel
interface flow and the penalty $Q^2$ give a dominating quadratic form,
with sector weights proportional to
$e^{-(1+k/m)Q^2/2}w_L(Q\rho_L)w_R(-Q\rho_R)$.
The zero and $\ell$ sectors have positive weights. In the following
identities $\operatorname{Cov}_q$ uses the centered neutral charge
law and $\E_s$ the shifted sector in
\eqref{physical-theory:charge-ratio}. Differentiating
\eqref{physical-theory:charge-ratio} gives
\[
 \begin{aligned}
 k^2\operatorname{Cov}_q((q,g),(q,g'))
 &=k(g,Ag')-\operatorname{Cov}((h,Ag),(h,Ag')),\\
 D_{Ag}\log w(s)&=k(g,\E_s(q-s)).
 \end{aligned}
\]
Total variance and shifted-sector covariance comparison therefore force
the conditional affine charge means to have zero limiting variance.
Here is the precise finite inequality, also needed when $k$ varies.
For the global affine shift $g=x_1/m$, let $g_B$ be its restriction
to $B=L,R,LR$, and put
$E_B=(g_B,A_Bg_B)$ and
$C_B=\operatorname{Var}_B(h,A_Bg_B)$. Write
\[
 \begin{split}
 \pi_m(Q)&=\frac{e^{-(1+k/m)Q^2/2}w_L(Q\rho_L)w_R(-Q\rho_R)}
 {\sum_j e^{-(1+k/m)j^2/2}w_L(j\rho_L)w_R(-j\rho_R)},\\
 b_Q&=-kQ/m+D_{A_Lg_L}\log w_L(Q\rho_L)
                   +D_{A_Rg_R}\log w_R(-Q\rho_R).
 \end{split}
\]
Thus $b_Q$ is $k$ times the conditional affine charge mean in the
dominating model. The face contribution is exactly $-kQ/m$, since
the two seam faces have affine values $(m-1)/m$ and $1$.
Precision comparison bounds its total charge variance by that in
the unmodified rectangle. Within each sector the cells are
independent, and their conditional charge variances dominate the
centered cell variances. The law of total variance, multiplied by
$k^2$, gives
\begin{equation}
 \operatorname{Var}_{\pi_m}(b_Q)
 \le k(E_{LR}-E_L-E_R)-C_{LR}+C_L+C_R.
 \label{physical-theory:two-cell-numerical}
\end{equation}
The right side tends to zero: the seam energy is $m^{-1}$ and the
three physical affine variances tend to $2a,a,a$. On the other hand,
the normalizer of $\pi_m$ is between $1$ and
$\sum_Qe^{-Q^2/2}$, while its $0$ and $\ell$ sectors have positive
limiting weights. Equation~\eqref{physical-theory:replica-PDE} gives
$-a\ell/2$ for each logarithmic derivative in $b_\ell$;
reflection and charge reversal on the right cell give the same sign.
Consequently $b_\ell\to-a\ell$ and $b_0=0$, contradicting
\eqref{physical-theory:two-cell-numerical} if $a>0$.
Every nonzero mean harmonic therefore tends to zero. The two-sided
square-root bound in \eqref{physical-theory:phase-shift} transfers
this conclusion to every bounded-energy face profile of the same
integer total.

Now tile a $Gm$-square by $G^2$ cells. For a neutral residual
charge profile $r$, put $Q_i=\sum_{x\in B_i}r_x$ and choose a linear
coarse flow $F(Q)$ with divergence $Q$. In cell $i$ let $F_{ie}$ be
the outward flux at face $e$ and $\rho_{i,e}$ the uniform probability
on that face. Subtracting the face fluxes leaves a neutral profile
$r_i^0=r|_{B_i}-\sum_e F_{ie}\rho_{i,e}$. The dual energy is bounded by
\[
 k\sum_i\mathcal E_{B_i}(r_i^0)
       +\frac{k}{m}\sum_eF_e^2+\sum_iQ_i^2.
\]
The internal minimizing flows and the parallel seam flows use disjoint
edges, so their displayed costs add. The last term is a nonnegative
penalty; the result is one dominating quadratic form, to which the
ambient comparison applies. Put
$\xi_i=\sum_{x\in B_i}\bar\rho_{Gm}(x)$. This vector is supported
on boundary cells, has total one, and satisfies
$\sum_i\xi_i^2\le C/G$. In the denominator a sector of integer
cell totals $n_i$ has $\sum_i n_i=0$ and $Q=n$; in the numerator
it has $\sum_i n_i=1$ and $Q=n-\xi$. Divide both sums by the
product of the cells' centered charge partition functions. The Bessel
weights factor exactly, and each sector is its Gaussian penalty and
seam factor times $\prod_iw_{B_i}(s_i)$, where $Q(s_i)=n_i$.
At fixed $G$ the penalties are dominated by summable Gaussian
sequences in $n$. Every nonzero denominator sector contains a cell
with nonzero integer total and hence tends to zero; its zero sector
is exactly one. Dominated convergence therefore makes the normalized
denominator tend to one.
Keep numerator sectors with one unit charge in any central cell.
In a sector of integer cell charges $n_i$, let $s_i$ be its incoming
face profile plus its prescribed exterior profile, of total $n_i$;
$H_{i,\pm j}$ are its face coefficients. A unit radial flow $V(x)=(x-x_0)/(2\pi|x-x_0|^2)$, corrected by a
smooth Neumann field of size $O(G^{-1})$, has face-flux odd and even
parts
$D_{ij}=(H_{i,+j}-H_{i,-j})/2$,
$S_{ij}=H_{i,+j}+H_{i,-j}$ satisfying
$\sum D_{ij}^2\le(2\pi)^{-1}\log G+C$ and
$\sum S_{ij}^2\le C$. Integrating over faces proves these discrete
bounds: the sum-integral error is summable, and the correction has
bounded energy. Prescribe this flow on the independent vectors
$e_{i_0}-\xi$, where $\xi$ is the exterior-face mass, and extend
linearly; one dominating form then works for all retained sectors.
Square symmetry kills the first variation of the total-one face
character. For $Y_i=s_i-n_i\bar\rho_i$, its Hessian bound and
reflection symmetry give
$\limsup\sum_i\operatorname{Var}\langle h,Y_i\rangle
\le a\sum D_{ij}^2+Ck\sum S_{ij}^2$.
The sector penalty is uniformly bounded and the interface cost
vanishes as $m\to\infty$ at fixed $G$. Consequently
\begin{equation}
 \liminf_{m\to\infty}\frac{z_{Gm}}{z_m}
       \ge cG^{2-a/(4\pi)},\qquad z_m=w_{B_m}(\bar\rho_m)>0.
 \label{physical-theory:charge-amplification}
\end{equation}
The exponent is positive for $0<a<8\pi$. Fix $G$ making this lower bound
larger than one and iterate, contradicting $z_m\le1$.

\emph{The zero-response case.}
At $a=0$ the same argument shows that some nonzero mean harmonic has
positive subsequential limit $r_\ell$: otherwise zero neutral
response makes every fixed face character differ by $o(1)$ from its
vanishing mean character, giving the same growth contradiction.
Extract all integer mean coefficients along that subsequence and
put $b(j)=r_j e^{-j^2/2}$.
The preceding trial form, now applied to a uniform total-$\ell$ source,
gives
\begin{equation}
 \liminf_{m\to\infty} w_{B_{Gm}}(\ell u_{Gm})
 \ge e^{\ell^2/(2G^2)}\frac{b^{*G^2}(\ell)}{b^{*G^2}(0)}
 \xrightarrow[G\to\infty]{}1.
 \label{physical-theory:convolution-amplification}
\end{equation}
Here $m$ runs through the extracted subsequence and $b^{*G^2}$ is the
$G^2$-fold convolution. Fourier inversion proves the ratio limit.
The normalization of $b$ is an even law with Gaussian tails, a
positive atom at zero and positive variance. At every point where
its Fourier transform has modulus one, the contributions to the
inversion integrals at $0$ and at $\ell$ have the same leading Gaussian
term, since $\ell$ belongs to the lattice span. The complementary arcs
decay exponentially. Meanwhile the adjacent-pair Laplace transform
tends to one by zero neutral response.

\emph{Uniform finiteness: the first limit.}
Suppose no $M,L$ exist. For each $n$ choose parameters $k_n,y_n$
and a base side $m_n\ge n$ satisfying the two hypotheses but failing
\eqref{physical-theory:free-tests} for every $1\le r,\ell\le n$.
Pass to $k_n\to\bar k\in[0,k_{\max}]$ and rescale by
$m_n^{-1}$, without rescaling the field. For each rational axis
rectangle retain a countable energy-dense list of neutral smooth,
face, volume and polygonal-gradient tests, together with all integer
mean phases. The phases are compact; the neutral observations have
the common Gaussian flow bounds. Diagonal extraction gives joint
laws, all fixed neutral moments, and locally uniform entire transforms
with any fixed integer phase inserted. The flow bound extends these
limits to the profiles used above. Each rectangle retains its own free
law; no compatibility of conditional boundary laws is assumed.

Write $a_R=\lim_n a_{Rm_n}(k_n,y_n)$ for integer $R\ge1$.
Cutting exactly into base squares has seam cost $O(m_n^{-1})$.
After this first limit, therefore,
\begin{equation}
 0\le a_R\le a_1\le8\pi-\delta.
 \label{physical-theory:limit-affine-deficit}
\end{equation}
The coefficient $a_R$ may depend on $R$. We do not identify it with
its large-$R$ limit at this stage.

Independent products of finitely many free cut cells converge and the
seam errors for a fixed partition vanish. We may therefore pass the
cutting inequalities, the partition squeeze for saturation and the
unconditional folding inequalities to the individual limiting rectangle
laws. Folding is performed on the finite lattice. Only its numerical
covariance inequality is passed to the limit, which requires no limiting
conditional trace law. For the replica inequalities the uniform fourth
moments justify passage to the limit.

Every fixed integer-total character needed on an enlarged rectangle
has a positive limiting lower bound. To see this, first cut off one
embedded base square and use its total-one face profile $s_n^0$.
Restoring the cut bonds increases precision, so its character in the
enlarged free rectangle is at least $z_*$. The midpoint inequality
gives $w(js_n^0)\ge z_*^{j^2}$: the sequence
$\log w(js_n^0)-j^2\log w(s_n^0)$ has nonnegative second
differences and vanishes at $j=0,1$. A prescribed profile of total
$j$ differs from $js_n^0$ by a neutral profile of bounded flow cost.
The square-root shift bound then gives its positive lower bound.
The bound may depend on the fixed enlarged rectangle and on $j$;
no uniform lower bound in $R$ is asserted.

At fixed cell number, the trial ratios have Gaussian-dominated sector
sums and a positive zero sector in the denominator. They pass by
dominated convergence. Source-jet convergence and the preceding
positivity pass their logarithmic derivatives. For the two-cell
comparison use \eqref{physical-theory:two-cell-numerical}, which
contains no division by $k_n$. Its sector probabilities converge
by Gaussian domination, and each fixed $b_Q$ converges by source-jet
convergence. Fatou's lemma applied to
\[
 \frac12\sum_{Q,Q'}\pi_m(Q)\pi_m(Q')
                         (b_Q-b_{Q'})^2
\]
passes the variance bound without a tightness assumption on the
unobserved charge means. These limits retain the required numerical
comparisons; they do not assign a lattice charge representation to a
limiting rectangle law.

\emph{Uniform finiteness: large scales inside the limit.}
Within the extracted family let $R\to\infty$. Cutting with the fixed-partition
inequalities and uniform flow bounds gives $a_\infty=\lim_{R\to\infty}a_R$. Rectangular saturation and
the Gaussian transform calculation now apply at this second limit.
Equation~\eqref{physical-theory:limit-affine-deficit} gives
$0\le a_\infty\le8\pi-\delta$. If $\bar k=0$, bare Gaussian
domination already gives $a_\infty=0$. Otherwise assume
$a_\infty>0$. A nonzero mean harmonic with positive subsequential
limit would satisfy the replica PDE and contradict the limiting
two-cell variance bound, exactly as above. If all such harmonics
vanish, the trial-form argument gives, for each fixed large $G$,
\[
 \liminf_{R\to\infty}\frac{z_{GR}}{z_R}
       \ge cG^{2-a_\infty/(4\pi)},
 \qquad z_R=\lim_n w_{B_{Rm_n}}(\bar\rho_{Rm_n})>0.
\]
The radial-flow estimates give $c>0$ independent of $G$. Choose
$G$ so the right side exceeds one and iterate in $R$. This
contradicts $z_R\le1$. Hence $a_\infty=0$ in all cases.

The zero-response argument now applies within this family. If every
nonzero mean harmonic vanished at large $R$, charge amplification
would give the same contradiction. Thus some integer $\ell$ has
a positive subsequential mean coefficient. Applying
\eqref{physical-theory:convolution-amplification} along that
subsequence makes the coefficient arbitrarily close to one at a
finite larger integer scale; the pair Laplace transform tends to
one since $a_\infty=0$. There are therefore finite integers
$R,\ell$ for which the limiting tests obey
\[
 1-w_R(\ell u_R)<\delta_{\rm ph}/2,\qquad
 \log\E_{P_R}e^{\tau\ell(\bar h_L-\bar h_R)}<1/2.
\]
Here the subscripts denote the extracted free laws at scale $R$.
Return only now to $n\to\infty$. Convergence of the character and
of the Gaussian-controlled pair Laplace transform makes the original
tests strict at side $Rm_n$ for all sufficiently large $n$.
Since then $R,\ell\le n$, this contradicts the chosen failures and
proves the existence of finite $M,L$.

\emph{The joint probability bound.}
Set $q=(p/5)^{40}$; choose $s<\min(1,d_0/4)$ so that
$(8\pi/s)e^{-d_0^2/(32s^2)}<q/4$, and take
$\delta_{\rm ph}=qs(1-\cos(d_0/4))/(16\pi)$,
$\tau=d_0^{-1}\log(8e/q)$.
Add independent variance-$s^2$ noises to the selected $\ell \bar h_i$
and wrap them modulo $2\pi$. Write $p_T$ for their target joint
density and $p_i$ for the free one-cell densities. At positive proper
mass and finite comb number, the wrapped Gaussian adds integer
coordinates $j_i$ with energy
\[
 \sum_i\frac{(v_i+2\pi j_i-\ell\bar h_i)^2}{2s^2}.
\]
After discretizing the real cores, this is a Gaussian lattice with
fixed displacement $v_i/(2\pi)$ in its new integer coordinates.
Restoring the cut bonds or adding a centered PSD pin increases
precision without changing that displacement. The ambient coset
comparison in Lemma~\ref{positive-cosine-laws:lattice-comparison}
therefore gives
\begin{equation}
 \frac{p_T(v)}{p_T(0)}\le\prod_i\frac{p_i(v_i)}{p_i(0)}.
 \label{physical-theory:wrapped-density-ratio}
\end{equation}
First remove the core mesh and comb at fixed proper mass and graph,
then remove the mass. The noise densities are bounded and continuous,
and integer $\ell$ makes them invariant under the released common
periods; hence the free quotient limit retains all root phases.
The cut components factor with their own full normalizers.

The wrapped-noise bound gives $\sup p_T\le s^{-|I|}$. Positive
Fourier coefficients give $p_i(0)\ge(2\pi)^{-1}$. In a free cell
the probability of a noisy phase exceeding $d_0/2$ is at most
\[
 \frac{\delta_{\rm ph}}{1-\cos(d_0/4)}
                  +2e^{-d_0^2/(32s^2)}.
\]
An original bad phase remains bad at the half threshold with
conditional probability at least $1/2$, independently for each
added noise. Integrating \eqref{physical-theory:wrapped-density-ratio}
over those noisy bad phases therefore bounds the original joint
probability by the $|I|$th power of
\[
 \frac{4\pi}{s}\left\{
 \frac{\delta_{\rm ph}}{1-\cos(d_0/4)}
                      +2e^{-d_0^2/(32s^2)}\right\}<q/2.
\]
For $j$ neighboring pairs with disjoint end cells, cut into their
free pair rectangles. Neutral Laplace comparison and the $2^j$
possible signs give $(2e e^{-\tau d_0})^j=(q/4)^j$.
Assign one of the five failure types to each specified bad cell;
one type contains at least a fifth, and a greedy choice keeps at
least a fortieth as disjoint pairs. The sum is bounded by
$5^{|S|}q^{|S|/40}=p^{|S|}$. For a true wall, first add exterior
precision $vI$ in a torus, retain all full block denominators, let
$v\to\infty$, then remove the auxiliary mass. All removed bonds
and pins are PSD. The same proof allows extra centered PSD pins;
continuous nontrivial averages put no mass on the fixed thresholds.
\end{proof}

\subsection{Pinning of the canonical family}
\label{physical-theory:canonical-pinning}

At one physical scale the finite criterion needs both a strict stiffness
deficit and a positive fundamental character. The first variation of the
canonical trajectory exhibits both. Its bulk preparation must, however,
be compared through an actual interaction collar with the weak-scale
preparation at the free boundary used in the test.

\begin{proposition}[Canonical pinning scales]
\label{physical-theory:canonical-pinning-scales}
On each compact positive endpoint interval, the hypotheses of
Proposition~\ref{physical-theory:finite-pinning} hold uniformly at one
sufficiently early physical scale. Its pinning blocks may be chosen
from a finite family of physical lengths comparable to
$a_{\rm exit}$, uniformly in the cutoff.
\end{proposition}
\begin{proof}
Choose a weak physical scale $r$ and a free square of side
$\Lambda r+O(\epsilon)$; localize the cosine and excess kinetic
term to an inner square, leaving a collar wider than the prefix
range. The localized precision and activity are smaller, so its
variance is an upper bound and its character a lower bound for the
homogeneous law. Let $C$ be the canonical bulk prefix,
$P_0=m^{-2}\mathbf1\mathbf1^{\top}$, and
$Q=A^\dagger-(I-P_0)C(I-P_0)\succeq0$. Haar translation absorbs
the constant part of $C$, so $\tau+\psi+\zeta$, with
$\psi\sim N(0,Q)$, $\zeta\sim N(0,C)$, is the exact free quotient.
Even reflection at the free walls leaves every prefix on the inner
support equal to the bulk prefix.

We record the interface argument once, since it is also used below
for free current cavities. Color all supports meeting a deleted
interaction boundary. In their analytic charge decomposition, neutral
nonconstant terms begin quadratically in increments and nonzero
charge has free critical dimension at least two; the weak
perturbation preserves the resulting strict interface gap. A face has $O(b)$ anchors, a
corner $O(1)$, so after grouping their homogeneous norm is at most
$C(1+\log b)^c/b<1$. The charge decomposition is bounded between
reserved analytic radii by the constant-direction contour shift;
scalars are extracted only in charge zero. The division-free
operation therefore gives an exact interface recursion
$B_{j+1}=L_jB_j+Q_j+N_j$ with
$\|L_j\|\le\theta<1$, $Q_j=O(v_j)$ and
$N_j=O(v_jB_j+B_j^2)$. Composing with the marginal inverse makes
its endpoint derivatives of order $q\le2$ contract by
$(\theta+Cv_*)(1+Cv_*)^q<1$. Induction bounds these derivatives
and their preparation memory. The retained derivative regulator has
bounded operator norm and trace $O(\Lambda^2)$, by even torus
compression and the reserved Gaussian estimate. Thus the complete
localized density and its two integrated smooth-shift responses have
uniform $O_\Lambda(v^2)$ Taylor remainders. Telescoping one covariance
or site/edge allocation difference gives a bound of
$C(\epsilon/r_j)^\nu n^p\gamma^n$ on a history of size $n$.
The exponential factor makes the sum over $n$ finite, and contraction
controls the sum over earlier scales uniformly. This proves the required arbitrary-cutoff
comparison, including residual phases, without assuming a homogeneous
flow at the interface.

The exact linear term of its scalar-factored reconstruction is
$d_\epsilon-\phi^{\top}A_E\phi/4+
(2/c_{\rm lat})\sum_S e^{-4\pi C(x,x)}\cos(\beta_*\phi_x)$.
For $g=x_1/m$, Cameron--Martin translation by a smooth extension
equal to $g$ on the interaction support gives
\begin{equation}
 \operatorname{Var}_{\rm loc}\langle h,Ag\rangle
 =8\pi(g,Ag)-4\pi v(g,A_Eg)+O_\Lambda(v^2),\qquad
 w_{\rm loc}(u_m)\ge c_\Lambda v.
 \label{physical-theory:canonical-free-input}
\end{equation}
Haar integration kills the charged linear term in the first formula.
For the second, its first harmonic is
$c_{\rm lat}^{-1}v\sum_S e^{-4\pi(C(x,x)+Q(x,x))}$;
the exact constant-prefix conversion is
$e^{\beta_*^2\lambda_0/(2m^2)}$, where $CP_0=\lambda_0P_0$.
The bulk diagonal $e^{-4\pi C(x,x)}\asymp(\epsilon/r)^2$,
$Q(x,x)\le C_\Lambda$ and $|S|\asymp(r/\epsilon)^2$ give the
strict lower bound. Choose $\Lambda$ so $(g,A_Eg)\ge e_0>0$,
then an earlier canonical section with $0<v_-\le v_r\le v_*$.
The variance deficit and positive mean character are uniform.
The flow from $u_m$ to $\bar\rho_m$ has energy at most $1/3$,
so \eqref{physical-theory:phase-shift} supplies $z_*>0$.
The criterion gives a finite family $a_\epsilon=j m_0(\epsilon)\epsilon$,
$m_0=\lfloor\Lambda r/\epsilon\rfloor$, with
$0<c\le a_\epsilon/a_{\rm exit}\le C$; the integer choice is never
differentiated.
\end{proof}

\subsection{Localization at the fundamental period}
\label{physical-theory:fundamental-localization}

The finite criterion allowed an auxiliary harmonic $\ell$. One further
finite enlargement gives $\ell=1$ from the canonical fundamental character,
identifies the well selected by the zero boundary and supplies the
moments for height convergence. Let $d_1(i,j)$ denote graph distance between blocks in the square tiling.
For block probabilities $u_i,u_j$, parallel row flows and the
two-dimensional Fourier annulus sum give
\[
 (u_i-u_j,A^\dagger(u_i-u_j))
 \le\min\{d_1(i,j)+2,\ C\log(2+d_1(i,j))\}.
\]
For the logarithmic bound the summand is bounded, after $q=mp$,
by $\min(1,|q|^{-2})\min(1,d_1^2|q|^2)/|q|^2$; dyadic annuli
below and above one give the displayed logarithm. Even reflection
gives the same free-square estimate. A fixed root has at most $64^n$
bad animals of size $n$. Thus a bounded-aspect-ratio rectangle fails
to have a good crossing with probability $Ce^{-cs}$, where $s$ is
its shorter side. Dyadic overlapping crossings connect good sites
within distance $Ct$ of any two prescribed cells, except on an
event of probability $Ce^{-ct}$. In a free square use the two
ancestor chains of its dyadic rectangular subdivision; on a torus
connect the terminal circuits by bounded-aspect-ratio strips.
Good neighbors have the same well label. Gaussian domination of
the short flows consequently gives
\begin{equation}
 \Pr(|\bar h_i-\bar h_j|>u)
 \le Ce^{-ct}+C(t+1)^2e^{-c'u^2/\log(t+2)}
 \le C e^{-cu^2/\log(2+u)}.
 \label{physical-theory:block-tails}
\end{equation}
Choose $t\asymp u^2/\log(2+u)$; if this exceeds the domain size,
the direct logarithmic flow proves the same result. All exponential
moments follow. Comparing to an exterior block and hard-pinning the
exterior transfers these bounds to each zero-extended Dirichlet
average and to every further centered PSD pin.

To eliminate the auxiliary $\ell$, embed a base block in a free square
containing arbitrarily many pinning blocks. Restoring bonds keeps its
mean character at least $c_b>0$. The difference between its average and the
whole mean has uniformly bounded variance. Indeed, a bounded number of
corridor flows reaches the pinning block, and \eqref{physical-theory:block-tails} controls the remaining
difference. The physical-variance bound in \eqref{physical-theory:phase-shift} now gives a uniform
fundamental mean coefficient $r_*>0$. In the compactness argument of the
criterion the limiting response is zero, so \eqref{physical-theory:convolution-amplification} applies with $\ell=1$ because
$b(0),b(1)>0$. A sequence failing the strict fundamental test at all bounded scales
would contradict this limit. One enlargement of the finite scale family
therefore suffices uniformly, and henceforth $\ell=1$.

With a zero wall, any good cell disconnected from the exterior zero
phase has a surrounding bad circuit of length at least four. Hence $\Pr(|\bar h_i|\ge d_0)\le p+C\sum_{n\ge4}n^2(64p)^n\le p+C_*p^4$.
The moments in \eqref{physical-theory:block-tails} are uniform as $d_0$ and $p$ decrease from their initial
values, giving
\begin{equation}
 \E|\bar h_i|^q\le d_0^q+C_{2q}^{1/2}(p+C_*p^4)^{1/2}.
 \label{physical-theory:small-block-moments}
\end{equation}
Given finite $q$ and a target accuracy, one finite
family of physical scales attains it uniformly over domains, cutoffs,
block positions and centered PSD pins. A partial boundary block keeps
its full denominator. For a point relative to its block average, the
Neumann-mode sum $\sum_{0\le a,b<m,\ (a,b)\ne(0,0)}(a^2+b^2)^{-1}=O(1+\log m)$ gives $\E e^{t(h_x-\bar h_B)}\le e^{Ck_{\max}t^2(1+\log m)}$.

\subsection{The selected representation and its spectral core}
\label{physical-theory:representation}

\begin{proposition}[Literal fields, moments and the common cyclic space]
\label{literal-fields:literal-derivative-fixed-box}
\label{literal-fields:fixedbox-height}
\label{literal-fields:monotone-selection}
\label{height-representation:compact-norm}
\label{height-representation:selected}
For compactly supported $L^2$ tests, the literal angular field has
joint arbitrary-cutoff limits in each fixed true box and then a unique
cofinal zero-wall volume limit. All fixed mixed moments converge.
On each compact support $K$, for nonzero $f$, $1\le p<\infty$
and $\lambda\ge0$,
\begin{equation}
 \|h(f)\|_p\le C_{K,p}\|f\|_2,
 \qquad \E e^{\lambda|h(f)|/\|f\|_2}\le C_{K,\lambda},
 \label{physical-theory:height-moments}
\end{equation}
uniformly also before either limit and under $L^2$-consistent lattice
sampling. The limiting law is translation invariant, reflection positive
and locally distribution valued. With the literal derivative
$D(v)=\langle\nabla\phi,v\rangle$,
$h(\nabla\cdot v)=-\beta_*D(v)$.
Positive-time height Fourier cylinders, their finite polynomials and
the bounded derivative Fourier algebra generate the same cyclic OS
space, denoted $\mathcal H_{\rm ht}=\mathcal H_\partial$.
\end{proposition}
\begin{proof}
At weak physical resolution $r$, Gaussian conditioning in the
unit-reference split $G=C+Q$ gives
$\zeta=CA_D\phi+\eta$, with
$\operatorname{Cov}\eta=C-CA_DC$. The physical tilt depends only
on $\phi$, so the identity persists under that tilt. Its high-field
proxy is $C+(Z_\epsilon^{-1}-1)CA_DC\preceq2C$, while the positive
spectral integral gives
\begin{equation}
 0\preceq C_\epsilon[0,r]\preceq Cr^2I
       \qquad(\epsilon\le r).
 \label{literal-fields:high-covariance-smallness}
\end{equation}
The bounded microscopic interval costs $O(\epsilon^2)$; integrating
the bounded periodized multiplier above it costs $O(r^2)$.
Thus each smooth high-field smear has $L^p$ norm $O(r)$;
summation by parts gives the same result for derivative smears.
At fixed $r$ the complete retained law converges in each fixed box
by Theorem~\ref{fixed-volume-integration:fixed-box},
with its positive denominator and Gaussian regulator. The triangle
inequality in finite-dimensional $W_p$ makes two cutoff laws differ
by at most $Cr$ in the limit. Letting $r\downarrow0$ proves their
unique joint limit, uniformly on compact parameter intervals.

For the volume-independent estimate, write $j_x=\epsilon^2f_x$,
$\|f\|_{2,\epsilon}^2=\epsilon^2\sum_x|f_x|^2$, and choose a full
pinning block $B$ near $K$. With $b=\sum_xj_x$,
\begin{equation}
 \begin{gathered}
 j^{\top}h=b\bar h_B+(j-bu_B)^{\top}h,\\
 |b|\le C_K\|f\|_{2,\epsilon},\quad
 \|j-bu_B\|_2\le C_K\epsilon\|f\|_{2,\epsilon}.
 \end{gathered}
 \label{height-representation:linear-decomposition}
\end{equation}
The neutral part has a flow of energy $C_K\|f\|_{2,\epsilon}^2$:
solve on a fixed containing square, whose Neumann gap is
$\ge c_K\epsilon^2$, and extend its flow by zero. Gaussian domination
and the block exponential moments prove
\eqref{physical-theory:height-moments} by H\"older. If the original
domain does not contain $B$, enlarge it first; centered-pin comparison
has the required direction. This estimate extends the smooth
fixed-box construction to compact $L^2$ tests and arbitrary consistent
samplings.

Each finite linear combination has a positive characteristic function
decreasing under domain enlargement. Its directed infimum is continuous
at zero by the second-moment bound. The characteristic-function
continuity theorem gives consistent limiting vector laws; higher
moments give uniform integrability, and the exponential bound gives
entire Fourier--Laplace transforms and moment determinacy. Cofinal
infima and opposite lattice translations prove translation invariance;
the sampling estimate removes the rounding. Reflected Fourier Gram matrices remain positive in both limits. On a
compact square a Laplacian basis satisfies $\E\sum_n(1+\lambda_n)^{-s}|h(e_n)|^2<\infty$ for $s>1$, realizing the
local random distribution. Summation by parts identifies the derivatives
also after reflection. A reflected forward difference is the backward
difference with its test shifted by one edge, and the shift error
vanishes in the kinetic norm.

To remove the height zero mode from the core, fix $p$ and use
\eqref{physical-theory:small-block-moments} at accuracy $\eta$.
If the finite possible block sides are $jm_0\epsilon$, use a side
divisible by every such $j$. Its mean is a convex combination of
the selected block means, so Minkowski keeps the same $L^p$ bound.
Place it in positive time, pass the fixed-box and volume limits, and
smooth its indicator using \eqref{physical-theory:height-moments}.
This gives a smooth positive-time probability $\rho_\eta$ with
$\|h(\rho_\eta)\|_p\le2\eta$.
For $g=f-(\int f)\rho_\eta$ choose a positive-time rectangle
containing its support. Put $a(x)=\int g(t,x)dt$ and choose a
time cutoff $\chi$ of integral one. Then
\[
 v_t(t,x)=\int_{-\infty}^t[g(s,x)-\chi(s)a(x)]ds,
 \qquad v_x(t,x)=\chi(t)\int_{-\infty}^x a(y)dy
\]
are compactly supported and $\operatorname{div}v=g$.
Hence $h(f)$ is an $L^p$ limit of literal derivative smears.
For a degree-$m$ polynomial use $L^{2m}$ approximation and H\"older;
replace each derivative by $\sin(tD(v))/t$ to obtain finite Fourier
sums. Characters use $|e^{ix}-e^{iy}|\le|x-y|$. Finally
$\|[F]\|_{\rm OS}\le\|F\|_2$; the reverse inclusion is the
derivative dictionary. All approximations preserve positive-time
clearance, although their supports may grow.
\end{proof}

\begin{proposition}[A surviving derivative vector]
\label{literal-fields:theorem-2}
A fixed smooth positive-time literal derivative insertion has squared
reflected norm bounded below and, at one fixed physical time $t_*>0$,
translated overlap divided by that squared norm at least $q>0$, uniformly in every
containing true rectangle and fine cutoff. It survives both ordered
limits and gives $Ma_{\rm exit}\le C$ in every representation
containing its class.
\end{proposition}
\begin{proof}
Choose nonconstant $f_0\in C_c^\infty(-1,1)$,
$f_L(x)=f_0(x/L)$, and the literal row derivative
$F_0=\sum_j f_L(\epsilon j)(\phi_{0,j+1}-\phi_{0,j})
=\ell_\epsilon^{\top}\phi$. Write $F_u$ for its translate to a lattice
row at physical time $u$, suppressing the fixed $\epsilon,L$.
In $(-4L,4L)^2$, split at $r=L/\Lambda$.
The prefix row sum is $O(r/\epsilon)$ and
$\|\ell_\epsilon\|_2^2=O(\epsilon/L)$, so the high contribution
is $O(\Lambda^{-1})$. The unit Gaussian variance tends to
\[
 q_f=\iint f'_0(x)G_{(-4,4)^2}((0,x),(0,y))f'_0(y)dxdy>0.
\]
Lattice convergence follows by spatial sine modes: repeated summation
by parts bounds mode coefficients by $C_R\sqrt{\epsilon/L}\,j^{-2R}$,
and the temporal diagonal Green value is at most $CL/(\epsilon j)$;
the tails are summable. At fixed $\Lambda$, retained Gaussian and
physical laws have common smooth-jet moments. On a large jet ball
their complete weak density is $1+O(\delta)$, since it has finitely
many scale-$r$ blocks. Both laws give the complement probability
$\le\eta$, so the actual normalizer lies between
$(1-\eta)^2$ and $(1+\eta)/(1-\eta)$ and the retained second
moment differs from its Gaussian value by $O(\sqrt\eta)$.
Choose successively $\Lambda$, $\eta$, the jet ball, $\delta$, and
an earlier fixed canonical section realizing $\delta$. Then
$\E_DF_u^2\ge c_{\rm loc}>0$ in every containing true rectangle.

For spatial Dirichlet $A_W$, the free strip row covariance is
\[
 \begin{aligned}
 G_W(n)&=e^{-n\kappa(A_W)}/\sqrt{A_W(A_W+4)},\\
 \kappa(\lambda)&=\operatorname{arcosh}(1+\lambda/2).
 \end{aligned}
\]
Since $(1-e^{-n\kappa})/(2\sinh\kappa)\le n/2$ and
$d(d^*d)^{-1}d^*$ is a projection, covariance comparison gives
\[
 c_{\rm loc}\le\E_DF_u^2\le C_{\rm loc},\qquad
 \E_DF_uF_v\ge c_{\rm loc}-C_{\rm inc}|u-v|/L.
\]
Choose $\chi\ge0$, $\int\chi=1$, supported in $(\sigma,2\sigma)$,
and $t_*>0$ with $C_{\rm inc}(t_*+4\sigma)/L<c_{\rm loc}/4$.
The squared reflected norm $N$ and translated overlap $K(t_*)$ of the smooth
insertion $\chi(t)f_L(x)\partial_1\phi$ then obey
$N\ge c_{\rm loc}/8$, $K\ge c_{\rm loc}/8$ and
$K/N\ge q=c_{\rm loc}/(2C_{\rm loc})>0$ at every fine cutoff.
The same quadrature mass multiplies numerator and denominator, so it
cancels in this ratio. Ordered convergence preserves this nonzero
vector. Its positive spectral measure implies $M\le-\log q/t_*$.
The finite exit atlas bounds $t_*/a_{\rm exit}$ above and away from
zero, giving a uniform upper bound for $Ma_{\rm exit}$.
\end{proof}

\subsection{Axial summability}
\label{physical-theory:axial-summability}

The moment bound on block heights leaves open a possible zero-energy
atom in the two-point function. The finite-domain contour bound gives axial summability after removal of the cutoff and passage to the selected-volume limit, in that order.
Let $\E_B$ denote the law with an added centered PSD precision $B$.
Write a zero-extended block average as $\bar h_i=2\pi n_i+u_i$,
$n_i\in\mathbb Z$, $-\pi\le u_i<\pi$. For a set $E$ of coarse
edges, nonzero $n_i-n_j$ forces both endpoints to be bad, hence
probability at most $p^{|E|/2}$. The adjacent-block flows have energy
at most two and overlap at most twice in each orientation. Thus
\[
 \E_B e^{\sum_E t_e(\bar h_i-\bar h_j)}\le e^{K\sum_Et_e^2/2},
 \qquad K=4k_{\max}.
\]
Gaussian integration of this bound and
$(n_i-n_j)^2\le(\bar h_i-\bar h_j)^2/(2\pi^2)+2$ give
$\E_B e^{\eta\sum_E(n_i-n_j)^2}\le M^{|E|}$, with
$\eta=\pi^2/(4K)$ and $M=e^{2\eta}\sqrt{4/3}$.
Cauchy--Schwarz therefore proves, for integers $r_e\ge1$ and $p$
small enough that $a_p>0$,
\[
 \begin{gathered}
 \Pr_B\{|n_i-n_j|\ge r_e\ (e\in E)\}
 \le e^{-a_p|E|-\eta\sum_Er_e^2/2}
 \le e^{-\kappa_p\sum_Er_e},\\
 a_p=-\log(M^{1/2}p^{1/4}),\qquad
 \kappa_p=\sqrt{2a_p\eta}\longrightarrow\infty.
 \end{gathered}
\]
All estimates include centered PSD pins and partial boundary cells.

The integer field is zero outside the finite domain. Round dual corners
by one fixed inside-preserving rule and take outer
contours of all positive and negative integer level sets. If
$N_\gamma$ is a signed contour's multiplicity and $L_\gamma$ its
length, the joint edge bound yields
$\Pr(N_\gamma\ge r_\gamma\ \forall\gamma)
\le\prod_\gamma e^{-\kappa_pL_\gamma r_\gamma}$:
an edge is crossed at most $|n_i-n_j|$ times over all levels.
For $z_\gamma=e^{-\kappa_pL_\gamma}$ this implies
\begin{equation}
 \E_B e^{\sum_\gamma t_\gamma N_\gamma}
 \le\prod_\gamma\frac{1-z_\gamma}{1-z_\gamma e^{t_\gamma}}
       \quad(t_\gamma\ge0, z_\gamma e^{t_\gamma}<1).
 \label{physical-theory:contour-product}
\end{equation}
For a finite contour family, expand in factorial moments and use
$\binom Nm=\sum_{r\ge m}\binom{r-1}{m-1}\mathbf1_{N\ge r}$;
Tonelli sums the geometric series; monotone convergence treats infinite
families when the product is finite. No contour independence is assumed.
There are at most $CL^2c_0^L$ length-$L$ contours enclosing a cell.
For a set $\Gamma$ with at most one cell in each column, a contour
encloses at most $L_\gamma$ of its sites. Consequently
$\log\E_Be^{\sum_\Gamma|n_i|}\le\delta(\kappa_p)|\Gamma|$,
where $\delta(\kappa)=4C\sum_{L\ge4}L^2c_0^Le^{-(\kappa-1)L}\to0$.
Also $|u_i|\le d_0+\pi\mathbf1_{\rm bad}$ and the prescribed-set
bound controls the exponential of the bad count. For $|w_i|\le1$,
$X=\sum_\Gamma w_i\bar h_i$, $t=1/(4\pi)$, Cauchy--Schwarz and
\eqref{physical-theory:comparison} give
\begin{equation}
 \operatorname{Var}_B X\le K_0|\Gamma|,
 \quad K_0=2d_0/t+t^{-2}\{\delta(\kappa_p)+p(e^{1/2}-1)\}.
 \label{physical-theory:line-variance}
\end{equation}
A single finite physical scale family makes $K_0$ as small as
prescribed, before choosing the domain, cutoff, pin or line.

Choose a cutoff subsequence on which that family's integer scale
index is constant, with limiting physical side $a>0$. For each fixed
$N$, place $N$ adjacent cells along either coordinate axis in a fixed
true box. Their sampled
indicators converge in physical $L^2$; first pass
\eqref{physical-theory:line-variance} through cutoff removal there,
then through selected volume. Translation invariance and association
give, for the square-average covariances $c_n\ge0$,
\begin{equation}
 Nc_0+2\sum_{n=1}^{N-1}(N-n)c_n\le K_0N,
 \qquad c_0+2\sum_{n\ge1}c_n\le K_0.
 \label{physical-theory:ordered-axial}
\end{equation}
Monotone convergence of $(1-n/N)_+$ gives the last inequality. A positive atom at
zero energy would give every buffered axial self-overlap a common lower
bound equal to its mass, contrary to summability. Only the square was
chosen by subsequence. Its law is fixed by fixed-box uniqueness
throughout this argument, which uses only finite-domain lattice laws.

\subsection{From one covariance to the whole cyclic space}
\label{physical-theory:covariance-spectral-promotion}
On the Fourier quotient physical translations form a positive symmetric
contraction semigroup. Reflection gives $\|U_tF\|_{\rm OS}^2=\langle F,U_{2t}F\rangle$. Iterating Cauchy--Schwarz
and using $\|U_sF\|_{\rm OS}\le\|F\|_2$ proves contraction and preservation of null vectors.
Continuity in the tests gives strong continuity, while $U_t=U_{t/2}^2$ gives
positivity. Denote the semigroup by $e^{-tH}$ with $H\ge0$ and the constant class by
$\Omega$.

\begin{lemma}[Covariance envelope]
\label{height-representation:association-envelope}
For square-integrable associated coordinates, if real $F,G$ have coordinate
Lipschitz constants $L_i,M_j$, then
\begin{equation}
 |\operatorname{Cov}(F,G)|\le
       \sum_{i,j}L_iM_j\operatorname{Cov}(X_i,Y_j).
 \label{physical-theory:association-envelope}
\end{equation}
\end{lemma}
\begin{proof}
Apply association to $(\sum L_iX_i)\pm F$ and
$(\sum M_jY_j)\pm G$ and add equal or opposite signs.
Truncation of increasing factors treats unbounded functions.
\end{proof}
\begin{lemma}[Entrywise covariance under Dirichlet enlargement]
\label{physical-theory:entrywise-domain-order}
For the same canonical bare data at a fixed cutoff, let
$D\subset D'$ be finite zero-Dirichlet domains. For nonnegative
tests $f,g$ supported in $D$,
\[
 0\le\operatorname{Cov}_{\epsilon,D}(h(f),h(g))
 \le\operatorname{Cov}_{\epsilon,D'}(h(f),h(g)).
\]
For fixed admitted Dirichlet rectangles, after cutoff removal in
both domains and then selected enlargement of $D'$, the continuum
laws therefore satisfy
\[
 0\le\operatorname{Cov}_{D,\rho}(h(f),h(g))
 \le\operatorname{Cov}_{\infty,\rho}(h(f),h(g)).
\]
The continuum assertion holds for compactly supported nonnegative
$L^2$ tests. Thus any upper bound for the selected covariance transfers to the
finite-domain association envelope. The comparison uses nonnegative
diagonal pins; it does not infer entrywise order from a
positive-semidefinite matrix inequality.
\end{lemma}
\begin{proof}
At fixed cutoff write the finite angular-field law as
\[
 \begin{gathered}
 d\mu(h)=Z_\mu^{-1}
 \exp\!\left\{-\frac12\sum_i a_i h_i^2
       +\sum_{\{i,j\}}J_{ij}h_ih_j+\sum_i w_i(h_i)\right\}\prod_i dh_i,\\
 J_{ij}\ge0,
 \end{gathered}
\]
where the quadratic form is positive definite and the $w_i$ are
bounded and even. For the canonical law $w_i(h)=2y_\epsilon\cos h$;
extra centered diagonal pins change only the $a_i$.
Put $h_i=r_i\sigma_i$, $r_i\ge0$, $\sigma_i\in\{-1,1\}$.
Conditionally on $r$, the signs form the zero-field ferromagnetic
Ising law with couplings $J_{ij}r_ir_j$.

For this finite Ising law, every spin-product expectation is
nonnegative, and the covariance of any two spin products is
nonnegative. The first assertion follows by expanding the positive
coupling exponentials and summing the independent signs. For the
second, use independent copies and set
$t_i=(\sigma_i+\sigma_i')/2$ and
$s_i=(\sigma_i-\sigma_i')/2$. Their weight is
$\exp\{2\sum_{\{i,j\}}J_{ij}r_ir_j(t_it_j+s_is_j)\}$ and
\[
 \sigma_A-\sigma_A'
 =2\sum_{U\subset A,\ |U|\ {\rm odd}}s_Ut_{A\setminus U}.
\]
The independent-copy formula for the covariance becomes a sum
of nonnegative monomial integrals: at each site
$(t_i,s_i)\in\{(\pm1,0),(0,\pm1)\}$, so mixed powers vanish,
odd powers cancel and even powers sum nonnegatively.

The radial density is a product of one-site factors times
$Z_{\rm Ising}(r)$. For $i\ne j$,
\[
 \partial_{r_i}\partial_{r_j}\log Z_{\rm Ising}(r)
 =J_{ij}\langle\sigma_i\sigma_j\rangle_r
       +\operatorname{Cov}_r(U_i,U_j)\ge0,
 \qquad U_i=\sum_zJ_{iz}r_z\sigma_i\sigma_z.
\]
The preceding spin inequalities give the sign, so the lattice
condition makes the magnitudes associated. One may first truncate
to a rectangular radial box and then remove the truncation using
the Gaussian tails. The function
$F_{ij}(r)=r_ir_j\langle\sigma_i\sigma_j\rangle_r$ is nonnegative
and increasing in every magnitude: coupling differentiation of
the spin expectation is a nonnegative spin-product covariance.
For $i=j$ it is simply $r_i^2$.

Introduce the diagonal pin $v\sum_{k\in A}h_k^2/2$ on the vertices $A$. All field means vanish
by global sign symmetry. Radial association therefore yields
\[
 \frac{d}{dv}\operatorname{Cov}_{\mu_v}(h_i,h_j)
 =-\frac12\operatorname{Cov}_{\mu_v^{\rm rad}}
       \left(F_{ij},\sum_{k\in A}r_k^2\right)\le0.
\]
One can
differentiate first with truncated observables and remove the truncation
by Gaussian tails. With $A=D'\setminus D$ and $v\to\infty$, Gaussian completion recovers the
normalized zero-Dirichlet law on $D$, with the original bonds and bare
data. This orders the lattice covariances entry by entry, and summation
with nonnegative weights orders the smears.

In the admitted rectangles take nonnegative $L^2$-consistent samplings.
Proposition~\ref{height-representation:selected} gives convergence of their mixed second moments as the
cutoff vanishes at fixed $D,D'$. Then let $D'$ exhaust the plane. Convergence
of the selected moments gives the second inequality, first for smooth
tests and then for the stated class by the same local $L^2$ bound. Applied
to the nonnegative envelope profiles of Lemma~\ref{height-representation:association-envelope}, this order transfers
the selected screening bound to each finite $D$. All lattice comparisons
take place at finite volume.
\end{proof}

Polarizing gives the estimate for complex cylinders. For smooth compact
profiles $g_i$ with coordinate Lipschitz constants $L_i$, the envelope $\sum_iL_i|g_i|$ is
bounded and compactly supported. A finite positive sum of translates of
a single square-average profile therefore dominates it. The lattice
covariance inequality passes to continuum smears, and reflection
Cauchy--Schwarz at the temporal midpoint bounds any spatially separated
term by the square's axial self-covariance. If the latter decays
exponentially at rate $\mu$, the spectral measure of each centered smooth
Fourier cylinder vanishes on $[0,\mu)$. These cylinders form a dense core.
Local $L^2$ continuity extends the exclusion to all compact $L^2$ profiles,
and boundedness of the spectral projection extends it to the closure.
Hence $\inf\operatorname{Spec}(H|_{\Omega^\perp})\ge\mu$.

In the converse direction, the spectral bound applies to the square
itself and gives its covariance decay. Its spectral measure is nonzero,
since otherwise the envelope would annihilate the surviving derivative
vector. Its lower edge thus equals the full spectral bottom. The Laplace
transform is log-convex, and for any $t_0>0$ the ratio of its values at
successive times separated by $t_0$ increases to $e^{-Mt_0}$, with $M$ the common
lower edge.

\section{The complete local observable space}
\label{sec:admission}
\label{source-admission:local-admission}
\label{mollifier-sources:section}

To admit the fixed-volume limits of the renormalized coefficients to the
selected Hilbert space, we compare them with mollified heights in the
reflected norm of the original measure. The mollified observables have
the same local normalization. Uniform control as the Dirichlet boundary
recedes comes from a transfer bound independent of the number of
repeated insertions and an estimate on the actual zero-boundary caps.

Fix a finite polarized source list $I$, closed under the required
reflected copies, independent higher columns, identity and contact
coordinates, and prescribed local mixing partners. If $B_I(\eta)$ is
its logarithmic insertion in the finite source algebra $\mathcal J_I$,
the observable associated with $S\subset I$ is
$O_S=[\eta_S]e^{B_I(\eta)}$. All comparisons below concern this
coefficient, including the products and contacts in its exponential.

\subsection{Mollified local fields}
\label{source-admission:mollified-packets}
Fix a marked subset $S$. Each template $\nu\in\mathcal L_S$ pairs a
field polynomial $P_\nu$ with a product of source profiles. Write
$J_{S,\nu}(x)=\prod_{i\in S}\nabla_\epsilon^{\gamma_{i\nu}}J_i(x)$
for that product. The microscopic local column is
\begin{equation}
 B_{\epsilon,S}=\epsilon^2\sum_x\sum_{\nu\in\mathcal L_S}
 c_{\epsilon,S\nu}(\rho,x)\,e^{iq_\nu h_x}
 P_\nu\big((\nabla_\epsilon^\alpha h)_x\big)\,J_{S,\nu}(x).
 \label{local-density-moments:source-density}
\end{equation}
Here the field differences in $P_\nu$ have $|\alpha|\ge1$.
For every fixed $p$ its first four coefficient derivatives have
$L^p$ norm $\le C\epsilon^{-P}$, uniformly in containing boxes.
Indeed the finite triangular inverse costs $L_I^n$ over
$n=O(|\log\epsilon|)$ levels, including its four derivatives.
Engineering and finite Wick conversions add powers and logarithms;
the vertex Wick factor has power growth since its diagonal covariance
is $O(\log\epsilon^{-1})$. A zero-sum finite stencil has a fixed
supported bond flow, so Gaussian domination gives
$\|\nabla_\epsilon^\alpha h_x\|_p\le C_{\alpha,p}\epsilon^{-|\alpha|}$.
H\"older and Minkowski handle the finite polynomial and spatial sum.
Finally $O_S=\sum_{\pi\in\operatorname{Part}(S)}\prod_{A\in\pi}B_A$;
the finite Leibniz rule includes every independent contact and identity
column. Fixed field derivatives and the prescribed compact metric
substitution only enlarge $P$. The four differentiated inverse bounds are those of
\eqref{source-convergence:local-inverse-jets}. These are
coefficient derivatives at fixed field, not derivatives of the law.

Let $\varrho$ be a real radial compactly supported smooth probability
density, put $\varrho_r(x)=r^{-2}\varrho(x/r)$, and set
$P_rh=\varrho_r*h$. The corresponding lattice rows are normalized to
mass one.
For an angular prefix covariance $C_s$, exact Gaussian translation
gives, before any quotient,
\[
 \E_{C_s}[e^{U(\psi+\zeta)}e^{i\langle f,\psi+\zeta\rangle}]
 =e^{i\langle f,\psi\rangle-\langle f,C_sf\rangle/2}
                   F_s(\psi+iC_sf),\qquad F_s=\E_{C_s}e^U.
\]
The finite-cutoff cosine is entire and the completed quadratic
precision positive, so contour translation on the Gaussian support
is legitimate. For
$f_x=q\varrho_r(x-\cdot)+\sum t_\alpha r^{|\alpha|}
\partial_x^\alpha\varrho_r(x-\cdot)$ and $s=c_Ir$,
\begin{equation}
 \|C_sf_x\|_{C_s^{R_I}}
 \le C_I(s/r)^2(|q|+\sum|t_\alpha|),\qquad
 |\langle f_x,C_sf_y\rangle|\le C_I.
 \label{mollifier-sources:shift-budget}
\end{equation}
The prefix row mass is $O(s^2)$; summation by parts moves its
derivatives onto the smooth row. Fixed clearance removes wall images.
Choose $c_I$ so the joint shift has norm less than half the analytic
radius. Collect the local and polymer factors meeting its $O(r)$
support into a marked union. Exterior factors keep their original
weights. The union occupies a bounded number of $s$-blocks, and
the exponential bound on polymer size sums the assignments of the
fixed marks. Differentiating in the $t_\alpha$ treats the
polynomials and retains cross contractions. The packet has integer
total charge under the common shift, although individual averaged
coordinates need not be periodic.

In \eqref{local-density-moments:source-density} substitute jets of $P_rh$ for the field jets, keeping the template
coefficients independent until insertion of the numerical profiles. The
free ordinary map has vertex diagonal $e^{-q^2\langle\varrho_r,C_s\varrho_r\rangle/2}>0$, triangular Wick contraction
and unit leading diagonal on averaged polynomial jets. These properties
give its inverse without inverting spatial averaging. For $a/r\ge2$ one has $\langle\varrho_r,C_a\varrho_r\rangle=4\log(a/r)+O(1)$
uniformly, by splitting the momentum integral below $a^{-1}$, between $a^{-1},r^{-1}$ and
above $r^{-1}$. Engineering restoration gives retained free eigenvalues $(a/r)^{\sigma_S-2q^2-d}\ge1$ up
to bounded triangular maps. After a fixed preparation to $a=A_Ir$ the
normalized complementary graph is $O(A_I^{-1})+O(\delta_g^2)$. Each unretained Taylor monomial
gains a degree and each free proper-subset fusion costs the grading
weight $\delta_g^2$. Fix $\delta_g$ first, then $A_I$, and finally the weak radius. The
interacting correction is $O(v_a)$, placing the graph and its ordinary
inverse in the common controlled ball.

Write the one-step linear blocks on ordinary and complementary
coordinates as $\left(\begin{smallmatrix}A&B\\C&D\end{smallmatrix}\right)$,
and let $H^+$ be the outgoing graph. The principal factor of the
linearized graph transform is $(D-H^+B)(A+BH)^{-1}$; the finite
nonlinear terms carry the reserved grading weights. Its strict gap
gives geometric forgetting
$q_I^j$. The finite inverse is Lipschitz there, and hence
\begin{equation}
 \|U_j\cdots U_{n-1}\|
 \le L_I^{n-j}\exp\{C_Iq_I^j/(1-q_I)\},\qquad q_I<1.
 \label{mollifier-sources:inverse-transient}
\end{equation}
Put $n_I=|I|$, $d_I=\dim\mathcal M_I$, $p_I=2d_I+2$ and
$L_I=B_I^{a_I}$. A triangular inverse and its variation use at
most $p_I$ inverse factors. Choose
$a_I<\min\{1/(2n_I),\gamma_I/(2p_I)\}$, where
$\theta_{\rm st}=B_I^{-\gamma_I}$.
Then normalized graph differences decay with exponent
$\kappa_I=\gamma_I-p_Ia_I>0$. The finite free transient changes
only the prefactor. Four endpoint derivatives add summable
$(j+1)^4q_I^j$ transients, without changing the ultraviolet loss.

At a fixed weak stop $R$ we use the mollifier's triangular inverse to
match its ordinary readout to the canonical one. The coefficients of the
resulting complete polynomial $\mathcal P_{r,I}$ are chosen at their cutoff limit, which
is independent of the containing box. The independent scalar history
starts at zero. The local value at zero field is retained even
initially. A collision of superficial degree $\omega$ has packet bound $|\mathcal K_r(f)|\le Cr^{-\omega-a_I}\|f(r\cdot)\|_{C^{R_I}}$. Put
$T_\omega f$ in the identity/contact coordinate, leaving a paired remainder $O(r^{1-a_I})$.
After at most $n_I-1$ proper-subset branchings the power $1-n_Ia_I$ is still
positive. Off its diagonal the fixed contact cutoff costs only $Cr^{-P}e^{-c/(\mu_{\rm ref}r)}$. This
bounds the scalar tail directly for the actual packet.

Use a rooted norm summing the local, stable, wall, ordinary coefficient,
paired contact-test and independent scalar seminorms with weight
$e^{\sigma|X|_R+\kappa\tau(X,z)/R}$, where $\tau(X,z)$ is the
minimum physical tree length joining its full generating support
and source anchors.
One marked root ranges only over the fixed support. The original
connected size reserve bounds its sum by
$C_I\sum_n(1+n)^{p_I}e^{-3\sigma_0n}$, independently of the domain;
scalars keep their generating support after field evaluation.
Let $X^{\rm can}(R)$ and $X^{\rm moll}_r(R)$ denote the canonical and
mollified source records at the common scale $R$. Let $d_0$ be the
distance of the source support from the wall. At $r\ll s<R$, graph
contraction, the scalar coboundary and this distance give
\begin{equation}
 \|X^{\rm can}(R)-X^{\rm moll}_r(R)\|_{\rm rooted}
 \le C\left[(s/R)^\eta
 +(R/s)^P
   \left\{(r/s)^\gamma+e^{-cd_0/s}\right\}\right].
 \label{physical-theory:record-error}
\end{equation}
Above $s$ both records use the same actual operation and its one-factor
telescope; below $s$ only the Taylor-subtracted scalar tail is small.
Choose $s/R=(r/R)^{\gamma/(P+\gamma+\eta)}$ to obtain
$C(r/R)^\nu$, $\nu=\eta\gamma/(P+\gamma+\eta)>0$.
Finite-cutoff matching uses its actual lattice inverse; replacing it
by the limiting coefficients adds $\omega_r(\epsilon)\to0$.
This modulus is local and uniform in auxiliary widths and packet
numbers. Thus write
$d_\epsilon(r)=C(r/R)^\nu+\omega_r(\epsilon)$.
Mixed lists are treated in the same finite algebra. Uniform fourth
derivative bounds give interior $C^k$ errors of order
$(r/R)^{\nu(1-k/4)}$, $k\le3$; if necessary, smooth the limiting
coefficients on a slightly larger parameter interval first.

\begin{lemma}[Matching of normalized local constructions]
\label{source-dilation:marked-recognition}
Two controlled local source constructions have identical complete
fixed-box limits if their unmarked histories converge to the same
canonical history, their full ordinary normalization data agree,
their initial independent scalar histories vanish, and they use the
same contact finite parts. Here ``controlled'' requires bounded local
graph preparation and the finite dimension/charge and positive
Taylor-subtracted scalar-tail estimates just used; ordinary matching
alone is insufficient.
\end{lemma}
\begin{proof}
At an intermediate scale $s$, graph contraction and a retarded
one-factor telescope compare ordinary and stable coordinates; an early
tail is contracted and a fixed recent window converges. The exact
scalar zero-evaluation coboundary retains both initial local values;
its independent tail below $s$ is $O(s^\eta)$. A source-attached
wall term costs $Cs^{-P}e^{-cd/s}$. Above $s$ scalar increments
depend on the current record, so the same finite-window comparison
applies to all contacts and forward wall records. The full error is
$C_{D,I,R}\{s^\eta+s^{-P}[e_\epsilon(s)+e^{-cd/s}]\}$ with
$e_\epsilon(s)\to0$. Take $\epsilon\downarrow0$ first, then
$s\downarrow0$; fixed-box integration has its original positive
denominator. The differentiated window bounds give the stated
parameter orders as well.
\end{proof}

\subsection{Kinetic completion and positive normalizers}
\label{source-admission:positive-normalizers}

Division by the physical partition function controls the volume factor
in the reflected-norm comparison. After exact completion of the excess kinetic term, positivity of the cosine weight permits the required lower bound on the partition function.
Let $v_R$ denote the scalar weak activity coordinate at scale $R$. On a
Dirichlet rectangle of side lengths at least $c_0R$, put $L=-\Delta_{\epsilon,D}$, $\kappa_0=\beta_*^{-2}$, $\kappa_\epsilon=Z_\epsilon\beta_*^{-2}$ and $\Delta\kappa=\kappa_\epsilon-\kappa_0$.
With angular prefix $C=\beta_*^2C_\epsilon[0,R]$ and positive cosine partition $F_Q$, exact
completion gives
\begin{equation}
 \begin{gathered}
 A=(I+\Delta\kappa LC)^{-1},\quad Q=CA,\\
 T=\kappa_\epsilon^{-1}L^{-1}-Q,\quad M_Q=-D^2\log F_Q(0).
 \end{gathered}
 \label{gaussian-normalization:compatible-completion}
\end{equation}
and the complete reference logarithm satisfies
\begin{equation}
 \begin{aligned}
 \mathcal H_R(\psi)&=c_{\rm kin}
       -\tfrac12\Delta\kappa\langle\psi,LA\psi\rangle
       +\log F_Q(A\psi),\\
 AM_QA&=-D^2\mathcal H_R(0)-\Delta\kappa LA.
 \end{aligned}
 \label{gaussian-normalization:phase-completion-identity}
\end{equation}

Marking the field twice in the complete rooted expansion adds only a
fixed polynomial in the support. The odd Dirichlet extension permits
local Sobolev estimates also at faces and corners, yielding $0\preceq M_Q\preceq Cv_RR^{-2}\sum_{j=0}^s(R^2L)^j$.
The bounded spectral maps $A^{\pm1}$ preserve the form bound. For $C(\lambda)=(1-\omega_R)/(\kappa_0\lambda)$ one finds

\[
 T(\lambda)=\frac{\kappa_0\omega_R(\lambda)}
 {\kappa_\epsilon\lambda[\kappa_\epsilon-\Delta\kappa\omega_R(\lambda)]}
 \le C_N\lambda^{-1}(1+R^2\lambda)^{-N}.
\]
In a rectangle of width $W$ and height $H$ the global phase inequality
implies
\[
 \det\!\big(I+T^{1/2}M_QT^{1/2}\big)^{-1/2}
 \le \frac{\E_T F_Q}{F_Q(0)}\le 1.
\]
To bound the logarithm of the determinant, count Dirichlet
eigenvalues and integrate the decreasing summand. The resulting bound $C(WH/R^2)v_R[1+\log(1/v_R)]$
is uniform in the aspect ratio. The radial integral is $\int_0^\infty\log(1+Cv_R/[t(1+t)])dt$, split at $v_R,1$.

We need the corresponding local version for a positive mask on $N$
buffered $R$-blocks. Fold one common mean in each connected component
by an integer $2\pi$ shift, using a smooth cutoff on its guard.
Phase ordering and the preceding Hessian give
$F_A(\psi)/F_A(0)\ge
e^{-Cv_R(N+A_0+B_0)}$, where $A_0$ is the sum of squared neutral
block means and the derivative form $B_0$ has trace $O(N)$.
For the retained block-mean vector $a$ and componentwise mean-zero
projection $P$, fine flows give
$\operatorname{Cov}(P a)\preceq C L_{\rm graph}^\dagger$.
A spanning tree doubled into a depth-first path gives
$\lambda_j(L_{\rm graph})\ge c(j-1)^2/N^2$ by min--max;
the pullback increases the norm and has energy at most twice the
graph energy. Hence
$\log\det{}'(I+v_RL_{\rm graph}^\dagger)\le CN\sqrt{v_R}$.
Hadamard--Fischer retains cross-component correlations and gives
\begin{equation}
 1\le F_A(0)/Z_A\le e^{C(\sqrt{v_R}+v_R)N}.
 \label{physical-theory:positive-normalizer}
\end{equation}
At a true zero wall use the grounded graph, with eigenvalue floor
$c(j-\tfrac12)^2/N^2$; no nonzero boundary trace is folded.

For the inner region $U$, guarded enlargement $V$, collar $B=V\setminus U$ and exterior
$E=D\setminus U$, consider $J=\log F_D(0)+\log F_B(0)-\log F_E(0)-\log F_V(0)$. A history cancels unless its support meets both $U$ and
$D\setminus V$. It must then cross the guard, at cost $Cv_aN(R/a)^Pe^{-cR/a}$. The rooted tree bound for
the last fixed-ratio massive shell has the same cancellation when the
shell is integrated before taking its numerical connected logarithm. We
obtain $0\le J\le Cv_RN$. Positive supermodularity and \eqref{physical-theory:positive-normalizer} now bound the physical
cavity ratio by $e^{-CN}F_D(0)/F_E(0)\le Z_D/Z_E\le F_D(0)/F_E(0)$.

\subsection{Repeated insertions and Dirichlet caps}
\label{source-admission:transfer-bounds}

Fix the cutoff. The transfer $T_W$ on a row of physical Dirichlet width $W$
has largest eigenvalue $\lambda_W$. For an insertion through a slab of physical
length $\ell$ the row-to-row operator $\mathcal A_W$ includes unmarked propagation
across the slab. Its norm is determined by repeating the insertion and
its reflection. A rotation interchanges the long direction of repetition
with the bounded footprint of the insertion, so the positive normalizer
bound controls the cost of each copy.

Fix the finite packet alphabet and its complete internal source
algebras, the support and buffer ratios, and a common weak scale $R$. Any
number of translated or reflected occurrences use the same finite normalization matrices, each with independent labels. The repetition
limit precedes cutoff removal. Accordingly, at the native cutoff and
fixed $r$, both the source--mollifier bound $d_\epsilon(r)=C(r/R)^\nu+\omega_r(\epsilon)$ and its cutoff threshold
are uniform in the repetition count and auxiliary width.

\begin{proposition}[Uniform insertion and cap estimates]
\label{physical-theory:quantitative-admission}
Let $\mathcal A_W$ be a complete packet's insertion through a
buffered time slab of length $\ell$ and bounded spatial footprint.
Put $Z_W(s)=T_W^{s/\epsilon}(0,0)$, with $s/\epsilon$ an integer.
All finite-cutoff time lengths below obey this convention. If the
insertion's complete stop-scale size is
$b$, including the local perturbation bound for separated copies
verified below,
then
\[
 \sup_{W\ge W_0}\|\lambda_W^{-\ell/\epsilon}\mathcal A_W\|
 \le Cb.
\]
For a fixed cap-smoothing length $a$ and $N>a+m$,
\begin{equation}
 B_{W,N,m}:=\lambda_W^{2m/\epsilon}
                   \frac{Z_W(2N-2m)}{Z_W(2N)}
 \le\exp\!\left\{C(1+W/R)\frac{m}{N-a-m}\right\}.
 \label{source-admission:local-cap-price}
\end{equation}
These finite-cutoff estimates are uniform in width and in the
number of repeated packets. Their constants depend on fixed
geometric ratios and the weak and source reserves.
\end{proposition}
\begin{proof}
\emph{Repetition at fixed cutoff.}
The buffers make $\mathcal A_W$ Hilbert--Schmidt at fixed cutoff and
width. The positive operator $T_W$ has simple leading eigenvalue
$\lambda_W$, and its buffered powers are trace class. If $\mu_j$ are
the eigenvalues of $\mathcal A_W^*\mathcal A_W$ in decreasing order,
then $(\sum_j\mu_j^n)^{1/n}\to\mu_1$; the same statement for
$T_W^{2\ell/\epsilon}$ gives the denominator below.
The repeated insertion gives the trace ratio
\begin{equation}
 a_n=\frac{\Tr[(\mathcal A_W^*\mathcal A_W)^n]}
             {\Tr T_W^{2n\ell/\epsilon}},\qquad
 \|\lambda_W^{-\ell/\epsilon}\mathcal A_W\|^2
       =\lim_{n\to\infty}a_n^{1/n}.
 \label{source-admission:local-repetition}
\end{equation}
After a quarter turn, the $2n$ copies occupy one slab of fixed width.
To bound the $n$th root we need a bound per insertion and a lower
bound for the partition function with its original Dirichlet caps.

\emph{The local perturbation bound for separated copies.}
We verify the stronger hypothesis in
Lemma~\ref{connected-expansions:unified-tree-count}, rather than use
its fixed-list constant $C_k$ with $k$ tending to infinity.
For one source error write
$E_S=[\eta_S](e^{B^1(\eta)}-e^{B^0(\eta)})$.
The exact switch is the idempotent matrix
\[
 T=\begin{pmatrix}0&1\\0&1\end{pmatrix},\qquad
 f(T)=\begin{pmatrix}f(0)&f(1)-f(0)\\0&f(1)\end{pmatrix}.
\]
Evaluation at $T$ is an algebra homomorphism. In particular its
off-diagonal entry, denoted by $\Delta F=F(1)-F(0)$, satisfies
$\Delta(FG)=F(0)\Delta G+(\Delta F)G(1)$.
Conjugation by $\operatorname{diag}(d^{-1},1)$ divides this entry alone
by $d$, so tensoring the switches computes exact mixed differences with one
$d_i^{-1}$ per distinguished error of size $d_i$. Each $d_i$ here equals $d_\epsilon(r)$. Leave the
diagonal norms at their original bounds. With the fixed internal nilpotent
coefficient norm tensored into the matrix norm, submultiplicativity
controls products, contractions and finite contact substitutions.
Coefficient extraction therefore acts on $e^{B^1}-e^{B^0}$, with every lower column and
its products retained. Replacing it by $e^{B^1-B^0}$ would change the observable.

Keep the generating support even if extraction shrinks the field carrier.
At the intermediate scale $s$ of \eqref{physical-theory:record-error}, classify histories by footprints met
before $s$. Single-footprint terms form that packet's exact one-factor
telescope, of complete stop norm at most $d_\epsilon(r)$. Assign terms lost on
excluding a second footprint to the two-footprint class before taking
absolute values; this preserves the signed cancellation.

A history meeting $k\ge2$ separated footprints has a connecting
tree of length at least $c\ell(k-1)$. The complete engineering loss
at its joining scale is at most $(Cs^{-P})^k$: the ordinary running
normalizations have already been summed, so there is no bare
$\epsilon^{-Pk}$ in this bound. Spend half the length reserve to get
\begin{equation}
 (Cs^{-P})^k e^{-c\ell(k-1)/(2s)}
 \le\bigl(Cs^{-P}e^{-c\ell/(4s)}\bigr)^k
 \le d_\epsilon(r)^k,
 \qquad k\ge2.
 \label{source-admission:separated-merger}
\end{equation}
The last inequality holds for all sufficiently small $r$, since
$s/R$ is a positive power of $r/R$ and the exponential dominates
every such power. Its choice is independent of $k$. The unspent
half of the length reserve pays the spatial endpoint sums and the
sum over birth scales.

Here is the local counting that makes these bounds independent of
the number of copies. Choose one coefficient radius $\eta_*>0$ in
each normalized packet algebra, after dividing its distinguished
coefficient by $d_\epsilon(r)$, or by $b_i$ for an ordinary packet.
At most a fixed number of footprints meet an enlarged $R$-block.
For the tails, the rooted exponential weight gives
\[
 \sup_x\sum_i e^{-c\,\operatorname{dist}(x,K_i)/R}<\infty,
\]
where $x$ ranges over $R$-blocks and $K_i$ is the $i$th footprint;
bounded packet density proves this by summing successive annuli.
Thus the normalized single-packet input at any block is at most
$C\eta_*$, with $C$ independent of the domain and packet count.
For the merger terms, join the internal trees to their Gaussian
forest before summing endpoints. The remaining length and size
reserves in \eqref{source-admission:separated-merger} pay exactly
the same block-degree factorials as the coloured-tree count in
Lemma~\ref{connected-expansions:unified-tree-count}. The fixed
internal algebra and bounded overlap add only a fixed factor per
occupied block. Summing the resulting geometric majorant, with
one $\eta_*$ per marked packet, bounds its local perturbation norm
by $C\eta_*$ as well. Choose $\eta_*$ so that this lies inside the
unused weak-ball margin. More explicitly, let $X_R(z)$ be the
complete stop-scale record with one label $z_i$ for each packet.
Let $\|\cdot\|_{\mathrm{loc},\mathbf b}$ be the maximum over
$R$-blocks of its rooted record sum, with the original size and
length weights and the coefficient weights of
\eqref{connected-expansions:packet-algebra-norm}. Scalar records
retain their generating roots in this sum. We have proved
\begin{equation}
 \|X_R(z)-X_R(0)\|_{\mathrm{loc},\mathbf b}
       \le C\eta_*,
 \label{source-admission:uniform-local-perturbation}
\end{equation}
with $C$ independent of the number and positions of the separated
copies, cutoff and containing box. This verifies the local hypothesis of
\eqref{connected-expansions:repeated-packet-bound} with one radius
for every copy.

Only the fixed complete packet alphabet is inverted. Disjoint stencils
kill local cross-packet contacts. Nonlocal returns enter the
connected-history sum with their labels and generating support. Above the
joining scale the same exact operations in the submultiplicative packet
algebra converge with the local, packet-count-uniform constants.

\emph{The numerator and its physical denominator.}
In \eqref{connected-expansions:repeated-packet-bound} each occupied Gaussian regulator enters once, with one unmarked
prefactor for the whole integral. The complete numerator record $\mathcal N_{N,D}$ for $N$
packets and actual unmarked scalar $E_D$ satisfy
\begin{equation}
 \int |\mathcal N_{N,D}|\,d\gamma_T
 \le e^{E_D}(Cb)^N e^{CN_R(D)}.
 \label{source-admission:repeated-numerator}
\end{equation}
Here $N_R(D)$ counts scale-$R$
blocks in the region; unequal packets give $C^N\prod_i b_i$ for $(Cb)^N$. Integrate the last
fixed-ratio centered Gaussian shell before taking its numerical connected
logarithm. Its vacuum clusters and \eqref{physical-theory:positive-normalizer} give $Z_D\ge e^{E_D-CN_R(D)}$ for the unmarked partition
$Z_D$ in the same Gaussian decomposition and scalar convention. Division
gives $(Cb)^N e^{C'N_R(D)}$: the potentially divergent $E_D$ cancels but marked scalars and
contacts persist. The rotated region has only $O(N)$ blocks.

In the short-buffer application the footprint shrinks with the buffer.
Write $a$ for its diameter and take $R\asymp a$. Summing over geometric birth
scales gives the merger estimate before error absorption,
\begin{equation}
 \sum_{s\le c a}(Cs^{-P})^k e^{-ca(k-1)/(2s)}
 \le(C'a^{-P})^k,\qquad k\ge2.
 \label{source-admission:shrinking-merger}
\end{equation}
Indeed, on
setting $u=a/s$ one has $(k-1)/k\ge1/2$. A fixed part of the exponential bounds $u^{Pk}$ by $C^k$,
and the remaining factor $e^{-c'u(k-1)}$ is summable on the geometric scales
uniformly in $k$. Increasing the fixed power for one packet to cover
profile derivatives preserves the normalized local count. The cost is
polynomial in $a^{-1}$, avoiding exponential dependence on the ratio of a
fixed macroscopic footprint to $a$.

\emph{The original caps.}
We make the scalar cancellation needed for caps explicit. Retain each
scalar's complete generating support, including preparation and
subtraction history. Its span-$d$ rooted bound is
$Ca_j^{-P}e^{-cd/a_j}$. The extra tree-length weight is
submultiplicative when supports join across a Gaussian edge, and
the exponential decay in support size bounds the weighted sum. Rechart templates by the common factor
$\lfloor R/\epsilon\rfloor\epsilon/R=1+O(\epsilon/R)$ and
average their origins over that lattice period. This convex average
of exact identities introduces no counting factor. Choose auxiliary
circumferences commensurate with that period; buffered physical lengths
can be rounded within fixed ratios. A nonwinding
translation class has exactly the circumference many translates on
a cylinder; on a rectangle its count is affine in the length, with
one contribution from each cap. Therefore every common bulk and
single-cap class cancels in any signed combination with zero area
and zero cap multiplicity. Survivors wind or reach two boundaries
and are summable by their span. Literal Gaussian and onsite constants
cancel as well. With $d_s$ sites per row and $n$ time steps, their
rectangle contribution is
$z_\epsilon d_s n+(\frac12\log Z_\epsilon+E_\epsilon\epsilon^2)d_s(n-1)$,
where $z_\epsilon$ is the scalar per site in the term proportional to
$n$. On a cylinder, both $n$ and $n-1$ are replaced by its length $k$.
The combination $(n-l)E_{\rm cyl}/k+E_l-E_n$ cancels them exactly.

Put $u_a=T_W^{a/\epsilon}\delta_0/\|T_W^{a/\epsilon}\delta_0\|$.
One has
\[
 \begin{aligned}
 |\langle u_a,u_N\rangle|^2
 &=\frac{Z_W(a+N)^2}{Z_W(2a)Z_W(2N)}\\
 &\ge e^{-C(1+W/R)}.
 \end{aligned}
\]
To see this, put $G=\kappa_\epsilon^{-1}L^{-1}$ and use the
smoothed split $T^*=GP^2$, $C^*=G(1-P^2)$, with
$P=\omega_R(L)$. The spectral formula following
\eqref{gaussian-normalization:phase-completion-identity} gives
$GP^2\preceq T$ at sufficiently fine cutoffs, where we may assume
$Z_\epsilon\le2$. Indeed the scalar inequality reduces to
$(1-\omega_R)(\kappa_0-\Delta\kappa\,\omega_R)\ge0$,
since $0\le\omega_R\le1$ and $\Delta\kappa\le\kappa_0$.
Thus $C^*\succeq Q$: it is the completed
prefix $Q$, rather than the uncompleted $C$, that is being compared.
Charge monotonicity therefore preserves its Hessian bound.
Write the retained field as $P\xi$, with covariance $G$ for $\xi$,
and pin only $\xi$ at the joining interface. If $\Delta_T\succeq0$
is the removed covariance and $T^*_{\rm cut}=T^*-\Delta_T$, global
phase semiconvexity gives
\[
 \det(I+\Delta_T^{1/2}M_{C^*}\Delta_T^{1/2})^{-1/2}
 \le\frac{\E_{T^*}F_{C^*}}{\E_{T^*_{\rm cut}}F_{C^*}}\le1,
 \qquad M_{C^*}=-D^2\log F_{C^*}(0).
\]
For the lower bound integrate
$F_{C^*}(x+z)\ge F_{C^*}(x)
e^{D\log F_{C^*}(x)z-\langle z,M_{C^*}z\rangle/2}$ in the
removed Gaussian variable. The remaining real-source Gaussian
factor is at least one. The upper bound follows from nonnegative
Fourier coefficients after centered integration.

In transverse Dirichlet modes the seam variance is
$O((\kappa_\epsilon\omega_j)^{-1})$ and the squared norm of its
harmonic extension is $O(\omega_j^{-1})$. Smoothing and the
completed Hessian bound therefore give for the pinning determinant
\[
 C\sum_{j\ge1}\log\{1+Cv_R(R\omega_j)^{-2}(1+R\omega_j)^{-q}\}
 \le C\sqrt{v_R}(1+W/R),\qquad \omega_j\ge cj/W.
\]
Hadamard's inequality is applied to the Gram matrix of mode columns;
no commutation of the Hessian with $L$ is required. Delete an
$O(R)$ activity strip at the seam. The latent pin and finite range
make the two remaining arms independent. Restoring the strip uses
the cavity bound. In the three pairings $(a,N)$, $(a,a)$ and
$(N,N)$, their positive arm factors cancel, and the scalar contrast
has zero area and cap count. The surviving scalar histories cross
both ends and their span sum costs only $C v_R(1+W/R)$.

For completeness, if the Dirichlet row has $M$ interior sites,
$W=(M+1)\epsilon$, its free frequencies are
\[
 \mu_j=4\sin^2\frac{\pi j}{2(M+1)},\qquad
 \omega_j=\epsilon^{-1}\operatorname{arcosh}(1+\mu_j/2)
       \ge cj/W,\quad 1\le j\le M.
\]
Writing $\lambda_j$ for the largest eigenvalue of the corresponding
one-mode row transfer, the Mehler formula is a constant times
$\lambda_j^{s/\epsilon}(1-e^{-2s\omega_j})^{-1/2}$ for the
zero-cap kernel at time $s$. Its width constant and $\lambda_j$
cancel in the overlap ratio, leaving exactly
\[
 \frac{\sqrt{(1-e^{-4a\omega_j})(1-e^{-4N\omega_j})}}
      {1-e^{-2(a+N)\omega_j}}
       \ge\sqrt{1-e^{-4a\omega_j}},\qquad N\ge a.
\]
The last inequality follows by writing
$p=e^{-2a\omega_j}$, $q=e^{-2N\omega_j}$ and using
$0\le q\le p<1$ and $1-q^2\ge(1-pq)^2$.
The decreasing function $-\log(1-e^{-x})$ has integral
$\pi^2/6$ on $(0,\infty)$. Summing it at $4a\omega_j$ bounds
the negative logarithm of the free overlap by $CW/a$, uniformly
in $M$. The determinant, strip restoration and free-mode bounds
together prove the overlap estimate.
At fixed width and cutoff the positive compact transfer projects
onto its simple ground state. Thus
$f(t)=\log\langle u_a,e^{-2tH_W}u_a\rangle$, with
$H_W=-\epsilon^{-1}\log(T_W/\lambda_W)$, is convex, decreasing
and bounded below by $-C(1+W/R)$, since the limiting ground-state
overlap has that lower bound. Also $f(0)=0$. Convexity gives
\[
 \frac{f(t)-f(t-m)}m
       \ge\frac{f(t-m)-f(0)}{t-m}
       \ge-\frac{C(1+W/R)}{t-m}.
\]
The identity
$\log B_{W,N,m}=f(N-a-m)-f(N-a)$ now proves
\eqref{source-admission:local-cap-price}. This ground-state limit
has proved a finite-cutoff inequality; it has not replaced the
selected physical order of limits.

\emph{Rotation and the periodic common coordinate.}
In \eqref{source-admission:local-repetition} the norm limit can be taken along the subsequence of repetitions
making its circumference commensurate with the scalar-allocation
period. Rotate this cylinder by a quarter turn. Its periodic circumference
is $L=2n\ell$, its temporal zero caps are distance $W$ apart, and
all $2n$ packets lie in one slab of fixed width $w$, with insertion
$\mathcal B_n$. If $t=(W-w)/2$, the exact rotated integral is
\[
 a_n=\frac{\langle T_L^{t/\epsilon}\delta_0,
                   \mathcal B_n T_L^{t/\epsilon}\delta_0\rangle}
                {Z_L(W)}.
\]
Here $T_L$ is the transfer on the periodic row after the rotation,
and $Z_L$ is its zero-cap kernel on the noncompact covering space.
Cauchy--Schwarz gives
$a_n\le C_{\rm cap}\|\lambda_L^{-w/\epsilon}\mathcal B_n\|$,
where $C_{\rm cap}=\lambda_L^{w/\epsilon}Z_L(W-w)/Z_L(W)$.

The periodic row has one noncompact common coordinate. Floquet
decomposition under $h\mapsto h+2\pi$ gives fibers $\theta$,
whose absolute kernels are dominated by the positive compact zero
fiber; continuity gives $\lambda_L=\|T_{L,0}\|$. The source packets
are common-shift periodic. Writing $\kappa$ for the angular stiffness,
a winding-$j$ bridge satisfies
$T_L^{s/\epsilon}(0,2\pi j\mathbf1)
\le e^{-2\pi^2\kappa L j^2/s}Z_L(s)$, so its quotient cap obeys
$1\le\widetilde Z_L(s)/Z_L(s)\le1+C\sqrt{s/L}$.
Take fixed buffered times $t_1>t_0$ and $\tau>0$, and put
$\alpha=(t_1-t_0)/\tau$. Scalar cancellation and the retained
normalizers give
\[
 \Xi_L=(\Tr T_{L,0}^{\tau/\epsilon})^{\alpha}
          \frac{Z_L(t_0)}{Z_L(t_1)}
 \le\Xi_L^{\rm free}e^{C(1+L/R)}.
\]
In this contrast the signed area is
$\alpha\tau+t_0-t_1=0$, and the two zero-cap terms have opposite
cap multiplicities. Hence every common bulk or single-cap scalar
history cancels. A surviving history winds around the time circle
or reaches both caps. Its fine-scale sum is at most
\[
 C(1+L/R)\sum_{a_j\le R}
       v_{a_j}(R/a_j)^P e^{-c\min(\tau,t_0)/a_j}
       \le C'(1+L/R).
\]
The final fixed-ratio Gaussian shell is integrated before the
numerical logarithm, and has the same rooted bound. All the
remaining integration regions have fixed buffered heights, hence
$N_R=O(1+L/R)$, so their positive normalizers cost
$e^{\pm C(1+L/R)}$. On the periodic quotient the common phase is
integrated over one $2\pi$ cell. Convolution of this phase with the
constant-mode part of the prefix Gaussian preserves Haar measure;
the local positive Gaussian splitting is therefore exact on the
quotient as well. The neutral means use the mean-zero determinant
in \eqref{physical-theory:positive-normalizer}. Covering zero caps
are compared with quotient caps by the winding bound above. These
observations justify the displayed comparison with the original
scalar normalization.

On a periodic row of $M=L/\epsilon$ lattice sites, the nonzero
frequencies are
\[
 \begin{gathered}
 \mu_j=4\sin^2(\pi j/M),\qquad 1\le j<M,\\
 \omega_j=\epsilon^{-1}\operatorname{arcosh}(1+\mu_j/2)
       \ge c\min(j,M-j)/L.
 \end{gathered}
\]
The two real coordinates of a conjugate Fourier pair are counted
separately. For the corresponding one-mode row transfer $T_j$, with
largest eigenvalue $\lambda_j$, the Mehler formula gives
$\Tr(T_j/\lambda_j)^{\tau/\epsilon}=(1-e^{-\tau\omega_j})^{-1}$;
the zero-cap formula above supplies its other factor. Thus all row
eigenvalues cancel from $\Xi_L^{\rm free}$, leaving the factors
\[
 (1-e^{-\tau\omega_j})^{-\alpha}
 \left(\frac{1-e^{-2t_1\omega_j}}{1-e^{-2t_0\omega_j}}\right)^{1/2}.
\]
The same integrable logarithm bounds the sum of their logarithms by
$CL\alpha/\tau+CL/(2t_0)$. The common mode contributes
$(\sum_{q\in\mathbb Z}e^{-\tau q^2/(2\kappa L)})^\alpha\sqrt{t_1/t_0}$,
also at most $e^{C(1+L/R)}$. Since
$\lambda_L\le(\Tr T_{L,0}^{\tau/\epsilon})^{\epsilon/\tau}$,
choose $t_0=W_0-w$, $t_1=W_0$ in the bound for $\Xi_L$.
This gives $C_{\rm cap}\le e^{C(1+L/R)}$ at one fixed buffered
$W_0$. The logarithm of the covering cap kernel is convex, so
$Z_L(W-w)/Z_L(W)$ is nonincreasing in $W$. The same bound holds
for every $W\ge W_0$.

We also need a normalized trace, rather than just an upper bound
for $\lambda_L$. The Rayleigh quotient of a quotient cap smoothed
for time $a/2$ gives
$\lambda_L^{\tau/\epsilon}\ge
\widetilde Z_L(a+\tau)/\widetilde Z_L(a)$.
For fixed buffered $s>0$ it follows that
\begin{equation}
 \Tr(T_{L,0}/\lambda_L)^{s/\epsilon}
 \le \Tr T_{L,0}^{s/\epsilon}
 \left[\frac{\widetilde Z_L(a)}{\widetilde Z_L(a+\tau)}\right]^{s/\tau}
 \le e^{C(1+L/R)}.
 \label{source-admission:periodic-normalized-trace}
\end{equation}
For the second inequality use the winding comparison to pass to
covering caps. Again the signed area and cap multiplicities vanish,
so the same scalar cancellation and retained normalizer bounds apply.
Its nonzero-mode log sum is bounded by $CL/s+CsL/(\tau a)$, and
its common factor is $[1+\sqrt{2\pi\kappa L/s}](1+\tau/a)^{s/(2\tau)}$.

\emph{The Hilbert--Schmidt bound and the repetition limit.}
The Hilbert--Schmidt square of each fiber of $\mathcal B_n$
is a periodic rectangle of size $L\times2w$ with $4n$ packets.
In winding sector $j$ the affine lift has energy
$\kappa\pi^2j^2L/w$. There are $O(1+L/R)$ stop blocks. The
derivative regulator is block-additive, so its coefficient can be
chosen once to consume a fixed fraction of this energy. The degree
per packet is fixed, although the total polynomial degree grows
with $n$. If that degree is at most $p$, its winding-dependent
factor is bounded by $(C(1+|j|)^p)^{4n}$. Since $L=2n\ell$,
another fixed fraction of the affine energy absorbs this factor
at a cost $C_1^{4n}$: use the finite supremum of
$(1+|j|)^{4p}e^{-c j^2}$ and then raise it to the $n$th power.
The remaining winding sum
$\sum_j e^{-c\kappa j^2L/w}$ is uniformly bounded for $n\ge1$.
Floquet phases have modulus one, so the bound is uniform in the
fiber parameter.

Form the complete $4n$-packet coefficient before this winding
estimate. Dividing \eqref{source-admission:repeated-numerator}
by the same positive partition gives explicitly
\[
 \sup_\theta\Tr(\mathcal B_{n,\theta}^*\mathcal B_{n,\theta})
 \le (Cb)^{4n}e^{C(1+L/R)}\Tr T_{L,0}^{2w/\epsilon}.
\]
Taking its square root and using
\eqref{source-admission:periodic-normalized-trace} at $s=2w$
therefore gives
$\|\lambda_L^{-w/\epsilon}\mathcal B_n\|
\le(Cb)^{2n}e^{C(1+L/R)}$.
Together with the rotated cap comparison, this bounds
$a_n\le(Cb)^{2n}e^{C'(1+2n\ell/R)}$. Take the $n$th root in
\eqref{source-admission:local-repetition}, at fixed native cutoff
and original width, and then a square root. The result is
$\|\lambda_W^{-\ell/\epsilon}\mathcal A_W\|
\le Cb\,e^{C'\ell/R}$, which is the stated bound because
$\ell/R$ is a fixed geometric ratio. For the shrinking packets in
\eqref{source-admission:shrinking-merger}, that ratio stays bounded
as $a\downarrow0$. All cutoff thresholds used in the local
perturbation estimate were independent of $n$ and $W$, so the
auxiliary repetition limit has exchanged neither physical limit.
\end{proof}

\subsection{Identification of the source and height spaces}
\label{source-admission:identification}

\begin{theorem}[Complete local admission]
\label{source-admission:local-gram-limits}
Every fixed complete source list has bounded positive-time physical
cylinder approximants chosen independently of the containing domain
and cutoff. All ordinary complete coefficients and joint source--source and
source--core Grams converge in
the original bounded-family coefficient/contact-test topology, and
the resulting source OS space is canonically the height/derivative
space, intertwining translations and preserving null vectors.
The approximation is uniform on large squares, their fixed bounded
dilates and the tall rectangles with width $W$, half-height $W^2/R$.
\end{theorem}
\begin{proof}
Apply the insertion bound to the complete source--mollifier switch,
with $b=d_\epsilon(r)$. Its exact zero-cap Gram is
$B_{W,N,m}\|A_{W,r}u_{N-m}\|^2$, where the slab and buffer time
$m$ is fixed. The cap bound is uniform on the stated geometries,
and hence
\begin{equation}
 \sup_{D\ \text{sufficiently large}}
 \limsup_{\epsilon\downarrow0\ \text{in}\ D}
 \|O_{\epsilon,S}-P_{r,S,\epsilon}\|_{\rm OS,D}^2
       \le C(r/R)^{2\nu}.
 \label{source-admission:local-admission-conclusion}
\end{equation}
At fixed $r$, derivative mollifier rows have compact neutral flows
of bounded energy; differences of nearby rows have energy
$O(|x-y|^2)$. Periodic factors are bounded and their increments
obey $|e^{iqu}-e^{iqv}|\le|q||u-v|$. Gaussian moments and
H\"older therefore justify fixed anchor quadrature and bounded
smooth truncation of the derivative polynomials, uniformly in domain
and cutoff. They produce the claimed bounded cylinders in ordinary
$L^2$, hence in OS norm. Choose the approximation first, then a
domain threshold, then the cutoff threshold for each such domain.

For a finite list write $\Psi$ and $F_r$ for its source and comparator
synthesis maps. A single fixed approximant bounds $\|\Psi\|$ uniformly. By the
error estimate, $F_r$ is Cauchy in the selected height space. From $\|\Psi-F_r\|\le e$ and
$\|\Psi\|\le B$ we get $\|\Psi^*\Psi-F_r^*F_r\|\le(2B+e)e$. Adjoining any other source or core list and taking its
finite Gram through the ordered limits gives an isometry that preserves
null spaces even for singular Grams. Rescaling the test slots and using
multilinearity recovers the original distribution seminorms. Applying
the comparison to translated lists gives intertwining. The range lies in
the height core and contains the protected derivative core, already
proved equal to the height core.
\end{proof}

\section{Screening and the physical mass}
\label{sec:screening}
\label{physical-theory:screening}
\label{physical-theory:screening-mass-section}

The cyclic space of the height law selected by pinning is, by local admission,
the full observable space. To obtain a positive mass on this space, consider a
strip with nonzero lift. A string of bad pinning blocks gives its transfer
operator a positive threshold. At critical Gaussian normalization a local current
identity transfers this bound to the neutral theory. Free cavities realize the
identity in the microscopic law with its original Dirichlet caps.

\subsection{A threshold in the lift sectors}
\label{physical-screening:charged-sectors}

On a strip of physical width $W$ denote by $T_{j,W}$ the finite-cutoff row transfer for
boundary values $0$ and $4\pi j$, $j\in\mathbb Z$. Put $\lambda_j(W)=\|T_{j,W}\|$. The ordinary zero-boundary transfer is $T_{0,W}$.

\begin{lemma}[Charged transfer threshold]
\label{physical-screening:charged-threshold}
At a sufficiently early positive canonical endpoint, there is a
physical constant $\mu>0$, independent of the strip width and
sufficiently fine cutoff, such that
$\lambda_j(W)\le e^{-\mu\epsilon}\lambda_0(W)$ for every $j\ne0$.
\end{lemma}
\begin{proof}
Fix fundamental pinning parameters $d_0<\pi/10$, $p<1/64$ and
a finite physical scale family $a=m\epsilon\in[a_-,a_+]$.
On a strip of width $W$, impose spatial boundary values $0,4\pi j$;
make time periodic with length $T$, a multiple of $a$. Let $Z_j(W,T)$
and $\lambda_j(W)$ be its partition function and positive transfer
ground eigenvalue. Apply pinning to the sector sum $Z_\Sigma=\sum_jZ_j$,
not to a fixed affine sector. To justify this, enlarge the cylinder
by exterior columns and add centered precision
\[
 \frac\eta2\sum_xh_x^2+\frac\eta2j^2
 +\frac v2\sum_{x\ \text{left}}h_x^2
 +\frac v2\sum_{x\ \text{right}}(h_x-4\pi j)^2.
\]
The physical free tests extend to this real--integer Gaussian
system by the same core-mesh and comb comparison. Hard-pinning
$v\to\infty$ makes the exterior columns constant, with equal
cosine weights in every integer sector; removing $\eta$ then gives
the proper sector-summed law. Cutting full selected cells or pairs
subtracts only PSD forms. Therefore the identical prescribed-set
bound holds for its complete block averages, including wall blocks.

Good neighbors share the same nearest well. A good path cannot join
the exterior labels $0$ and $2j$ when $j\ne0$. Planar crossing
duality on the cylinder supplies a bad sup-connected circuit winding
around time, of at least $T/a$ cells. Root it on a fixed temporal
row and count animals to obtain
$\Pr_\Sigma(j\ne0)\le C(1+W/a)e^{-cT/a}$.
The exact affine decomposition $h=4\pi jx/W+\varphi$ and positive
Fourier phase maximum give
$Z_j/Z_0\le e^{-\kappa_\epsilon(4\pi)^2j^2T/(2W)}$, hence
$Z_\Sigma/Z_0\le1+C\sqrt{W/T}$.
Taking the temporal root at fixed cutoff and width removes these
polynomial factors and proves
\begin{equation}
 \lambda_j(W)/\lambda_0(W)\le e^{-\mu\epsilon},
 \qquad j\ne0,\quad\mu=c/a_+>0.
 \label{physical-theory:sector-gap}
\end{equation}
The Dirichlet quadratic form confines the row, so its bounded cosine
tilt has a compact positive transfer with simple top eigenvalue and
trace-class positive powers. Widths smaller than a fixed pinning
length satisfy the same type of bound directly from the affine
energy estimate. This establishes a charged threshold, not yet the
neutral gap.

\end{proof}

\subsection{The boundary current identity}
\label{physical-screening:critical-currents}

The local identity is computed in the critical free theory. Its boundary
phase matters: the zero-Dirichlet cap must be invariant under the
current charge, and the charged pair must have the same normalization
as the neutral derivative.

Use $z=t+ix$, $\partial=(\partial_t-i\partial_x)/2$ and
$\bar\partial=(\partial_t+i\partial_x)/2$. The critical action is
$S_0=(16\pi)^{-1}\int|\nabla h|^2$ and the chiral covariance is
$\langle\varphi(z)\varphi(w)\rangle=-2\log(z-w)$.
Only for the auxiliary current sectors take the $4\pi$ compact
extension; its zero-lift simply connected Dirichlet law is unchanged.
For $h=\varphi_L+\varphi_R$, with
$J^\pm=J^1\pm iJ^2$, choose
\[
 \begin{aligned}
 J_L^3&=\tfrac i2\partial\varphi_L,  & J_L^\pm&=:e^{\pm i\varphi_L}:,\\
 J_R^3&=-\tfrac i2\bar\partial\varphi_R, & J_R^\pm&=:e^{\mp i\varphi_R}:.
 \end{aligned}
\]
For electric charge $e$ and lift $2\pi n$, the chiral weights are
$p_L=e+n/4$, $p_R=e-n/4$;
left and right raising currents have $(e,n)=(1/2,2),(-1/2,2)$.
Both change the spatial lift by $4\pi$. At stiffness $Z/(8\pi)$
their dimension is $(Z+Z^{-1})/2=1+(Z-1)^2/(2Z)$.
Dirichlet doubling gives $h=\varphi(z)-\varphi(\bar z)$.
Since $(z-\bar z)^{-2}=-|z-\bar z|^{-2}$ whereas the real
Dirichlet vertex expectation is positive, its flat-normalized
cocycle gives the singlet
\begin{equation}
 \mathcal K=\sum_aJ_L^aJ_R^a
       =\tfrac14:\partial h\,\bar\partial h:-:\cos h:_{\rm CFT}.
 \label{physical-theory:singlet}
\end{equation}
Thus $+\lambda\int\mathcal K$ has the calibrated tangent
$Z-1=\pi\lambda$, $2y=\lambda/\sqrt{\kappa_{\rm lat}}$,
$v=2\pi\lambda$, $x=-v$, where
$\kappa_{\rm lat}=e^{-4\gamma_E}/64$.

At a temporal cap let $U$ translate the open right half-row by $4\pi$.
Symmetric ordering assigns half this jump to the endpoint. The
corresponding endpoint commutator $[h,\widetilde h]=4\pi i$,
$U=e^{i\widetilde h/2}$ gives, in electric-then-shift ordering,
$J_L^+=e^{ih/2}U$, $J_R^+=-e^{-ih/2}U$.
Their charge combination $2i\sin(h/2)U$ annihilates the zero cap;
their spatial combination is $2\cos(h/2)U$.
Equivalently $J_Ldz=J_Rd\bar z$ at every Dirichlet boundary,
including $dz=i\,dx$, $d\bar z=-i\,dx$ on the temporal cap.
Wick contraction gives the level-one OPEs
$J^3J^3\sim[2(z-w)^2]^{-1}$,
$J^3J^\pm\sim\pm J^\pm/(z-w)$ and
$J^+J^-\sim(z-w)^{-2}+2J^3/(z-w)$.
Their contour charges act by $\mathfrak{su}(2)$; boundary gluing
cancels the Ward contour, the zero cap is invariant, and
$\mathcal K$ is a singlet. Every two-current tensor with any finite
number of singlet insertions is therefore proportional to
$\delta_{ab}$. For $K^a=J_L^a-J_R^a$,
$K^3=(i/2)\partial_t h$, so its reflected charged pair is exactly
twice its Cartan pair. These free-boson conventions agree with
\cite[\S\S2--3]{GaberdielRecknagelWatts2002} and the zero-Dirichlet
member in \cite[\S3]{GaberdielRecknagel2001}; the calculation here
fixes the phases needed for the physical field.

\subsection{Comparison through free cavities}
\label{physical-screening:cavity-comparison}

Delete the cosine and the excess kinetic term in a guarded square.
The cavity interior is Gaussian at critical stiffness. For a given
trace the neutral current differentiates its harmonic extension. The
charged current also changes the trace winding. The regulator comparison
will use this charged kernel together with its exterior continuation.

\begin{lemma}[Free cavity kernels and their continuation]
\label{physical-screening:cavity-kernels}
Fix finitely many disjoint guarded cavities of size $\delta$ in a
Dirichlet rectangle $D$, with positive mutual and boundary clearances.
At critical stiffness $\kappa_0=1/(8\pi)$ their point-split currents
define Gaussian boundary kernels
$\mathcal B^a_{\epsilon,\delta}$, $a=3,+,-$.
The Cartan kernel preserves the ordinary trace space. The charged
kernels map it to the affine trace spaces with circulation $\pm4\pi$.
After the flat current normalization, their conditional kernels and
each prescribed finite number of field derivatives have a common
Gaussian regulator bound, uniformly as $\epsilon\downarrow0$ at
fixed $D,\delta$. They converge to the corresponding critical
current kernels. Their exterior continuations are independent of
the choice of a cut with the same endpoint and boundary germs.
\end{lemma}
\begin{proof}
Let $\xi$ be the ordinary boundary trace of one cavity, and let
$H_{\epsilon,\delta}[\xi]$ be its discrete harmonic extension.
Writing $\mathcal G_{\epsilon,\delta}(\xi)$ for the uninserted
critical Gaussian cavity kernel, conditional integration gives
\begin{equation}
 \mathcal B^3_{\epsilon,\delta}(\xi)
 =\mathcal G_{\epsilon,\delta}(\xi)
       \frac i2\partial_t^\epsilon
                     H_{\epsilon,\delta}[\xi](x).
 \label{physical-screening:cartan-kernel}
\end{equation}
Here $x$ is the insertion center and $\partial_t^\epsilon$ is the
symmetric finite difference used in its point splitting.

In the prescribed ordering the right-half-row translation $U$ and electric
factors give the spatial raising kernel its factor $(e^{ih/2}+e^{-ih/2})U$. Reflection gives the
lowering kernel as its adjoint. The charges $\pm1/2$ occur with a lift change $4\pi$,
outside the integer-vertex source calculus.

In the punctured free cavity choose the discrete Hodge representative
$\omega_{\epsilon,\delta}$ with circulation $4\pi$, zero
divergence and zero normal component at the boundary. The charged
trace is $\xi$ plus the corresponding winding trace. With this
affine identification, summation by parts gives
$\langle\omega_{\epsilon,\delta},du\rangle=0$ for every ordinary
single-valued fluctuation $u$. Its magnetic energy is consequently
deterministic. If $G_{\epsilon,\delta}$ is the zero-boundary angular
covariance in the cavity, its magnetic and electric singular factors
are respectively
\[
 \exp\{-\tfrac12\kappa_0\|\omega_{\epsilon,\delta}\|^2\},
 \qquad
 \exp\{-\tfrac18G_{\epsilon,\delta}(x,x)\}.
\]
Their logarithmic exponents are
\begin{equation}
 \frac{\kappa_0(4\pi)^2}{4\pi}=\frac12,
 \qquad
 \frac{(1/2)^2}{4\pi\kappa_0}=\frac12.
 \label{physical-screening:cavity-dimensions}
\end{equation}
The product therefore has order $\epsilon/\delta$. Multiplication
by $\epsilon^{-1}$ and the fixed flat shape constant leaves a
finite kernel of dimension one. After division by
$\mathcal G_{\epsilon,\delta}$ and the affine trace identification,
its remaining electric factors are phases of the harmonic trace,
combined with the cocycle already fixed by the boundary calculation.
For real traces they have bounded modulus. Their field derivatives
are products of smooth harmonic observations; the Cartan ratio in
\eqref{physical-screening:cartan-kernel} is one such observation.
The inner guard separates these observations from the trace boundary.
The Poisson-kernel bounds at this fixed separation and the reserved
Gaussian derivative regulator thus give the asserted bounds and
cutoff convergence. In particular the winding trace must be retained:
putting this kernel on the zero-winding trace space would undo the
orthogonal decomposition and introduce an artificial endpoint term.

To continue the kernel on the lattice, join the oppositely charged cores by a
dual cut $\gamma$ with signed integer edge cochain $\sigma_\gamma$. Take the exterior covariant
increment to be
\[
 d_\gamma h=dh-4\pi\sigma_\gamma.
\]
Thus $d_\gamma h$ enters the kinetic form and the potential stays $\cos h$. Moving the cut by an
integer vertex function $k$ amounts to the change of variables
\begin{equation}
 (h,\sigma_\gamma)\longmapsto
          (h+4\pi k,\sigma_\gamma+dk).
 \label{physical-screening:cut-gauge}
\end{equation}
Neither action term changes, and the electric phases $e^{\pm ih/2}$ are invariant. For a
common true-wall germ choose $k=0$ there to keep the Dirichlet caps. In row
transfer language this is the sector change $0\to1$ of $U$, reversed by the opposite
insertion. Away from the two free cores closedness gives smooth local
representatives of a fixed affine field differing by $4\pi$ constants. At fixed $D,\delta$
their derivatives and Dirichlet energy are finite. These charts serve the
ultraviolet comparison only: the charged kernel changes trace space and is no
local integer vertex.
\end{proof}

\begin{proposition}[Fixed-cavity current comparison]
\label{physical-screening:fixed-cavity-comparison}
Fix a smooth positive-time test $F$, and choose a cavity size $\delta$
below its support clearances. At a sufficiently early canonical
endpoint there are normalized cavity insertions
$O^a_{\epsilon,\delta}(F)$, with a common local scalar and a common
ordinary current normalization, having the following properties.
Write $C^{33}_{\delta,D}(s)$ for the reflected Gram of the Cartan
insertion and its physical-time translate by $s$, and
$C^{+-}_{\delta,D}(s)$ for the corresponding charged reflected pair.
Whenever the cavity slabs are disjoint and contained in a fixed
Dirichlet rectangle, cutoff removal gives
$C^{+-}_{\delta,D}(s)=2C^{33}_{\delta,D}(s)$.
The Cartan insertion has bounded local cylinder approximants and
converges in the selected reflected norm to
$(i/2)\partial_t h(F)$ as $\delta\downarrow0$, with the quantitative
bound \eqref{physical-theory:neutral-cavity}. The approximation is
uniform on squares and on the tall rectangles admitted in
Theorem~\ref{source-admission:local-gram-limits}.
\end{proposition}
\begin{proof}
\emph{A regulator that preserves this identity.}
At resolution $r$ impose wall clearance $r$ and mutual separation at least $r$ on
the centers. In the signed gas
\[
 \mathfrak Z^{\rm hc}_{r,D}(\lambda;\mathcal O)
 =\sum_n\frac{(-\lambda)^n}{n!}
   \int_{\Omega_{n,r}}\left\langle\mathcal O
                   \prod_{i=1}^n\mathcal K(x_i)\right\rangle_{0,D}
                                      d^{2n}x.
\]
the integration region $\Omega_{n,r}$ contains the permitted configurations of $n$ centers,
and $\langle\cdot\rangle_{0,D}$ denotes critical zero-Dirichlet expectation. Packing terminates the
series, wall clearance removes the boundary one-point singularity, and each
summand obeys the invariant tensor identity. Point split, then integrate the
independent fields at the allowed centers for the positive continuum prefix of
range $<r/4$. The exact packet is $k_r(\psi;x)=\frac14:\partial\psi\bar\partial\psi:_{Q_r}
-a_{\rm fr}r^{-2}\cos\psi+c_{r,D}(x)$, with retained covariance $Q_r$, flat prefix
normal-ordering constant $a_{\rm fr}>0$ and local scalar contraction $c_{r,D}(x)$ kept inside the
insertion.

Multiply the extended product by the hard-core indicator. Occupancy at most
four per $r$-cell permits an exact finite sum inside each cell. Expand only
between neighbors: there are at most sixteen particle edges per pair and
boundedly many neighbors, hence connected activity $(C|\lambda|)^{|X|}$. The estimate $r^2|D^mk_r|\le C_m(1+\mathcal E_{B^*})$
requires one reserved Gaussian regulator per occupied cell for its local form
$\mathcal E_{B^*}$. Replacing the one-cell factor with its local exponential costs $O(\lambda^2)$.
Projection and quadratic preparation leave errors $O(t^2)$ in the stable record and
$O(t)$ on the even wall, with $t=2\pi\lambda$.

Above $r$ use the canonical full limiting map. Denote the marginal coordinates
by $g_j=(x_j,v_j)$, the displacement from the invariant graph by $R_j=K_j-H(g_j)$, and the marginal and
stable outputs by $(f_0,p_0)$. Its graph defect $G(g,K)=p_0(g,K)-H(f_0(g,K))$ vanishes for $K=H(g)$. Integrating along
the segment from $H(g_j)$ to $K_j$ yields $R_{j+1}=A_jR_j$, where
\[
 A_j=\int_0^1 D_K G\big(g_j,H(g_j)+sR_j\big)\,ds,
 \qquad \|A_j\|\le\vartheta<1.
\]
In coordinates $u=-x$ consider the normal displacement $d_j$ from the outgoing curve
$u=\varphi(v)$. Its recursion reads
\[
 d_{j+1}=(1-\ell_bv_j+O(v_j^2))d_j+e_j,
 \quad |e_j|\le Cq^jv_j^2,\quad d_1=O(t^2),\quad\ell_b=\log b.
\]
The cone clock bounds the forward multiplier by $Cv_k/v_j$ and gives $|d_j|\le Ct^3/v_j$. At $r_N=Rb^{-N}$,
continuity between the lower and upper reciprocal-clock trials permits the choice
$v_N=\rho$ with $t\asymp(N+\rho^{-1})^{-1}$. For tangential error, run backward from this common endpoint using
the multiplier $C(v_j/v_k)^2$. This gives $|g_j-\bar g_j|\le Ct^3/v_j$ and $\sum_{j\le N}|g_j-\bar g_j|\le Ct$, using $\sum_{k\ge j}v_k^{-2}\le Cv_j^{-3}$ and $\sum_jv_j^{-1}=O(t^{-2})$. Stable and wall
contractions identify the records on each fixed physical window. Scalars cancel
separately within each regulator's normalized ratio. Fixed-box Gaussian
regulators integrate the nonscalar comparison. Positivity of the canonical
denominator ensures that the auxiliary denominator is eventually nonzero even
though the auxiliary gas is signed.

\emph{The masked, decorated comparison.}
Use the same $C_4$-symmetric cavities and inner guards in both
regulators. In the singlet gas omit centers in the masked region;
in the native law delete both the cosine and the excess kinetic
term there. The critical Gaussian part is retained. The current
kernels are those of Lemma~\ref{physical-screening:cavity-kernels}.
They are integrated against the exterior density, with the original
unmasked partition function in the denominator.

Normalize all three components by the same complete one-cavity scalar
$e^{-E_{\epsilon,\delta}}$ and the same ordinary Cartan matching
factor $q_{\delta,\epsilon}$. Scalars joining distinct cavities remain
in the multipacket record. No containing-box expectation enters this
normalization. Let the resulting operators be $O^a_{\epsilon,\delta}(F)$.
The one-cavity owner and the flat Gaussian kernel fix these factors
before the containing box is chosen. In the auxiliary regulator use
the same local prescription.

Here is the extension of the unmarked comparison required by a
charged pair. Fix $D,\delta$ and the separated insertion supports,
and choose an intermediate scale $s_{\rm uv}$ smaller than their clearances
and than the free guards. A connected ultraviolet history of diameter
less than a fixed multiple of $\delta$ lies in a chart of the
punctured exterior, unless it surrounds a cavity. In a chart the
operation is the same local Gaussian integration, evaluated on the
ordinary field plus a fixed smooth lift $w$. On overlaps the lifts
differ by $4\pi$ and \eqref{physical-screening:cut-gauge} identifies
the records exactly. A history which cannot be put in such a chart
has diameter at least $c\delta$; its unused support decay bounds its
contribution below $s_{\rm uv}$ by
\begin{equation}
 C_{D,\delta}s_{\rm uv}^{-P}e^{-c\delta/s_{\rm uv}}.
 \label{physical-screening:wrapping-tail}
\end{equation}
Indeed a history born at scale $a_j<s_{\rm uv}$ must contain at least
$c\delta/a_j$ connected blocks. Reserve part of its exponential
size decay for this lower bound and sum the remaining rooted
history bound; the geometric scale sum is bounded by the right side
of \eqref{physical-screening:wrapping-tail}, with a smaller $c$.

A smooth lift changes the interaction to $\cos(h+w)$ and contributes a linear kinetic cross
term with $dw$. Keep this deterministic affine term in the Gaussian completion once
at reconstruction. The constant phases $e^{inw}$ have unit modulus, so constant
translation is isometric in the analytic charge norm. Subtract the constant part:
the scale-$a_j$ profile jet is $O_{D,\delta}(a_j/\delta)$ and its Taylor remainder below $s_{\rm uv}$ sums to $C_{D,\delta}s_{\rm uv}/\delta$. Apply
the even-mask recursion before evaluating the profile. With $\tau_wF(h)=F(h+w)$ and profile $w_j$ in
scale-$a_j$ coordinates, the full operation $\mathcal M_j$, including projection and scalar
zero-evaluation, becomes
\[
 \mathcal M_j^w
   =\tau_{w_{j+1}}\mathcal M_j\tau_{-w_j}.
\]
Intermediate translations cancel, leaving the even-interface contraction
conjugated only at its endpoints. Thus no step recharges the background or
projects away its linear terms. On the reserved field radii at any fixed stop
below $\delta$ the nonconstant profile is bounded. Undoing the conjugacy restores both
original normalizations with their scalar conversions.

The fractional electric charge causes no further interacting
ultraviolet subtraction. Its point singularity is inside the exactly
free core and has already been removed in
\eqref{physical-screening:cavity-dimensions}. At fixed $\delta$
the remaining boundary germ is a smooth observation at that scale.
One can see this without an integer-charge expansion. A smooth
radial probability density in the free inner guard averages every
harmonic extension to its value at the center. Conditional Gaussian
integration represents $e^{\pm iH[\xi](x)/2}$ by the corresponding
smooth electric smear, with its finite Gaussian variance factor;
differentiating the averaging density gives the Cartan observation.
The discrete harmonic kernels have the same limit and uniform
guarded bounds. Write $\Gamma_j^h=\beta_*^2\Gamma_j$ for the angular
version of the covariance shell, rescaled to its physical length
$a_j$. For any of these fixed smooth profiles $f_\delta$,
the high shells satisfy
\[
 \sum_{a_j<s_{\rm uv}}\langle f_\delta,\Gamma_j^hf_\delta\rangle
 \le C\|f_\delta\|_2^2\sum_{a_j<s_{\rm uv}}a_j^2
 \le C_\delta s_{\rm uv}^2.
\]
The $L^2$ operator bound for $\Gamma_j^h$ is $Ca_j^2$.
Consequently this germ has a convergent Gaussian
tail even for electric coefficient $1/2$. At a fixed positive stop
below $\delta$ it is kept as a boundary kernel and integrated
directly. No integer-vertex recognition assertion is being applied
to it, and no charged source flow down to zero cavity size is needed.

For completeness, the comparison can be expressed at the level of
the actual decorated integrals, with their fixed physical separation
suppressed in the notation. Let $e_\epsilon(s_{\rm uv})$ and
$e^{\rm hc}_r(s_{\rm uv})$ be the nonscalar record errors at the intermediate
scale for the native and hard-core regulators, respectively, relative
to their common limiting history. They tend to zero at each fixed
$s_{\rm uv}$ as their cutoffs are removed. Telescope at the first
unequal local operation. A difference in the fixed window above
$s_{\rm uv}$ costs its window error times $(\delta/s_{\rm uv})^P$.
Below that window the common free-core data cancel; a remaining
stable or Taylor-subtracted scalar tail gains
$(s_{\rm uv}/\delta)^\eta$. Any early interacting history which
reaches an inner current must cross its free guard, and has the
same exponential bound as \eqref{physical-screening:wrapping-tail}.
The exact zero-evaluation identity retains its scalar if it joins
different cavities. Adding the already bounded profile tails gives,
at the fixed positive stop,
\begin{equation}
 \begin{aligned}
 |C^{ab}_{\epsilon,\delta,D}
                -C^{ab,\rm hc}_{r,\delta,D}|
 \le{}& C_{D,\delta,F}\bigg\{
       (s_{\rm uv}/\delta)^\eta+s_{\rm uv}/\delta\\
 &+(\delta/s_{\rm uv})^P\big[
          e_\epsilon(s_{\rm uv})+e^{\rm hc}_r(s_{\rm uv})\big]\\
 &+(\delta/s_{\rm uv})^P e^{-c\delta/s_{\rm uv}}
                         \bigg\}+o_\epsilon(1).
 \end{aligned}
 \label{physical-screening:decorated-comparison}
\end{equation}
This is needed only for the two-current components $ab=33,+-$
and for their unmarked normalizer. The small last term is the
free-core harmonic-kernel approximation at fixed $\delta$.
The scalar in each regulator is factored only from its own
unmarked density. All mask scalars, including those whose generating
support joins different cavities, remain in the decorated numerator.

For the integrals in \eqref{physical-screening:decorated-comparison} apply fixed-box reconstruction (Theorem~\ref{fixed-volume-integration:fixed-box}). With fixed
reserve the common derivative regulator has a Gaussian $p$th moment for some $p>1$,
and the boundary kernels have conjugate moments. An affine shift costs a finite
factor depending on $D,\delta$: for its exterior one-form $\mathfrak a_\delta$, admitted bounded kinetic
coefficient $\kappa$ and any $\zeta>0$,
\[
 |\kappa\langle\mathfrak a_\delta,\nabla\psi\rangle|
 \le \zeta\|\nabla\psi\|_2^2
       +\frac{\kappa^2}{4\zeta}\|\mathfrak a_\delta\|_2^2.
\]
The energy is finite at fixed $D,\delta$. Choose $\zeta$ within the fixed-stop Gaussian
reserve to absorb the Cameron--Martin linear term once. H\"older then integrates
the absolute decorated differences, including signed or complex numerators,
independently of positivity of the shifted cosine expansions. The unmarked
physical denominator is bounded away from zero at this stop, hence so is the
convergent auxiliary one eventually. The normalized bound vanishes on taking $\epsilon,r\downarrow0$
at fixed $D,\delta,s_{\rm uv}$, then $s_{\rm uv}\downarrow0$. Constants may depend on $D,\delta$. Only the following neutral
approximation needs uniform removal of $\delta$.

The comparison just proved matches the auxiliary one-cavity scalar
and Cartan factor to the native ones; there is no
independent charged rescaling or separately adjustable pair scalar.
Every auxiliary two-current tensor is invariant before integration
and before division. Applying
\eqref{physical-screening:decorated-comparison} to its two sides
therefore gives
\begin{equation}
 C^{+-}_{\delta,D}(s)=2C^{33}_{\delta,D}(s)
 \label{physical-screening:fixed-cavity-ward}
\end{equation}
after cutoff removal in each fixed rectangle, for each fixed cavity
size and separated pair of slabs.

\emph{The neutral approximation in its own norm.}
For the neutral difference insert an exact masked/unmasked idempotent
switch with the current's nilpotent mark. After the one-cavity scalar
is retained, its nonconstant interface difference at $\delta$ is
$O(v_\delta)$ in the dimension-one packet norm. A block of side
$r$ carries engineering coefficient $rF$. The stable forcing below
$\delta$ sums as $\sum_{a_j<\delta}(a_j/\delta)^\eta v_{a_j}
\le Cv_\delta$. The only dimension-one neutral odd outputs are the
two gradients; $C_4$ and reflection make their map scalar, so
$q_\delta=1+O(v_\delta)\ne0$. Undifferentiated height is nonperiodic,
the identity is even, and the first charged odd field has dimension two.
A single cosine cannot return a neutral current to its charge sector;
the derivative's Gaussian dimension is exactly one. Its multiplier
is therefore $1+O(v_j^2)$, whose product to $R$ is bounded since
$\sum_{\delta\le a_j\le R}v_j^2\le Cv_R$.
Thus the complete error at $R$ is bounded by
$C_{F,R}[v_\delta+(\delta/R)^\nu]$.
For $k\ge2$ copies separated by a fixed buffer $\ell$, an early
joining history spans at least $c\ell(k-1)$. Part of its exponential
decay gives $(C\delta^{-P})^ke^{-c\ell(k-1)/\delta}$, bounded by
the $k$th power of the single-copy error bound. The remaining decay
sums the histories.
The finite-cutoff modulus is uniform in auxiliary widths and repetitions.
The quantitative admission argument consequently gives
\begin{equation}
 \limsup_{D\uparrow\mathbb R^2}\,\limsup_{\epsilon\downarrow0}
 \|O^3_{\epsilon,\delta}(F)-(i/2)\partial_t h(F)\|_{\rm OS,D}
 \le C_{F,R}[v_\delta+(\delta/R)^\nu]\longrightarrow0.
 \label{physical-theory:neutral-cavity}
\end{equation}
The inner limit holds $D$ fixed. The outer limit includes the selected
tall rectangles. At fixed $\delta$ the neutral packet has bounded
cylinder approximants. Its admission is independent of the charged
comparison: the former is uniform in the containing domain and then
allows $\delta\downarrow0$, whereas
\eqref{physical-screening:fixed-cavity-ward} was obtained at each fixed
$D,\delta$. These are the two different uniformity statements needed
for the spectral argument.

\end{proof}

\subsection{Screening of the selected neutral state}
\label{physical-screening:neutral-state}

\begin{theorem}[Physical screening]
\label{physical-screening:theorem}
At one sufficiently early positive endpoint there is $\mu>0$ such
that a centered square-average vector $v_B$ in the selected height
space satisfies
\begin{equation}
 \mathbf1_{[0,\mu)}(H)v_B=0,\qquad
 0\le\langle v_B,e^{-tH}v_B\rangle\le\|v_B\|^2e^{-\mu t}.
 \label{physical-screening:screening-conclusion}
\end{equation}
Consequently the common source, height and derivative space has a
strictly positive spectral gap in the original physical time.
\end{theorem}
\begin{proof}
\emph{The finite-cutoff charged Gram.}
Let $\mathcal A_{\epsilon,\delta,W}$ be the charged slab kernel,
including its two-coupling masks, the free current kernel and the common
local normalization. It maps the row space with boundary values
$(0,0)$ to the row space with boundary values $(0,4\pi)$. These are
copies of $L^2$ of the interior row variables, with the sector recorded
in the adjacent transfer operators. The insertion is the half-row
shift from Lemma~\ref{physical-screening:cavity-kernels}; its reflected
partner is the Hilbert-space adjoint. A finite smear is an integral of
such kernels with the same sector change.

Let $\ell$ denote the length of each insertion slab, $s$ the length of the empty
middle interval and $N$ the half-height of the long rectangle. Both outer caps
give the same smoothed vector
\[
 b_{W,N,s}=T_{0,W}^{(N-\ell-s/2)/\epsilon}\delta_0,
\]
Here $\delta_0$ imposes the actual zero row, and $N-\ell-s/2$ exceeds the fixed cap-smoothing
length. Choose all finite-cutoff lengths on the time lattice. In the notation $Z_W(t)=T_{0,W}^{t/\epsilon}(0,0)$
of the admission theorem, the exact charged reflected Gram is
\begin{equation}
 C^{+-}_{\epsilon,\delta,D}(s)
 =\frac{\langle b_{W,N,s},
       \mathcal A_{\epsilon,\delta,W}^*
       T_{1,W}^{s/\epsilon}
       \mathcal A_{\epsilon,\delta,W}b_{W,N,s}\rangle}
       {Z_W(2N)}.
 \label{physical-screening:charged-gram-transfer}
\end{equation}
Its denominator is the original uninserted physical partition function in the
neutral sector with no cavity mask.

\emph{Comparison of the long and short rectangles.}
Choose $s_0>0$ keeping the short-interval cavities separated. Shorten only the empty
interval by $\Delta=s-s_0\ge0$: the cap vector and both kernels stay fixed, with denominator $Z_W(2N-\Delta)$.
Gaussian convolution conjugated by the positive row weight is positive
self-adjoint. The spectral theorem and \eqref{physical-theory:sector-gap} give
\[
 T_{1,W}^{s/\epsilon}
 \preceq\lambda_1(W)^{\Delta/\epsilon}T_{1,W}^{s_0/\epsilon},
 \qquad
 \lambda_1(W)^{\Delta/\epsilon}
       \le e^{-\mu\Delta}\lambda_0(W)^{\Delta/\epsilon}.
\]
Substitution in \eqref{physical-screening:charged-gram-transfer} leaves the cap ratio of \eqref{source-admission:local-cap-price}, yielding
\begin{equation}
 C^{+-}_{\epsilon,\delta,D}(s)
 \le e^{-\mu\Delta}B_{W,N,\Delta/2}
                 C^{+-}_{\epsilon,\delta,D'}(s_0).
 \label{physical-screening:long-short}
\end{equation}
independently of any cutoff-uniform insertion norm.

\emph{The selected neutral limit.}
Remove the cutoff in both rectangles at fixed $\delta$. Use \eqref{physical-screening:fixed-cavity-ward} on both sides of \eqref{physical-screening:long-short} and
cancel the factor of two. For $N=W^2/R$ and fixed $s,\delta$ the logarithmic cap cost is at most
\[
 C(1+W/R)\frac{\Delta/2}{W^2/R-a-\Delta/2}\longrightarrow0,
\]
with fixed cap-smoothing length $a$. This tall-rectangle limit removes the
separation-dependent exponential cost of a square before the spectral limit.

Approximate the fixed-$\delta$ Cartan kernel in its reflected norm by bounded physical
cylinders uniformly along both exhaustions. Positive character domain order
squeezes each cylinder law between inscribed and circumscribed squares.
Compact-support moment bounds pass products, and removing the approximation
identifies the tall-rectangle and square-selected cavity values. Local admission
bounds the short-separation Gram at fixed $\delta$, giving for each $s\ge s_0$ in the selected
representation
\begin{equation}
 0\le\langle[O^3_\delta(F)],e^{-sH}[O^3_\delta(F)]\rangle
 \le C_{\delta,F,s_0}e^{-\mu(s-s_0)}.
 \label{physical-screening:neutral-spectral-bound}
\end{equation}
Take the width limit at fixed separation before sending $s$ to infinity.

Positivity of the full spectral measure of the neutral vector now excludes mass
on $[0,\mu-\eta]$ for every $\eta>0$, since multiplying its Laplace transform bound \eqref{physical-screening:neutral-spectral-bound} by $e^{\mu s}$ would
otherwise give a contradiction. Hence $\mathbf1_{[0,\mu)}(H)[O^3_\delta(F)]=0$, including at zero energy. For the
literal current vector $v_F=[(i/2)\partial_t h(F)]$, boundedness of the projection gives
\[
 \|\mathbf1_{[0,\mu)}(H)v_F\|
 \le\|v_F-[O^3_\delta(F)]\|\longrightarrow0
\]
by \eqref{physical-theory:neutral-cavity}. Thus cavity removal follows both cutoff removal and selection of the
neutral infinite-volume state.

With a positive-time buffer the height vector belongs to the generator domain,
and its time derivative multiplies the spectral measure by $E^2/4$. This excludes
height spectrum in $(0,\mu)$ without dividing at zero. Smooth $L^2$ approximation gives the
same exclusion for the square selected in \eqref{physical-theory:ordered-axial}, where the zero atom was
independently removed. This proves \eqref{physical-screening:screening-conclusion}. The one-square covariance envelope extends
the gap to the entire height/derivative core, identified with the source space by
complete admission. The charged sector enters only through finite-cutoff transfer
inequalities and fixed-box cavity identities. The infinite-volume law used
throughout this last step is the selected neutral one.
\end{proof}

\subsection{Physical dilation and the source cocycle}
\label{physical-theory:dilation}

Dilation acts simultaneously on the endpoint parameter and the complete source
record. Its cocycle identifies the dilated OS representations and determines how
the physical generator transforms.

\begin{proposition}[Regular physical dilation and the complete source cocycle]
\label{height-representation:smooth-action}
\label{source-dilation:source-cocycle}
The canonical endpoint action $T_\lambda$ admits a $C^2$ orbit
coordinate $s(\rho)>0$ with nonzero derivative and
$s(T_\lambda\rho)=\lambda s(\rho)$. There is a
box-independent compatible finite-jet substitution
$\mathscr R_{\lambda,\rho}$, with bounded $C^2$ coefficients on
compact admissible sets, such that
\begin{equation}
 \begin{aligned}
 \mathcal Z_{D,T_\lambda\rho}[J]
 &=\mathcal Z_{\lambda D,\rho}[\mathscr R_{\lambda,\rho}J],\\
 \mathscr R_{\lambda\mu,\rho}
 &=\mathscr R_{\mu,\rho}\circ\mathscr R_{\lambda,T_\mu\rho}.
 \end{aligned}
 \label{source-dilation:source-law}
\end{equation}
It preserves zero source, label restriction, reflection, conjugation
and the definitions of earlier operator blocks. Its inverse is
$\mathscr R_{\lambda^{-1},T_\lambda\rho}$. All endpoints and
finite normalization windows must remain in their admitted weak
domains. On the common physical Hilbert spaces it induces a
surjective isometry satisfying
\begin{equation}
 U_\lambda e^{-tH_{T_\lambda\rho}}
   =e^{-\lambda tH_\rho}U_\lambda,\qquad
 U_\lambda\Omega_{T_\lambda\rho}=\Omega_\rho.
 \label{physical-theory:hamiltonian-scaling}
\end{equation}
\end{proposition}
\begin{proof}
The canonical continuous action of
Proposition~\ref{canonical-dilations:action} has fixed-$t$ $C^2$ diffeomorphisms
$D_t$ in its Abel coordinate, with
$D_{t+\ell_b}(q)=D_t(q)+1$, $\ell_b=\log b$.
Their derivatives and inverse derivatives are locally bounded jointly:
on larger compact intervals the closed sets bounding their $C^2$
norms cover the time interval; closedness follows from $C^0$
continuity and distributional lower semicontinuity. Baire gives a
bounded subinterval. Writing a small time as the difference of two
times there and using the group law propagates the bounds near zero,
then finite compositions cover any compact time interval. Thus
\[
 \Psi(q)=\ell_b^{-1}\int_0^{\ell_b}D_t(q)dt
 \quad\hbox{is }C^2,\qquad
 \Psi'>0,\qquad \Psi(D_sq)=\Psi(q)+s/\ell_b.
\]
Differentiation under the integral is justified by these bounds;
derivatives are measurable as limits of difference quotients.
Also $\Psi(q+1)=\Psi(q)+1$, so $\Psi$ is a global diffeomorphism.
With the Abel coordinate $\Phi$ of
Proposition~\ref{canonical-dilations:action}, the function
$s(\rho)=\exp\{\ell_b\Psi(\Phi(\rho))\}$ has the stated scaling law. The argument works in every finite $C^k$
class when the fixed-time maps have that class.

Under $y=\lambda x$ the microscopic source density gains $\lambda^{-2}$, an order-$k$ derivative gains $\lambda^k$
and the profiles become $J(y/\lambda)$ at fixed field and period. Unmarked recognition gives
the history at $\rho$. In ordinary normalization common ultraviolet prefixes cancel in
order, leaving finite physical windows between references. At weak scale $R$ the
ordinary readouts $N$ and $N_\lambda$ of the original and rescaled sources at $\rho$ give $\mathscr R_{\lambda,\rho}=N^{-1}N_\lambda$ on
their admitted finite symbolic image. Both undergo identical transport on changing
$R$. The inverse is finite triangular; stable and marked-wall records propagate
forward.

Convert Wick and kinetic references, every mixing partner and prescribed contact
finite part. For $T$ extended on all proper diagonals, the full-diagonal correction
is
\[
 c_\alpha(\lambda)=T[(\chi_{\rm ct}(\mu_{\rm ref}r)
             -\chi_{\rm ct}(\mu_{\rm ref}r/\lambda))r^\alpha/\alpha!],
 \qquad |\alpha|\le\omega.
\]
where $r$ is the relative coordinate and $\omega$ the divergence degree. Proper-subset
corrections come first, and the test vanishes near the full diagonal, avoiding
unknown full-subset contacts. On compact positive dilation intervals the packet
and Taylor-remainder bounds persist. The initial independent scalar stays zero;
local identity values remain ordinary coordinates. Recognition of the full records
gives \eqref{source-dilation:source-law}. Exact composition before integration and faithfulness of the ordinary
inverse give the cocycle and symmetries.

Finite-window comparison gives $C^2$ dependence on $\rho$ at fixed $\lambda$ and joint
continuity. For joint $C^2$ coefficients remove only the profile pullback in orbit
coordinate $q=\log s(\rho)$. The finite lifted matrices obey $A_{t+s}(q)=A_s(q)A_t(q+s)$. Baire again gives locally
uniform fixed-$t$ $C^2$ bounds for $A_t$ and its inverse. With $B(q)=A_{q-q_0}(q_0)$ initially continuous,
$A_t(q)=B(q)^{-1}B(q+t)$. Choose smooth $\psi$ near $q_0$ with $V=\int\psi(s)B(s)ds$ invertible. The uniform bounds make $\int\psi(q+t)A_t(q)dt=B(q)^{-1}V$
of class $C^2$, hence $B$ and then $A_t(q)$ jointly are $C^2$.

The literal selected height law satisfies
$h_{T_\lambda\rho}(x)\overset{\rm law}=h_\rho(\lambda x)$:
dilation takes a cofinal zero-wall exhaustion to another one.
Its action on the positive-time Fourier core preserves reflected
Grams and commutes with reflection. Inverse dilation makes the
resulting isometry onto, and physical time scales by $\lambda$.
This proves \eqref{physical-theory:hamiltonian-scaling}, including
generator domains. Complete admission extends the same identification
to the full source quotient.
\end{proof}

At the Gaussian endpoint the protected derivative transforms as
$D(f)\mapsto D(\lambda^{-1}f(\cdot/\lambda))$, and the charge-$n$
vertex as $V_n(f)\mapsto V_n(\lambda^{2n^2-2}f(\cdot/\lambda))$.
For a polynomial $P$ of derivative jets in $\phi$ units, the complete
Wick conversion is
\[
 P\longmapsto e^{(n\beta_*)^2\Delta C_{00}/2}
 \exp\{-\tfrac12\Delta D:\partial_z^2
                  -in\beta_*\Delta c\cdot\partial_z\}P.
\]
Here $\Delta C_{00}$, $\Delta D$ and $\Delta c$ are respectively
the reference differences of the field, derivative-jet and mixed
covariances. The first factor is the vertex scaling above; the other
terms include identity mixing. Covariance differences add and contact
cutoff differences telescope under composition. Metric jets are first
expanded by their fixed local substitution and then transformed;
that substitution is not inverted. In the calibrated coordinate $a=\mathcal A(\rho)$, the dilated
parameter is $\mathcal A(T_\lambda\rho)$.

\subsection{The physical mass and its ultraviolet scale}
\label{physical-mass:mass-coordinate-section}

\begin{theorem}[The common mass and dimensional transmutation]
\label{physical-mass:mass-theorem}
The complete source, height and derivative Hilbert spaces are
canonically identical, with constant vacuum $\Omega_\rho$ and
physical generator $H_\rho\ge0$. For
$M(\rho)=\inf\operatorname{Spec}(H_\rho|_{\Omega_\rho^\perp})$,
\begin{equation}
 0<c_-\le M(\rho)a_{\rm exit}(\rho)\le c_+<\infty.
 \label{physical-theory:mass-matching}
\end{equation}
The constants are uniform on the fixed compact positive interval.
For every fixed positive-time polynomial in the declared renormalized
smears and complete contacts,
$|C_\rho^F(t)|\le A_F e^{-c_-t/a_{\rm exit}(\rho)}$ for $t\ge0$,
with $A_F$ uniform on its bounded test family. The rate is common to
the entire algebra; no common amplitude over degrees or source lists
is asserted. With $g_\epsilon=c_{\rm lat}y_\epsilon$,
\begin{equation}
 \begin{aligned}
 \log\frac{a_{\rm exit}}{\epsilon}
 &=\frac{1}{g_\epsilon}+O\!\left(\log\frac{1}{g_\epsilon}\right),\\
 \log\frac{1}{\epsilon M(\rho)}
 &=\frac{1}{g_\epsilon}+O\!\left(\log\frac{1}{g_\epsilon}\right).
 \end{aligned}
 \label{physical-theory:mass-clock}
\end{equation}
\end{theorem}
\begin{proof}
On the common space identified by admission, screening gives a gap at one
endpoint. By \eqref{physical-theory:hamiltonian-scaling} it scales as $M(T_\lambda\rho)=\lambda M(\rho)$. Compactness of the dilation interval and
finiteness of the exit atlas give the lower estimate in \eqref{physical-theory:mass-matching}; the surviving
derivative certificate bounds the same threshold from above. For the centered
vector $\psi_F$ the spectral theorem yields $C_\rho^F(t)=\langle\psi_F,e^{-tH_\rho}\psi_F\rangle
\le\|\psi_F\|^2e^{-M(\rho)t}$ with norms uniformly bounded by joint
admission on the declared family. The ultraviolet clock in Theorem~\ref{arbitrary-cutoffs:allcutoff-theorem} gives the first
formula. The second differs by subtraction of $\log[M(\rho)a_{\rm exit}(\rho)]$, which mass matching bounds,
so the leading coefficient is again one. Here the generator is taken after
cutoff removal followed by the selected volume limit.
\end{proof}

\subsection{Selected values and finite-volume scale responses}
\label{scale-responses:selected-values}

\paragraph{Uniform selected values.}
The rescaled height test $f_\lambda(y)=\lambda^{-2}f(y/\lambda)$ is continuous in common-support $L^2$ on compact
positive dilation intervals. By the height moment bound, selected characters and
moments then depend continuously on the parameter. The continuous fixed-box
characters decrease to the continuous selected limit. Dini's theorem gives
uniform convergence in volume on compact parameter intervals, and common
exponential moments pass each fixed polynomial moment. This argument includes
reflected and translated Grams. For smooth tests, $f_\rho$ is $C^2$ in $L^2$. Differentiation
under the height moment bound yields $\partial_\rho\chi=\E[ih(f'_\rho)e^{ih(f_\rho)}]$ and $\partial_\rho^2\chi=
\E[(ih(f''_\rho)-h(f'_\rho)^2)e^{ih(f_\rho)}]$. Uniform local admission passes
the values and their distribution bounds to complete sources. The selected source
cocycle gives $C^2$ dependence of these limiting values. Convergence of the
derivatives of fixed-volume laws requires its own argument.

\begin{proposition}[The exact two scale responses]
\label{scale-responses:wall-identities}
\label{scale-responses:wall-defects}
Put $\mathfrak b(\rho)=\partial_tT_{e^t}\rho|_0$,
$r_\rho(J)=\partial_t\mathscr R_{e^t,\rho}J|_0$ and
$\mathscr D_\rho F=D_JF[r_\rho]$. For the complete normalized
finite-box jet $F_D$, define
$\mathscr W_DF_D=\partial_tF_{e^tD}|_0$. Then
\begin{align}
 \mathfrak b F_D'&=(\mathscr D_\rho+\mathscr W_D)F_D,\\
 \mathfrak b^2F_D''+\mathfrak b\mathfrak b'F_D'
 &=\{(\mathscr D_\rho+\mathscr W_D)^2
                       +\mathfrak b\partial_\rho\mathscr D_\rho\}F_D.
 \label{physical-theory:wall-identities}
\end{align}
In the topology of the uniform selected value bounds, convergence
of the original first and second parameter derivatives is equivalent
to the vanishing of, respectively,
\begin{equation}
 \mathscr W_DF_D,
 \qquad
 (\mathscr W_D^2+2\mathscr D_\rho\mathscr W_D
                       -\mathfrak b'\mathscr W_D)F_D.
 \label{physical-theory:wall-defects}
\end{equation}
\end{proposition}
\begin{proof}
The identity
$F_{e^tD}(\rho;K)=F_D(T_{e^t}\rho;\mathscr R_{e^t,\rho}^{-1}K)$
proves existence of these particular domain-scaling derivatives by
fixed-box $C^2$ regularity. Differentiating the cocycle twice gives
$\partial_t^2\mathscr R|_0=D_Jr_\rho[r_\rho]
+\mathfrak b\partial_\rho r_\rho$, proving the displayed identities.
Box independence makes $\mathscr D_\rho$ commute with
$\mathscr W_D$. For the selected values the same calculation has
$\mathscr W=0$. Subtract those identities and eliminate $F_D'$
from the second. The local terms converge by the bounded-family
value bounds, and $\mathfrak b$ is bounded away from zero on the
compact interval, giving exactly \eqref{physical-theory:wall-defects}.
The later extensive-response theorem proves this vanishing.
\end{proof}

For clarity, differentiating the actual finite-cutoff law uses its
whole-box score $A=-\partial_\rho S$, modulo an unmarked scalar:
$A=-Z_\epsilon'\phi^{\top}A_D\phi/2
+2y_\epsilon'\sum_D\cos(\beta_*\phi)$.
For any complete coefficient $O$,
\[
 \begin{aligned}
 \partial_\rho\E O&=\E O'+\operatorname{Cov}(O,A),\\
 \partial_\rho^2\E O&=\E O''+2\operatorname{Cov}(O',A)\\
 &\quad+\operatorname{Cov}(O,A')+\kappa(O,A,A).
 \end{aligned}
\]
The last cumulant includes all normalization subtractions. These
renormalized combinations are controlled; separate bare summands
need not have limits. An extensive score is not a compact metric
insertion, and the scale identities above are not local geometric
Ward identities.

\section{Gaussian observations with matching traces}
\label{sec:observations}
\label{response-framework:architecture}

To differentiate the expectation of a local observable with respect to
the endpoint parameter, we must
compare several source regions inside their original domain. Consider
first one interface. Given its trace, let $u$ and $v$ be the positive
conditional partition functions on the inside and outside. An insertion
$O$ on the inside replaces $u$ by $u_O$. For the periodic sources used
here, $u$ and $u_O$ depend on the ungrounded component means only modulo
$2\pi$. Fold these means modulo $2\pi$, keeping every neutral difference,
and let $\bar v$ be the conditional expectation of $v$ given the folded
trace. Using the free trace measure for
expectations and $L^2$ norms gives
\[
 p=\frac{\bigl(\mathbb E[u\bar v]\bigr)^2}
         {\|u\|_2^2\,\|\bar v\|_2^2},\qquad
 |\langle O\rangle_D|
 \le \frac{\|u_O\|_2}{\sqrt p\,\|u\|_2}.
\]
Indeed, the physical expectation is the quotient of $\mathbb E[u_O\bar v]$ by $\mathbb E[u\bar v]$. An inside
bound suffices provided $p$ stays above a positive constant. To estimate $p$
we need all three gluings, inside to inside ($II$), inside to outside ($IE$) and
outside to outside ($EE$). The overlap is bounded in Proposition~\ref{extensive:normalizer}. For the squared
insertion Theorem~\ref{extensive:comparison} gives the comparison with its normalizer, retaining the same
unmarked scalar.

The dimension-zero interface term $\cos(h_1-h_2)$ could occur if fine traces fluctuated
independently on the two sheets. We enforce equality by conditioning on nested
observations with local, trace-preserving duals. The positive covariance
increments have vanishing harmonic moments to control their long-range part
across the interface. In the local regulators keep the neutral differences
separate from the unbounded component means.

Fix $k$ source regions and merge overlapping guards into bounded rectangles,
leaving distant rectangles separate. Across their artificial boundary $\Gamma$ write
$I$ for the inside and $E$ for the connected outside. A rectangle may meet one
true Dirichlet wall or two perpendicular walls. Impose common lower bound $R_0>0$
on face lengths, corner charts and unrelated boundary clearances, and upper
diameter bound $L_0$. Only the angles $\pi/2$ and $3\pi/2$ occur. Fix this geometry, the
finite source/contact lists and the derivative orders before choosing $R\le cR_0$.

Use the nearest-neighbor Dirichlet Laplacian $L_s=-\Delta_\epsilon$ of side $s$ with every hole
present, and summation weights $\epsilon^2$ in the bulk and $\epsilon$ on the trace. For $\kappa$ in a
fixed compact positive interval all estimates are uniform in the containing
square, separations, $R\le cR_0$ and ultraviolet depth. The fixed-domain cutoff
threshold $\epsilon_D$ may depend on the domain and its compact family of rounded region
positions. The selected squares must leave open zero-data segments on two
opposite true walls. The count of occupied $R$-blocks alone grows with $L_0/R$.

For the Dirichlet-to-Neumann operator $\Lambda_s$ on $\Gamma$ take the outward normal on side
$s\in\{I,E\}$, so its form is harmonic extension energy. The Gaussian energy pairing $\mathcal E(f,g)$
includes stiffness. For the original free field $\psi$ let $\gamma_\Gamma$ denote restriction
to $\Gamma$; its trace $\gamma_\Gamma\psi$ has covariance $\Sigma_B=[\kappa(\Lambda_I+\Lambda_E)]^{-1}$ in the trace pairing.

The Gaussian replica is specified as follows. Let $H_s$ be the actual
harmonic extension from $\Gamma$ on sheet $s$, and let $U$ insert
one constant on each of the $k_u\le k$ ungrounded components. Set
$S=(U^*\Sigma_B^{-1}U)^{-1}$. The field on sheet $s$ is
\[
 h_s=\eta_s+H_s(b+U\delta_s),\qquad
 b\sim N(0,\Sigma_B),\quad \delta_s\sim N(0,S).
\]
Here the killed fields $\eta_s$, the common trace $b$, and the private
shifts $\delta_s$ are independent.
Repeated sheet types have independent killed and private coordinates
and share $b$. Keep every $\delta_s$ entirely in the retained field.
The periodic
constraint $\delta_1-\delta_2\in2\pi\mathbb Z^{k_u}$ precedes weak interface integration; its positive normalization
follows below.

At fixed derivative order $m$ choose $1/2<\alpha<\alpha_0<2/3$. The norm at a corner $v$ records a common
value and the increments
\begin{equation}
 |f(v)|+\sup_{r\in[\epsilon,a]}(a/r)^\alpha
       \sum_{j=0}^m r^j\|D_\epsilon^j(f-f(v))\|_{\infty,A_r}.
 \label{extensive:corner-norm}
\end{equation}
Here the radii are dyadic and $A_r$ denotes the fixed fat annuli. For the
corresponding quadratic form, sum the squared scaled Sobolev norms with a fixed
reserve of derivatives in place of taking the supremum. On ordinary patches take
scaled $C^{m,\gamma}$ norms with $0<\gamma<1$. Write $\mathcal E_a$ for the resulting field space. At these scales
the integrated fluctuations have matching traces on adjacent sheets. Killed
fields have zero trace.

\begin{theorem}[Localized positive Gaussian decomposition]
\label{extensive:gaussian}
\label{covariance-bounds:package-verification}
\label{gaussian-observations:retained-neutral}
For any prescribed finite spatial and tree-weight orders, there are
nested finite linear observation spaces and positive covariance prefixes
$C_a$, $B_0\epsilon\le a\le R$, with the following properties.

Their increments and retained complement reconstruct the exact bare
Markov replica covariance. Each sheet type has identical self
covariance in $II,IE,EE$, the relative fine trace is zero, and all
activity masks use this same decomposition. Write $\Gamma_a$ for a positive innovation between two consecutive
scales of the fixed-ratio hierarchy. For physical $a$-blocks $X,Y$, let $\|\cdot\|_\pi$ denote the
projective tensor norm on $\mathcal E_a(X)\widehat\otimes_\pi\mathcal E_a(Y)$.
Then
\begin{equation}
 \sup_X\sum_Y(1+d(X,Y)/a)^M
 \|\Gamma_a(X,Y)\|_\pi
 \le C_M.
 \label{extensive:covariance-row}
\end{equation}
The tensor directions have matching trace. A positive forest
interpolation, restricted to occupied guarded blocks with bounded
overlap, has regulator covariance operator norm $C$ and trace $CN$.
The retained field has these same bounds for derivative and corner
increment forms. Its unwrapped component means remain separate.

Boundary innovations on a guard at distance $d_0$ from $\Gamma$ have
projective norm $C_M(a/d_0)^M$. At a flat interface patch distance
$d\ge Ca$ from a defect, reflected prefix kernels differ by
$C_M(a/d)^M$ in each prescribed weighted projective norm, including
diagonal entries. At an ungrounded point their diagonal is
\begin{equation}
 C_a^{ss}(x,x)=(2\pi\kappa_\epsilon)^{-1}\log(a/\epsilon)+O(1),
 \label{extensive:critical-diagonal}
\end{equation}
with the original pinned/image convention at true walls. Consequently
charge attenuation over an interval has its exact bare-stiffness power
and two bounded endpoint Wick factors. On a cluster of diameter
$O(\sigma)$, the pinned interval shift
$(C_A-C_\sigma)(x,v)-(C_A-C_\sigma)(v,v)$ has bounded scaled
regular or corner jets, uniformly in $A/\sigma$.
\end{theorem}

\begin{proof}
\emph{The observation space.}
Prescribe independently on each region all harmonic-polynomial boundary
data through degree $p$, with odd reflection at true walls. Let $P_\mu$
be their interior polynomials, $h_\mu$ their exterior extensions, and
$\ell_\mu=\kappa(\Lambda_I+\Lambda_E)P_\mu$ their full DtN weights.
Near a vertex $h_\mu-P_\mu$ has zero data on both rays; its flux is
$\ell_\mu$. Thus its local powers are $t^{n\pi/\omega-1}$, with
no boundary-data resonance. On regular cells use
$\mathcal P_q+\operatorname{span}\{\ell_\mu\}$ and match $J$ jets
at endpoints, where $q\ge\max(p-1,2J+1)$. On the first corner pair
share the coefficients of $\ell_\mu$ between the two rays and keep
independent polynomial parts. These spaces are nested by restriction.
A corner germ can be stopped at a transition cell by subtracting the
Hermite polynomial with zero left jet and the germ's right jet.
No products of a polynomial with a singular flux are required.

We first establish uniform estimates for the full normalized span.
Locally write $h_\mu=\operatorname{Re}F_\mu$. Choose a multiple $n$ of four with $n>p$.
The boundary polynomial disappears from $F_\mu^{(n)}$,
which reflects to the polygon double. On its orientation cover, $dz$
is global and the corner coordinates are $z-v=w$ or $w^3$.
The cover has bounded topological complexity. The flux has at most
a simple pole in $w$, and $D_z$ increases a corner pole order by
at most three. Continuation of $F'_\mu$ changes it by a polynomial
of degree at most $n-2$. Order these polynomials before the fluxes
and delete dependent generators. Every independent prefix has
unipotent triangular monodromy, so its Wronskian is a nonzero
single-valued meromorphic function. Its pole degree is bounded by
$C[j+3j(j-1)/2+pj]$, independently of the coefficients and distant
geometry. Analytic independence makes it nonzero, as follows by
reducing the Taylor series to distinct leading orders. Single-valuedness
removes logarithms at the corners.

A nonvanishing $P$-valent function $W$ on a disk of radius $r$ obeys
$|D\log W|\le C(P+1)/r$. To see this, maximize
$(1/2-|z|)|g'(z)|$ for $g=\log W$ on the unit disk. Cauchy's estimate
and Rouch\'e's theorem give an inverse of $g$ over an image disk of
radius at least $|g'(0)|/128$. A disk containing $P+1$ points separated
by $2\pi i$ contradicts $P$-valency of $W$. Rescaling and Cauchy
give all fixed higher logarithmic derivatives. The divisor bound
therefore controls the coefficients of the exact annihilating equation
away from the finitely many zeros and poles. In fact, with $W_j$
the Wronskian of the first $j$ generators and $W_0=1$, its
first-order factors are
\[
 A_j=D-D\log(W_j/W_{j-1}),\qquad A_d\cdots A_1 f=0.
\]
In a cell of length $a$, choose a fat interval, a circle of comparable radius and connecting
paths staying $ca$ from those points. The number of excluded points is bounded, so the
paths have length $O(a)$. The equation controls its initial jet by the
interval $L^2$ norm: normalize the jet, use the nonzero norm of its
finite Taylor polynomial, and make the integral remainder smaller on
a fixed short interval. Gronwall along the paths, followed by Cauchy
on the circle, proves
\begin{equation}
 a^j\|D^jf\|_{\infty,I}\le C_ja^{-1/2}\|f\|_{2,I},\qquad
 \|f\|_{2,I}\le C\|f\|_{2,I_{\rm fat}}.
 \label{extensive:observation-inverse}
\end{equation}
At a reentrant vertex $wf(w)$ is holomorphic. The same argument
on $|w|\asymp a^{1/3}$ gives
$|f(t)|\le Ca^{-1/2}\|f\|_{2,(ca,da)}(a/t)^{1/3}$.
Hence corner mass in $(0,\eta a)$ is at most $C\eta^{1/3}$ of
the full mass. Convex vertices give stronger bounds.

Suppose a nontrivial real linear combination of the full fluxes agreed with a
polynomial on one face. Analytic continuation would make the corresponding
exterior complex derivative polynomial and the real exterior harmonic function a
global harmonic polynomial. This polynomial vanishes on open segments of two
opposite true walls and hence on both lines. Odd reflection makes it periodic in the normal direction;
a periodic polynomial is constant and its zero boundary value makes
it zero. Independence of the prescribed boundary data proves the
claim. Complex relations split into real ones since the fluxes on
the selected ray are real. Thus the potential Gram on a fixed inner
annulus is positive definite, even though its smallest eigenvalue
may depend on $D$.

\emph{Exact lattice realization.}
On fixed polynomial spaces choose a right inverse $Q_0$ for $\Delta$.
After scaling the cell size $a$ to one, put $h=\epsilon/a$.
Then $\Delta_hQ_0=I+N_h$, where $N_h$ lowers degree and is nilpotent,
so $Q_h=Q_0\sum_j(-N_h)^j$ is an exact uniformly bounded lattice
right inverse. Here the Taylor estimate is applied to a function $u$
that is lattice harmonic on a fixed enlargement of the cell, including
every difference stencil used below. In this paragraph $D_\epsilon^j$
denotes the collection of order-$j$ lattice differences, each divided
by $\epsilon^j$; $\Delta_\epsilon$ is the symmetric nearest-neighbor
Laplacian. If $T_pu$ is its Newton--Taylor polynomial, let
\[
 P_{p,\epsilon}u
   =T_pu-Q_{\epsilon,a}\Delta_\epsilon T_pu,
\]
where $Q_{\epsilon,a}$ is the rescaling of $Q_h$ to physical
coordinates. Then $\Delta_\epsilon P_{p,\epsilon}u=0$ and, on the
inner cell $U$ with fixed enlargement $\widetilde U$,
\[
 \|D_\epsilon^j(u-P_{p,\epsilon}u)\|_{\infty,U}
 \le C_{p,j}a^{p+1-j}
       \|D_\epsilon^{p+1}u\|_{\infty,\widetilde U},
 \qquad 0\le j\le p+1.
\]
Indeed $\Delta_\epsilon T_pu=-\Delta_\epsilon(u-T_pu)$ on
the required stencils, so the Newton remainder and the bounded
polynomial right inverse give this estimate. Averaging over the wall
reflection group preserves odd data. Separately, correcting a
continuum harmonic polynomial by the same right inverse gives a basis
converging to its continuum basis, since its lattice Laplacian is
$O(h^2)$ after rescaling. The lattice-harmonic polynomial space has
dimension $2p+1$.

A polynomial flux $p(x)$ has an exact odd Cauchy extension: set
$q_0(k)=k$ and choose odd $q_j$ with
$\Delta_kq_j=-q_{j-1}$, $q_j(0)=q_j(1)=0$ for $j\ge1$. Then
$\epsilon\sum_j\epsilon^{2j}q_j(y/\epsilon)
\Delta_{x,\epsilon}^jp(x)$ has zero trace and first normal
difference $p$. A lattice observation flux is therefore the flux of
a zero-ray harmonic function plus this Cauchy polynomial on the
$6\pi$ reflection cone. Its potential $V$ obeys
$(R_{3\pi}-I)(R_{2\pi}-I)V=0$. Continuum limits have ball mass
\[
 M(r)=\sum_{\lambda\in\Lambda} b_\lambda r^{2\lambda+2},
 \qquad b_\lambda\ge0,\qquad
 \Lambda=\{2n/3:n\ge1\}\cup\{1,2,\ldots\}.
\]
Angular integration from the zero ray and the preceding inverse estimate
compare annular potential mass with $r^3$ times the fat-interval flux norm. In
the ball of radius $r/4$ the positive expansion gives mass at most $4^{-10/3}M(r)$. The
annulus thus carries a fixed fraction of the ball mass, yielding uniform
continuum outer doubling. For fixed $D$ the finite potential family converges:
its Gram bound $A_D\le C_*B_D$ passes to the lattice as soon as each matrix error falls
below $\delta\lambda_{\min}(B_D)$. Only this cutoff threshold depends on the domain; the transferred
constant is uniform.

Fix $\mu$ off the displayed spectrum with $Q=2^{2\mu+2}>2C_*$. If the doubling ratio could cross
$Q$ inwards on scales $a\ge B_0\epsilon$ for arbitrarily large fixed $B_0$, rescale a sequence
with $\epsilon/a\to0$ and $M(1)=1$. Discrete Caccioppoli and compactness of the piecewise-affine
interpolants give a harmonic cone limit satisfying $M(2)\le Q$ and $M(1/2)\le Q^{-1}$. Positivity of
the expansion implies $1=M(1)^2\le M(1/2)M(2)\le1$. Equality leaves a single exponent $\mu$, excluded by
choice. Hence one doubling bound reaches $B_0\epsilon$ with no losses accumulated over
shells. The same compactness gives flux observability $M(a)\le Ca^3\|f\|_{2,(ca,da)}^2$ by harmonic Cauchy
uniqueness: a nonzero limit cannot have both value and flux zero on a ray.
Iterated discrete Caccioppoli and Fourier $H^2$
Sobolev give $\|D^jv\|_\infty\le C_ja^{-j-1}\|v\|_2$,
and hence \eqref{extensive:observation-inverse} on the lattice.
For regular cells at distance $r\gg a$ from a corner, compactness and
unique continuation first compare cone mass to mass in a face-centred
disk of radius $cr$; the ordinary integer-spectrum version propagates
its doubling down to $a$. This covers intermediate face scales.

The exact lattice barrier supplies endpoint nonconcentration. For
$1/2<\beta<2/3$, $r^\beta\sin(2\theta/3)$ is superharmonic outside
a fixed microscopic disk. Add the killed-wedge Green potential of its
finite negative defect. The negative-power barrier bounds that correction
by $Cr^{-\beta}\sin(2\theta/3)$ outside the disk; inside, every return
has a fixed chance to hit a killed ray. Thus the repaired barrier is
comparable to the original positive one. Comparison and reflected
interior estimates give
$|\epsilon^{-1}V(t,\epsilon)|\le
Ca^{-2}M(4a)^{1/2}(\max(t,\epsilon)/a)^{\beta-1}$.
The part odd under $R_{3\pi}$ is a bounded-degree polynomial and has
the stronger linear bound. Summing the squares gives lattice corner
mass $C\eta^{2\beta-1}$ for $\eta\ge\epsilon/a$ and the
endpoint-slot bound. Both incident
edge fluxes couple to the single physical corner value.

\emph{Poisson approximation of arbitrary order.}
The finite observation list must approximate arbitrary corner solutions,
not just its own span. In an infinite wedge, a bump $b$ on one ray
has $n$th coefficient
$\omega^{-1}\int b(t)t^{-n\pi/\omega-1}\,dt$.
Narrow bumps near distinct $\ell t_j$, $1<t_j<2$, give an invertible
generalized Vandermonde matrix. Their harmonic interior extensions
are smooth at vertices by odd reflection. Approximate their holomorphic
potentials in $C^{K+1}$ by polynomials matching all vertex jets through
$K$. Such approximation follows by inward dilation, Cauchy integration
on a larger rectangle, approximation of the outside pole kernels by
geometric-series polynomials, and finite Hermite correction. At true
walls project onto the odd polynomial subspace. For $K+1>2N/3$ the
weighted moment errors are small. The required functions are $h_E-P$:
noninteger coefficients are unchanged, whereas at exponent $2m$ the
opposite interior orientation changes the row factor to
$2/(3\pi)+2/\pi=8/(3\pi)$, still nonzero. At a grounded junction
the two right-angle coefficients likewise add. Finite-domain and
interior arc corrections have $n$th coefficient at most $CR_0^{-2n/3}$.
Choosing $\ell/R_0$ and then the approximation tolerance small gives
uniformly invertible actual probes. Compactness of the allowed local
shapes fixes one degree $p$ for the entire family.

On the square wedge arc let $T_bf=(Hf)(\cdot/b)$. Its low modes
have eigenvalues $b^{-2n/3}$, and the complement after $N$ modes
has norm $C_Nb^{-2(N+1)/3}$. Lattice transfers converge in operator
norm: their kernels are $h^{-1}G_h(x,z+hn_{\rm in})$. The lattice
potential kernel bounds the off-diagonal Green function uniformly;
summation by parts and uniqueness identify its limit. Reflected
interior estimates give convergence of the first normal difference
uniformly on the separated arcs, including the convex endpoints,
where the kernel vanishes. Uniform kernel convergence implies
Hilbert--Schmidt convergence.
Choose $N$ and $2N/3<\sigma<2(N+1)/3$ so that $\sigma$ exceeds
the prescribed decay orders. Then take $b$ large. In the low/high splitting successive
lattice transfers have blocks $(A_j,B_j;C_j,D_j)$ close to
$\operatorname{diag}(A,D)$, uniformly for $a_j/\epsilon\ge B_0$.
The exact probe space is a graph $X_j$ satisfying
\[
 X_{j+1}=(C_j+D_jX_j)(A_j+B_jX_j)^{-1},\quad
 w_{j+1}=(D_j-X_{j+1}B_j)w_j.
\]
Here $\|A^{-1}\|=b^{2N/3}$ and $\|D\|\le\tfrac14b^{-\sigma}$.
The invertible probe matrix makes $\|X_0\|\le1/2$ at a fixed
small radius; sufficiently fine lattice convergence starts the same
graph. Choosing the operator error small gives $\|X_j\|\le1/2$
and $\|w_{j+1}\|\le b^{-\sigma}\|w_j\|$. After a fixed inward
buffer, the barrier and interior differences give flux remainder
$Ca^{-1/2}(a/t)^\sigma\|u(t\cdot)\|_2$, with all fixed scaled jets.
The construction restarts at any later $t$ and commutes with source
derivatives. Hermite correction joins it to ordinary face Taylor
approximation. Thus every needed Poisson weight has an actual lattice
observation approximant, with no per-shell discretization loss.

\emph{Smooth duals and conditional covariance.}
Use dyadic subdivisions of each complete face and rectangle; cell
lengths lie in $(a/2,a]$. Fix a cardinal $C^J$ lattice reconstruction
exact on degree-$q$ polynomials, but retain the original quadrature in
every observation. Restriction gives literal nesting at arbitrary
endpoint phases. Endpoint Hermite functions and orthonormal zero-jet
bubbles give a stable local basis $w_i$ for the trace space $W_a$.
Let $T_a$ vanish within $\eta a$ of vertices and equal one beyond
$2\eta a$. Corner nonconcentration makes
$G_{ij}=\langle w_i,T_aw_j\rangle$ uniformly positive. Its banded
inverse has exponential decay, since
$G^{-1}=c\sum_{n\ge0}(I-cG)^n$ with $\|I-cG\|<1$.
Consequently
\[
 R_a^\Gamma f=\sum_{i,j}T_aw_i(G^{-1})_{ij}\langle w_j,f\rangle
\]
is smooth, exact on all moments, exponentially local and zero on
their common kernel. The bulk spline space has the identical
construction with a cutoff removing an $\eta a$ boundary tube:
the removed mass is at most $C(\eta+\epsilon/a)$ by the polynomial
inverse. Denote its map by $R_a^b$ and a smooth normal collar lift
by $E_a$. Lift the reconstructed trace first, then reconstruct the
remaining bulk data.
\[
 g=E_a R_a^\Gamma(\gamma_\Gamma f),\qquad R_af=g+R_a^b(f-g).
\]
Its two dual energy norms are $O(a^{-1/2})$ for trace rows and
$O(a^{-1})$ for bulk rows normalized in $L^2$.

Let $\mathcal F_a$ be the sigma algebra generated by the finite
observations at resolution $a$, and let $\mathcal K_a$ annihilate
these observations in Dirichlet energy. Put
$\Sigma_a=\operatorname{Cov}(\psi\mid\mathcal F_a)$ and
$B_a=\gamma_\Gamma\Sigma_a\gamma_\Gamma^*$. Local Poincar\'e
and trace inequalities give $\Sigma_a\preceq Ca^2I$, $B_a\preceq CaI$.
For $v\in\mathcal K_a$, the kernel of $R_a(\chi v)$ contains
$\chi(y)-\chi(x)$, so
$\|\nabla R_a(\chi v)\|_2\le
Ca\|\nabla\chi\|_\infty\|\nabla v\|_2$, also with exponential
local weights. Testing the constrained inverse equation with
$(I-R_a)(e^{2\theta d/a}v)$ and absorbing $C\theta$ proves
exponential energy localization. Conditional expectation is orthogonal
projection: the nested $\Sigma_a$ and their differences are positive.
An innovation's Euler equation has smooth bulk and paired DtN forcing,
whose coefficients are its energy pairings with the exact duals.

These forcings give precisely the corner regularity in
\eqref{extensive:corner-norm}. Decompose the total normalized paired
flux on the reflection cone into its $R_{3\pi}$-even zero-ray part
and its bounded-degree odd polynomial part; observability bounds both
without separating nearly cancelling original coefficients. The former
is realized by the zero-ray harmonic function outside and zero inside.
For a polynomial jump $p$, the lattice single layer
$\epsilon\sum_y p(y)\chi(y)[A_\epsilon(x-y)-A_\epsilon(-y)]$
is $O(r(1+|\log r|))$: split at $|y|=2r$, using the integrable
near logarithmic difference and $r\int_{2r}^1dt/t$ outside.
Subtract its full-line Cauchy polynomial on each annulus to obtain all
one-sided jets. A smooth bulk Poisson particular leaves a harmonic
remainder through the vertex, controlled by Poincar\'e. Thus a unit
innovation satisfies
\begin{equation}
 |z(x)-z(v)|+\sum_{j=1}^m r^j|D^jz(x)|
       \le C(r/a)^{\alpha_0},\qquad |z(v)|\le C.
 \label{extensive:corner-innovation}
\end{equation}
Harmonic extension has the same bound: its difference from the
auxiliary field has zero rays and smooth forcing, and annular
particulars sum with exponent $2-\alpha_0>0$.

\emph{Killed prefixes and the physical map.}
Write $A_s=\epsilon^2L_s$, whose spectrum lies in $(0,8]$. Take an even nonnegative Schwartz function
$\varphi$ with smooth Fourier support in $[-1,1]$ and normalization $\int_0^\infty u\varphi(u)\,du=1$, and define
\begin{equation}
 \begin{aligned}
 W_t(\lambda)&=\sum_{n\in\mathbb Z}
 \varphi(t\arccos(1-\lambda/2)-2\pi nt),\\
 Q_a^s&=\frac{\epsilon^2}{3\kappa}
 \int_0^{a/(c_{\rm fr}\epsilon)}tW_t(A_s/3)\,dt.
 \end{aligned}
 \label{killed-covariances:killed-spectral}
\end{equation}
By Poisson summation $W_t$ is a Chebyshev polynomial of degree at most $t$. The
resulting matrix has finite range. It equals the bulk or half-plane matrix
wherever their path neighborhoods coincide. The sum defining it is nonnegative,
and $\int_0^\infty tW_t(\lambda)dt=1/\lambda$ follows from $\lambda=4\sin^2(\omega/2)$ and $\sum_n(\omega-2\pi n)^{-2}=[4\sin^2(\omega/2)]^{-1}$. This proves the exact identity $Q_\infty^s=\kappa^{-1}L_s^{-1}$.

Schwartz decay gives the following bounds
for the killed covariance and for its retained tail $T_R^s=Q_\infty^s-Q_R^s$.
\[
 Q_{Ba}^s-Q_a^s\preceq C_Na^2(1+a^2L_s)^{-N},\qquad
 T_R^s\preceq C_N\kappa^{-1}L_s^{-1}(1+R^2L_s)^{-N}.
\]
For the killed heat kernel,
$p_t(x,x)\le C(t+\epsilon^2)^{-1}\mathbb P_x(\tau>t/3)^2$,
by splitting the semigroup into three pieces. A positive wedge barrier
$r^{\alpha_0}\cos(\beta(\theta-\omega/2))$,
$\alpha_0<\beta<\pi/\omega$, repaired on its finite lattice core
as above, gives exit probability $C(r/R)^{\alpha_0}$ and mean exit
time $CR^2(r/R)^{\alpha_0}$. Hence survival at time $t$ is
$C(r/\sqrt t)^{\alpha_0}$ in the local chart. The heat representation
of the spectral envelope yields
$\operatorname{Var}((a^2L_s)^j\zeta_a(x))\le
C_j(r/a)^{2\alpha_0}$. Iterated local energy/Sobolev estimates give
\eqref{extensive:corner-innovation} with zero vertex value. Annular
quadratic traces sum as $\sum_m2^{-2m(\alpha_0-\alpha)}$.
Extra derivatives give projective Fourier expansions on each block;
finite range bounds their rows.

Two sheets share only the harmonic extension of the trace at these
scales. With $s,t$ indexing sheet copies (even when their types agree),
the physical covariance prefix before stiffness completion is
\begin{equation}
 C_a^{st}=\delta_{st}Q_a^s+H_sB_aH_t^*.
 \label{extensive:master-prefix}
\end{equation}
Its positive complement is the sum of the killed tails, the harmonic
image of the remaining trace covariance, and the retained private
constant lifts. This proves exact reconstruction and the sheet identities.
For a boundary innovation $\mathcal B$, put $P=I-R_{Ba}^\Gamma$.
Then $\mathcal B=P\mathcal BP^*$ exactly. The Poisson approximants
proved above imply
\begin{equation}
 \|(H_sP)_{XA}\|\le C_M(1+d(X,A)/a)^{-M}.
 \label{extensive:poisson-row}
\end{equation}
Near a face this is the one-sided Poisson estimate on finite smooth
jets; at a corner split increments into annuli and sum the wedge
barrier. At distance $d\gg a$, subtract the actual observation
approximant before pairing. A unit trace cell has $L^2$ norm
$O(\sqrt a)$ and its corner flux remainder is
$O(a^{-1/2}(a/d)^M)$, giving the stated power.

For distant regions exact multipoles also matter. Fill every hole
harmonically. The filled field has source
$\kappa(\Lambda_I+\Lambda_E)b$, whose harmonic-polynomial moments
vanish for $b=Pf_A$ by self-adjointness. The projection is block
diagonal between regions, so $b$ remains on the input region.
At a target distance $d$, choose a contour of radius in $[d/4,d/2]$
staying $c_kd$ from every region, excluding their finitely many radial
intervals. On it the reciprocity identity is
$H_Eb(y)=\kappa\langle b,(\Lambda_I+\Lambda_E)
G_D(y,\cdot)\rangle$.
Subtract the exact lattice harmonic Taylor polynomial on each region.
The Green derivative bound $|D^jG_D|\le C_jd^{-j}$, $j>0$,
leaves data $C(R_0/d)^L$. This bound follows from Fourier contour
shift for heat times $t\ge\epsilon^2$, giving
$C_jt^{-1-j/2}e^{-cd/\sqrt t}$, and the factorial jump-count tail
for shorter times; rectangular images and the first Dirichlet
eigenvalue make it independent of $D$. The maximum principle bounds
the remainder extension by the same number. Its local flux
approximation supplies another $(a/R_0)^M$. Outside the contour
all other region data are zero, so the maximum principle propagates
the product to the target. Taking $L\ge M$ proves
\eqref{extensive:poisson-row} also there. Differentiated interior,
reflected and corner estimates give its jets.

Both weighted row and column sums of \eqref{extensive:poisson-row}
are finite. Convolving them with the exponentially localized auxiliary
trace covariance proves \eqref{extensive:covariance-row} and the guard
bound. Localize all adjacent sheets by cutoffs with one common trace.
Any ambient projective series can be projected onto matching directions:
replace each trace by its average and extend the difference by a bounded
normal collar map, or annular angular interpolation at a corner.
This projection fixes the covariance and changes its norm by a constant.
Positive forest Schur matrices have diagonal one and entries bounded
by one. On occupied blocks each guard occurs only with its fixed
multiplicity; the same row and diagonal estimates give operator norm
$C$, trace $CN$, independently of the number of forest copies.

\emph{Retained fields and the critical coefficient.}
For the retained auxiliary mean, dual forcing coefficients are
$\mathcal E(z,\rho_i)$; exponential dual localization bounds their
local squares by $Ce_X(z)$ with $\sum_Xe_X(z)\le C\mathcal E(z,z)$.
Subtracting a local mean and repeating the particular-solution estimate
gives regular derivatives and corner increments
$C(r/R)^{\alpha_0}e_X(z)^{1/2}$. Harmonic extension minimizes energy;
subtract its boundary increments and use the zero-ray barrier to obtain
the same local bounds. For the retained killed tail, set
\[
 e_{X,R}(z)=\|\nabla z\|_{2,X^+}^2+
     \sum_{j=1}^{m+1}R^{4j-2}\|L_s^jz\|_{2,X^+}^2.
\]
The local wedge Green kernel satisfies
$\|G(x,\cdot)\|_2\le CR(r/R)^{\alpha_0}$: for $|y-v|>2r$
sum shell bounds $C(r/|y-v|)^{\alpha_0}$; within $2r$ integrate
the square of $1+\log_+(r/(|x-y|\vee\epsilon))$.
Thus a Poisson particular obeys
$|p(x)|\le CR(r/R)^{\alpha_0}\|L_sw\|_2$.
Apply this to $R^{2j}L_s^jz$, and use Caccioppoli on nested
annuli, to bound all corner jets by
$C(r/R)^{\alpha_0}e_{X,R}(z)^{1/2}$; ordinary patches follow from
local elliptic estimates after mean subtraction.
The spectral tail envelope gives
$\sum_Xe_{X,R}((T_R^s)^{1/2}u)\le C\|u\|_2^2$.
This proves the retained operator bound; squared evaluation norms
and the convergent annular sum prove trace $CN$. Private constant
lifts have covariance
$S=[\kappa U^*(\Lambda_I+\Lambda_E)U]^{-1}$.
Their harmonic energy covariance is bounded since
$\Lambda_s\preceq\Lambda_I+\Lambda_E$; their rank is at most $k$.
This controls their neutral forms while leaving common means unbounded.

For the auxiliary diagonal, subtract a retained unit-mass spline
average near $x$ for the upper bound. The neutral lattice potential
kernel gives $(2\pi\kappa)^{-1}\log(a/\epsilon)+O(1)$.
For the lower bound condition additionally outside a buffered local
rectangle. Choose its sides away from the finitely many cell endpoints
by $\eta a$, so every clipped observation cell has a fat subcell.
The dual construction on those pieces makes the normalized observation
Gram $Q\succeq cI$. Its point-to-row covariance vector is bounded:
bulk rows pair with a local logarithm, and corner rows use
$\int_0^1t^{\beta-1}(1+|\log t|)dt<\infty$.
Regression removes only $O(1)$ from the local killed-square diagonal.
For a physical sheet at boundary distance $d\le a$, the killed prefix
has diagonal $(2\pi\kappa)^{-1}\log((d\vee\epsilon)/\epsilon)+O(1)$.
Indeed its local wedge shells above $d$ sum to a constant by the
barrier; the actual-side Green differs by a bounded harmonic correction.
Uniformity of that correction follows by comparison with the exterior
of a fixed disk inside the hole, using the difference of two lattice
potential kernels with a Kelvin-image pole; true walls use odd images.

The harmonic part supplies
$(2\pi\kappa)^{-1}\log(a/(d\vee\epsilon))+O(1)$.
For its upper bound use
the decomposition into the harmonic extension and the independent killed
field. With $\phi_a$ the spline average and $G_s$ the killed covariance,
\[
 \operatorname{Var}\big((H_s\gamma_\Gamma\psi)(x)-\psi(\phi_a)\big)
 =\operatorname{Var}\big(\psi(x)-\psi(\phi_a)\big)
   -G_s(x,x)+2G_s(x,\phi_a).
\]
The last term is bounded by finite-range shell summation and the
bounded killed tail at distance $O(a)$ from the boundary.
For the lower bound, in the local conditioned rectangle the actual
Poisson kernel dominates the locally killed one. Positivity of the
trace Green kernel bounds its variance below by the difference of the
two local Green diagonals; the remaining finite observation regression
again costs $O(1)$. At the vertex the common trace is the auxiliary
field; at $d\ge a$ the killed bulk prefix already supplies the whole
logarithm. This proves its exact coefficient, including pinned images.

Consequently the Wick conversion for charge $q$ is exactly
\[
 e^{-q^2(C_A^{ss}-C_a^{ss})(x,x)/2}
 =(a/A)^{q^2/(4\pi\kappa)}e^{-q^2(b_A(x)-b_a(x))/2},
 \qquad |b_a|\le C.
\]
Store its two endpoint factors. For a pinned interval, use
\eqref{extensive:corner-innovation} and covariance Cauchy--Schwarz
on each fixed-ratio shell separately. Its contribution is
$C(r/t)^{\alpha_0}$; summing $t\ge\sigma$ is bounded. For reflection,
the local half-plane singularities in
$H_I(x,\cdot)-H_E(\mathcal Rx,\cdot)$ cancel exactly. Approximation
in $W_t$ gives arbitrary $(t/s)^M$ for the remaining smooth kernel.
Apply Cauchy--Schwarz shellwise to the difference of covariance products,
including their diagonals, and sum before any cumulative point
variance. Combining with the arbitrary spatial row bound by geometric
mean retains both the defect gain and a summable row.

\emph{Exact bare completion.}
The preceding construction at reference stiffness $\kappa_0$ is
completed with $\delta=Z_\epsilon-1$, $\alpha_\epsilon=(1+\delta)^{-1}$,
$\kappa_\epsilon=\kappa_0Z_\epsilon$. For $G^0=C+T$, write $L=(G^0)^{-1}$ for its precision. Completing
the quadratic form $\delta\langle h,Lh\rangle/2$ gives
\[
 \begin{aligned}
 Q&=C(I+\delta L C)^{-1},\qquad A=(I+\delta C L)^{-1},\\
 c_{\rm kin}&=-\tfrac12\log\det(I+\delta L^{1/2}C L^{1/2}).
 \end{aligned}
\]
The completed Gaussian integral is
\[
 \begin{aligned}
 &\mathbb E_C e^{-\delta\langle\phi+\zeta,L(\phi+\zeta)\rangle/2
                         +V(\phi+\zeta)}\\
 &\qquad=e^{c_{\rm kin}-\delta\langle\phi,L A\phi\rangle/2}
                \mathbb E_Qe^{V(A\phi+\xi)},
 \end{aligned}
\]
where the quadratic expression is interpreted as its bilinear form. The normalized conditional covariance
is an energy projection, so its completion is exactly
$\alpha_\epsilon\Sigma_a$. On sheet $s$ the full killed covariance at
reference stiffness is $G_s^0=(\kappa_0 L_s)^{-1}$. Set
$\omega_a=I-(G_s^0)^{-1}Q_a^s$. Then
\[
 A_a=\alpha_\epsilon\sum_{n\ge0}
       (\delta\alpha_\epsilon)^n\omega_a^n,\qquad
 \widehat Q_a-\widehat Q_b=A_a(Q_a-Q_b)A_b.
\]
The smooth finite-range kernels $\omega_a$ and $Q_a\omega_a$ have
the established weighted bounds. For fine cutoff $|\delta|$ is small;
the distance weights of an $n$-fold product grow only polynomially in
$n$, so the Neumann series converges in the weighted norms. This gives positive completed increments and the same estimates.
The completed physical prefix is
$\delta_{st}\widehat Q_a^s+\alpha_\epsilon H_sB_aH_t^*$;
the killed retained complement is
$\alpha_\epsilon(G_s^0-Q_a^s)(I+\delta(G_s^0)^{-1}Q_a^s)^{-1}$.
It has the same spectral envelope. The smooth endpoint correction is
$O(|\delta|)$, proving \eqref{extensive:critical-diagonal} with the
actual bare coefficient.

The physical inverse mean map is
$h_s\mapsto h_s+\delta Q_a^s\kappa_0\mathcal L_{s,b}h_s$,
where $\mathcal L_{s,b}$ is the local Laplacian with its actual
boundary value $b$. It preserves the common trace and is finite range.
On a fat chart, subtract the harmonic extension $e$ of that trace:
$Q_a^s\kappa_0\mathcal L_{s,b}h_s=(I-\omega_a)(h_s-e)$.
Harmonic extension is bounded in the same $C^{m,\gamma}$ space:
subtract the degree-$m$ boundary polynomial and sum the Poisson
remainder $t^{m+\gamma}$; the singular integral is
$\int_0^a t^{\gamma-1}dt$. Its difference splits at the displacement
and is $O(h^\gamma)$. At corners sum boundary annuli with a barrier
exponent between $\alpha$ and $2/3$. Thus both mean maps are
$I+O(|\delta|)$ on one fixed field space without derivative loss.
Subtract constants and use local Poincar\'e for the derivative regulator.
Under the exact quadratic conjugacy the canonical weak record carries its full
scalar and source dictionaries, with intermediate endpoint changes cancelling.
Even the relative lifts enter the replica compensation as Gaussian data. The
positive-cosine inequalities thus hold for the completed law with no omitted
stiffness error multiplied by an ultraviolet logarithm.
\end{proof}

\section{Replica comparison with the physical normalizer}
\label{sec:replicas}

Two quantities are needed for the physical comparison, the overlap $p$ of
Section~\ref{sec:observations} and the inside partition $\mathbb E[u^2]$ normalizing the squared insertion.
Removing the interaction in an interface collar gives a Gaussian region across
which the positive conditional integrals on either side can be compared. On
restoring the interaction, cancel the scalar terms shared by the replicas before
taking absolute values.

Admit the complete local coefficients of integer vertices and of polynomials in
positive-order derivatives of $h=\beta_*\phi$, including their prescribed scalar and contact
columns. A common $2\pi$ shift preserves these sources. Fix the entire list, its
sublists and its mixed comparison switches. Constants can depend on the list size
$k$, with no uniformity as $k\to\infty$. Literal nonperiodic height cylinders occur in the
separate comparison core.

\begin{proposition}[Connected interface and scalar estimates]
\label{extensive:interface}
\label{interface-expansion:rr-restoration-scalar}
With the covariance of Theorem~\ref{extensive:gaussian}, the complete
nonconstant interface record satisfies $\|K_j\|\le Cv_j$, uniformly
also for literal activity masks. For a collar containing $N_\Gamma$
occupied stop blocks, let $v_{st}$ be its complete zero-field
restoration ratio in replica $st$. Then
\begin{equation}
 |2\log v_{IE}-\log v_{II}-\log v_{EE}|
       \le CN_\Gamma v_R.
 \label{extensive:restoration}
\end{equation}
For four masks $U+B+O,B,U+B,B+O$, where $B$ separates $U$ from
$O$ by $cR$, their zero-field phase partitions satisfy
\begin{equation}
 0\le\log\Phi_{U+B+O}(0)+\log\Phi_B(0)
       -\log\Phi_{U+B}(0)-\log\Phi_{B+O}(0)
       \le CN_\Gamma v_R.
 \label{extensive:four-mask}
\end{equation}
All scalars are the complete scalars of the same Gaussian hierarchy.
Only their displayed contrasts are bounded.
\end{proposition}

\begin{proof}
For the covariance row \eqref{extensive:covariance-row} the forest certificate in the local expansion lemma
follows directly from the operator and trace bounds: for $Q=\|J\zeta\|^2$ and $K=J\Gamma J^*$ one has $\mathbb E e^{cQ}=\det(I-2cK)^{-1/2}\le C^N$
at fixed sufficiently small $c$. Attach a boundary birth root to each new
interface atom and keep it through extractions into the bulk. Then the two
Schur bounds in \eqref{extensive:poisson-row} control both birth sums, row and column, by $Cv_j$. Away from
the interface, histories built solely from the original killed bulk operation
reproduce the calibrated canonical trajectory; the new atoms supply its
inhomogeneous correction.

At a corner decompose $f=c+u$ with $u(v)=0$, counting the common value $c$ only once. The
increment form obeys
\[
 Q_{a,\alpha}(u)=\sum_{r\le a}(a/r)^{2\alpha}
             \sum_{j=0}^m r^{2j-2}\|D^ju\|_{2,A_r}^2,
 \qquad Q_{a,\alpha}\le CB^{-2\alpha}Q_{Ba,\alpha}.
\]
Including $c$ in the norm makes this corner space a Banach algebra; localize by
$c+\chi u$. Subtract the noise at the vertex in the gradient cross term to get
\[
 2\left|\int\Delta f\,[\zeta-\zeta(v)]\right|
 \le\eta\int r^2|D^2f|^2+
          C\eta^{-1}\int r^{-2}|\zeta-\zeta(v)|^2.
\]
The annular trace is finite since $\alpha<\alpha_0$, and the shrinking puncture contributes
zero in the limit. The boundary-gradient form is bounded since $\alpha>1/2$. Grouping
absorbs the Young factor without increasing the unit coefficient of the
background gradient energy. In the neutral sector common shifts eliminate $c$
and inversion eliminates the linear increment. Upon extracting the identity,
the unmarked neutral remainder is quadratic, gaining $B^{-2\alpha}$. A marked linear term gains
$B^{-\alpha}$ in its own graded record. At grounded corners $c=0$.

A fine flat term restricted to a coarse corner must be summed over annuli. For
a neutral term of $p\ge2$ increment factors, degree $\ell>d$ and root dimension $d=1$ or $2$
this costs
\[
 B^{-p\alpha}\sum_{m\le B}m^{p\alpha-\ell+d-1}
       \le C(1+\log B)(B^{d-\ell}+B^{-p\alpha}).
\]
At cluster diameter $B^\theta a$ split the sum at $m=B^\theta$. The resulting exponents are $d-\ell+\theta\ell$ and
$-p\alpha+(p\alpha+d)\theta$, bounded in the bulk by $-1+3\theta$ and $-2\alpha+(2\alpha+2)\theta$ and on a face by $-1+2\theta$ and $-2\alpha+(2\alpha+1)\theta$,
respectively. All are negative for $\theta<1/3$. The reference action keeps the bulk
marginal terms; induced degree-two interface terms have root dimension one.
Matching the relative trace excludes the dimension-zero face term $\cos(h_1-h_2)$.

For charged terms split a short cluster's Gaussian interval at
$B^\theta a$. The pinned shift from Theorem~\ref{extensive:gaussian}
is bounded on the remaining interval, and the capped contour
\eqref{local-source-operation:contour} gives exponent
$2d(q)/Z_\epsilon$, where $d(q)=q^2$ for $|q|\le1$ and
$d(q)=2|q|-1$ otherwise. The exact quadratic charge exponent is
used only on the finite Wick basis. The analytic weight on the charge
coefficients makes their sum finite. Keep both endpoint Wick factors.
Once $B$ is fixed, $B^{C|Z_\epsilon-1|}$ tends to one, preserving
the strict contraction for fine cutoff.

For a cluster of $N$ blocks the exponential decay in $N$ absorbs
the polynomial number of possible anchors. For a long link or a
generating graph of length $ta_j$, with $t\ge B^\theta$, reblocking
changes its length weight by a factor at most
\[
 \left(\frac{1+t/B}{1+t}\right)^{L_H}
 \le 2^{L_H} B^{-\theta L_H}.
\]
Here $L_H$ is the length-weight exponent. Counting the fine anchors
contributes at most $CN^2B^2$, and rescaling the field norm contributes
at most $B^P$, as in Appendix~\ref{app:connected-expansion}.
Thus $\theta L_H>P+2$ gives decay in $B$ in both the birth-root row
sum and the current-block column sum. Fix these weight exponents first,
then take $B$ large and the weak-coupling ball small. The recursion $\|K_{j+1}\|\le\vartheta\|K_j\|+Cv_j$ has $\vartheta<1$, hence $\|K_j\|\le Cv_j$
by stable convolution. All four masks have these same even interfaces and only
finitely many corner intersections.

Keep the sheet label through contractions, extractions and scalar
compensation. Because the self covariance of a sheet is replica independent,
turning off its partner recovers the single-sheet operation. For the
accumulated scalar $E_{st}$ of sheets $s,t$ all single-sheet terms, including
subsequent returns to the scalar, consequently cancel in $2E_{IE}-E_{II}-E_{EE}$. A surviving
connected graph uses an interaction on each sheet. The quadratic expansion
bound makes its forcing and stable record $O(v_j^2)$; marked coefficients still
include orders zero and one.

For a boundary root at distance $s$ from the nearest defect distinguish
complete generating graphs longer than $cs$ from the rest. The long graphs gain
$(1+s/a_j)^{-p}$ from the length weight. For a short graph the reflected templates agree in
a common flat chart: telescope its covariances and use the reflection bound,
diagonals included, to recover that decay. Retain the label assigned at the
first changed covariance under all subsequent transport, counting it once.
Displacing the root changes its distance to the defect by at most the recorded
path length, so the same path weight controls the displaced root.

For $p\ge2$ the root contributes at most
$Cv_j^2\min\{1,(a_j/s)^2\}$. Along a face the sum per corner is
bounded by a constant times $\sum_{m\ge1}m^{-2}$. A remote defect
at clearance $L_0$ along a perimeter of length $P_\Gamma$ contributes
at most $(P_\Gamma/a_j)(a_j/L_0)^2$ to the root count. For roots
off the interface, the transverse and distant covariance bounds
in Theorem~\ref{extensive:gaussian} are summable over the possible
locations. Summing over scales with $\sum_{a_j\le R}v_j^2\le Cv_R$
proves \eqref{extensive:restoration}.

For the four-mask contrast, every graph missing $U$ or $O$ cancels
before absolute values. A survivor spans $cR$, so its scale-$a_j$
sum is at most $CN_\Gamma v_j(R/a_j)^p(a_j/R)^L$, with $L>p$.
This is summable. Integrate the final positive interval $[cR,R]$
before taking the numerical logarithm. It has fixed row and regulator
constants, and the local gas criterion depends on $v_{cR}$ rather
than on $v_RN_\Gamma$. Its numeric connected terms retain the same
mask cancellation. Ginibre supermodularity gives the lower sign in
\eqref{extensive:four-mask}.
\end{proof}

\subsection{Positive normalization and complete source errors}
\label{positive-replicas:fold}

\begin{proposition}[Positive normalization]
\label{extensive:normalizer}
\label{positive-replicas:affinity-theorem}
\label{normalized-sources:rr-positive-cavity}
\label{normalized-sources:rr-affinity-normalizer}
The exact folded inside/outside affinity is bounded below by $c_k>0$.
The deleted-collar replicas satisfy
$(Z_{IE}^{\rm del})^2\ge c_k Z_{II}^{\rm del}Z_{EE}^{\rm del}$,
and collar restoration satisfies
$e^{-CN_\Gamma}v_{st}Z_{st}^{\rm del}\le Z_{st}^{\rm full}
\le v_{st}Z_{st}^{\rm del}$.
The inside-replica partition is between $c_ke^{E_{\rm rep}}$ and
$C_ke^{E_{\rm rep}}$, where $E_{\rm rep}$ is its actual complete
scalar. These constants are uniform in containing domain, separation
and fine cutoff, with no weak-coupling requirement on the surviving
interacting arms after collar deletion. They may depend on the fixed
collars and stop $R$; the normalization bounds may deteriorate with
$N_R$. The weak-coupling threshold remains independent
of the number $N_R$ of occupied stop blocks.
\end{proposition}

\begin{proof}
\emph{Fold and replica.}
Condition on all original traces at once, assigning half of each
trace-site cosine to each side. With free trace law $\mu_0$, let
$u,v$ be the positive inside and outside conditional integrals, and
let $u_O$ be the possibly complex marked inside integral. Fold one
common trace value per ungrounded component modulo
$2\pi$, retaining all neutral differences, and put
$\bar v=\mathbb E[v\mid\text{folded trace}]$. Periodicity gives
the identity from Section~\ref{sec:observations}. Applying
Cauchy--Schwarz to its numerator gives, with $p$ as defined there,
\begin{equation}
 \langle O\rangle_D
   =\frac{\mathbb E[u_O\bar v]}{\mathbb E[u\bar v]},\qquad
 |\langle O\rangle_D|^2
   \le \frac{\mathbb E[|u_O|^2]}{p\,\mathbb E[u^2]}.
 \label{extensive:folded-cauchy}
\end{equation}
Let $U$ insert component constants and let $\Sigma_B$ be the trace
covariance. Choose a linear section of the quotient of the trace
space by $\operatorname{ran}U$, and write $b=\beta+Ua$, with
$\beta$ in this section. Conditional on the neutral trace coordinate
$\beta$, the constant coordinate $a$ has law $N(m(\beta),S)$, where
$S=(U^*\Sigma_B^{-1}U)^{-1}\succeq cI$: disjoint collar plateaus
realize a constant vector with energy at most $C|a|^2$.
Its wrapped density $w_S$ is bounded above and below because it is
the convolution of $w_{cI}$ with a probability law. The exact folded
replica density is
$\delta_{2\pi\mathbb Z^{k_u}}(a_1-a_2)/w_S(a_1-m(\beta))$
relative to two conditional Gaussian copies. It is uniformly
comparable to the normalized positive comb. Add an independent
$N(0,S)$ common shift to the two copies. Their common covariance
increases from $S/2$ to $3S/2$; the density ratio is at least
$3^{-k_u/2}\ge 3^{-k/2}$, while positive Fourier coefficients bound unmarked
partitions above by their old values. This gives the Gaussian
representation $h_s=\eta_s+H_s(b+U\delta_s)$ of
Section~\ref{sec:observations}, with common trace $b$ and independent
private shifts.
The comb constrains $\delta_1-\delta_2$. For a marked square,
integrate the two conditional inside fields first: periodicity makes
their product $|u_O|^2\ge0$, so the density comparison applies.
Grounded components use zero-boundary Poincar\'e and are not folded.

\emph{Affinity through an empty collar.}
Delete the collar activities, and cut inside the resulting Gaussian
collar. The two surviving interacting arms have their actual positive
amplitudes $F_s(X_s)$, with no weak-coupling restriction. Their cut
covariances are
\[
 \Sigma_s=A_s+K_{ss}+D_s,\qquad
 K_{st}=H_s\Sigma_BH_t^*,\qquad
 D_s=(H_sU)S(H_sU)^* .
\]
Here $A_s$ is the independent killed covariance. On neutral trace data let $L$
be the interior DtN form. A fixed collar extension bounds the exterior
form by $CL$. Schur complementation gives
$\Sigma_B\preceq C(L^{-1}_{\rm neu}+USU^*)$:
the neutral covariance is at most $L^{-1}$, and the conditional mean
covariance is at most $CS$ since
$\Lambda_{c\beta}L^{-1}\Lambda_{\beta c}\preceq CS^{-1}$.

Cut off a neutral harmonic extension to vanish at the old trace and
equal itself at the new cut. Its energy is bounded by the neutral
trace energy; therefore its cut covariance is at most $CA_s$.
Its relative trace is bounded as well: point and gradient evaluation
variances on the separated collar are $C$ and $CR^{-2}$, whose cutoff
energy integrates to $C$ per occupied collar block. The private
constant term has rank at most $k$. Consequently
the normalized shared covariance
$J_{st}=\Sigma_s^{-1/2}K_{st}\Sigma_t^{-1/2}$ satisfies
$\|J_{st}\|\le q<1$ and $\operatorname{Tr}J_{ss}\le C$.
Conditional on both cuts, the comb difference has covariance at least
$c'I$: change its private shifts by $\pm a/2$ and cancel their cut
values by killed collar extensions of energy $C|a|^2$. Thus the comb
multiplies the cut joining kernel by bounded positive factors.

For whitened cuts $x_s$, put $R_s=x_s^{\top}J_{ss}x_s$. Gaussian block
inversion and the determinant bound
$-\log\det(I-J_{st}^*J_{st})\le C_q\operatorname{Tr}(J_{st}^*J_{st})$ give
$e^{-C(1+R_s+R_t)}\le k_{st}\le e^{C(1+R_s+R_t)}$.
Under the complete positive-cosine self law, Gaussian domination
\eqref{positive-cosine-laws:pinned-gaussian-domination} gives
$\mathbb E[R_s]\le\operatorname{Tr}J_{ss}\le C$. Choose fixed balls
where both cuts carry at least three quarters of each self law.
There $k_{ss}\le C_B$ and $k_{IE}\ge c_B>0$. Writing
$m_s=\int_{B_s}F_s\,d\mu_s$ gives
$m_s\ge(3Z_{ss}^{\rm del}/4C_B)^{1/2}$ and
$Z_{IE}^{\rm del}\ge c_Bm_Im_E$. Hence
\begin{equation}
 (Z_{IE}^{\rm del})^2\ge c_kZ_{II}^{\rm del}Z_{EE}^{\rm del}.
 \label{extensive:deleted-affinity}
\end{equation}
The constants depend on occupied collars, not the size of the arms.

\emph{Restoration and the identical scalar.}
For one replica let $\Phi_A(\psi)=\mathbb E_{C_R}e^{V_A(\psi+\zeta)}$.
The connected estimate, differentiated twice at zero after the final
fixed-ratio Gaussian interval, bounds
$-D^2\log\Phi_A(0)$ by $Cv_R$ times the local Sobolev form including
block means. This is a bound on the complete numeric logarithm;
source-free extracted scalars have zero Hessian. The global phase
inequality \eqref{positive-cosine-laws:phase-bound} transports it to
all real matching-trace phases. Charge-covariance monotonicity bounds
every masked curvature by the full one.

Take the comb limit at fixed cutoff before estimating the interface.
Gauge the integer differences throughout each active component, use
identical cutoffs on its sheets, and fold one common phase. This keeps
the gauged field in the linear matching-trace space and gives
$\Phi_V(\psi)/\Phi_V(0)\ge
e^{-Cv_R(N_\Gamma+Q_{\rm means}+Q_{\rm jets})}$.
The derivative form has bounded covariance operator and trace
$CN_\Gamma$. Neutral means are dominated by a coarse graph inverse
using fixed local energy flows. For a connected graph with $N$ vertices,
a spanning-tree traversal repeats every edge twice, and the min--max
principle compares its nonzero Laplacian eigenvalues with those of
a path of length at most $2N$, hence with $c(j/N)^2$. Therefore
$\sum_{j=1}^{N-1}\log(1+Cv_RN^2/j^2)\le C\sqrt{v_R}N$.
The compact common phase is already bounded. Hadamard--Fischer
handles correlations between distinct components. The determinant
lower bound also holds under the positive comb decoration: completing
a nonnegative quadratic form replaces Gaussian covariance $Q$ by
$Q_T\preceq Q$, and the positive Fourier series gives
$Z(Q_T)\ge Z(Q)$. Thus the decorated negative quadratic integral
is at least $\det(I+Q^{1/2}TQ^{1/2})^{-1/2}$, also after the
fixed-cutoff comb limit. We obtain
\[
 e^{-C(\sqrt{v_R}+v_R)N_\Gamma}\Phi_V(0)
       \le Z_V\le\Phi_V(0).
\]
This does not put the possibly large mean covariance into a positive
exponential regulator.

Ginibre supermodularity and phase ordering give
\[
 \frac{Z_{U+B}}{Z_B}\le\frac{Z_{U+B+O}}{Z_{B+O}}
       \le\frac{\Phi_{U+B+O}(0)}{\Phi_{B+O}(0)}.
\]
Together with \eqref{extensive:four-mask} these prove
\begin{equation}
 e^{-CN_\Gamma}v_{st}Z_{st}^{\rm del}
       \le Z_{st}^{\rm full}\le v_{st}Z_{st}^{\rm del}.
 \label{extensive:cavity}
\end{equation}
Combine \eqref{extensive:restoration}, \eqref{extensive:deleted-affinity}
and \eqref{extensive:cavity}, then restore the bounded fold comparisons.
The actual $p$ in \eqref{extensive:folded-cauchy} is bounded below.
On the fixed inside region the same proof yields
\begin{equation}
 c_ke^{E_{\rm rep}}\le Z_{\rm rep}=\mathbb E[u^2]
                       \le C_ke^{E_{\rm rep}}.
 \label{normalized-sources:rr-9-3}
\end{equation}
Here $E_{\rm rep}$ is the actual complete two-sheet scalar, including
kinetic, interface, corner and compensation terms. The terminal numeric
log is $O(v_RN_R)$. Keep its complete unmarked weak factors
$U_{\mathcal F}$ together with all marked occupied regulators.
The rooted majorant sums these complete factors with a bound
$\exp(CN_R+cQ_{\rm jets})$ in the same graded norm.
For
derivative and corner quadratic forms take the coefficient below $1/(4K)$, with $K$
the uniform covariance operator bound, reserving the same margin for the
forest. The Gaussian determinant now gives $\int|U_{\mathcal F}|\prod_vG_v\,d\gamma_{\mathcal F}
\le e^{CN_R}C_0^m$ on $m$ occupied blocks without
changing the factored scalar $E_{\rm rep}$. This joint estimate is what the local
expansion lemma requires beyond the unmarked partition bound. Integer phases
have modulus one and common phases range over compact tori; the weak threshold
is consequently independent of $N_R$ even for $N_R\asymp(L_0/R)^2$.
\end{proof}

\begin{theorem}[Normalized simultaneous source comparison]
\label{extensive:comparison}
\label{normalized-sources:rr-complete-packet-comparison}
The actual finite-box complete coefficients admit the prescribed local
physical comparators $P_r$, fixed before the containing domain, with
\begin{equation}
 |\langle O\rangle_D-\langle P_r\rangle_D|\le C_k(r/R)^\nu
 \label{extensive:normalized-error}
\end{equation}
after cutoff removal in each fixed box. The estimate holds for every
required sublist and mixed switch coefficient. At lattice cutoff, with
$E_i=O_i-P_{r_i}$ and $d_i=C(r_i/R)^\nu+\omega_{r_i,D}(\epsilon)$,
where $\omega_{r_i,D}(\epsilon)\to0$ at fixed $r_i,D$, the exact
inside conditional integrals satisfy
\begin{equation}
 \frac{\mathbb E[|u_{\boldsymbol E}|^2]}{\mathbb E[u^2]}
       \le C_k\prod_{i\in\mathcal A}d_i^2
                     \prod_{i\notin\mathcal A}M_i^2.
 \label{extensive:conditional-errors}
\end{equation}
The denominator is the actual complete inside-replica normalizer.
Here $\mathcal A$ indexes error packets and $M_i$ is the complete
norm of each remaining packet. The comparators are the periodic
mollifier and derivative cylinders of the preceding local matching
construction, including their complete contact columns.
\end{theorem}

\begin{proof}
\emph{The complete source switch.}
Two triangular switches in the full nilpotent source algebra have off-diagonal
product $(O_1-P_{r,1})(\overline O_2-\overline P_{r,2})$, including each scalar and contact coefficient before absolute values.
At scale $s=R(r/R)^\beta$ below clearance $d_0$, a source-connected history differing from the
ordinary killed prefix uses a guard boundary covariance or reaches the interface.
Its generating support gives
\begin{equation}
 \|\Delta\mathcal P_{s,S}\|
 \le C_S\sum_{a_j\le s}a_j^{-P_S}(a_j/d_0)^M
 \le C_Ss^{-P_S}(s/d_0)^M.
 \label{normalized-sources:rr-10-2}
\end{equation}
Choose $M$ with $\beta(M-P_S)\ge |S|\nu$ for all sublists. Together with ordinary complete matching this
gives $d_i$ per switch. Extraction preserves the support graphs of vacuum
dependencies meeting the guard, covering both early common links and later contact
returns. Later boundary integrations cost at most $\exp(C\sum_{a_j\le R}(a_j/d_0)^M)$. Rescale the switches by
their factors. Products, Gaussian maps, local subtractions and scalar coboundaries
preserve tensor grading. By local expansion and the retained regulator estimate a
switch $E=O-P_r$ of error $d=C(r/R)^\nu+\omega_{r,D}(\epsilon)$ satisfies
\begin{equation}
 \mathbb E[|u_E|^2]\le C_kd^2e^{E_{\rm rep}}.
 \label{normalized-sources:rr-10-3}
\end{equation}
For joint lists reconstruction gives the product of squared factors with exactly
the scalar of \eqref{normalized-sources:rr-9-3}, proving \eqref{extensive:conditional-errors} by division. Treat bounded-region contacts as one
joint coefficient and condition on the union once, with one constant coordinate
per component. This avoids both a separation-dependent hull and repeated
denominators. Apply \eqref{extensive:folded-cauchy}, telescope the fixed comparator list and remove the cutoff
at fixed $r,D$ to get \eqref{extensive:normalized-error}.
\end{proof}

\section{Extensive parameter responses}
\label{sec:responses}

An endpoint derivative inserts the action throughout the domain.
To avoid the possible volume factor in local bounds, split the action into
packets, compare them in the normalized law, and sum connected terms
by screening with one source packet fixed. We obtain the fourth
derivative directly in a fixed box. Interior interpolation then takes
the first two derivatives to the selected infinite-volume limit.

Choose a smooth physical partition $\sum_b\chi_b=1$ on all sites
and original wall bonds. Label the profile that equals one on the
entire source/contact guard by $\chi_0$. All other profiles vanish
there. The profiles vary at the fixed
length $L_0$, independently of the containing domain. Put
$r_x(\mathbf t)=\rho+\sum_bt_b\chi_b(x)$, write $r_e$ for its prescribed interpolation on an edge $e$, and use the literal log weight relative to the fixed
critical Gaussian,
\begin{equation}
 -\frac12\sum_e[Z_\epsilon(r_e)-1](\nabla_e\phi)^2
       +2\sum_xy_\epsilon(r_x)\cos(\beta_*\phi_x).
 \label{extensive:literal-action}
\end{equation}
The original source injection has parameter $\rho+t_0$ on its flat
guard. On $t_b=t$, this is exactly the original $\rho+t$ law,
including its boundary bonds; hence
$(\sum_b\partial_{t_b})^q=\partial_\rho^q$ on the diagonal.
The Gaussian observation basis is held fixed: actual parameter
derivatives are complete action marks in this fixed reference law.

\begin{theorem}[Complete parameter responses]
\label{extensive:responses}
\label{parameter-responses:action-packets}
\label{parameter-responses:rr-fourth-response}
For every fixed complete source list, its prescribed partners and
source-generator enlargement, the literal action coefficients through
four labels have the local bounds of Theorem~\ref{extensive:comparison}
and a direct continuous cutoff limit in each fixed original square.
After that cutoff limit the actual normalized coefficient $F_D$ is
$C^4$ in $\rho$. On compact positive intervals $J_0\Subset J_1$,
uniformly over the selected squares and their bounded dilates,
\begin{equation}
 \sup_D\max_{0\le q\le4}\|\partial_\rho^qF_D\|_{J_1,I,\mathcal B}
       \le C_{I,\mathcal B,J_1}.
 \label{parameter-responses:final-C4}
\end{equation}
Here $I$ is the fixed source list and $\mathcal B$ a bounded family
of tests. The norm is the coefficient seminorm on this family,
uniform in the indicated parameter interval. If $d_D\to0$ bounds the complete value
errors on $J_1$ for the generator-enlarged family, then
$F_\infty\in C^2(J_0)$ and
\begin{equation}
 \|\partial_\rho^qF_D-\partial_\rho^qF_\infty\|_{J_0,I,\mathcal B}
       \le C_qd_D^{1-q/4},\qquad q=1,2.
 \label{parameter-responses:final-response-passage}
\end{equation}
The two original domain defects satisfy
\begin{equation}
 \|B_{D,1}F_D\|\le C(d_D^{3/4}+d_D),\qquad
 \|B_{D,2}F_D\|\le C(d_D^{1/2}+d_D).
 \label{extensive:wall-limits}
\end{equation}
All limits remove the cutoff in a fixed box first. No convergence of
the selected fourth derivative or rate for $d_D$ is asserted.
\end{theorem}

\begin{proof}
\emph{Local action coefficients.}
In the polarized algebra of at most four labels the complete canonical input
satisfies $|\partial_\rho^qv_j|\le Cv_j^2$, $1\le q\le4$ and the corresponding stable and even-wall bounds. At a
deterministic root $x_B$ freeze the summed input at $\rho+\sum t_b\chi_b(x_B)$ to obtain $\mathcal C_j[\chi]$. Store the
pure-parameter scalar before subtracting the frozen record. Naturality at
constant parameters gives
\[
 E_{j+1}=\Phi_j(\mathcal C_j[\chi]+E_j)-\Phi_j(\mathcal C_j[\chi])
       +\mathcal F_j[\chi].
\]
At root $x$ the forcing $\mathcal F_j$ is the difference between $\Phi_j(\mathcal C_j[\chi])$ and $\Phi_j(\mathcal C_j[\chi(x)])$. Telescoping
the inputs gives one profile difference bounded by $C\min(1,a_j\ell/L_0)$ per term, where
$\ell$ is graph length in scale-$a_j$ units. For $0<\eta\le1$ this is at most $C(a_j/L_0)^\eta\ell^\eta$. The
length weight sums $\ell^\eta$. Allocations of up to four derivatives grow
polynomially in block count, absorbed by its exponential weight. The
canonical derivative bound gives $\|\mathcal F_{j,S}\|\le C(a_j/L_0)^\eta v_j^2$ for every nonempty action-label set $S$.

At a true wall freeze the even-wall jet including boundary bonds. A
source-to-transition graph spans the guard, gaining any power of the ratio of
$a_j$ to its width.

The dimension-two action block has zero engineering loss and interaction
correction $O(v_j)$. By triangularity in label degree, $v_{n-q}\le Cv_n/(1+cqv_n)$ bounds its propagator by
$C\exp(C\sum_{k=j}^{n-1}v_k)\le C(1+n-j)^P$. Duhamel at $a_n\asymp R$ gives $\|E_{n,S}\|\le Cv_n^2\sum_{q\ge0}(1+q)^Pb^{-\eta q}$. Induction controls proper subsets, and the
initial defect vanishes with its factor $(\epsilon/L_0)^\eta$.

Pure action contacts have superficial degree two at every order. Their
retained nonconstant columns are kinetic and integer-charge columns. Profile
derivatives raise dimension, and relevant profile-only columns are scalar. At
source contacts the profile is constant, so the counterterm is the derivative
of the prescribed contact in proper-subset order. For a guard crossing choose
the length weight above the list's largest superficial degree: the resulting
power of $a_j$ absorbs every contact loss.

In fixed volume the source construction already gives $C^3$ regularity.
For the fourth derivative, stop first at resolution $\delta$. Transport
leaves a positive power of $\delta/R$ in both tails, the one from profile
variation and the one from Taylor subtraction at contacts. For the frozen
four-jet we can remove the cutoff directly above $\delta$: only a finite
window of lattice kernels and exact moment inverses has to converge
there. Now take $\delta\downarrow0$. The four action coefficients are
continuous, and the last is the fourth derivative.

Write $\eta$ for the source labels. The complete scalar without $\eta$ factors out:
\[
 Z_\epsilon(\mathbf t;\eta)
 =e^{C_\epsilon^{\rm vac}(\mathbf t)}\,\widetilde Z_\epsilon(\mathbf t;\eta).
\]
Division by $Z_\epsilon(\mathbf t;0)$ cancels it, preserving all source-bearing identity and contact
columns in the derivatives of the normalized law corresponding to the log
weight \eqref{extensive:literal-action}.

\emph{Anchored summation.}
For Theorem~\ref{normalized-source-theorems:reusable-response} with $q_0=4$, the fixed-box coefficient limit was proved above. By Theorem~\ref{extensive:comparison} every
required sublist admits simultaneous approximation. An observable bundle and
four action marks require five packets, or $k_0+4$ if the original packets are
translated separately. For the comparators, finite integer phases and
polynomials in normalized derivative smears, the coefficients and coordinate
Lipschitz constants are bounded by $Cr^{-P}$. Gaussian domination bounds the clipping
error by $Cr^{-P}e^{-cK^2}$. Integration by parts expresses each signed derivative-height test
as a difference of nonnegative tests, so the association envelope (Lemma~\ref{height-representation:association-envelope}), the
finite-domain entrywise comparison of Lemma~\ref{physical-theory:entrywise-domain-order} and the independent screening estimate
in the selected state give
\begin{equation}
 |\operatorname{Cov}_D(P_{A,r,K},P_{B,r,K})|
       \le Cr^{-P}(1+K)^Pe^{-\mu d}.
 \label{parameter-responses:rr-12-2}
\end{equation}
Uniform moments at a single comparator resolution pass by simultaneous
comparison to the complete packet moments. This verifies the remaining
hypotheses of the response theorem. With $r=r_0e^{-\zeta d}$ and $K^2=Md$ it gives exponential packet
factorization. Cutting a longest edge in a minimum spanning tree gives
exponential decay of the connected terms. In \eqref{normalized-source-theorems:reuse-polarized-response} we can therefore sum over every
action position with an original source label fixed. The normalized logarithm
has already lost its parameter-only components, giving \eqref{parameter-responses:final-C4} without a volume
factor. At fixed volume the action partition is finite. The continuous
fourth-order coefficient limit passes the successive identities of the
fundamental theorem of calculus, giving classical $C^4$ regularity.

\emph{Selected response limits.}
The interior finite-difference estimate of the same theorem, applied to the
value modulus, gives \eqref{parameter-responses:final-response-passage}. Set $H_D=F_D-F_\infty$. For the generators $\mathfrak b$ and $\mathscr D_\rho$ in Proposition~\ref{scale-responses:wall-identities} write $L_\rho=\mathscr D_\rho^2+\mathfrak b\,\partial_\rho\mathscr D_\rho
-\mathfrak b'\mathscr D_\rho$ as
in Section~\ref{sec:transfer}, where $\partial_\rho\mathscr D_\rho$ differentiates only the coefficients and profile substitutions
of $\mathscr D_\rho$. By the exact source/domain cocycle,
\[
 \begin{aligned}
 B_{D,1}F_D&=\mathfrak b H_D'-\mathscr D_\rho H_D,\\
 B_{D,2}F_D&=\mathfrak b^2 H_D''-L_\rho H_D.
 \end{aligned}
\]
These are the defects $\mathscr W_D$ and $\mathscr W_D^2+2\mathscr D_\rho\mathscr W_D-\mathfrak b'\mathscr W_D$ of Proposition~\ref{scale-responses:wall-identities}. The value modulus on the
generator-enlarged list controls the remaining source terms, yielding \eqref{extensive:wall-limits}. The
second-order conclusion concerns the complete defect. Both local admission and
screening are inputs proved independently of this response argument.
\end{proof}

\section{Euclidean covariance and relativistic reconstruction}
\label{sec:reconstruction}
\label{sec:relativistic-theory}

To reconstruct a relativistic theory from the gapped source representation with
positive Hamiltonian, we recover continuous rotations from a radial auxiliary
law and bound the growth of the correlation distributions with the number of
insertions by a uniform repeated-insertion estimate. A common dense
chronological core identifies the reconstructed representation with the one
carrying the physical gap.

Our conventions are $h=\beta_*\phi$ and $D_\mu(f)=-\beta_*^{-1}h(\partial_\mu f)$, with $\beta_*^2=8\pi$. We work on a fixed compact positive
interval of endpoint parameters and remove the cutoff before taking the selected
Dirichlet exhaustion. For every finite field list, the coefficient comparison
includes its conjugates, its required descendants and its prescribed contacts.

\subsection{Stability of the microscopic preparation}

Matching the auxiliary regulator requires its actual initial density. Beyond the
quadratic marginal flow we must keep the free linear shape, its quadratic
returns and the independent scalar history. Both radial regularization and the
higher harmonics in Section~\ref{regulator-universality:universality-section} use this comparison.

\begin{lemma}[Stability under a decaying preparation change]
\label{continuum:preparation-comparison}
Suppose an actual microscopic preparation has an invertible ordinary
two-coordinate map with positive activity derivative and the allowed
stable and even-wall bounds on a sufficiently small common graph ball.
After retaining its actual free linear shape and quadratic memory,
suppose its complete operations differ from the canonical limiting
operation by $O(\sigma^j)$ in the reserved finite jets at age $j$,
$\sigma<1$. The free linear shape may itself have such a decaying
defect. Calibrate its exact ordinary
free multiplier so that every reduced marginal map has linear jet $I$.
Then its true forward graphs converge to the canonical graph, and
its reduced maps $f_j$ admit conjugacies
\begin{equation}
 \|H_j-H_\infty\|_{C^8,w}+\|f_j-f_0\|_{C^8}\le C\sigma_1^j,
 \quad c_{j+1}f_j=f_0c_j,
 \quad c_j(g)-g=O(\sigma_2^j|g|^2),
 \label{continuum:preparation-stability}
\end{equation}
with bounded $C^{10}$ norms and $\sigma_1,\sigma_2<1$.
The bare choice
\begin{equation}
 (1-Z,y)=\mathcal P^{-1}c_0^{-1}f_0^{-N}p_\rho
 \label{continuum:matched-preparation}
\end{equation}
tunes the actual law to the canonical history. For complete source
comparison, assume the targets agree and the histories satisfy the bounded-preparation,
locality and contact-tail hypotheses of marked recognition, with the
established anomalous inverse loss and zero independent initial marked
scalar. If their retained laws satisfy the original Gaussian domination,
then their
complete normalized fixed-box limits agree, including all mixed Grams.
The original fixed-reference response can be imposed exactly by its
own local inverse. All conclusions retain the available four-jet
bounds and interior $C^3$ fixed-box comparison.
\end{lemma}
\begin{proof}
Compare the two forward graph equations at the same ordinary
argument. After $r\le8$ differentiations, their highest-order
difference is multiplied by
$q_r=(\theta+C\eta)(1-C\eta)^{-r}<1$.
All lower-order differences occur triangularly, and the inhomogeneous
operation error is $O(\sigma^j)$. Two unused derivatives bound the
change in the inverse arguments. A fixed weighting of this triangular
system therefore gives the graph estimate in
\eqref{continuum:preparation-stability}. The quadratic memory satisfies
the same retarded convolution, so its initial discrepancy is forgotten
at a geometric rate.

The summable conjugacies for these reduced maps come from Theorem~\ref{arbitrary-cutoffs:allcutoff-theorem}. This gives $c_j$ and
the remaining bounds in \eqref{continuum:preparation-stability}. With the bare pair \eqref{continuum:matched-preparation} the ordinary orbit agrees
exactly, including the effect of the higher terms in the limiting map. At a wall
the initial discrepancy contracts. The forcing converges on every fixed recent
window, and the tail beyond that window is geometric. Throughout the comparison
we track the scalar histories separately.

For a marked list $I$ let $d_j$ measure the discrepancy in the entire unmarked input
and $e_j$ the discrepancy between the marked graphs. The recurrence $e_{j+1}\le q_Ie_j+C_Id_j$, with $q_I<1$,
gives a convergent retarded convolution. There is only a finite factor $\exp(C_I\sum_j\sigma_1^j)$ in the
ordinary inverse product, and the ultraviolet exponent in the contact Taylor
remainder stays positive. We may therefore invoke Lemma~\ref{source-dilation:marked-recognition}. Its scalar comparison
involves $Q_a(0)+\sum s_j$ even if the ordinary normalization data already coincide.

In a fixed box, integrability of the retained Gaussian regulator and the lower
bound on the true partition denominator give the comparison of normalized
coefficients and mixed Grams. Partial shells of bounded length suffice for
arbitrary cutoffs. On a compact endpoint interval we start on one common earlier
weak section and continue for finitely many steps. The compact-reference readout
converges with its first derivative to the original nondegenerate readout.
Inverting it in the fixed reference box fixes the original reference value,
independently of the target domain. The stated regularity follows by
differentiating the recurrences.
\end{proof}

\subsection{Radial matching and rotations}
\label{radial-preparation:radial-auxiliary-matching}

For rotations beyond the square symmetries of the lattice, we match to a
positive radial regularization and use domain order. Including the exterior
collar makes the interaction a single functional of the zero-extended field for
all nested domains.

Take a nonnegative radial $\varrho\in C_c^\infty(B_1)$ with $\int\varrho=1$ and set $\varrho_r(x)=r^{-2}\varrho(x/r)$ and $P_rh=\varrho_r*h$. Denote extension by zero
by $h^0$ for fields on $D$, and write $G_D=(-\Delta_D)^{-1}$. If $Z_{r,D}$ is the partition function at
stiffness $Z$ and activity $u$, the radial measure is
\begin{equation}
 d\nu_{r,D}^{Z,u}(h)=Z_{r,D}^{-1}
 \exp\left\{2ur^{-2}\int_{\mathbb R^2}
                   [\cos(P_rh^0)-1]\,dx\right\}
 d\gamma_{\beta_*^2Z^{-1}G_D}(h).
 \label{radial-preparation:auxiliary-law}
\end{equation}
Write its characteristic functional as $\chi_{r,D}(f)=\int e^{ih(f)}d\nu_{r,D}(h)$. The integral of the interaction
extends over $D+B_r$, exterior collar included, and the exponent takes values
between $-4ur^{-2}|D+B_r|$ and zero. This bound gives convergence of smooth Gaussian
approximants for fixed $r>0$ in each bounded domain.

\begin{proposition}[Radial matching]
\label{radial-preparation:auxiliary-preparation}
\label{radial-preparation:auxiliary-axis-matching}
There are domain-independent $Z_r(\rho)>0,u_r(\rho)>0$ for which
\eqref{radial-preparation:auxiliary-law} has the canonical fixed-box
height and derivative limit in every true axis rectangle, jointly
with all finite moments and the original reference normalization.
\end{proposition}
\begin{proof}
The comparison begins before cutoff removal. In a finite Gaussian
approximation, decompose $G=C+T$ into its ultraviolet prefix and
retained covariance, set $A=G^{-1}$ and $s=1-Z$, and complete the
kinetic square. The resulting covariance, mean and quadratic terms are
\begin{equation}
 \begin{aligned}
 C_s&=C(1-sAC)^{-1},& M_s&=(1-sCA)^{-1},\\
 Q_s&=sA(1-sCA)^{-1},& T_s&=T+\frac{s}{1-s}TAT.
 \end{aligned}
 \label{continuum:kinetic-completion}
\end{equation}
The high-field determinant cancels in the normalized functional
whose coefficients are being compared. At any positive retained
resolution the covariance correction is smoothing; each mollified
row has finite covariance and depends analytically on $s$.
Thus the completion supplies a finite preparation without requiring
a Radon--Nikodym comparison of the rough limiting Gaussian fields.

At a preparation scale $cr$, partition the anchor integral into
blocks. Each block activity is $O(|u|)$ in the fixed analytic norm;
its enlarged support meets a bounded number of blocks. For wall rows
use the exact Fourier identity
$\mathbb E_{C_s}e^{if(M_s\psi+\zeta)}
=e^{-C_s(f,f)/2}e^{if(M_s\psi)}$;
field derivatives act on the bounded-mass mollifier row, not on a
truncated covariance shift. This also gives a uniform complex-kinetic
bound: $\Re C_s\succeq cC$, while covariance Cauchy--Schwarz bounds
the imaginary mean by
$\frac c4 C(f,f)+C|\Im s|^2\langle\psi,ACA\psi\rangle$.
Before using the regulator, localize the latter form to the row
support by expanding $M_s-I=\sum_{k\ge0}s^{k+1}(CA)^{k+1}$.
Finite range gives successive support enlargements and the smaller
fixed $s$-disk gives an exponentially summable tail.
Writing $Q_a$ for the retained covariance at the current preparation
scale, $CA=1-AQ_a$ is uniformly bounded on the same little H\"older
spaces, since $AQ_a$ is a smooth low-pass kernel.  Insert smooth block
partitions of unity with fixed overlap between consecutive factors of $(CA)^k$ and retain each
resulting neighboring-block path as its generating support.  Writing $A_{\rm pol}$ for the polymer size weight, there are
at most $D^k$ paths and each occupies at most $c(k+1)$ blocks. Their
support-weighted sum is bounded by
$C\sum_{k\ge1}(|s|D L A_{\rm pol}^{c})^k$ for a fixed local operator
constant $L$. Choose the complex $s$-disk after $A_{\rm pol},D,L$ are fixed.  Gradient localization follows from
$(CA)^k\phi=Cd^*(dCd^*)^{k-1}d\phi$, so this decomposition introduces
no undifferentiated height term.
 The exponential weight in the support size absorbs this tail once
per occupied support, independently of the number of signed rows in $f$. The connected activity majorant
now constructs an actual analytic weak record and removes the temporary
Gaussian cutoff at fixed $r$.

Integration leaves the free linear vertex $2ur^{-2}e^{-Q_r(a)/2}\cos(P_r\psi)$. To compute its scale factor, use
the unit-field prefix multiplier $|p|^{-2}k(a|p|)$, where $k(s)=\int_0^{s/(c_{\rm fr}\sqrt3)}v\varphi(v)\,dv$. Here $k(0)=0$, $k(\infty)=1$ and $k'$ decays
rapidly. The normalization $\widehat\varrho(0)=1$ gives
\begin{equation}
 Q_r(a)=4\int_0^\infty|\widehat\varrho(q)|^2k(aq/r)\frac{dq}{q}
       =4\log(a/r)+c_\varrho^{(0)}+O((r/a)^\kappa).
 \label{radial-preparation:auxiliary-free-diagonal}
\end{equation}
Indeed, on differentiating the integral with respect to $\log(a/r)$, we may use $\widehat\varrho(q)=1+O(q^2)$ and
the rapid decay of $k'$ to obtain $4+O((r/a)^2)$. Integration recovers \eqref{radial-preparation:auxiliary-free-diagonal} with the same
fixed derivative bounds. In particular $d_j=(a_j/r)^2e^{-Q_r(a_j)/2}$ approaches a positive limit $d_\infty$ at a
geometric rate. Since $P_r1=1$, the ordinary free multiplier is exactly $d_{j+1}/d_j$ with the
returns of the complementary shape included. In the activity coordinate $(d_\infty/d_j)v_j^{\rm raw}$ the
marginal linear jet is $I$ and the limiting chart agrees with the original one.

At $a_j=Arb^j$, radiality gives $P_r\psi-\psi=O((r/a_j)^2)$
in current-scale jets. Keep this actual linear complementary shape
and the actual quadratic memory in the polynomial preparation.
The remaining initial body is cubic; a large fixed $A$ makes the
linear graph slope small. Its quadratic memory difference obeys
$e_{j+1}^{(2)}\le q e_j^{(2)}+C\sigma^j$, $q<1$; hence it also
decays geometrically. The operation differs only through these
explicit templates, so its full one-defect norm and reserved jets
have the rate required by Lemma~\ref{continuum:preparation-comparison}.
The ordinary microscopic Jacobian is
$\operatorname{diag}(2,c_\varrho)$, $c_\varrho>0$.
Wall and exterior-collar anchors enter the even-wall record, of size
$O(|g|)$: its pinned nonconstant part starts quadratically, giving
the original face and corner gains. Scalar evaluations are retained.
Lemma~\ref{continuum:preparation-comparison} supplies the actual
tuning and canonical retained limit. The Gaussian domination argument
applies to arbitrary positive cosine rows. Its high-field error for
each smooth height or derivative smear is $O(R)$ in every fixed
$L^p$, as in Proposition~\ref{literal-fields:literal-derivative-fixed-box}.
Remove $r$ at fixed retained $R,D$, then $R$. This identifies the
literal laws and their moments.
\end{proof}

The two bounding axis squares give the same limit and thus identify the rotated height law. For
composite fields the prescribed finite parts need not form irreducible tensor
multiplets. We keep those finite parts in determining the rotation action.

\begin{theorem}[Euclidean covariance in the original dictionary]
\label{euclidean-covariance:radial-domain-order}
\label{euclidean-covariance:selected-height-rotations}
\label{euclidean-covariance:complete-source-rotations}
The selected height law is Euclidean invariant. Every finite local
field multiplet has a bounded, spatially constant finite conversion
$\widehat{\mathcal O}=C_I\mathcal O$ to a covariant tensor frame.
In the original frame its separated correlations transform by
$A_I(R)=C_I^{-1}\Gamma_I(R)C_I$. On complete source jets the action is
\begin{equation}
 \mathfrak U_I(R)=\mathfrak A_I\Gamma_I^{\rm src}(R)
                              \mathfrak A_I^{-1},\qquad
 \mathcal Z_\rho[\mathfrak U_I(R)J]
                      =\mathcal Z_\rho[J],
 \label{euclidean-covariance:original-contact-rotation}
\end{equation}
Here $\Gamma_I(R)$ is the usual tensor representation of $R\in O(2)$,
and $(\Gamma_I^{\rm src}(R)J)(x)=\Gamma_I(R)^{-\top}J(R^{-1}x)$,
extended by the tensor and derivative rules to complete jets.
Moreover $C_I=(D\mathfrak A_I(0))^*$, with the formal adjoint including
integration by parts. In particular the literal derivative transforms
by $A_I(R)D_\mu=R_{\mu\nu}D_\nu$. The finite substitution includes every contact,
fixes the identity and literal derivative, commutes with the original
reflection and square symmetries, and is coherent under list inclusion.
Its coefficients and inverse are bounded and $C^3$ in the parameter.
\end{theorem}
\begin{proof}
For $D\subset D'$, the Dirichlet variational principle gives
$G_D\preceq G_{D'}$ after extension by zero. The two radial
interactions, including their collars, are restrictions of one
functional. The positive-cosine comparison of
Corollary~\ref{covariance-comparison:cosine-covariance-order} therefore
gives $0<\chi_{r,D'}(f)\le\chi_{r,D}(f)$, while radiality gives
$\chi_{r,RD}(f)=\chi_{r,D}(f\circ R)$.
For $D_L=(-L,L)^2$ the inclusions
$D_{L/2}\subset RD_L\subset D_{2L}$ trap the rotated
characteristic functional between those in two axis squares.
First remove $r$ in these fixed squares by radial matching; only
then let $L\to\infty$. Both bounds tend to the same selected law,
so $\chi_\infty(f\circ R)=\chi_\infty(f)$.
Polarization and the established moment bounds identify the joint
height and derivative laws. Together with translations, this gives
their Euclidean covariance.

\label{euclidean-covariance:intrinsic-marked-frame}
For composite fields, consider the connected marked germ at a weak
physical scale $a$, retaining its preparation, stable and marked
scalar terms. Differentiate in the declared field and source variables
and project the constant field onto charge $q$. Translation invariance
separates the common anchor from a relative distribution
$k_{a,S,q}$. Its intrinsic local moments, on the finite polynomial
module modulo integration by parts, are
\begin{equation}
 \mathfrak m_{a,S,q}(p)
   =\lim_{L\to\infty}\langle k_{a,S,q},p(z)\chi(z/L)\rangle,
 \qquad \chi\text{ radial and }\chi=1\text{ near }0.
 \label{euclidean-covariance:intrinsic-marked-moment}
\end{equation}
On a support of $n$ blocks, the polynomial jet grows at most as
$C_In^{p_I}$. The rooted bound decays as $\gamma_I^n$, with
$\gamma_I<1$, so the moment sum converges.
Hence these moments exist, are bounded, and are independent of the
cutoff and polymer allocation. Center changes are finite polynomial
jet changes; their same left factor in the readout and its dual matrix
cancels. On an independent local basis, the weight-triangular free
matrix is invertible. Normalize it once. Its interacting diagonal is
$I+O_I(\eta)$; proper-subset outputs are inverted successively.
This constructs a bounded finite ordinary inverse $\mathfrak N_{a,I}^{-1}$.

\label{euclidean-covariance:source-frame-conversion}
Fix a sufficiently early scale $a$ with intrinsic tensor targets. Keep the
earlier columns fixed and transport the original targets to $a$ by their cocycle.
For the given list denote the original and covariant target maps at that scale
by $\Lambda^{\rm orig}$ and $\Lambda^{\rm cov}$. The coordinate change satisfies
\begin{equation}
 \mathfrak N_{a,I}\circ\Lambda^{\rm orig}\circ\mathfrak A_I=\Lambda^{\rm cov},
 \qquad B_{\epsilon,I}^{\rm cov}=B_{\epsilon,I}^{\rm orig}\circ\mathfrak A_I.
 \label{euclidean-covariance:finite-frame-substitution}
\end{equation}
The two finite inverses in the first identity determine $\mathfrak A_I$ uniquely. The second
identity holds already at finite cutoff, with transported target converging to
the intrinsic one. By uniqueness the conversion commutes with translations,
reflection, conjugation and restriction. Translating arbitrary prescribed test
jets shows that the coefficients of a translation-invariant local finite
differential polynomial are constant. Protected normalization fixes the
derivative column. The bounded moment map and finite inverse preserve the given
$C^3$ regularity.

The radial continuum operation commutes with $O(2)$, as do the intrinsic moments and
radial contact forests. The normalized marked graph is unique, so in the
covariant frame it carries the tensor action. Its radial mollifier realization
and joint source admission give the same conclusion. Conjugation by the exact
finite substitution gives \eqref{euclidean-covariance:original-contact-rotation} with its group law. When the points are separated
the nonlinear contact columns vanish. The matrix in each field slot then reduces
to $C_I^{-1}\Gamma_I(R)C_I$. By compactness of $O(2)$, the matrices and their inverses are uniformly
bounded. The original charged fields can still transform among components.
\end{proof}

\subsection{Growth of local fields}

Reconstruction requires a bound on the order of the limiting correlation
distributions as the list of fields grows. We use the uniform repeated-insertion
estimate. Constants may depend on the finite field alphabet, but are uniform in
the number of occurrences.

\begin{lemma}[Uniform local bounds]
\label{local-field-bounds:height-factorial}
\label{local-field-bounds:one-species-buffer}
\label{local-field-bounds:separated-product-bound}
The selected height is a random tempered distribution, and for one
fixed continuous Schwartz seminorm $q$,
\begin{equation}
 |\mathbb E h(f_1)\cdots h(f_n)|\le K^nn!\prod_iq(f_i).
 \label{continuum:height-factorial}
\end{equation}
Fix a finite completed field alphabet $\mathcal A$. There are fixed
$a_*>0$ and $C,P,r$ such that, for $0<a\le a_*$, an original lattice
insertion of $O(f)$ supported
in an $a$-square, with time buffers comparable to $a$, defines a slab operator
$\mathcal A_W(f)$ of time length $\ell_a\asymp a$. Let $\lambda_W$
be the leading eigenvalue of the row transfer at this cutoff.
For $W\ge C_0a$,
\begin{equation}
 \lambda_W^{-\ell_a/\epsilon}\|\mathcal A_W(f)\|
 \le Ca^{-P}p_{a,r}(f),
 \qquad p_{a,r}(f)=\max_{|\alpha|\le r}
                      a^{|\alpha|}\|\partial^\alpha f\|_\infty.
 \label{local-field-bounds:buffer-bound}
\end{equation}
The scale $a$ is fixed before the cutoff threshold. Consequently, for every number $n$ of
chronologically separated such squares,
\begin{equation}
 |S_{O_1\cdots O_n}(f_1\otimes\cdots\otimes f_n)|
 \le C^na^{-Pn}\prod_i p_{a,r}(f_i).
 \label{local-field-bounds:separated-products}
\end{equation}
The constants are independent of $n$ and of empty separations.
\end{lemma}
\begin{proof}
Proposition~\ref{height-representation:compact-norm} bounds
$\|h(g)\|_n$ by $Cn\|g\|_2$ for tests in a fixed square.
Apply a bounded-overlap partition into unit squares and sum with
Cauchy--Schwarz. For $s>1$ this gives
$\|h(f)\|_n\le C_sn\|(1+|x|^2)^{s/2}f\|_2$.
H\"older's inequality and $n^n\le e^nn!$ give
\eqref{continuum:height-factorial} with one fixed Schwartz seminorm.
To realize the field on $\mathcal S'$, expand in a Hermite basis.
The $L^2$ estimate makes a sufficiently negatively weighted square
sum of its coefficients integrable. That sum is finite almost surely
and defines a tempered distribution; continuity in $L^2$ identifies
its test-function pairings with the original $h(f)$.

Take $a_*$ below the weak section of the alphabet and every fixed source reference.
At the master weak scale $R\asymp a$ with the fixed buffer ratios, the anchored estimate
bounds the complete packet norm by $B_a(f)\le Ca^{-P_0}p_{a,r}(f)$, descendants included. By Proposition~\ref{physical-theory:quantitative-admission} the norm
bound $CB_a$ holds at every width and for arbitrarily many reflected copies. Apply
\eqref{source-admission:uniform-local-perturbation} to the fixed alphabet before increasing the number of copies. Its constants
depend only on the alphabet and buffer ratios. The estimate \eqref{source-admission:shrinking-merger} covers every
birth-scale merger at the cost of a fixed extra power of $a^{-1}$. In extracting the
caps the only ratios entering the constants are $\ell_a/R$ and the footprint size
divided by $R$. These bounds give \eqref{local-field-bounds:buffer-bound}. For a macroscopic test, partition first
into $a$-squares at a further fixed power cost in $a^{-1}$.

For separated footprints the microscopic cross-contact coefficient vanishes. The
joint coefficient is consequently a chronological product of insertion kernels
and unmarked transfer contractions. At finite volume its bound includes the
Dirichlet cap factor $B_{W,N,m}$ of \eqref{source-admission:local-cap-price}. Remove the cutoff, then choose $N=W^2/R$. At fixed $m$,
$R$ and cap-smoothing length, the logarithm of the cap factor has upper bound $C(1+W/R)m/(N-b_{\rm cap}-m)$,
tending to zero as $W\to\infty$. Here $b_{\rm cap}$ is the cap-smoothing length. Local admission
identifies the limiting coefficient with the selected one, giving \eqref{local-field-bounds:separated-products}.
\end{proof}

\begin{proposition}[The reconstruction growth condition]
\label{schwinger-growth:fixed-list-tempered}
\label{schwinger-growth:full-source-linear-growth}
Every fixed complete list is tempered, including its contacts. For
each finite alphabet there are $A,B,s<\infty$ such that
\begin{equation}
 \begin{gathered}
 |S_{a_1\ldots a_n}(F)|\le A^n(n!)^B|F|_{sn},
 \quad F\in\mathcal S_0((\mathbb R^2)^n),\\
 |F|_p=\max_{|\alpha|\le p}\sup_X
               (1+|X|^2)^{p/2}|\partial^\alpha F(X)|.
 \end{gathered}
 \label{schwinger-growth:linear-growth}
\end{equation}
Here $\mathcal S_0$ consists of Schwartz tests flat to every order on
all pair diagonals. Constants are independent of $n$.
\end{proposition}
\begin{proof}
For a fixed list, merge neighboring contact regions and apply
Theorem~\ref{normalized-sources:rr-complete-packet-comparison} on
translated unit product boxes. The resulting bound is uniform in the
translations. Expanding a test in a Fourier tensor series on each box
and summing with polynomial spatial weights proves continuity in a
finite Schwartz seminorm, including the fixed contact extension.

To keep the bound uniform in the number $n$ of insertions, take an even integer
$m>r+2$ and an integer $q>P+2$, with $r$ and $P$ as in \eqref{local-field-bounds:buffer-bound}. On an $a$-product box, Fourier
expansion and \eqref{local-field-bounds:separated-products} bound the coefficient by $C^na^{-Pn}\max_{|\alpha_i|\le m}
a^{|\alpha|}\|\partial^\alpha F\|_\infty$ whenever one direction separates
all the buffered slabs. Pair distances greater than $C_0n^2a$ suffice: each pair
excludes a set of angles of length $O(n^{-2})$, and for sufficiently large $C_0$ the fewer
than $n^2/2$ such sets cannot cover the circle. Conversion of tensor frames
contributes at most another factor $C^n$.

Put $a_j=a_*2^{-j}$ and denote the least pair distance by $\delta(X)$. Use a partition of Gevrey order
two on the region $C_0n^2a_j\le\delta(X)\le4C_0n^2a_j$, then divide it into product cells of side $a_j$. After scale
normalization, its derivatives through order $mn$ cost at most a fixed factorial
power in $n$. Indeed the Leibniz rule for the $O(n^2)$ pair cutoffs gives a bound $C^{k+1}(k!)^2(Cn^2)^k$
for $k\le mn$. This is at most $C^n(n!)^b$ for a suitable $b$.

Flatness of $F$ on each diagonal permits Taylor expansion to order $qn$ at a
nearest pair diagonal. This gives $a_j^{qn}$, a fixed additional factorial power, and $|F|_{sn}\prod_i(1+|x_i|^2)^{-2}$
for sufficiently large fixed $s$. Only the nearest pair moves, through a distance
$O(n^2a_j)$, so adjusting the spatial weights costs a polynomial in $n$. Summing the
product cells costs $(Ca_j^{-2})^n$. The contribution at scale $j$ is at most $C^n(n!)^{B'}a_j^{(q-P-2)n}|F|_{sn}$. The choice
$q>P+2$ makes the sum over $j$ finite, and one fixed scale suffices away from the
diagonals. Flatness gives convergence of the partition in the original Schwartz
topology, so the sum equals the distribution already constructed. Diagonal
contacts annihilate $\mathcal S_0$. Thus the $E0'$ condition of \cite[Section~IV.1, equation~(4.1)]{OsterwalderSchrader1975} holds on tests flat at the diagonals. General tests retain the original contact extensions.
\end{proof}

\subsection{A common chronological core}

Products of literally normalized derivatives give a common dense set for the
full source representation and the cyclic representation reconstructed from
correlations at distinct points.

\begin{lemma}[A common chronological core]
\label{relativistic-reconstruction:thin-strip-height}
\label{relativistic-reconstruction:height-chronological-density}
\label{relativistic-reconstruction:slab-derivative-approximation}
\label{relativistic-reconstruction:flat-core-density}
Products of literal derivative smears in strictly separated compact
positive-time slabs are dense in the selected source OS space.
More generally, a continuous fixed $n$-linear map with bound
$C\prod_i\|Y_i\|_p$ on height smears is approximated by its sum
over separated time slabs, with omitted sum $O(r)$ at mesh $r$.
\end{lemma}
\begin{proof}
First let $f$ be supported in a fixed spatial rectangle and a time
strip of length $r$. We claim
$\|h(f)\|_p\le C_pr\|f\|_\infty$.
Choose a fixed interior mass-one profile $b$ and put $m_f=\int f$.
The neutral test $f-m_fb$ has Neumann-flow energy at most
$Cr^2\|f\|_\infty^2$. To see the scale factor, integrate the
logarithmic Green singularity in the two time variables and use
$\log_+(|X-Y|^{-1})\le\log_+(|x-y|^{-1})$; the last expression
is integrable in the two spatial variables. Gaussian current
domination bounds the neutral smear, and
$|m_f|\le Cr\|f\|_\infty$ bounds the remaining term $m_fh(b)$.
This proves the claim.

Use a smooth time partition of mesh $r$ for each test. An $n$-tuple with
neighboring intervals contributes at most $O(r^n)$ to the bounded $n$-linear
functional. There are $O(r^{-(n-1)})$ tuples of this kind, so discarding them costs $O(r)$ in
all. The rest can be chosen with compact, strictly separated supports. Applying
this multilinear estimate to height polynomials gives their chronological
approximations. The exponential height moments give polynomial density in the
Fourier-cylinder core, and ordinary $L^2$ convergence suffices for OS convergence
since the reflected norm is bounded by the ordinary $L^2$ norm.

In a fixed open slab choose a compact mass-one profile $b$ even in space, with
well-separated spatial translates $b_j$. By the physical gap and quarter-turn
symmetry the centered $h(b_j)$ have exponentially summable covariances. Their average
uses the profile $b^{(N)}=N^{-1}\sum_{j=1}^Nb_j$, for which $\|h(b^{(N)})\|_2=O(N^{-1/2})$. Interpolating with the uniform higher moment
bounds gives convergence to zero in every finite $L^p$. The neutral test $g_N=f-(\int f)b^{(N)}$ has a
compact divergence primitive within the slab: take $\int\psi=1$ there, set $a_N(x)=\int g_N(t,x)dt$, and put
\[
 v_t(t,x)=\int_{-\infty}^t[g_N(s,x)-\psi(s)a_N(x)]ds,
 \qquad v_x(t,x)=\psi(t)\int_{-\infty}^xa_N(y)dy.
\]
Then $\nabla\cdot v=g_N$ and $h(g_N)=-\beta_*D(v)$. Approximate each factor of a chronological monomial within its
original slab in $L^{2n}$. Telescoping the product gives ordinary $L^2$ convergence.
Passing through the source/height Gram isometry of Theorem~\ref{source-admission:local-gram-limits} proves density in the
source space.
\end{proof}

\subsection{Relativistic fields and the spectrum}

\begin{theorem}[The common local relativistic theory]
\label{relativistic-reconstruction:full-field-reconstruction}
\label{relativistic-spectrum:relativistic-gap}
All declared finite alphabets have compatible Wightman realizations on
the original selected source Hilbert space, with its vacuum $\Omega$
and physical generator $H$. Their finite words span one dense invariant
domain. Smeared fields are closable, conjugation gives their adjoints
on that domain, and spacelike-separated bosonic fields commute there.
With Minkowski metric $\eta=\operatorname{diag}(1,-1)$ and
$U(t,x)=e^{i(tH-xP)}$, let $\Sigma_\rho$ be the support of the
joint spectral measure restricted to $\Omega^\perp$. Then
\begin{equation}
 \Sigma_\rho\subset\{(E,p):E\ge\sqrt{p^2+M(\rho)^2}\},\qquad
 M(\rho)=\inf_{(E,p)\in\Sigma_\rho}\sqrt{E^2-p^2}>0.
 \label{continuum:invariant-gap}
\end{equation}
The vacuum is unique. The original derivative upper witness retains
its norm and overlap, so $M(\rho)<\infty$ and
$0<c_-\le M(\rho)a_{\rm exit}(\rho)\le c_+<\infty$ on compact
parameter intervals. The selected Euclidean height law is ergodic
under translations and extremal among translation-invariant laws.
\end{theorem}
\begin{proof}
A completed finite alphabet satisfies permutation symmetry,
reflection positivity in the original Grams, Euclidean covariance in
the finite tensor frame, and the clustering estimate of
Theorem~\ref{parameter-responses:rr-fourth-response}.
Equation~\eqref{schwinger-growth:linear-growth} supplies the remaining
growth hypothesis. Apply the corrected reconstruction theorem
\cite[Section~IV.1 and Remark~5 on p.~283]{OsterwalderSchrader1975}
to obtain its tempered Wightman representation.

Map a positive-time polynomial to its source coefficient in the physical
representation. On coincidence-free product tests this map preserves the
reflected form. Fourier tensor approximation extends the equality to joint
tests. Since the range contains the dense chronological derivative core, the
induced isometry is onto. Translating its tests intertwines physical time
evolution in the two spaces, so both the generators and their domains
correspond.

\label{representation-transport:gram-transport}
Indeed a map preserving reflected forms descends through the null spaces, and
density of its image gives a unitary on the completions. Under the identity $RT_t=T'_{ct}R$,
uniqueness of the generator gives $UHU^{-1}=cH'$. For two nested alphabets the restricted
Wightman functions agree by uniqueness of analytic continuation, as do their
maps on the common core. Taking the directed union gives the stated domain and
locality. Inclusion of the conjugate field in the adjoint proves closability.
Complete Euclidean contacts and source vectors retain their original Grams. An
arbitrary Euclidean contact extension has no unique Lorentzian time-ordered
interpretation from reconstruction alone.

The gap identifies $\ker H=\mathbb C\Omega$. If the joint spectral support contained a nonzero null
point, Lorentz covariance would put its boosts in the support at arbitrarily
small positive energies. For a timelike point, a boost to rest shows that its
invariant mass is at least $M=\inf\operatorname{Spec}(H|_{\Omega^\perp})$. Conversely $\sqrt{E^2-p^2}\le E$ bounds the infimum of the invariant
masses above by the infimum of the energies. This gives \eqref{continuum:invariant-gap} for the spectral
support, without assuming an isolated particle shell.
\label{relativistic-spectrum:identified-upper-witness}
For the normalized actual derivative vector of Proposition~\ref{literal-fields:theorem-2} the overlap is $\langle\psi_F,e^{-t_*H}\psi_F\rangle\ge q>0$. Its
spectral measure is preserved, so $q\le e^{-t_*M}$ gives the same upper bound and exit-scale
comparison.

\label{relativistic-reconstruction:selected-law-ergodicity}
\label{relativistic-reconstruction:selected-extremality}
The gap gives exponentially integrable time covariance for any centered bounded
height cylinder. Hence its time average in ordinary $L^2$ has norm $O(T^{-1/2})$. By cylinder
density every translation-invariant random variable is constant. The bounded
Radon--Nikodym densities of an invariant convex decomposition are translation
invariant, so they too are constant and the component laws coincide. This proves
extremality for the selected Dirichlet vacuum. Other boundary or charged sectors
are outside the claim.
\end{proof}

\subsection{Dilation and the physical mass coordinate}
\label{physical-scale:physical-scale}

Physical dilation now gives a smooth coordinate on the family of
reconstructed theories, namely the mass itself.

\begin{theorem}[Complete dilation and the mass coordinate]
\label{physical-scale:selected-source-scale}
\label{physical-scale:full-source-unitary-scale}
\label{physical-scale:mass-coordinate}
For admitted endpoints, the original complete source cocycle satisfies
\begin{equation}
 \mathcal Z_{T_\lambda\rho}[J]
  =\mathcal Z_\rho[\mathscr R_{\lambda,\rho}J],\qquad
 U_{\lambda,\rho}H_{T_\lambda\rho}U_{\lambda,\rho}^{-1}
  =\lambda H_\rho,
 \label{physical-scale:full-generator-scale}
\end{equation}
where $U_{\lambda,\rho}$ is unitary and preserves the vacuum and
complete source Grams. Hence
\begin{equation}
 M(T_\lambda\rho)=\lambda M(\rho),\qquad
 M(\rho)=M(\rho_*)\exp\!\left(\int_{\rho_*}^{\rho}
                                  \frac{ds}{\mathfrak b(s)}\right).
 \label{physical-scale:mass-coordinate-eq}
\end{equation}
Here $\mathfrak b(\rho)=\left.\frac{d}{dt}T_{e^t}\rho\right|_{t=0}>0$
is the generator of the endpoint action. Thus $M$ is a positive,
finite, strictly increasing $C^2$ coordinate on this chart.
\end{theorem}
\begin{proof}
Cofinality of the dilated squares passes \eqref{source-dilation:source-law}
to the selected limit, including contacts. The semigroup identity of
Proposition~\ref{height-representation:smooth-action} acts on the generators
identified in Theorem~\ref{relativistic-reconstruction:full-field-reconstruction}.
It transports the full spectral measure. Hence along the smooth orbit
$\rho(t)=T_{e^t}\rho_*$ the mass is $e^tM(\rho_*)$. Since
$\rho'(t)=\mathfrak b(\rho(t))>0$, differentiation along the orbit
gives $\mathfrak b M'=M$ without differentiating a spectral infimum.
Integration gives \eqref{physical-scale:mass-coordinate-eq}. In the chart
$\alpha=\mathcal A(\rho)$ the generator becomes
$\mathfrak b_\alpha=\mathcal A'\mathfrak b$.
\end{proof}

\label{physical-scale:prescribed-mass}
To prescribe any mass $M_*>0$, choose physical units once for the entire family.
Starting at mass $M_0$, put $\lambda=M_*/M_0$ and set
\begin{equation}
 \begin{aligned}
 Z_\epsilon^{(\lambda)}&=Z_{\lambda\epsilon}^{(0)},&
 y_\epsilon^{(\lambda)}&=y_{\lambda\epsilon}^{(0)},\\
 E_\epsilon^{(\lambda)}&=\lambda^2E_{\lambda\epsilon}^{(0)},&
 a_0^{(\lambda)}&=a_0/\lambda .
 \end{aligned}
 \label{physical-scale:rescaled-bare-family}
\end{equation}
The change of coordinates $x\mapsto\lambda x$ identifies the finite lattice actions down to the
vacuum term, with a single transport of all source references. In the
dimensionless compact readout the squared-test integral scales by $\lambda^{-2}$ and the
gradient energy is invariant, leaving the readout unchanged. The selected
generator becomes $\lambda H$, while $a_{\rm exit}$ is multiplied by $\lambda^{-1}$. The choice of units gives
neither a numerical mass--coupling constant nor an unrestricted mass range
within a bounded weak chart.

\section{Interaction and regulator universality}
\label{sec:interaction-universality}

Even a Gaussian field can have a mass gap. For a single smear of the linearly
normalized derivative field, screening suppresses the variance while the leading
growth of the real-source exponential moment retains its ultraviolet Gaussian
normalization. This discrepancy detects interaction. We also identify the
continuum theory for a specified class of local periodic perturbations of the
microscopic action.

\subsection{A non-Gaussian derivative witness}
\label{derivative-interaction:interaction-section}

For real $U\in C_c^\infty(\mathbb R^2)$ write
\[
 X(U)=D(\nabla U)=\phi(-\Delta U),\qquad
 E(U)=\int_{\mathbb R^2}|\nabla U|^2\,dx.
\]
The observable $X(U)$ is thus a linear smear of the original derivative field. The
canonical activity $v_r(\rho)$ at weak physical scale $r$ tends to zero as $r\downarrow0$, uniformly
on compact positive endpoint intervals. This smallness will control real
backgrounds of unrestricted absolute value.

\begin{lemma}[Localized background logarithm]
\label{derivative-interaction:bounded-gradient-logarithm}
Fix the cutoff $\epsilon$ and the domain $D$. Write $\mathcal U_r$
for the interaction after integrating the unit-reference Gaussian
prefix to scale $r$. For a real
background $\psi$ compactly supported away from the wall, assume
$\max_{k=1,2}r^k\|\nabla^k\psi\|_\infty\le h_*$.
Then
\begin{equation}
 \left|\log\frac{\mathcal U_r(\psi)}{\mathcal U_r(0)}\right|
 \le C\{v_r+\delta_\epsilon(r)\}N_r(\operatorname{supp}\psi),
 \qquad \delta_\epsilon(r)\xrightarrow[\epsilon\downarrow0]{}0.
 \label{derivative-interaction:bounded-gradient-log-bound}
\end{equation}
Here $N_r$ counts the blocks in a fixed dependency enlargement.
The bound imposes no restriction on the absolute field mean.
\end{lemma}
\begin{proof}
For an admissible background the local factor $e^{V_B}(1+W_B^\sharp)$ has positive sign and logarithm
of order $O(v_r)$. The derivative regulator costs at most $e^{Ch_*^2|X|_r}$, the stable body has size
$O(v_r^3)$ and the even-wall history has size $O(v_r)$, up to the respective cutoff errors.
Take $h_*$ and the weak radius small enough for the block-number exponential decay
to absorb $e^{Ch_*^2|X|_r}$. Divide out the positive local factors. The connected-polymer
criterion then gives the rooted bound $C(v_r+\delta_\epsilon)$, also with any fixed support moment.

In \eqref{derivative-interaction:bounded-gradient-log-bound}, the two logarithms have identical contributions from any cluster whose
full dependency support misses $\operatorname{supp}\psi$. Root the other clusters where this support
meets the background. Summing there costs $N_r$, even for clusters that reach the
wall. The field-independent scalars cancel before taking absolute values.
\end{proof}

\begin{theorem}[Interaction in the literal derivative sector]
\label{derivative-interaction:free-laplace-rate}
\label{derivative-interaction:derivative-nongaussianity}
\label{derivative-interaction:separated-derivative-witness}
For every real compactly supported smooth $U$,
\begin{equation}
 \lim_{|t|\to\infty}t^{-2}\log\mathbb E_\rho e^{tX(U)}
       =\tfrac12E(U),
 \label{derivative-interaction:free-laplace-limit}
\end{equation}
uniformly on compact positive endpoint intervals. On each such
interval there are a fixed $U$ and a finite $t_*>0$ for which
\begin{equation}
 \log\mathbb E_\rho e^{t_*X(U)}
   -\tfrac12t_*^2\mathbb E_\rho X(U)^2
                  \ge\tfrac18t_*^2E(U)>0.
 \label{derivative-interaction:finite-derivative-witness}
\end{equation}
For each admitted $\rho$, some finite even order $n\ge4$ has a
nonzero connected correlation of literal derivative components on
mutually separated compact time slabs. The order need not be four.
\end{theorem}
\begin{proof}
We first determine the large-source coefficient. At finite cutoff,
write $A=A_D$ for the counting-coordinate precision,
$G=A^{-1}$, $s=1-Z_\epsilon$, and $U_\epsilon$ for the sampled
test. The literal discrete variable is
$X_\epsilon=U_\epsilon^\top A\phi$, and its discrete Dirichlet energy is
$E_\epsilon(U)=U_\epsilon^\top AU_\epsilon\to E(U)$.
Normalize the positive-cosine convolution at zero by writing
$R_Q(x)=F_Q(x)/F_Q(0)$. With $Q=G/Z_\epsilon$ and
$x=tU_\epsilon/Z_\epsilon$, Gaussian translation gives
\[
 \mathbb E e^{tX_\epsilon}
       =e^{t^2E_\epsilon(U)/(2Z_\epsilon)}R_Q(x).
\]
Positive phase comparison gives $R_Q(x)\le1$.

For the lower bound take a weak physical scale $r\asymp(1+|t|)^{-1}$, leaving its fixed
proportionality constant to be chosen below. During cutoff removal hold $t$, $r$
and $D$ fixed. For the actual prefix $C\preceq G$, completing the square as in \eqref{continuum:kinetic-completion} gives
\[
 \log\frac{\mathcal U_r(\psi)}{\mathcal U_r(0)}
  =\tfrac12\psi^\top Q_s\psi+\log R_{C_s}(M_s\psi).
\]
The spectrum of $B=A^{1/2}CA^{1/2}$ is contained in $[0,1]$. The scalar inequality $b/(1-sb)\le 1/(1-s)$ on this
interval implies $C_s\preceq Q$, and Corollary~\ref{covariance-comparison:phase-covariance-monotonicity} gives $R_Q(x)\ge R_{C_s}(x)$. The desired argument for this prefix
convolution is obtained by taking $\psi=(1-sCA)x$, for which $M_s\psi=x$.

Finite range gives
$\operatorname{supp}\psi\subset\operatorname{supp}U+B_{Cr}$.
In the bulk, $\|C\|_{\infty\to\infty}\le C(r/\epsilon)^2$.
Move finite differences onto the smooth test to obtain
\[
 r^k\|\nabla^kCAU_\epsilon\|_\infty
       \le Cr^{k+2}\|U\|_{C^{k+2}},\qquad k\le2.
\]
The proportionality constant in $r\asymp(1+|t|)^{-1}$ can
therefore be chosen so that the background satisfies the preceding
lemma. The quadratic correction obeys
$|\psi^\top Q_s\psi|\le C|s|t^2E_\epsilon(U)$ and tends to zero at
fixed $t$. Since the support meets at most $C_U(1+r^{-2})$
blocks, cutoff removal followed by volume selection gives
\begin{equation}
 \tfrac12t^2E(U)-C_Uv_r(1+r^{-2})
 \le\log\mathbb E e^{tX(U)}\le\tfrac12t^2E(U).
 \label{derivative-interaction:free-laplace-sandwich}
\end{equation}
The same upper estimate at $2t$ supplies uniform integrability in
the field limits. Now let $|t|\to\infty$. The error divided by
$t^2$ is $O(v_r)$, which proves
\eqref{derivative-interaction:free-laplace-limit}.

\label{derivative-interaction:height-L2-covariance}
For unit-square pieces $f_k$ of a height test, the local $L^2$ bound and the gap
along either axis give
\[
 |\mathbb E\phi(f_k)\phi(f_l)|
       \le Ce^{-c|k-l|}\|f_k\|_2\|f_l\|_2.
\]
Summing the kernel gives $\operatorname{Var}\phi(f)\le C\|f\|_2^2$. Take $U_0$ with $E_0=E(U_0)>0$ and put $U_L(x)=U_0(x/L)$. In two dimensions
$E(U_L)=E_0$, while
\[
 \operatorname{Var}X(U_L)
       \le CL^{-2}\|\Delta U_0\|_2^2.
\]
A single choice of $L$ makes the variance at most $E_0/2$ throughout the
interval. By uniformity in \eqref{derivative-interaction:free-laplace-limit}
we may fix a finite $t_*$ at which the log Laplace transform is at least
$3t_*^2E_0/8$. Subtract the quadratic variance term to obtain
\eqref{derivative-interaction:finite-derivative-witness}.

The sub-Gaussian estimate makes $\mathbb E e^{zX}$ entire. Were all
higher cumulants zero, the logarithm near zero would be quadratic. The
identity theorem would then give the Gaussian moment generating function
on the whole plane, contradicting the strict inequality at the finite
source just chosen. Field-sign symmetry eliminates the odd orders, so a
nonzero cumulant occurs at a finite even order $n\ge4$. At that order we have
\[
 |\kappa_n(Y_1,\ldots,Y_n)|\le C_n\prod_i\|Y_i\|_n.
\]
By Lemma~\ref{relativistic-reconstruction:flat-core-density} separate the height factors in time, then approximate each by literal
derivatives in its slab. Continuity preserves nonvanishing in some finite
approximation and hence in at least one derivative-component correlation.
Separation excludes contacts.
\end{proof}

\subsection{Universality for finite local periodic perturbations}
\label{regulator-universality:universality-section}

Fix an integer $N\ge2$, a finite $A$, and $a=(a_2,\ldots,a_N)\in[0,A]^{N-1}$. The perturbed action on the original
nearest-neighbor Dirichlet lattice is
\begin{equation}
 \begin{aligned}
 S^a_{\epsilon,D}
 &={}\tfrac Z2\phi^\top A_D\phi
  +2y\sum_x(1-\cos\beta_*\phi_x)\\
 &\quad+2y^2\sum_{n=2}^Na_n\sum_x(1-\cos n\beta_*\phi_x)
  +E^a_\epsilon\epsilon^2|D|.
 \end{aligned}
 \label{regulator-universality:regulator-action}
\end{equation}
Apply the original reference prescription to this action to fix its vacuum term.
The microscopic coefficients $a_n$ stay fixed as $Z$ and $y$ are tuned. Nonnegativity
of the harmonic coefficients preserves the positive comparison estimates for the
physical limit.

\begin{theorem}[Universality in the verified regulator class]
\label{regulator-universality:verified-regulator-universality}
Each action \eqref{regulator-universality:regulator-action} admits
positive domain-independent bare functions $Z_\epsilon^a,y_\epsilon^a$
and the original complete source normalization such that its
fixed-box continuum coefficients and reflected Grams equal the
canonical ones. The equality retains the interior $C^3$ fixed-box
comparison and uniform fourth-derivative bounds, and hence the
selected values and their proved $C^2$ parameter dependence.
The reconstructed local theories are canonically unitarily equivalent,
with the same mass and local fields. In common physical units,
matching the mass and the complete field/contact convention identifies
the endpoint. A single global rescaling gives the corresponding
statement at every prescribed positive mass.
\end{theorem}
\begin{proof}
\label{regulator-universality:harmonic-preparation}
The perturbation acts at individual sites, preserving reflection positivity. Its
nonnegative Bessel Fourier coefficients give Gaussian moment domination at every
positive precision, also in the presence of added pins or retained regulators.
At the bounded first scale the number of sites per block is bounded. Together
with $|\cos(n\beta_*\phi)|_{a_f}\le e^{n\beta_*a_f}$ this bounds the norm of the additional analytic activity by $C_{N,A}y^2$. The
resulting $O(y^2)$ seed in the initial stable record has vanishing constant and linear
jets. Thus the ordinary microscopic Jacobian is still $\operatorname{diag}(2,c_{\rm lat})$. At the wall the new
seed is even and of order $O(y^2)$, with the original pinned-field Taylor gains.
Retain every scalar coefficient. The coupling jets of the seed are $O(|g|^{\max(2-k,0)})$, and
differentiation in the fixed regulator labels remains bounded.

\label{regulator-universality:seed-forgetting}
\label{regulator-universality:regulator-tuning}
After preparation the harmonics belong to the input density of the original
operation. Its free fundamental multiplier and calibration persist. By Lemma~\ref{continuum:preparation-comparison} the
extra seed contracts, and tuning by \eqref{continuum:matched-preparation} includes the finite ultraviolet clock
shift. Odd higher harmonics may break activity-sign parity, which the proof
does not use. For the fixed perturbation and cutoff write $g=c_{\rm lat}y_\epsilon^a$. Telescoping
reciprocals gives
\[
 1-Z_\epsilon^a=-\tfrac12g+O(g^2),\qquad
 \log(a_{\rm exit}/\epsilon)=g^{-1}+O(\log(1/g)).
\]

\label{regulator-universality:regulator-source-hypotheses}
For the fixed complete list $I$, transport its original stencil embedding and
target along the actual history:
\begin{equation}
 B^a_{\epsilon,I}=\iota_{\epsilon,I}
                 (P^{a,\epsilon}_{R,0})^{-1}\mathcal C_{R,I}.
 \label{regulator-universality:regulator-source-injection}
\end{equation}
Only the finite triangular ordinary map is inverted. For this fixed harmonic
list the exponential block weight absorbs the extra analytic factor per
occupied block on the stable ball.

At superficial degree $\omega\ge0$ the original contact Taylor subtraction sends $Cr^{-\omega-a_I}$ to
$Cr^{1-a_I}$. Negative integer degrees already have positive exponent. Choosing $a_I<1/(2|I|)$
leaves a positive scalar-tail power in every proper-subset forest. The
summable unmarked discrepancy in $e_{j+1}\le q_Ie_j+C_Id_j$ preserves the marked inverse loss. The
injection retains its identity outputs. Its independent initial marked scalar
is zero. New charged corrections have dimension above one, leaving the
derivative's principal coefficient exactly one.

Thus Lemma~\ref{source-dilation:marked-recognition} identifies the source histories through the stated parameter orders,
with the full scalar telescope starting at the initial zero-field evaluation.
Division by the positive partition functions identifies normalized fixed-box
coefficients, including the finite metric substitutions of Section~\ref{stress-and-trace:section}. A common
Dirichlet exhaustion gives equal selected coefficients and Grams. The
identity on source vectors descends to an onto vacuum- and
translation-preserving unitary. The chronological core identifies all local
fields, hence the masses agree. Strict monotonicity in \eqref{physical-scale:mass-coordinate-eq} recovers the
endpoint. The rescaling \eqref{physical-scale:rescaled-bare-family} preserves each $a_ny^2$ at prescribed mass.

Constants may depend on the fixed harmonic and source lists.
Signed amplitudes, infinitely many harmonics, and changes to the
Gaussian lattice geometry are outside the proved class.
\end{proof}

\section{Field equations and geometric stress}
\label{sec:local-identities}
\label{local-currents:current-eom-section}
\label{stress-and-trace:section}

Changing the microscopic field or metric before cutoff removal yields local
identities for the fields with the contact extensions used in the
construction. We include the change and the ordinary sources in the same
finite labelled algebra. Its mixed coefficients keep the quadratic
variations, identity terms and contacts on every proper partial diagonal.

Denote the normalized selected generating jet by $\mathcal Z_\rho[J]$ and set $W_\rho[J]=\log\mathcal Z_\rho[J]$ at the fixed
finite source order. At a weak reference scale the intrinsic moment map \eqref{euclidean-covariance:intrinsic-marked-moment}
and its inverse fix the ordinary local coefficients. For each of the two
deformations we identify the remaining parts of the physical insertion by
bounding the preparation, obtaining a positive ultraviolet power for the
contact remainder, contracting the stable differences and summing the actual
scalar increments from zero independent initial marked scalar.

\subsection{Local field translations}

Fix a smooth real shift $\xi$ supported in the interior of the physical domain.
In counting coordinates at finite cutoff let $A=A_D$ and $G=A^{-1}$. Take an ultraviolet
prefix $0\preceq C\preceq G$. In these coordinates $C$ equals the covariance operator in the
physical pairing divided by $\epsilon^2$. Write $R_C=1-CA$ and
\[
 \begin{aligned}
 \Psi_0(\phi;J)&=\exp\{(1-Z_\epsilon)\phi^\top A\phi/2
          +2y_\epsilon\sum_x\cos(\beta_*\phi_x)
          +B_{\epsilon,I}[J](\phi)\},\\
 \Psi_C(\phi;J)&=\mathbb E_C\Psi_0(\phi+\zeta;J).
 \end{aligned}
\]
Integration over the prefix gives the translated density
\begin{multline}
 \mathbb E_C\!\left[e^{-(\phi+\zeta)^\top A\xi-\xi^\top A\xi/2}
                   \Psi_0(\phi+\zeta+\xi;J)\right]
 \\
 =e^{-\phi^\top A\xi-\xi^\top AR_C\xi/2}
                   \Psi_C(\phi+R_C\xi;J).
 \label{local-currents:shifted-density-formula}
\end{multline}
The Gaussian translation formula $\mathbb E_C[e^{-\zeta^\top A\xi}f(\zeta)]
=e^{\xi^\top ACA\xi/2}\mathbb E_Cf(\zeta-CA\xi)$, valid also for singular $C$, gives \eqref{local-currents:shifted-density-formula} upon
application to the translated interaction. Thus the first action score after
the prefix is $\phi^\top A\xi-D_\phi\log\Psi_C[R_C\xi]$. Its independent second-order scalar comes from the exterior
quadratic term.

\begin{proposition}[Admission of the complete translation jet]
\label{local-currents:shifted-density}
\label{local-currents:shift-reference-data}
The raw shifted injection
$-S_\epsilon(\phi+\xi)+S_\epsilon(\phi)+B_{\epsilon,I}[J](\phi+\xi)$
admits a unique local source substitution $\mathscr S_\rho(\xi,J)$,
with $\mathscr S_\rho(0,J)=J$.
At a fixed weak reference $a$, put $F_C=\log\Psi_C$ on the
positive bounded-field germ. The logarithm of the shifted density
relative to the unshifted, source-free density is
\[
 L_{\xi,J}(\phi)=-\phi^\top A\xi-\tfrac12\xi^\top AR_C\xi
            +F_C(\phi+R_C\xi;J)-F_C(\phi;0).
\]
Its moments determine the cutoff source coordinates by the finite
normalization equation
\begin{equation}
 \bigl(\mathfrak N_{a,\epsilon}\circ\Lambda^{\rm orig}_{a,\epsilon}\bigr)
       \bigl(\mathscr S_{\epsilon,\rho}(\xi,J)\bigr)
       =\mathfrak m_{a,\epsilon}(L_{\xi,J}).
 \label{local-currents:shift-readout-formula}
\end{equation}
The two normalization maps on the left are invertible
on the ordinary source coordinates. The division-free connected
reconstruction gives the same readout. The maps are coherent under
list restriction,
reflection and conjugation. They retain the supplied three convergent
fixed-box parameter derivatives and fourth bounds.
\end{proposition}
\begin{proof}
By retained smoothing, all required finite field derivatives of $R_C\xi$ are
bounded, the range is $O(a)$ and the cutoff limit is common. The exterior scalar
also has a finite limit, with $0\le\xi^\top AR_C\xi\le\xi^\top A\xi$. In \eqref{local-currents:shifted-density-formula} applied to the full circle-product
reconstruction, each field derivative costs a fixed analytic-radius reserve
and the graded regulator bounds each exterior gradient polynomial. The
number of mark placements contributes a polynomial in the support size,
summable with the unused exponential weight.

At the microscopic scale use the finite expansions of
$e^{iq\beta_*(\phi+\xi)}$ and
$(\nabla^\alpha\phi+\nabla^\alpha\xi)^m$.
They give bounded kinetic and phase packets in the dimension-two
translation module. A smooth shift of the derivative regulator is
controlled by Young's inequality, costing
$e^{Ca_j^2\|\nabla\xi\|_\infty^2}$ per early block.
For contact bounds, shift the complete Taylor forest before estimating
it: neutral collisions cancel the common phase, while a charged
collision retains that phase in its output vertex. Taylor expansion
of the remaining smooth differences has the same first omitted power
as the original forest. Commutators with its projectors are finite
local contacts. Relative moments of degree $m$ cost
$C a_j^m(1+n+\ell/a_j)^m$ on a history of $n$ blocks and span
$\ell$. The support and history bounds have moments of that order,
including for contributions from earlier stable terms. Thus the
independent remainder still has a positive ultraviolet power.

A field-independent shift coefficient, including
$-Z_\epsilon\xi^\top A\xi/2$, belongs to the local identity column
of $Q_0$. Its value is retained; the independent initial scalar is
zero. The readout \eqref{local-currents:shift-readout-formula} is
performed on the field/profile germ before inserting numerical
profiles. It consequently defines a finite local differential
polynomial in those profiles. The convolution $R_C\xi$ is only an
intermediate term in that readout, and all complementary Taylor
terms remain in the stable record.

For the shifted history and the matched local source, complete recognition
bounds the joint error in the ordinary, stable, wall and scalar records at
scale $\delta$ by $C[\delta^{\eta_I}+\delta^{-P_I}
(e_\epsilon(\delta)+e^{-cd(D)/\delta})]$. Remove the cutoff before taking $\delta\downarrow0$. The parameter and coherence
assertions follow from the differentiated bounds and finite inverses.
\end{proof}

\begin{theorem}[Field equation and topological current]
\label{local-currents:shift-ward}
\label{local-currents:top-ward}
The complete normalized functional satisfies
\begin{equation}
 \mathcal Z_\rho[\mathscr S_\rho(\xi,J)]=\mathcal Z_\rho[J],\qquad
 \mathscr S_\rho(\xi,\mathscr S_\rho(k,J))
                     =\mathscr S_\rho(\xi+k,J).
 \label{local-currents:shift-invariance}
\end{equation}
Define the infinitesimal source action and its zero-source part by
$s(\xi,J)=\partial_t\mathscr S_\rho(t\xi,J)|_{t=0}$,
$e(\xi)=-s(\xi,0)$ and $v(\xi,J)=s(\xi,J)+e(\xi)$.
Both the full and the connected functional obey
\begin{equation}
 DF(J)[e(\xi)]=DF(J)[v(\xi,J)],
 \qquad F\in\{\mathcal Z_\rho,W_\rho\},
 \label{local-currents:eom-ward-functional}
\end{equation}
where $DF(J)$ denotes the derivative in the source variables.
The insertion specified by $e$ is
\begin{equation}
 \begin{aligned}
 \mathcal E_\rho&=-Z_E(\rho)\partial_\mu D_\mu
              +\lambda_E(\rho)V_{\sin},\\
 V_{\sin}&=\frac{\mu_{\rm ref}^2}{2i}(V_1-V_{-1}),\quad
 Z_E(0)=1,\quad\lambda_E(0)=0.
 \end{aligned}
 \label{local-currents:finite-eom-block}
\end{equation}
Writing $\mathcal V_{\xi,A}=[\eta_A]v$ for a contact coefficient,
the identity for the insertions $O_I$ reads
\[
 S(\mathcal E(\xi),O_I)=
     \sum_{\varnothing\ne A\subseteq I}
     S(\mathcal V_{\xi,A}(J_A),O_{I\setminus A}).
\]
The force
$\mathcal F_\rho=(1-Z_E)\partial_\mu D_\mu+
\lambda_EV_{\sin}$ is nonzero. With
$\varepsilon_{01}=1$, the current
\begin{equation}
 j_\mu^{\rm top}=\frac{\beta_*}{2\pi}\varepsilon_{\mu\nu}D_\nu
 \label{local-currents:top-current}
\end{equation}
obeys $\partial^\mu j_\mu^{\rm top}=0$ on the common Wightman
domain after the prescribed tensor Wick rotation.
Its complete Euclidean contacts are the derivative commutators of
the original extension, specified below.
\end{theorem}
\begin{proof}
At finite cutoff the substitution $\phi\mapsto\phi+\xi$ fixes the Dirichlet boundary and has
unit Jacobian, so the normalized integral is invariant. The admission
proposition carries invariance through the ordered limits. Two raw shifts
compose by addition. By continuity in the reserved norms and uniqueness of
the local inverse, their source substitutions obey the group law in \eqref{local-currents:shift-invariance}.
Differentiation at the identity with the labelled Leibniz rule gives \eqref{local-currents:eom-ward-functional}. The
coefficient indexed by $I$ is precisely the sum over nonempty subsets.

Modulo integration by parts, the dimension-two square-scalar
block that is odd under field sign has basis
$\partial_\mu D_\mu,V_{\sin}$.
The other tensors and the even identity/cosine are excluded by
symmetry. The two coefficients in
\eqref{local-currents:finite-eom-block} are the linear readout of
\eqref{local-currents:shift-readout-formula}. This agrees directly
with the lattice first score
$Z_\epsilon D_\epsilon(\nabla\xi)
+2\beta_*y_\epsilon\sum_x\xi_x\sin(\beta_*\phi_x)$.
For a second shift profile $k$, its second variation includes the
kinetic scalar and
$2\beta_*^2y_\epsilon\sum_x\xi_xk_x\cos(\beta_*\phi_x)$.
The admission proposition retains both, together with all mixed
source contacts. The protected derivative normalization gives the
stated free endpoint value.

\label{local-currents:contact-commutator}
For the conservation contacts, first extend the proper-subset diagonals. In
one relative collision variable $r$ write $T_m$ for Taylor projection through
degree $m$ and set $W_mf=f-\chi T_mf$, $R_mt(f)=t(W_mf)$, with $\chi(r)=\chi_{\rm ct}(\mu_{\rm ref}r)$ and $T_{-1}=0$.
Distributional differentiation, together with
$\partial_\ell T_{m+1}=T_m\partial_\ell$, gives
\begin{equation}
 (\partial_\ell R_mt-R_{m+1}\partial_\ell t)(f)
 =-\sum_{|\alpha|\le m+1}
 t\left((\partial_\ell\chi)\frac{r^\alpha}{\alpha!}\right)
                                \partial^\alpha f(0),\quad m\ge-1.
 \label{local-currents:single-contact-commutator}
\end{equation}
For $m<-1$ the commutator vanishes. Differentiate exceptional finite parts
literally. On an ordered forest the product rule inserts this commutator at
each factor, retaining all factors on both sides. The coefficients exist
because they are evaluated away from the current collision once lower
subsets have been extended.

Commuting lattice differences give the off-diagonal curl identity
$\partial_\mu D_\nu-\partial_\nu D_\mu=0$. If $S^\circ$ denotes
its off-diagonal kernel and $\mathcal R$ the fixed forest, then
\begin{equation}
 \partial_x^\mu S_{j_\mu^{\rm top},I}
 =\frac{\beta_*}{2\pi}\sum_{\mu,\nu}\varepsilon_{\mu\nu}
       [\partial_{x^\mu},\mathcal R_{\nu,I}]S_{D_\nu,I}^\circ
       +\mathcal C_I^{\rm fin}.
 \label{local-currents:top-contact-identity}
\end{equation}
Here $\mathcal C_I^{\rm fin}$ is the derivative of the prescribed
exceptional finite parts; all remaining contacts are computed by
\eqref{local-currents:single-contact-commutator} and the product
commutator rule. Reconstruction of the identity away from the
diagonals gives operator conservation on the common domain.
\label{local-currents:nonzero-force}
Suppose the force were zero. The continued field equation and
curl identity would imply $\Box D_\nu=0$. Thus the Fourier
support of $D_\nu(x)\Omega$ would lie on $p^2=0$.
Field-sign symmetry removes its vacuum component, while the mass
gap places the remaining support in $p^2\ge M^2$.
Every derivative two-point function would then vanish, contradicting
the nonzero derivative witness. The force is therefore nonzero.
This argument concerns the declared vacuum algebra, which contains
no disorder field or soliton endpoint.
\end{proof}

\subsection{Metric variations and their Gaussian conversion}

Consider a Weyl change in two dimensions that preserves the scalar
covariance. The Gaussian measure can still have the nonzero second variation
\eqref{stress-and-trace:liouville-contact-conversion}. Thus geometric stress requires both the kinetic and cosine terms of the
retained metric polynomial and the variations of the measure extended to
coincidence. The latter contribute to the Ward identities.

Take jets of arbitrary fixed finite order at $g=\delta$ in smooth directions
supported away from the true boundary. Our volume element is $dv_g=\sqrt{\det g}\,dx$, and the
round sphere has positive scalar curvature. Neither convergence in metric
order nor reflection positivity for general nonlinear metric changes enters
these jet calculations.

\begin{proposition}[Complete metric admission]
\label{stress-and-trace:spectral-metric-jets}
\label{stress-and-trace:gaussian-contact-lemma}
\label{stress-and-trace:metric-admission}
The selected theory admits coherent finite geometric metric/source
jets. At a fixed weak reference $a$, their expression in the
original source coordinates is
\begin{equation}
 q_I(g,J)=(\mathfrak N_{a,I}^{\rm orig})^{-1}
                    \mathfrak N_{a,I}^{\rm met}(g,J),
 \qquad q_I(\delta,J)=J.
 \label{stress-and-trace:metric-conversion}
\end{equation}
The readout includes the complete Gaussian metric conversion and all
local identity/contact columns. Its complete source realization has
the geometric construction's actual scalar history with the original
flat denominator. Mixed jets commute and respect restrictions, earlier
fields, reflection, conjugation and the source cocycles. Parameter
regularity remains at the established finite orders.
\end{proposition}
\begin{proof}
\emph{Covariant shells.}
Use the spectral shape fixed in the flat construction to define
\begin{equation}
 C_g[a,b]=\frac13\int_{a/c_{\rm fr}}^{b/c_{\rm fr}}
                 r\varphi(r\sqrt{-\Delta_g/3})\,dr,
 \qquad C_g[0,\infty)=(-\Delta_g)^{-1}.
 \label{stress-and-trace:covariant-shell}
\end{equation}
These unit-reference shells are positive and additive, commute
with geometric pullback, and have finite propagation range.
To differentiate them on one fixed Hilbert space, put
$P_g=(\det g)^{1/4}(-\Delta_g)(\det g)^{-1/4}$ on $L^2(dx)$.
A shell has the form $a^2F(a^2P_g)$.
For $p$ labelled metric directions, its derivative is a sum over
ordered partitions $(S_1,\ldots,S_k)$ of those labels. In Fourier
variables, each term contains
\[
 a^{2+2k}F^{[k]}(a^2|p_0|^2,\ldots,a^2|p_k|^2)
                         \prod_{j=1}^kP_{S_j}(p_{j-1},p_j),
\]
where $p_0,\ldots,p_k$ are momenta,
$P_{S_j}=D_g^{S_j}P_g|_{g=\delta}$ has total differential degree
at most two, and $F^{[k]}$ is the $k$th divided difference.
The Hermite--Genocchi formula expresses $F^{[k]}$ as a simplex
integral of $F^{(k)}$. If
$|p_0|\ge2\sum_j|p_j-p_{j-1}|$, rapid decay of $F^{(k)}$
absorbs the momentum factors. Otherwise the Fourier moments of the
smooth metric coefficients do so. Momentum integration by parts gives
any fixed off-diagonal decay and the required endpoint derivatives,
hence the block-pair projective row bounds.
Finite propagation first restricts the coefficient functions to the
relevant neighborhoods. Subtracting their value at a root gains a
factor $a$ times a fixed positive-derivative seminorm. The endpoint
volume-density factors have the same estimates.

At positive flat covariance a variation $D$ inserts $D_C\mathbb E_CF[D]=\frac12\mathbb E_C(D:D_\phi^2F)$. Thus $p$ metric labels
cost at most $2p$ reserved field derivatives, whose projective directions and
factorials are summed by the block-Cauchy/tree estimate. Signed variations
enter solely as insertions; the positive flat forest integral uses each
merged regulator once.

Use the radial auxiliary law with proper-radius mollifier $P_{r,g}$,
interaction $2u_rr^{-2}\int[\cos(P_{r,g}h^0)-1]dv_g$, and the
already matched flat pair $(Z_r,u_r)$. Normal-coordinate ODEs give
bounded finite mollifier jets at scale $r$. Differentiating the full
kernel measure preserves $P_{r,g}1=1$, so every nonempty metric mark is
\[
 (D_g^SP_{r,g})\phi(x)
   =\int D_g^S[p_{r,g}(x,y)dv_g(y)]\,[\phi(y)-\phi(x)].
\]
The derivative regulator controls the surviving increment from $x$ across the
mollifier support. For constant metrics the cutoff length is fixed in proper
coordinates. The resulting affine pullback is exact and differs from
changing the bare fugacity at fixed coordinate radius. For ordinary sources
and their collision forests we use covariant derivatives, normal coordinates
and volume Jacobians with the original flat intrinsic targets.

\emph{Gaussian measure and determinant.}
Let $w=\sqrt{\det g}$,
$A_g^{\rm den}=-\partial_\mu(wg^{\mu\nu}\partial_\nu)$ and
$K_g=A_g^{\rm den}-A_0$, where $A_0=-\Delta$ is the unit-reference
precision. In a finite Gaussian approximation, put
$M_g=(1+CK_g)^{-1}$, $C_g^{\rm nat}=M_gC$. Exact completion gives
\begin{multline}
 \mathbb E_C\!\left[e^{-\langle\phi+\zeta,K_g(\phi+\zeta)\rangle/2}
                         \Psi_g(\phi+\zeta)\right]
 \\
 =\det(1+CK_g)^{-1/2}e^{-\langle\phi,K_gM_g\phi\rangle/2}
                \mathbb E_{C_g^{\rm nat}}\Psi_g(M_g\phi+\zeta).
 \label{stress-and-trace:metric-completion}
\end{multline}
\begin{equation}
 (A_0+K_g)^{-1}-C_g^{\rm nat}
 =(1+G_0K_g)^{-1}(G_0-C)(1+K_gC)^{-1}.
 \label{stress-and-trace:retained-metric-resolvent}
\end{equation}
The resolvent factors must be kept in the displayed order. The interaction $\Psi_g$
includes $(1-Z_r)\int dv_g\,g^{\mu\nu}\partial_\mu\phi\partial_\nu\phi/2$ and hence retains the physical stiffness. At each finite metric
order \eqref{stress-and-trace:retained-metric-resolvent} contains a smoothing retained covariance. The difference from the
retained covariant spectral split is smoothing as well. To see its finite
range, subtract the two complements from their common full inverse. Their
difference is the negative prefix difference. At order $p$, a coefficient of $C_g^{\rm nat}$
has at most $p+1$ finite-range factors $C$ separated by local precision
variations. Every fixed derivative of the spectral prefix also has finite
propagation.
In the finite metric algebra the covariance-change exponential truncates. To
bound its mean and precision chains let $C_a=C[0,a]$, $K=d^*kd$ and $\nabla_a=ad$. In a flat bulk chart put
\[
 T_a=dC_ad^*,\qquad U_a=a^{-1}C_ad^*.
\]
For every fixed $s=R+1/2$, uniformly in the lower cutoff,
\[
 \|T_au\|_{c^{s-1}_a(B)}
       \le C_s\|u\|_{c^{s-1}_a(B^+)},\qquad
 \|U_au\|_{c^s_a(B)}
       \le C_s\|u\|_{c^{s-1}_a(B^+)}.
\]
The enlargement $B^+$ is fixed and contains the propagation neighborhood. We use
the little H\"older space $c^s=c^{R,1/2}=\mathfrak h^{R,\alpha_{\rm H}}$ from Section~\ref{sec:source-calculus}. For the external metric profiles
take fixed seminorms through integer order $R_I\ge R+3$, increasing this order if the
prescribed mixed kernel derivatives require it. Ordinary field directions
have order $R$ throughout.

To prove these estimates, the flat spectral prefix has multiplier
$|p|^{-2}k_{\rm fr}(a|p|)$, where $k_{\rm fr}(t)=O(t^2)$ at zero, $k_{\rm fr}(t)\to1$, and the
derivatives of $1-k_{\rm fr}$ decrease rapidly at infinity.  After rescaling,
the multipliers of $T_a$ and $U_a$ are respectively
\[
 \frac{\xi_\mu\xi_\nu}{|\xi|^2}k_{\rm fr}(|\xi|),
       \qquad \frac{-i\xi_\mu}{|\xi|^2}k_{\rm fr}(|\xi|).
\]
They are smooth at zero and have orders zero and minus one.  On the
dyadic annulus $|\xi|\asymp2^j$, their localized convolution kernels
have $L^1$ norms at most $C_s$ and $C_s2^{-j}$, respectively: rescale
the annulus and integrate by parts more than twice.  The inhomogeneous low-frequency kernels
have bounded $L^1$ norms. Weighting by $2^{j(s-1)}$ and $2^{js}$ proves the asserted bounds, then
smooth density gives their little H\"older versions. Cardinal lifting
transfers the annular bounds to the lattice. Use finite Gaussian
approximations for the algebra and the positive smooth shell cutoffs for
uniform H\"older constants; arbitrary sharp Fourier cutoffs are outside this
argument. Finite propagation allows restriction to $B^+$ first.

Multiplication in the H\"older algebra gives a bound on the same space,
\[
 \|C_aKf\|_{c^s_a(B)}
 =\|U_a(k\nabla_af)\|_{c^s_a(B)}
 \le C_s\|k\|_{c^{s-1}_a(B^+)}
              \|\nabla_af\|_{c^{s-1}_a(B^+)}.
\]
Differentiating the chain $L=(C_aK_1)\cdots(C_aK_\ell)$ gives
\[
 \nabla_aL\phi=T_ak_1T_ak_2\cdots T_ak_\ell\nabla_a\phi.
\]
Each factor preserves the H\"older index. Ordinary analytic directions of
unit $c^s_a$ norm cost at most an exponential in chain length. On real backgrounds
the local Sobolev bound modulo constants gives a fixed polynomial in $E_a+\sum_{k=2}^{R+2}Q_{a,k}$,
absorbed by the positive source-degree regulator increment. No height
regulator or extra spatial derivative is needed.

Reflect the field oddly and $k$ tensorially on a true rectangle. Vanishing of
the metric directions in a boundary collar gives smooth coefficient
extensions preserving odd fields. Compression from the doubled flat torus
costs fixed reflection constants, including at corners since the odd
reflections commute. This uses the stated boundary compatibility.

At metric order $p$ no mean chain has more than $p$ precision factors and no
differentiated natural prefix more than $p+1$ covariances. Thus only finitely
many background polynomial factors enter the flat-metric mean derivatives.
Allocate ordinary field derivatives from the fixed strong/weak radius
reserve, absorbing the polynomial number of allocations. Each extra
allocation spends a metric label; unmarked steps use no further spatial or
analytic reserve. Varying covariances costs at most $2p$ extra functional field
derivatives by the block-pair projective bound. The forest integral still
uses positive flat covariance and one actual regulator per merged support.

Finally $\|T_a\|_{2\to2}\le1$ gives
\[
 |\langle\phi,K_1(C_aK_2)\cdots(C_aK_\ell)\phi\rangle|
 \le\prod_{i=1}^{\ell}\|k_i\|_\infty\,\|d\phi\|_2^2.
\]
This bounds the quadratic completion factors. The finitely many bounded
chain factors preserve smoothing of a retained covariance in the split
discrepancy. Local Fourier expansion with its fixed endpoint derivatives
gives the $c^s\widehat\otimes_\pi c^s$ kernel bound. At any fixed metric order only finitely many fixed
smooth profile seminorms occur.

The scalar field has covariance $(A_g^{\rm den})^{-1}$. The geometric determinant, on the other
hand, is the determinant of $P_g=w^{-1/2}A_g^{\rm den}w^{-1/2}$. For a metric direction $k$ write $\ell_k=D_k\log w$ and denote
multiplication by this function by $M_{\ell_k}$. At the flat metric,
\begin{equation}
 D_kP=D_kA^{\rm den}-\tfrac12(M_{\ell_k}A_0+A_0M_{\ell_k}).
 \label{stress-and-trace:half-density-variation}
\end{equation}
In each Gaussian chain or loop the extra precision in
\eqref{stress-and-trace:half-density-variation} collapses an adjacent
covariance, since $A_0G_0=1$. External collapses cancel the scalar
endpoint-density variations; internal collapses are local contacts.
Higher precision variations give the corresponding local quadratic
and higher variation terms. The two Gaussian constructions therefore
agree away from partial diagonals, also with arbitrary remaining
cylindrical field probes.

Extend proper diagonals recursively. With $p\ge2$ pure metric labels,
the scaling degree is at most $2p$, while the full relative diagonal
has codimension $2(p-1)$. The remaining difference is consequently
a distribution supported there of derivative order at most two:
\begin{equation}
 D_H=\sum_{|\alpha|\le2}c_{\alpha,H}\partial^\alpha\delta,
 \qquad c_{\alpha,H}=(-1)^{|\alpha|}
                 D_H[\chi(z)z^\alpha/\alpha!].
 \label{stress-and-trace:gaussian-local-contact}
\end{equation}
Subtract the test function's Taylor polynomial on that support to obtain \eqref{stress-and-trace:gaussian-local-contact}
with the stated coefficients. For mixed ordinary labels the Taylor degree is
the one prescribed for those labels. The coefficients describe bulk
contacts, leaving the common smooth wall part in its record. At singleton
order the free readout fixes the quadratic coefficient and the local
heat/Wick subtraction fixes the identity coefficient. Recursion between the
two defined extensions now determines every contact.

In the finite approximation the identity
$(\det Q_g/\det Q_0)^{1/2}\mathbb E_{Q_g}F
=\mathbb E_{Q_0}[e^{-\psi^*(Q_g^{-1}-Q_0^{-1})\psi/2}F]$
generates closed loops with the same quadratic source exponential as
open chains. The above contact map therefore leaves zero independent
initial marked scalar. Its identity with arbitrary cylindrical probes
allows insertion inside the bounded, mollified interaction before
cutoff removal. The finite rows converge at fixed $r$; only afterwards
is the physical limit $r\downarrow0$ taken.

For a direct normalization check, write $g=e^{2\sigma}\widehat g$,
$\det\widehat g=1$. The scalar heat coefficient $R/6$
\cite[Lemma~2.4.3(b) and Section~4.8]{Gilkey84} gives the
area-subtracted free conversion
\begin{equation}
 W^{\rm free}[g]-W^{\rm free}[\widehat g]
 =\frac1{24\pi}\int
        (|\nabla_{\widehat g}\sigma|^2+R_{\widehat g}\sigma)dv_{\widehat g}.
 \label{stress-and-trace:liouville-contact-conversion}
\end{equation}
It follows by integrating
$\delta_\omega W^{\rm free}=(24\pi)^{-1}\int\omega R_gdv_g$.
This local measure/contact term is part of the preceding map, counted
once; it retains the nonzero second Weyl jet even when the scalar
covariance is conformally unchanged.

\emph{Local projection and scalar summation.}
Average the connected kernel, including stable and preparation terms, over
its roots. Pair relative normal coordinates with the finite polynomial module
using volume Jacobians and parallel transports. The resulting bounded natural
projector gives $F=\iota_g\mathscr P_gF+(1-\iota_g\mathscr P_g)F$ with full generating support in the remainder. Reserved
moments absorb the fixed polynomial in $1+n+\ell/a$ from root changes and chart
derivatives. Histories leaving the coordinate ball stay in the remainder,
bounded by exponential size weight or any prescribed fixed row-decay order. A
sufficiently large permitted grouped step makes restriction gain times
norm-equivalence constant less than one, preserving the stable gap.

The ordinary projector leaves the scalar history undetermined. For the
connected interaction-prefix germ $\Gamma_{g,j}$, the increment $\Gamma_{g,j+1}(0)-\Gamma_{g,j}(0)$ eliminates the
accumulated independent scalar. The root density of this increment is
bounded by $Ca_j^{-2-p\eta}$. Up to geometric weight two the only natural scalar terms are
\begin{equation}
                  a_j^{-2}A_jdv_g+B_jR_gdv_g.
 \label{stress-and-trace:vacuum-primitives}
\end{equation}
The coefficients depend on the canonical state at $a_j$. First metric derivatives
vanish in normal coordinates; curvature spans the scalars linear in second
derivatives \cite[Lemma~2.4.6]{Gilkey84}. Fix the volume term by the metric vacuum convention. For
interior variations $\int_D R_gdv_g=0$: the positive square root of $g$ gives a global oriented
orthonormal frame, Cartesian near the boundary. Its connection $\omega$ satisfies $R_gdv_g=2d\omega$.
Stokes' theorem, with no corner term, cancels the pure curvature primitive
before absolute scale summation. The third-order remainder obeys
\begin{equation}
 Ca_j^{-2-p\eta}a_j^3
 \sum_{\mathfrak h\,{\rm rooted}}w_{\mathfrak h}(1+n+\ell/a_j)^{m+3}
                         \le C_Ia_j^{1-p\eta}.
 \label{stress-and-trace:complete-scalar-third-moment}
\end{equation}
Here $w_{\mathfrak h}$ is the rooted-history majorant, and $n$ and
$\ell$ are its block number and support span. Choose $p\eta<1/2$.
Mixed terms such as $R_gJ(x)$ are retained
local contacts, not canceled topologically. Their remainder uses the
declared larger contact degree. The relative free determinant has the
same area/curvature subtraction from its heat expansion
$\Delta\operatorname{Tr}e^{-tP}
=\Delta\operatorname{Area}/(4\pi t)+(24\pi)^{-1}\int R_gdv_g+O(t)$;
where $\Delta$ denotes subtraction of the flat expression. Its
finite nonlocal remainder is retained.

For a constant metric, all the data are affine pullbacks of the
flat construction. Every other geometric forcing has a positive
scale power. Choose the finite anomalous loss so that
$\sum_jC_Ia_j^{\nu_I-|I|\eta}<\infty$.
The early tail then vanishes and the fixed scale window
converges, jointly for ordinary, wall, contact and independent scalar
records by induction on subsets. In the ordinary coordinates, inversion
gives \eqref{stress-and-trace:metric-conversion}. Gaussian conversion and the scalar bounds complete the recognition
hypotheses. Normalized transfer now identifies the physical functionals with
their original flat denominator. In one finite labelled algebra, fixing
proper subsets ensures commutation of mixed derivatives and compatibility of
restrictions. Higher metric order needs larger fixed analytic reserves; the
required regularity of the physical dilation orbit is the same at each
order.
\end{proof}

\subsection{Ward identities and the local trace}

We vary the inverse metric and choose the positive sign for the
action-conjugate field,
\[
 T_{\mu\nu}=\frac2{\sqrt{\det g}}\frac{\delta S_R}{\delta g^{\mu\nu}},
 \qquad\mathcal O_\rho=+\partial_\rho S_R.
\]
Localize the endpoint derivative of $S_R$, including its source prescription.
This produces $\mathcal O_\rho$ from the kinetic, cosine, source and counterterm derivatives
of the original one-parameter family, without further inhomogeneous tuning.
The local readout and metric vacuum convention fix its identity coordinate
before the trace comparison; Theorem~\ref{parameter-responses:action-packets} fixes the others.

\begin{theorem}[Complete geometric and trace identities]
\label{stress-and-trace:geometric-ward}
\label{stress-and-trace:local-trace}
The stress tensor is a symmetric admitted local field. For every
compact interior vector field $\xi$,
\begin{equation}
 (\delta_g[\mathcal L_\xi g]+D_J[\mathcal L_\xi^{\rm src}J])
                  W^{\rm met}[g,J]=0.
 \label{stress-and-trace:ward-functional}
\end{equation}
The source action is geometric pullback conjugated by the complete
substitution $q_I$. At flat metric this is
$\int\partial_\mu\xi^\nu\langle T_{\mu\nu}\rangle_Jdx
+D_JW[\mathcal L_\xi^{\rm src}J]=0$.
Its mixed derivatives give all translation and diffeomorphism contacts.
In the original periodic, field-even dimension-two module the flat trace is
\begin{equation}
 \Theta=T^\mu{}_{\mu}
   =\mathsf B(\rho)\mathcal O_\rho+c_{\mathbf1}(\rho)\mathbf1
                                +\mathcal C_{\rm ct},
 \qquad \mathsf B=-\mathfrak b.
 \label{stress-and-trace:trace-sign}
\end{equation}
The scalar and contact coefficients are fixed by the finite maps and
the scalar telescope below; they are not freely assigned to make the
identity hold. No assertion $c_{\mathbf1}=0$ is made.
\end{theorem}
\begin{proof}
Simultaneously pulling back metric, field, domain and sources preserves the
auxiliary integral: the proper mollifier, covariance, determinant,
counterterms and rooted projection all commute with this change of
variables. Metric admission transfers the equality to the original
functional. Twice differentiating there gives both $D_J^2W[q_{k_1},q_{k_2}]$ and $D_JW[q_{k_1k_2}]$. At the flat
metric $\delta g^{\mu\nu}=-(\partial^\mu\xi^\nu+\partial^\nu\xi^\mu)$; the minus sign from $e^{-S_R}$ then gives \eqref{stress-and-trace:ward-functional} with the displayed sign. The
first metric coefficient belongs to the admitted dimension-two module, hence
to the common Wightman representation. Free metric variation sets its
kinetic normalization, the vacuum convention its identity component.

Use the passive scale coordinate $u(\mu)=T_{\mu_0/\mu}\rho$, rescale the proper reference length with
it, and carry the operator references back by the complete source cocycle.
Differentiation in $\log\mu$ gives
$\mathsf B(u)=-\mathfrak b(u)$.
Under a new parameter chart $v=f(u)$, the transformations
$\mathsf B^v=f'\mathsf B$ and $\mathcal O_v=\mathcal O_u/f'$
leave their product invariant.
If $\mathscr P_c$ denotes geometric source pullback under
$x\mapsto cx$, then the conversion from metric to original
source coordinates is
\begin{equation}
                 \mathscr C_{c,\rho}
                    =\mathscr R_{c,\rho}^{-1}\mathscr P_c.
 \label{stress-and-trace:constant-weyl-source-map}
\end{equation}
Its coordinate pullbacks cancel, leaving local normalization and
contact coefficients. For the complete unmarked-plus-sourced germ
$\mathfrak G_a$ and its intrinsic ordinary readout $\mathfrak N_a$,
constant-metric affine change and ordered-prefix cancellation give
\begin{equation}
 \mathfrak N_a\mathfrak G^{\rm met}_{\rho,a}[c^2\delta,J]
 =\mathfrak N_a\mathfrak G^{\rm flat}_{T_c\rho,a}
                         [\mathscr C_{c,\rho}J]
                          +\mathfrak v_{c,\rho,I}[J].
 \label{stress-and-trace:constant-weyl-record}
\end{equation}
We keep the source-zero value of the germ so its unmarked parameter
derivative includes the bulk action score.

The scalar $\mathfrak v$ must be constructed separately from the
ordinary comparison. Express the geometric history $M$ and the
transported flat history $A$ in the same operation and source
convention. Their ordinary preparations include the full
$\mathscr R^{-1}\mathscr P$, its identity outputs, and the
quadratic Gaussian contact conversion. Write $L^{\rm in}_{c,r,I}$
for the remaining central local conversion columns and finite
vacuum-convention difference. This includes the measure term
\eqref{stress-and-trace:liouville-contact-conversion} once.
Place these columns in $Q_0$ with zero independent scalar, and
remove them from the subsequent $Q$ records in the expression below.
At fixed cutoff $r$, write $\Delta E_a=Q^M_{a,r}(0)-Q^A_{a,r}(0)$
and $\Delta s_j=s^M_{j,r}-s^A_{j,r}$ for the differences of the
zero-field values and scalar increments. With
$\operatorname{Loc}_{\mathbf1,I}$ the fixed local identity/contact
projection, the scalar difference is
\begin{equation}
 \mathfrak v_{c,\rho,I}
 =\lim_{r\downarrow0}\operatorname{Loc}_{\mathbf1,I}
 \left[L^{\rm in}_{c,r,I}+\Delta E_a+\sum_{a_j<a}\Delta s_j\right].
 \label{stress-and-trace:weyl-scalar-telescope}
\end{equation}
The input targets are fixed and $\mathfrak v_{1,\rho,I}=0$. Apply the prescribed subtraction to the area term and
cancel the pure curvature primitive, retaining the mixed curvature terms in
the local projection. Split at a physical scale $d$. Below $d$, Taylor
subtraction gives the positive powers in \eqref{stress-and-trace:complete-scalar-third-moment} and in the contact forest. Above
$d$, affine pullback and ordered-prefix cancellation identify the locally
converted ordinary data. On this finite window, stable contraction and
telescoping send the generated scalar discrepancy to zero as $r\downarrow0$. Taking $d\downarrow0$
leaves only the explicit conversion in \eqref{stress-and-trace:weyl-scalar-telescope}, with no further nonlocal scalar.
Here the independent initial scalar must be zero and the wall scalars keep
their own record. This gives \eqref{stress-and-trace:constant-weyl-record} also for the scalar histories of the full
arrays.

Define the finite Weyl and action-source polynomials
$\mathscr T_I[\sigma,J]=D_gq_I(\delta,J)[2\sigma\delta]$ and
$\mathscr A_{\rho,I}[\sigma,J]$, the fixed ordinary realization of
$\mathcal O_\rho(\sigma)$. Let $\Pi_{{\rm ct},I}$ retain mixed
and collision columns, including profile derivatives, but exclude the
source-free bulk columns $\sigma\mathcal K_{\alpha\alpha}$,
$\sigma V_{\cos}$, $\sigma\mathbf1$, where
$\mathcal K_{\mu\nu}=[\partial_\mu\phi\partial_\nu\phi]_R$ and
$V_{\cos}=\mu_{\rm ref}^2(V_1+V_{-1})/2$.
Let $v_{\mathbf1}(t)$ be the identity component of
$\mathfrak v_{e^t,\rho,I}$ in the same finite readout. With the
action-identity convention fixed as above, the polynomial generating
$\mathcal C_{\rm ct}$ and the identity coefficient are, respectively,
\begin{equation}
 \Pi_{{\rm ct},I}(\mathscr T_I-\mathsf B\mathscr A_{\rho,I}),
 \qquad c_{\mathbf1}(\rho)=v_{\mathbf1}'(0).
 \label{stress-and-trace:local-weyl-contact-map}
\end{equation}
These are computable coefficients of independently constructed maps.

For the Weyl variation $\delta_\sigma g_{\mu\nu}=2\sigma g_{\mu\nu}$,
our convention gives $\delta_\sigma S_R=-\int\sigma\Theta$;
thus differentiating the log weight inserts $+\Theta$.
Differentiate \eqref{stress-and-trace:constant-weyl-record} at
$c=e^t=1$. The endpoint derivative is
$\mathfrak b\partial_\rho$ and inserts
$-\mathfrak b\mathcal O_\rho$. This fixes the beta-function sign
and the three source-free bulk coefficients.

Take intrinsic polynomial moments before integrating over fields. For a Weyl
profile constant near the root, the support and history-tail estimates
recover the constant-profile moments as the transition annulus recedes.
There is no even dimension-one vector in the module to pair with a first
profile derivative. Higher derivatives and source overlaps occupy the mixed
and collision columns of \eqref{stress-and-trace:local-weyl-contact-map}. With both action and metric scalar histories
controlled, complete recognition gives the identity in distributions.

The only possible retained curvature improvement is
$a\int R_gFdv_g$, with
$T^{\rm imp}_{\mu\nu}=2a(\delta_{\mu\nu}\Delta-
\partial_\mu\partial_\nu)F$ and $\Theta^{\rm imp}=2a\Delta F$.
Its dimension-zero periodic even $F$ is constant, so it vanishes.
There is likewise no nontrivial allowed dimension-zero field
redefinition giving a bulk EOM term; $\phi\mathcal E_\phi$ lies
outside the periodic module. Larger declared schemes transport those
redundant fields and contacts together. Changing
$\mathcal O_\rho\mapsto\mathcal O_\rho+a(\rho)\mathbf1$
changes $c_{\mathbf1}$ by $-\mathsf B a$, preserving the identity.
\end{proof}

The retained metric polynomial by itself has variation
\[
 T^{\rm ret}_{\mu\nu}=Z_R\mathcal K_{\mu\nu}
 -\delta_{\mu\nu}\bigl(Z_R\mathcal K_{\alpha\alpha}/2+
 \lambda_c V_{\cos}+e\mathbf1\bigr),\qquad
 \Theta^{\rm ret}=-2\lambda_c V_{\cos}-2e\mathbf1.
\]
Conversion by \eqref{stress-and-trace:metric-conversion} adds the normal-ordering, action-source, quadratic-variation
and identity contacts for geometric stress. Metric variation defines $Z_R$; the
independent field-equation calculation defines $Z_E$.

\section{Consequences and scope}
\label{sec:discussion}

The two lengths entering the mass law have independent definitions. One is
the exit length of the nonlinear weak trajectory selected by compact
physical normalization. The other is the correlation length determined by
the positive spectral threshold of the literal Hamiltonian. In the selected
representation their ratio stays between uniform positive finite bounds,
giving the leading exponential dependence of the mass on microscopic
activity.

This threshold applies to all local sources because heights, derivatives and
complete sources span the same cyclic space. For this identification
normalized transfer uses positive conditional comparisons, complete source
matching, bounds on repeated insertions and the true Dirichlet caps.
Extensive response uses, in addition, screening in the same state.

Interaction is detected there by derivative exponential moments. The
finite-harmonic comparison and geometric identities use the same field
normalization and contact prescription and need no comparison to another
continuum model. The existence and structure of isolated mass shells, and
the exact mass--activity constant, remain open independently of this
construction.

\appendix
\section{Positive periodic comparison}
\label{app:positive-comparison}

In finite dimensions pair field coordinates in the ordinary way and allow
singular Gaussian covariances. Denote the Gaussian and its expectation by $\gamma_m=N(m,Q)$
and $\gamma_m(f)=\int f\,d\gamma_m$. For $z_i\ge0$ let $V(h)=\sum_i z_i\cos h_i$. With $\zeta\sim\gamma_0$
define $F_Q(x)=\E e^{V(x+\zeta)}$ and $M_Q=-D^2\log F_Q(0)$.
The Bessel expansion gives $M_Q=\E_\pi[kk^\top]\succeq0$ for the
integer charge law $\pi(k)\propto e^{-k^\top Qk/2}\prod_i I_{k_i}(z_i)$.

\begin{proposition}[Positive periodic comparison]
\label{positive-cosine-laws:phase}\label{positive-cosine-laws:phase-character}
\label{positive-cosine-laws:tilt-minimum}\label{positive-cosine-laws:positive-characters}
The following inequalities hold for the actual normalized laws:
\begin{equation}
 D^2\log F_Q(x)\succeq-M_Q,\qquad
 e^{-x^\top M_Qx/2}\le F_Q(x)/F_Q(0)\le1.
 \label{positive-cosine-laws:phase-bound}
\end{equation}
If $\chi_m(k)$ is the characteristic function of
$d\mu_m\propto e^Vd\gamma_m$, then
$\chi_0(k)>0$ and $|\chi_m(k)|\le\chi_0(k)$ for every real $m,k$.
Let $\mu_0^{\,J}$ be the tilt of $\mu_0$ by $e^{J^\top h}$.
For centered linear forms $F,G$,
\begin{align}
 \Cov_{\mu_0^{\,J}}(h)&\succeq\Cov_{\mu_0}(h),
 \label{positive-cosine-laws:tilt-order}\\
 \Cov(F^2,G^2)&\ge2(\E FG)^2,
 \label{positive-cosine-laws:fourth}\\
 \Cov(\cos F,\cos G)&\ge0,\qquad 0<\E\cos F\le1.
 \label{positive-cosine-laws:positive-cosine-covariance}
\end{align}
Unsubscripted expectations and covariances here are taken in $\mu_0$.
Cosine characters increase under centered positive precision and
nonnegative activity increments. For common bare data and $D\subset D'$
with a charge supported in $D$,
\begin{equation}
            0<\E_{D'}\cos(f\cdot h)\le\E_D\cos(f\cdot h).
 \label{positive-cosine-laws:character-domain-order}
\end{equation}
\end{proposition}
\begin{proof}
\emph{Two and four replicas.} For independent $u,v\sim N(0,Q/2)$,
$K(a,b)=\E e^{2\sum_i z_i\cos(a_i+u_i)\cos(b_i+v_i)}$ is a Gram
kernel. Indeed, set $f_p(a)=\E\prod_i\cos(a_i+u_i)^{p_i}$. In the
power series for $K$, the coefficient of $f_p(a)f_p(b)$ is
$\prod_i\bigl((2z_i)^{p_i}/p_i!\bigr)\ge0$.
Replica rotation and Cauchy--Schwarz give
\[
 F_Q(x)^2F_Q(y)^2\le F_Q(0)^2F_Q(x+y)F_Q(x-y).
\]
Differentiate twice in $y$ at zero and integrate along $tx$ for the
lower bound in \eqref{positive-cosine-laws:phase-bound}; positive Fourier
coefficients give its upper bound.

Write $Z_m=\gamma_m(e^V)$ and $N_m(k)=\gamma_m(e^{V+ik\cdot h})$.
Take two replicas of mean $m$ and two of mean zero, and apply the
orthogonal four-by-four Hadamard transform. Their new means are
$(m,m,0,0)$. If
$a_{p,q}(h)=\prod_i\cos(h_i/2)^{p_i}\sin(h_i/2)^{q_i}$, expansion
of the transformed four weights gives exactly
\[
 \begin{aligned}
 &|N_0(k)|^2Z_m^2-|N_m(k)|^2Z_0^2\\
 &\quad=2\sum_{p,q}\prod_i\frac{(4z_i)^{p_i+q_i}}{p_i!q_i!}
 \gamma_m(a_{p,q})^2\,\gamma_0(\sin(k\cdot h)\,a_{p,q})^2\ge0.
 \end{aligned}
\]
The series has a finite absolute majorant and the positive Bessel expansion
gives $N_0(k)>0$. Expansion of the character modulus at $k=tu$ yields the covariance
minimum. Completing the square turns a real tilt into mean $QJ$, so the second
derivative of $\Var_{tG}F$ at its minimum gives \eqref{positive-cosine-laws:fourth}.

\emph{The stronger character inequality.} The lattice-Gaussian
inequality~\cite[Corollary 2.2]{RegevStephensDavidowitz2017} transfers
to the positive-cosine law by Gaussian combs. First suppose $Q\succ0$
and all activities are positive, and put $y_x=z_x/2$. For
$m>\max_x y_x$ and $t_{m,x}=2\log(m/y_x)$, Poisson summation gives
\[
 \begin{split}
 \Theta_{m,x}(h)
 &=\sum_{q\in\mathbb Z}(y_x/m)^{q^2}e^{iqh}\\
 &=\sqrt{2\pi/t_{m,x}}
             \sum_{n\in\mathbb Z}e^{-(h-2\pi n)^2/(2t_{m,x})},\\
 \Theta_{m,x}(h)^m&=e^{2y_x\cos h+O_y(m^{-1})}.
 \end{split}
\]
The last estimate holds uniformly on the real line and on each fixed complex
strip. Attach $m$ integer leaves $n_{x,a}$ to each real core $h_x$, giving joint energy
\[
 \frac12h^\top Q^{-1}h
       +\sum_{x,a}\frac{(h_x-2\pi n_{x,a})^2}{2t_{m,x}}.
\]
On restricting each core to $(2\pi/r)\mathbb Z$ we obtain a positive quadratic form on a
full-rank lattice. A linear change of coordinates puts it in the form of the
lattice inequality. With $m$ fixed, Gaussian Riemann sums converge as the core
mesh vanishes, $r\to\infty$, including after any fixed polynomial moment or real tilt
is inserted. Summing out the leaves gives the comb product. Its uniform
comparison with $e^V$ and the coercive Gaussian tail justify $m\to\infty$. The scalar
Poisson factors cancel on normalization. Allow zero activities by continuity
and replace $Q$ by $Q+\eta I$ with $\eta\downarrow0$ for singular covariances. Keeping the geometry
fixed throughout these limits yields
\[
 (\E\cos F)^2(\E\cos G)^2
 \le(\E\cos F\cos G)^2-(\E\sin F\sin G)^2.
\]
The two character means and $\E\cos F\cos G$ are positive, which gives the sign of the
cosine covariance. Replacing $G$ by $tG$ and expanding at zero gives $\Cov(\cos F,G^2)\le0$.
Differentiating normalized expectations then gives monotonicity in the
precision and activity. Alternatively, conditional on an exterior field, the
Gaussian mean is harmonic and the centered interior character bounds the
conditional character modulus. Averaging gives \eqref{positive-cosine-laws:character-domain-order} with the respective domain
normalizers. Continuity allows zero activities.
\end{proof}

\paragraph{Scores and covariance changes.}
Let $G$ be a proper reference covariance and let $A=Ph$, with $P$ of
full row rank. Under the Gaussian law, $A$ has covariance $G_A=PGP^\top$.
Given $A=a$, the field has mean $\Psi a$ and covariance $Q$, where
$\Psi=GP^\top G_A^{-1}$ and $Q=G-\Psi G_A\Psi^\top$.
At phase $\theta$ and tilt $J$ the observation density is
\[
 p(a)\propto e^{-a^\top G_A^{-1}a/2+J^\top\Psi a}
               F_Q(\Psi a+\theta+QJ).
\]
Thus
$D^2(-\log p)\preceq H_0=G_A^{-1}+\Psi^\top M_Q\Psi$.
For $s=-\nabla\log p$, integration by parts gives
$\E s=0$, $\Cov(A,s)=1$ and $\Gram(s)=\E D^2(-\log p)\preceq H_0$.
At zero phase and tilt,
$e^{-a^\top H_0a/2}\le p(a)/p(0)\le e^{-a^\top G_A^{-1}a/2}$.
The ball $B_r$ of radius $r=\sqrt{2\Tr\Cov(A)}$ about zero contains
at least half the mass, so $p(0)\ge 1/(2\operatorname{vol}B_r)$.

\begin{corollary}[Covariance order]
\label{covariance-comparison:cosine-covariance-order}
\label{covariance-comparison:phase-covariance-monotonicity}
For $Q_2\succeq Q_1$ and a fixed positive periodic weight on real
linear rows, its centered characters satisfy
$0<\chi_{Q_2}(k)\le\chi_{Q_1}(k)$. Its phase partitions satisfy
$F_Q(x)/F_Q(0)\le F_{Q+R}(x)/F_{Q+R}(0)$ for $R\succeq0$.
These conclusions include arbitrary real charges, singular observation
matrices and compact integrals of smooth rows with positive amplitudes.
\end{corollary}
\begin{proof}
Decompose $x=x_1+m$ into independent parts with covariances $Q_1$ and $Q_2-Q_1$. The weight for
averaging the conditional character inequality is $Z(m)d\gamma(m)/\E Z(m)$. An extra row $k$ of zero
activity permits arbitrary charges. Positive quadrature extends the
comparison to compact integrals of rows, and the empty Poisson configuration
gives strict positivity of the character. For the phase ratio take square
roots in the replica midpoint inequality, use the arithmetic--geometric mean
bound, then average over $y\sim N(0,R)$.
\end{proof}

\begin{lemma}[Lattice charges and quantitative response]
\label{positive-cosine-laws:lattice-comparison}
\label{positive-cosine-laws:charge-monotonicity}
\label{positive-cosine-laws:differential}
Let $\mathcal L$ be a lattice in a finite-dimensional Euclidean space,
possibly of lower rank, and let $A\succ0$. Put
\[
 \rho_A(x)=\sum_{\ell\in\mathcal L}
                 e^{-(\ell+x)^\top A(\ell+x)/2}.
\]
The Gaussian on the coset $\mathcal L+x$ has covariance at least that
on $\mathcal L$. Consequently, for $B\succeq0$,
\begin{equation}
 D_B\log\frac{\rho_A(x)}{\rho_A(0)}\le0,
 \qquad D_x^2\log\rho_A(x)\succeq D_x^2\log\rho_A(0).
 \label{positive-cosine-laws:ambient-ratio}
\end{equation}
For the centered law, increasing $A$ decreases covariance and every
real linear Laplace transform, and
$D_B\Sigma_A\preceq-\Sigma_AB\Sigma_A$ for $B\succeq0$.
Conditioning on a homogeneous linear constraint decreases covariance.
For $Q\succeq0$ and $y_x\ge0$, the charge law
$\nu(n)\propto e^{-n^\top Qn/2}\prod_x I_{n_x}(2y_x)$,
increasing $y_x$ increases covariance and Laplace transforms, while
increasing $Q$ decreases them; homogeneous integer constraints
again decrease covariance.

For the continuous law of precision $K$ and activities $y_x>0$, put
$\mathsf C=\E\phi\phi^\top$, $c_x=\E\cos\beta_*\phi_x$,
$s_x=\E[\phi\sin\beta_*\phi_x]$ and $\mathsf R=\mathsf C^{-1}-K$.
Let $e_x$ be the coordinate vector at $x$ and $D_B$ the directional
derivative of the precision along $B\succeq0$. Then
\begin{align}
 -\partial_{y_x}\mathsf C&\succeq2s_xs_x^\top/c_x,&
 \partial_{y_x}\mathsf R&\succeq
 \frac{(\mathsf R e_x)(\mathsf R e_x)^\top}{2\beta_*^2y_x^2c_x},\nonumber\\
 D_B\mathsf C&\preceq-\mathsf CB\mathsf C,&D_B\mathsf R&\succeq0.
 \label{positive-cosine-laws:precision}
\end{align}
Also $D_Bc_x\ge s_x^\top Bs_x/(2c_x)$.
\end{lemma}
\begin{proof}
\emph{Cosets and precision.}
The lattice covariance minimum is
\cite[Corollary 3.2]{RegevStephensDavidowitz2017} after a linear change
on the lattice span. A displacement $A$-orthogonal to that span
contributes only a Gaussian scalar and does not alter covariance.
If $X$ denotes the point of $\mathcal L+x$, direct differentiation gives
\[
 \begin{aligned}
 D_B\log\frac{\rho_A(x)}{\rho_A(0)}
 &=-\tfrac12\Tr B(\E_xXX^\top-\E_0XX^\top),\\
 D_x^2\log\rho_A(x)&=-A+A\Cov_x(X)A.
 \end{aligned}
\]
The centered mean vanishes. The second-moment difference in the first line
is therefore the sum of the covariance difference and the outer product of
the shifted mean with itself, both positive semidefinite. This gives \eqref{positive-cosine-laws:ambient-ratio} also
for lower-rank lattices with fixed displacements outside their span.

Completing the square turns a real tilt of the centered law into a
translated coset Gaussian. Its covariance dominates the untilted covariance,
and hence its second moment dominates the centered second moment.
Differentiate the logarithm of the normalized Laplace transform along $B\succeq0$ to
obtain the Laplace comparison. For centered linear forms $F,G$ the minimum of $t\mapsto\Var_{tG}F$
occurs at zero, where its second derivative gives $\Cov(F^2,G^2)-2\Cov(F,G)^2\ge0$. Rank-one decomposition
of $B$ now yields $D_B\Sigma_A\preceq-\Sigma_AB\Sigma_A$. Every nonempty fiber of a homogeneous linear constraint
is a coset of the central fiber, so its conditional covariance dominates
that of the central fiber. Total covariance gives the claimed effect of
conditioning.

\emph{The Bessel weights.}
At each site introduce $m$ integer leaves, with $P$ the summation map. The leaf
precision $\diag(2\log(m/y_x))+P^\top QP$ repeats each site's diagonal entry $m$ times. Increasing $y_x$ lowers
the diagonal precision, while increasing $Q$ adds a positive form. A
homogeneous integer constraint on the aggregate acts on $P$ times the leaves.
Thus the lattice comparisons hold already at finite leaf number.

For one site the aggregate Fourier series is
\[
 \left(\sum_{q\in\mathbb Z}(y_x/m)^{q^2}e^{iqu}\right)^m
                    \longrightarrow e^{2y_x\cos u}.
\]
Convergence on every fixed complex strip gives coefficient convergence.
Nonnegativity and convergence of the total masses give $\ell^1$
convergence; the real exponential generating functions bound the
tails with every polynomial weight. Consequently aggregate charge
moments, real tilts and products over the finitely many sites converge
to those of $\prod_xI_{n_x}(2y_x)$. The factors
$e^{-n^\top Qn/2}\le1$ preserve this convergence. Restricting to
a homogeneous integer constraint also preserves it, and its denominator
has a strictly positive zero configuration. This passes all normalized
charge comparisons, including singular $Q$. Finally let any zero
activities tend to zero. No translation of a Bessel weight is used.

For the continuous response, the quadratic coefficient of the stronger
character inequality gives
$-\Cov(F^2,\cos\beta_*\phi_x)
\ge(\E[F\sin\beta_*\phi_x])^2/c_x$.
Normalized differentiation proves the first bound. Integration by parts
at site $x$ gives $\mathsf C K e_x+2\beta_* y_x s_x=e_x$. Hence
$\mathsf C^{-1}s_x=(2\beta_*y_x)^{-1}\mathsf R e_x$.
Differentiating the inverse proves the second bound. Decompose $B$ into squares and use
\eqref{positive-cosine-laws:fourth} for its precision response.
\end{proof}

\begin{proposition}[Gaussian domination and retained regulators]
\label{positive-cosine-laws:retained-subgaussian}
For $K=ZA_D+P$, $P\succeq0$, the centered positive-cosine law satisfies
\begin{equation}
             \E e^{t^\top\phi}\le e^{t^\top K^{-1}t/2},\qquad
             \Cov(\phi)\preceq K^{-1}.
 \label{positive-cosine-laws:pinned-gaussian-domination}
\end{equation}
In the exact unit-reference split $G=A_D^{-1}=C+Q$, the retained field is
\begin{equation}
 \psi=QA_D\phi+\eta,\qquad \eta\perp\phi,\qquad
 \Cov(\eta)=Q-QA_DQ=C-CA_DC\succeq0.
 \label{positive-cosine-laws:retained-pullback}
\end{equation}
Its sub-Gaussian covariance proxy is
\begin{equation}
 \begin{aligned}
 \Sigma_\psi&=QA_DK^{-1}A_DQ+Q-QA_DQ\\
       &\preceq Q+(Z^{-1}-1)QA_DQ,\\
 \E e^{t^\top\psi}&\le e^{t^\top\Sigma_\psi t/2}.
 \end{aligned}
 \label{positive-cosine-laws:retained-domination}
\end{equation}
For $T\succeq0$ set $S=T^{1/2}\Sigma_\psi T^{1/2}$. If $\|S\|<1$, then
\begin{equation}
 \E e^{\psi^\top T\psi/2}
 \le\det(1-S)^{-1/2}.
 \label{positive-cosine-laws:regulator-determinant}
\end{equation}
For the literal derivative smear, uniformly over domains and centered pins,
\begin{equation}
 \begin{gathered}
 \E e^{tD_\epsilon(f)}\le e^{c_Zt^2\|f\|_{L^2}^2/2},\\
 \E\left|\prod_{i=1}^mD_\epsilon(f_i)\right|^2
 \le2^{m+1}m!c_Z^m\prod_i\|f_i\|_{L^2}^2,\qquad
 c_Z=\sup_\epsilon Z_\epsilon^{-1}.
 \end{gathered}
 \label{positive-cosine-laws:derivative-subgaussian}
\end{equation}
\end{proposition}
\begin{proof}
Complete the square and apply $F_Q(x)\le F_Q(0)$ for the first bound.
Gaussian conditioning proves \eqref{positive-cosine-laws:retained-pullback};
reweighting only $\phi$ leaves its independent noise unchanged.
Use the first bound and $K^{-1}\preceq Z^{-1}G$ for its proxy.
Gaussian linearization of $e^{\psi^\top T\psi/2}$ gives the determinant
by Tonelli. For derivatives the coefficient is $\epsilon d^*f$ and
$dA_D^{-1}d^*$ is an orthogonal projection. Cell averaging contracts
$L^2$; sub-Gaussian tail integration and H\"older give the moments.
The OS norm is bounded by the same $L^2$ moment. Point samples have
the same limit because their discrete $L^2$ error is $O(\epsilon)$.
\end{proof}

\section{Connected expansions with local derivative allocation}
\label{app:connected-expansion}
\label{connected-expansions:tree-section}

For the polymer norm used above, record the field support $X$ and the internal
tree $T_X$ on its occupied blocks. A further graph $\mathcal H$ may record the generating
supports and contractions. Exclusion and the regulator use $X$ alone. Under
local extraction $\mathcal H$ keeps its physical edge routes, birth roots, colours and
source labels even when $X$ shrinks.

\begin{lemma}[Connected expansion with local derivative allocation]
\label{connected-expansions:unified-tree-count}
\label{interface-expansion:rr-ordinary-operation}
\label{massive-source-integration:block-analytic}
Suppose activities have analytic radius $r$, size reserve $A^{-|X|}$,
and a nonnegative internal-tree kernel of row sum $K_0$. Suppose Gaussian
edges have tensor-projective row sum $K_G$, and every positive forest
interpolation on $m$ occupied blocks satisfies the joint bound
\[
 \int |U_{\mathcal F}|\prod_{v=1}^mG_v\,d\gamma_{\mathcal F}
       \le B_*H^m.
\]
Here $U_{\mathcal F}$ is the remaining unmarked reconstruction. The
joint integral has a single prefactor $B_*$ and contains each occupied
regulator once.
All tensor directions lie in the admitted local field spaces, with bounded
support-enlargement overlap. Then the connected expansion converges
normally when the size reserve and weak input are chosen with a strict
margin. At scale $a$, give a graph of total length $\ell$ the weight
$w_a(\ell)=(1+\ell/a)^{L_H}$. The expansion preserves fixed finite
source jets and this graph weight if the covariance rows remain
summable with the additional edge weight $w_a$ as well as the
current-tree weight. The expansion and its
nonlinear derivative have, respectively, quadratic and linear bounds
in the weak input for terms joining at least two base activities.

More explicitly, after the internal-tree sum, let each unmarked vertex
contribute $\eta A_0^d d!$ and each of $k$ marked vertices contribute
$c_iw_iA_0^d d!$ at degree $d$. If the resulting edge kernel has row
sum $K$, put $L=\max(1,A_0^2KH)$ and $q=8L\eta<1$. A rooted
connected marked sum is bounded by $C_k\prod_i c_iw_i$, where
\[
 C_1=H[1-8\log(1-q)],\qquad
 C_k=\frac{H8^kL^{k-1}(k-2)!}{(1-q)^{k-1}}\quad(k\ge2).
\]
For a full marked coefficient, multiply $B_*$ by the sum of these
products over partitions of the marked list. Every component has a mark.

There is also an estimate uniform in the number $N$ of separated packets.
Suppose the packets $P_i$ of norms $b_i$ occur with bounded overlap and
their complete local perturbations belong to the same weak ball in the
nilpotent norm with weights $\eta_*/b_i$, for a fixed $\eta_*>0$.
The overlap bound, analytic radius and unused size reserve are fixed
before $N$. The weak-ball condition is a bound at each occupied block,
with the same rooted edge sums as for the unmarked activities; it is
not a bound on the sum over the entire domain. Source-bearing scalars
are included with their generating roots. We state the weighted norm
for $b_i>0$; the resulting bound extends to zero sizes by continuity.
Then their complete joint reconstructed coefficient, after its original
unmarked scalar is factored, is bounded by
\begin{equation}
 B_*C^N\prod_{i=1}^N b_i.
 \label{connected-expansions:repeated-packet-bound}
\end{equation}
This last assertion requires uniform local perturbation bounds; it is
stronger than the fixed-$k$ estimate. It is an unnormalized estimate.
\end{lemma}

\begin{proof}
If $d_b$ derivative directions are assigned to block $b$, bounded overlap
gives $\|\sum z_if_i\|\le C\max_b\sum_{i\mapsto b}|z_i|$.
Cauchy's formula on $|z_i|\le r/(Cd_b)$ and
$d_b^{d_b}\le e^{d_b}d_b!$ give
\begin{equation}
 |D^dF[f_1,\ldots,f_d]|
 \le (C/r)^d\,\|F\|_{r,G}G\prod_b d_b!.
 \label{connected-expansions:block-cauchy-bound}
\end{equation}
Thus endpoint choices stay attached to their blocks.

Apply the positive Gaussian forest formula
\cite[Theorems~III.1 and~IV.5]{AR}. The internal trees of an unordered
partition of $X$, together with the Gaussian tree joining its components
at specified block endpoints, form exactly one tree on $X$ with two
edge colours. Conversely, deleting its Gaussian edges recovers that
partition and all endpoint choices. If $N=|X|$ and $j$ edges are
Gaussian, there are $j+1$ base components. Their weak factors give
$\nu^{j+1}$; summing colours and positions leafwise therefore gives
$\nu(K_0+\nu K_G)^{N-1}$. The $1/n!$ for $n$ base components first
cancels their permutations. Only then label the $N-1$ nonroot occupied
blocks and divide by $(N-1)!$. Each derivative end is incident to its
actual block, so \eqref{connected-expansions:block-cauchy-bound} costs at
most $\prod_v\deg_T(v)!$. Lemma~\ref{tree-summation:tree-sum} gives
\[
 \sum_{T\text{ on }[N]}\prod_v\deg_T(v)!
 =(N-2)!\binom{3N-3}{N-2}\le(N-2)!8^N.
\]
This count applies for $N\ge2$; a singleton contributes one.
The rooted majorant is consequently
\[
 CB_*\nu\sum_{N\ge1}
  \left[\frac{Ce^tH(K_0+\nu K_G)}A\right]^N.
\]
Terms with a Gaussian edge have colour sum
$\nu[(K_0+\nu K_G)^{N-1}-K_0^{N-1}]$, which is quadratic in $\nu$.
For a fixed number of sources, the number of allocations grows only
polynomially in $N$. The geometric majorant remains summable after
multiplication by this polynomial, so one may differentiate termwise
to each prescribed finite order. The terminal logarithm uses the same
criterion after Gaussian integration, when its activities are numbers.

With $m=n+k$ vertices, leaf integration supplies $K^{m-1}$. The degree count leaves $H8^kL^{k-1}(n+k-2)!q^n/n!$,
which sums to $C_k$; for $k=1$ the nonsingleton terms are $8Hq^n/n$ and the singleton is $H$.
Keep the finite set of marked components together until applying the joint
regulator bound, so the prefactor $B_*$ occurs once.

For the uniform repeated-packet assertion, give each complete packet
one additional commuting label $z_i$, with $z_i^2=0$, and put
\begin{equation}
 \left\|\sum_{S\subseteq[N]}F_Sz_S\right\|_{\mathbf b}
 =\sum_{S\subseteq[N]}\|F_S\|
                  \prod_{i\in S}\frac{\eta_*}{b_i}.
 \label{connected-expansions:packet-algebra-norm}
\end{equation}
Give each coefficient its original analytic or scalar norm. Label the whole
packet only after forming its source coefficient, exponential products and
contacts. To extract $z_S$ from a product, sum over $S=A\sqcup B$. The triangle inequality
then makes \eqref{connected-expansions:packet-algebra-norm} submultiplicative with constant one, uniformly in $N$ and also
for a fixed finite internal contact algebra at each $z_i$.

The coloured-tree count and block-Cauchy estimate use only that product
inequality. Their constants therefore do not count dimensions of the
coefficient algebra. By the local weak-ball hypothesis the geometric ratio
is uniformly below one for every $N$. At the first packet anchor the rooted
connected sum is bounded by a constant. Summation over anchors costs $CN$ by
the footprint and overlap bounds. No volume factor or factorial in $N$
appears, including for contact and scalar outputs, which keep their source
roots.

Exponentiating the connected marked components bounds their joint
coefficient norm by $e^{CN}$. In the forest integral use the joint regulator
estimate on all components together. It gives $B_*$ once, with one regulator for
each occupied block, independent of the component count; there is no
separate $B_*$ per component. Extraction of $z_1\cdots z_N$ costs $\eta_*^{-N}\prod_i b_i$. Absorbing $e^C/\eta_*$ in \eqref{connected-expansions:repeated-packet-bound} proves
the bound with no factorial division of packet labels.

Only finite labelled polynomials occur, without any analyticity requirement
on numerical source strengths. Substituting $k=N$ in the earlier fixed-$k$
estimate would leave $(N-2)!$. To include $b_i=0$ assume the local bound uniformly for $b_i+\delta$
with $\delta>0$, as holds automatically for a packet of zero norm. Now pass to $\delta\downarrow0$ in
the coefficient bound.

If new edges $E$ join graphs $\mathcal H_i$, let $\mathcal H$ be their
union together with these edges. Write $\mathcal L$ for total graph
length and $\ell_e$ for an edge length. The graph estimate follows from
\[
 w_a\bigl(\mathcal L(\mathcal H)\bigr)
 \le\prod_i w_a\bigl(\mathcal L(\mathcal H_i)\bigr)
     \prod_{e\in E}w_a(\ell_e).
\]
Use weighted convolution to sum graph labels before the forest
count. In a local extraction keep $\mathcal H$ and give the output
its actual target support, attaching any new target by a bounded
local edge. The zero-sum compensation identities hold coefficient by coefficient
as polynomial identities in this algebra. Historical vertices have no exclusion
conditions or extra regulators. Reblocking may keep the old routes.
A change of representatives adds at most $CN$ segments of length
$CBa$ and costs a fixed polynomial in $N$.

Both interface norms carry weight $L_H$. Root at the first interface input for
the birth-root row estimate, at the input containing the prescribed block
for the current-block column estimate. Subsequent interface inputs sum by
the column norm in either rooting.
For reblocking by a factor $B>1$, fix $0<\theta<1$. Histories of length below $B^\theta a$ have $O(NB)$
face roots or $O(1)$ corner roots near a fixed coarse block. Sum all
supports once at each root. For longer histories,
\[
 \frac{1+t/B}{1+t}\le2B^{-\theta}\quad(t\ge B^\theta),
\]
where $t$ is the history length in units of $a$. There are at most
$CN^2B^2$ fine anchors, and the engineering rescaling contributes $B^P$.
Thus the ratio of the new and old weights leaves a factor at most
$C_{L_H}N^2B^{P+2-\theta L_H}$ after summing over these anchors.
Choosing $\theta L_H>P+2$ gives decay in $B$ in both norms with the
same exponent $L_H$. Short-history restriction gains are supplied by the
application.

Finally, on compensated stable coordinates the free derivative has
norm at most $\theta_0<1$. Choose the weak ball so that the nonlinear
contribution to this norm is at most $C\nu<1-\theta_0$.
If the initial stable record is $K_{\rm in}$,
exact variation of constants reads
\[
 K_j=T_{j-1}\cdots T_0K_{\rm in}
       +\sum_{l<j}T_{j-1}\cdots T_{l+1}F_l,
 \qquad \|T_j\|\le\theta<1.
\]
For complete forcing atoms of size $v_l^p$, the canonical monotone
ultraviolet bounds give
$\|K_j\|\le\theta^j\|K_{\rm in}\|+Cv_j^p\sum_{m\ge0}\theta^m$.
Any fixed age polynomial is also summable. One changed covariance is
labelled when its forcing atom is created and transported thereafter.
The canonical marginal flow is already summed in these input atoms;
it is not re-expanded into bare ultraviolet histories.
\end{proof}

\begin{corollary}[Massive integration and the normalized numeric gas]
\label{massive-source-integration:massive-stop}
Let the centered Gaussian covariance on a bounded-degree block graph
satisfy the exponential projective row bound
\[
 \sup_b\sum_c e^{m d(b,c)}\|\Gamma_{bc}\|_\pi\le K_m,
 \qquad
 \mathbb E_{\Gamma^{(T)}}\prod_iG(X_i,\zeta_i)
       \le C_0^{\sum_i|X_i|}
\]
for every positive forest interpolation on disjoint base polymers,
with each original regulator included once. Suppose the exact circle
product has activities $\|F(X)\|_{r,G}\le\nu A^{-|X|}$, including
singleton local factors, and complete finite marked activities with
the same anchored size bounds, together with their marked logarithmic
scalars. For a strict size reserve and sufficiently small $\nu K_m$,
Gaussian integration gives an exact numerical hard-core gas
\[
 \begin{aligned}
 \mathbb E_\Gamma U
 &=\sum_{\{Y_i\}\,\mathrm{disjoint}}\prod_i\mathfrak K(Y_i),\\
 \sup_b\sum_{Y\ni b}e^{t|Y|+m\tau(Y)}|\mathfrak K(Y)|&\le C\nu.
 \end{aligned}
\]
Its numeric connected logarithm converges. Division by its own unmarked
partition subtracts exactly the vacuum clusters. A complete connected
marked coefficient joining supports at distance $L$ gains $e^{-m'L}$
for any fixed $0<m'<m$, with the fixed-list test constants.
\end{corollary}

\begin{proof}
Apply the lemma with $B_*=1$, $H=C_0$ and exponentially weighted edges. Forest
components at zero cross covariance are independent. Grouping their occupied
unions gives the stated numerical gas, with rooted bound $C\nu\sum_{N\ge1}[Ce^tC_0(K_0+\nu K_G)/A]^N$ and $K_G\le CK_m$. The strict
margin yields convergence of this series and then of the ordinary numerical
cluster sum. In the finite nilpotent source algebra choose a coefficient
radius $\delta$ for which the marked input norm is at most $\nu$. The lemma then
applies with $2\nu$, and coefficient extraction introduces the finite factor $\delta^{-|S|}$.
Add the anchored marked scalars with all their contacts. Normalization by
the actual partition cancels the unmarked logarithm. A connected support
joining marks at distance $L$ has tree length $\tau\ge L$, giving $e^{-m\tau}\le e^{-m'L}e^{-(m-m')\tau}$. Its second factor
is summable in the rooted estimate. Taking the logarithm after Gaussian
integration means that repeated clusters contain numerical factors without
repeating any field regulator. A deterministic mean of cost $e^{cR^2|X|}$ changes only
the fixed size reserve.
\end{proof}

\section*{AI USE DECLARATION}

The author conceived the research programme and supplied central
strategies for pinning and state selection
(Section~\ref{sec:pinning}), charged-to-neutral screening
(Section~\ref{sec:screening}), and non-Gaussianity
(Section~\ref{derivative-interaction:interaction-section}). The author's earlier inverse-square Ising papers
\cite{BhattacharyaCriticalJump2026,BhattacharyaCriticalRelations2026} provide methodological precedents for
the use of physical response, logarithmic interface estimates and finite
criteria in the present pinning strategy.
OpenAI's ChatGPT contributed substantive intermediate proof ideas,
calculations and technical arguments, particularly to the ultraviolet
and source constructions
(Sections~\ref{sec:renormalization}--\ref{sec:transfer}),
observable admission (Section~\ref{sec:admission}), Gaussian observation,
replica and response estimates
(Sections~\ref{sec:observations}--\ref{sec:responses}),
reconstruction, universality and local identities
(Sections~\ref{sec:reconstruction}--\ref{sec:local-identities}),
and Appendices~\ref{app:positive-comparison}
and~\ref{app:connected-expansion}. It also assisted with technical
completion of the author's strategies and extensively with exposition.
These contributions developed iteratively under the author's direction.
The author manually verified the
mathematics, including AI-assisted arguments, and takes full
responsibility for its content, references and attribution.

\bibliographystyle{aomplain}
\bibliography{references}
\end{document}